\documentclass[a4paper,12pt]{article}
\pdfoutput=1
\usepackage{geometry}
\usepackage{amsmath}
\usepackage{amsfonts}
\usepackage{amssymb}
\usepackage{mathtools}
\usepackage{ytableau}
\usepackage{verbatim}
\usepackage[hidelinks]{hyperref}
\usepackage{color}
\usepackage{graphicx}
\usepackage{tikz}
\usetikzlibrary{decorations.pathreplacing}
\usepackage{multirow}
\usepackage{empheq}
\usepackage{tensor}

\numberwithin{equation}{subsection}

\newcommand{\bea}{\begin{eqnarray}}
\newcommand{\eea}{\end{eqnarray}}
\newcommand{\tr}{\text{Tr}}
\newcommand{\str}{\text{Str}}
\newcommand{\cof}{\text{Coeff}}
\newcommand{\Mult}{\text{Mult}}
\def\cN{ \mathcal{N} }
\def\cP{ \mathcal{P} }
\def\mR{ \mathbb{R} }
\def\mC{ \mathbb{C} }

\def\bC{ \mathbb{C} }
\def\bP{ \mathbb{P} }
\def\bp{\mathbf{p}}
\def\bq{\mathbf{q}}

\def\tC{\widetilde{C}}

\newcommand{\cF}{ \mathcal{F}} 
\newcommand{\cO}{ \mathcal{O}} 
\newcommand{\cA} {\mathcal{A} } 
\newcommand{\cX} {\mathcal{X} } 
\newcommand{\cY} {\mathcal{Y} } 
\newcommand{\cJ} {\mathcal{J} } 
\newcommand{\cH} {\mathcal{H} } 
\newcommand{\cU} {\mathcal{U} } 

\newcommand{\cR} {\mathcal{R} }

\newcommand{\fu}{\mathfrak{u}}
\newcommand{\fsu}{\mathfrak{su}}

\newcommand{\bX}{\mathbb{X}}
\newcommand{\bZ}{\mathbb{Z}}

\newcommand{\cI}{ \mathcal{I}} 
\newcommand{\cM}{ \mathcal{M}} 
\newcommand{\cG}{ \mathcal{G}} 

\newcommand{\hN}{\widehat{N}}

\newcommand{\tred}{ \textcolor{red}} 
\newcommand{\Diag}{ {\rm Diag} } 
\newcommand{\Sym}{ \operatorname{Sym} }
\newcommand{\im}{ \operatorname{Im} }
\newcommand{\Ker}{ \operatorname{Ker} } 
\newcommand{\Span}{ \operatorname{Span} }

\begin{document}

\begin{flushright}
	QMUL-PH-20-04\\
\end{flushright}

\bigskip

\begin{center}
	
	{\Large\bf Quarter-BPS states, multi-symmetric functions and set partitions 
	}

	\bigskip
	
	{
		Christopher Lewis-Brown $^{a,*}$ and Sanjaye Ramgoolam$^{a,b,\dag}  $}
	
	\bigskip
	$^{a}${\em School of Physics and Astronomy} , {\em  Centre for Research in String Theory}\\
	{\em Queen Mary University of London, London E1 4NS, United Kingdom }\\
	\medskip
	$^{b}${\em  School of Physics and Mandelstam Institute for Theoretical Physics,} \\   
	{\em University of Witwatersrand, Wits, 2050, South Africa} \\
	\medskip
	E-mails:  $^{*}$ c.h.lewis-brown@qmul.ac.uk,
	\quad $^{\dag}$s.ramgoolam@qmul.ac.uk

	\begin{abstract}
		
\noindent	
We give a construction of general holomorphic quarter BPS operators
in $ \cN=4$ SYM at weak coupling  with $U(N)$ gauge group at finite $N$. 
 The construction employs  the M\"obius inversion formula for set partitions, applied to 
 multi-symmetric functions, alongside computations in the group algebras of symmetric groups. We present a computational algorithm which produces an orthogonal basis for the physical inner product on the space of  holomorphic  operators. The basis is labelled by a $U(2)$ Young diagram, a $U(N)$ Young diagram and an additional plethystic multiplicity label.  
  We describe precision counting results of quarter BPS states which are expected to be reproducible from dual computations with giant gravitons in the bulk, including a symmetry relating sphere and AdS giants within the  quarter BPS sector. In the case $n \leq N$ ($n$ being the dimension of the composite operator) the   construction is analytic,  using multi-symmetric functions and $U(2)$ Clebsch-Gordan coefficients. Counting and correlators of the  BPS operators can be encoded in a two-dimensional topological field theory based on permutation algebras and equipped with appropriate defects.

	\end{abstract}
	
\end{center}

\vskip.2cm

\newpage 

\setcounter{tocdepth}{2}
\tableofcontents

\section{Introduction}

Half-BPS operators in $ \cN =4$ SYM with $ U(N)$ gauge group are related, via the AdS/CFT correspondence \cite{malda,gkp,witten}  to extremely rich physics including Kaluza-Klein gravitons, giant gravitons \cite{mst,gmt,hhi} and LLM geometries \cite{LLM}. These operators are constructed from gauge invariant holomorphic  functions of a single complex matrix $X = \phi_1 + i \phi_2$, where $\phi_1, \phi_2$ are two of the six hermitian scalars in the adjoint of the gauge group. The two-point function of a holomorphic with an anti-holomorphic operator gives an inner product on the space of BPS operators. An  orthogonal  basis of operators labelled by Young diagrams $R$, with length no greater than $N$ was constructed in \cite{CJR2001}.  The Young diagram $R$ is identified by the Casimirs of a $U(N)$ symmetry in the theory. Three-point functions were computed in terms of Littlewood-Richardson coefficients \cite{CJR2001}.  This allowed a general identification of gauge invariant operators dual to giant gravitons in the AdS spacetime, extending the results in \cite{BBNS}. An important guide to the precise identification of giant graviton geometries dual to particular Young diagrams, was the { \it stringy exclusion principle}  (SEP), which refers to the fact that certain states disappear from the Hilbert space of the theory as $N$ changes to $N-1$\cite{malstrom}. States corresponding to the Young diagrams with $N$ rows are precisely all the states which disappear from the Hilbert space as the rank of the gauge group changes from $N$ to $ N-1$. Recent computations of correlators  of giant gravitons in the AdS space-time have demonstrated agreement between CFT and AdS \cite{BKYZ1103,CDZ1208,Lin2012,KSY1507}.  Despite significant progress on the quarter-BPS and  eighth-BPS sectors, both in the construction of CFT operators and in the construction of quantum states in the dual space-time,  a comparable level of understanding of AdS/CFT in this sector remains an important challenge.

From  consideration of the quantization of  moduli spaces  of giant  gravitons which are large in the $S^5$ directions on $ AdS_5 \times S^5$, it has been argued that the quantum states arising are those of an $N$-boson system in a harmonic oscillator \cite{Mikhailov,Beasley,BGLM06}. From consideration of moduli spaces of giants which are large in the $ AdS_5$ directions, the same Hilbert space has been arrived at \cite{MS06}. Yet another perspective comes from the strong coupling matrix dynamics leading to commuting $ N \times N$ matrices \cite{Ber0507}. The counting of quarter and eighth BPS states has been derived by calculating the quarter and eighth-BPS cohomology \cite{KMMR07}. 

The construction of quarter BPS operators from the gauge theory side has been developed in 
\cite{Ryz0111,dHRyz0301,dHRyz0109,Ryz0307}. At zero field theory coupling, the quarter BPS states are general holomorphic 
operators built from two complex matrices $ X $ and $ Y $. A subspace of these operators 
is annihilated by the one-loop dilatation operator and forms the weak coupling quarter BPS space.  An important outcome of 
these papers is that the weak coupling quarter BPS operators form the orthogonal subspace, in the free field inner product, 
of the  operators which contain commutators $ [ X , Y ] $ within a trace. This a well-defined characterisation of the 
quarter-BPS operators at finite $N$. 
 
 An  orthogonal basis for the free field inner product, in terms of Young diagrams, was found in \cite{BHR0711,BHR0806}, which was covariant under the $U(2)$ symmetry of the quarter BPS sector or $U(3)$ for eighth-BPS, and valid at finite $N$. Closely related orthogonal bases were obtained in \cite{Brauer,BCD0801,BCD0805} and the relation between the different bases in terms of free field conserved charges was elucidated in \cite{EHS}. In  the  free-field $U(2)$ covariant constructions of quarter BPS operators  \cite{BHR0711,BHR0806} having a total of $n$ copies of $X$ and $Y$, the labels consist of a Young diagram $R$ with $n$ boxes and columns no longer than $N$, a Young diagram $ \Lambda $ with $n$ boxes and columns no longer than $2$, along with a label $ \tau$ which  runs over the multiplicity of trivial $S_n$ irreps in $ R \otimes R \otimes \Lambda $. The Young diagram label $ \Lambda $ is also a  representation of the global symmetry $U(2)$. 
 The construction of weak-coupling quarter-BPS operators based on this new understanding of the finite $N$ inner product was  further developed in \cite{Brown1002,CtoC}. The finite $N$ construction of quarter BPS operators was given in terms of 
 a projector $ \cP_N $ in $ \mC ( S_n)$, which projects to the intersection of two subspaces of $ \mC ( S_n)$ \cite{CtoC}. One subspace is associated with the symmetrised traces at large $N$, another with the finite $N$ cutoff on the free field basis.

 In \cite{PasRam1204},  the eighth-BPS Hilbert space  of $N$ bosons in a 3D harmonic oscillator obtained from  quantizing the space of holomorphic polynomials in $ \mC^3$ was used to give a labelling of eighth-BPS states by a $U(3)$ Young diagram, along with a $U(N)$ Young diagram and an additional multiplicity label constructed from the pair of  Young diagrams. We will refer to this multiplicity label as a plethystic multiplicity label, due to the role of plethysm operations in its definition. By restricting to the  quarter BPS states there is a labelling using a $U(2)$ Young diagram, a $U(N)$ Young diagram and a plethystic multiplicity label. In another development, a special class of quarter BPS operators at weak coupling was found to be related to Brauer algebra constructions \cite{YusukeQuarter}. 

 There has not been so far, a general  construction  of quarter-BPS  operators at weak coupling and finite $N$, which includes a $U(2)$ Young diagram label alongside a $U(N)$ Young diagram label.  In this paper, we will  address this open problem and give a basis of operators which are quarter-BPS at weak coupling, orthogonal with respect to the free field inner product, and labelled by a $U(2)$ Young diagram, a $U(N)$ Young diagram, alongside an associated multiplicity label depending on these two Young diagrams. The virtue of having a $U(N)$ Young diagram is that the disappearance of states upon a   reduction of $N$ to $ N-1$ can be directly expressed in terms of this Young diagram - the disappearing states as $N$ is reduced to $ N-1$ are precisely the ones corresponding to Young duagrams with  exactly $N$ rows. We may therefore describe our basis as an SEP-compatible (SEP $=$ stringy exclusion principle) basis which is also $U(2)$ covariant.

The key ingredient which allows us to find a manifestly SEP-compatible $U(2)$ covariant construction of quarter BPS states is the mathematics of multi-symmetric functions \cite{Vaccarino2007,Domokos,Procesi}.  When gauge invariant functions of two matrices $ X ,  Y$ are evaluated on diagonal matrices $ X = \Diag ( x_1 , x_2 , \cdots ,x_N )$ and $ Y = \Diag ( y_1 , y_2 , \cdots , y_N)$, we get polynomials which are invariant under the $\sigma \in S_N$ acting simultaneously as 
\begin{align}
\begin{aligned}
x_i & \rightarrow x_{ \sigma (i) } \\
y_i & \rightarrow y_{ \sigma (i) } 
\end{aligned}
\end{align}
These polynomials are called multisymmetric functions. More generally, we can have variables $x^{a}_i$ with $ a \in \{ 1, 2, \cdots , M \}$ and $ i \in \{ 1, 2, \cdots , N \}$. Polynomials invariant under simultaneous $S_N$ permutations of all the $M$ vectors are more general { \it multi-symmetric functions}. There is a rich mathematics associated with changing between different bases of multi-symmetric functions for any $M$ which is  relevant in this paper and is controlled by an underlying structure of {\it set partitions}.

The paper is organised as follows. Section \ref{section: zero coupling} reviews the connection between half BPS operators and symmetric functions before covering the construction of orthogonal bases of half and quarter BPS states at zero coupling. This is based on the use of permutation algebras and Schur-Weyl duality relations between permutation representation theory and the representation theory of $U(N)$. Section \ref{section: weak coupling} is an introduction to the key mathematical tools and physical concepts  we will use in this paper. In section \ref{sec:MSSP}  use the combinatorics of set partitions to derive results on the transformation between two bases for multi-symmetric functions. The first is the trace basis. Elements of this basis set are obtained by specifying a trace structure for matrices $X , Y$, or more generally $ X^1 , \cdots , X^M$ and specialising to diagonal matrices. Another basis is the multi-symmetric monomial basis, which allows a simple description of finite $N$ cut-offs. In section \ref{sec:Counting}, we  start from the observation  that every vector partition ${\bf p}$ defines an associated partition $ p$, which is invariant under the action of the $U(2)$ transformations which interchange $ X ,Y$. We use  results on  plethysms of $SU(2)$ representations to obtain detailed expressions for refined multiplicities depending on a  pairs of Young diagram $ \Lambda , p $, where $ \Lambda $ is a $U(2)$ Young diagram and $p$ is a Young diagram constrained to have no more than $N$ rows, which we refer to as a $U(N)$ Young diagram. In section \ref{sec:Construction} we describe an algorithm for producing a basis of operators labelled by the pair of Young diagrams $ ( \Lambda , p  ) $ alongside the appropriate multiplicity label. The basis is orthogonal under the free field inner product.  In section \ref{sec:Projectors} we elucidate the  vector space geometry within $ \mC ( S_n )$,  involving the interplay between a projector for the $U(2)$ flavour symmetry, a projector for the symmetrisation of traces $ \cP$ and an operator $ \cF_N$ whose kernel implements finite $N$ constraints. This discussion allows us to show that the counting and two-point correlators of quarter-BPS operators at weak coupling can be expressed in terms of observables in two-dimensional topological field theory based on permutation group algebras with appropriate defects.

\section{Review of zero coupling: symmetric functions, half-BPS operators and the quarter-BPS covariant basis}
\label{section: zero coupling}

The half-BPS sector is composed of multi-traces of a single complex matrix $X$. We diagonalise $X$ in terms of its eigenvalues
\begin{equation}
X = 
\begin{pmatrix}
x_1 & 0 & 0 & \dots & 0 \\
0 & x_2 & 0 & \dots & 0 \\
0 & 0 & x_3 & \dots & 0 \\
\vdots & \vdots & \vdots & \ddots & \vdots \\
0 & 0 & 0 & \dots & x_N
\end{pmatrix}
\label{X as a diagonal matrix}
\end{equation}
Thus any multi-trace of $X$ can instead be written as a function of the eigenvalues  $x_1, x_2, \dots, x_N$. These functions must be completely symmetric in the $N$ variables. The theory of such functions has long been studied in mathematics, and they have many interesting properties \cite{Macdonald}. In this section we review the basic definition, three of the commonly used bases and give the connection to the half-BPS sector of $\cN = 4$ super Yang-Mills.

Since correlators of single matrix multi-traces are not renormalised as we go from zero coupling to weak coupling to strong coupling \cite{EHSW0004}, these bases remain half-BPS in all regimes.

After covering the half-BPS sector, we then review the covariant basis \cite{BHR0711,BHR0806} for the free field quarter-BPS sector. This involves two complex matrices $X$ and $Y$, in general non-commuting, and therefore we cannot express generic multi-traces purely in terms of their eigenvalues as we did in the half-BPS sector.

For the quarter-BPS sector, there is a step-change as we transition from zero coupling to weak coupling. After turning on interactions, some linear combinations of multi-traces recombine into long non-BPS multiplets, and therefore the covariant basis introduced here is only a basis for the quarter-BPS sector in the free theory. The remainder of the paper focuses on finding a generalisation of the covariant basis for the quarter-BPS sector at weak coupling.

\subsection{Half-BPS operators and symmetric functions}
\label{section: half-bps}

Symmetric functions are defined as polynomials in the $N$ variables $x_1, x_2, x_3, \dots, x_N$ that are invariant under all 
permutations of the $x_i$. More explicitly, given a polynomial $f(x_1, x_2, \dots, x_N)$, $f$ is a symmetric function if
\begin{equation}
f(x_1, x_2, \dots, x_N) = f(x_{\sigma(1)}, x_{\sigma(2)}, \dots, x_{\sigma(N)})
\label{permutation invariance for symmetric functions}
\end{equation}
for all $\sigma \in S_N$, the group of permutations of $\{1,2,\dots, N\}$.

We can take the infinite $N$ limit of this definition by defining symmetric functions as formal power series in infinitely 
many variables $x_1, x_2, \dots$. To return to the finite $N$ case (or to reduce a symmetric function in $M > N$ variables to 
one in $N$ variables), we can set $x_{N+1} = 0, x_{N+2} = 0, \dots$.

Physically, symmetric polynomials in $N$ variables are the wavefunctions associated to $N$ bosons moving in one dimension. 
The variable $x_i$ corresponds to the position of the $i$th boson, and the symmetric nature of the function is exactly the 
condition on the wavefunction that the state must be symmetric between identical bosons.

There are several different bases for the ring of symmetric functions, of which we will look at three. In each of these bases, each basis element consists of polynomials of a single degree, $n$, and the basis for the degree $n$ subspace is labelled by the partitions of $n$, which we now describe. 

A partition $p$ of $n$ is a set of weakly decreasing positive integers $\lambda_1 \geq \lambda_2 \geq \dots \geq \lambda_k$ which sum to $n$. We call $\lambda_j$ the components of $p$, $k = l(p)$ the length of $p$ and write $p = [\lambda_1, \lambda_2, \dots, \lambda_k]$. We denote partitions of $n$ by $p \vdash n$.

An alternative way of defining a partition $p$ is to give the number, $p_1$, of $\lambda_j$ equal to 1, the number, $p_2$, of $\lambda_j$ equal to 2, etc. We write
\begin{equation}
p = < p_1, p_2, \dots >
\label{partition from multiplicities}
\end{equation}
We call the $p_i$ the multiplicities of $p$, and this is the multiplicity notation. The $p_i$ satisfy
\begin{equation}
\sum_i i p_i = n
\end{equation}
A partition $p \vdash n$ can be thought of visually using a Young diagram. This consists of a set of boxes arranged in rows. The first row has length $\lambda_1$, the second $\lambda_2$, etc. For example for $p = [4,2,1]$ the associated Young diagram is
\begin{equation}
\begin{gathered} \ydiagram{4,2,1} \end{gathered}
\end{equation}
We will in general use the same notation for $p$ and the associated Young diagram. When it is necessary to make a distinction, we will use $Y(p)$ for the Young diagram.

For a partition $p = [\lambda_1, \lambda_2, \dots]$, the conjugate partitions is defined by $p^c = [\mu_1, \mu_2, \dots]$ with $\mu_i$ defined so that
\begin{align}
\begin{aligned}
\lambda_j & \geq i & \quad \text{if } j & \leq \mu_i \\ 
\lambda_j & < i & \quad \text{if } j & > \mu_i
\end{aligned}
\end{align}
Intuitively, we transpose the Young diagram of $p$, so that rows of $p$ become columns of $p^c$ and vice versa.

\subsubsection{Monomial basis}
\label{section: monomial basis}

We start with the monomial basis. Given a partition $p = [\lambda_1, \lambda_2,\dots, \lambda_k]$ of $n$, take the monomial
\begin{equation}
x_1^{\lambda_1} x_2^{\lambda_2} \dots x_k^{\lambda_k}
\end{equation}
and then add all distinct permutations of it to form a symmetric function. So for example if we take $p = [3,1,1]$ (and use $N=3$ for simplicity), the associated monomial basis element is
\begin{equation}
m_{[3,1,1]} = x_1^3 x_2 x_3 + x_1 x_2^3 x_3 + x_1 x_2 x_3^3
\end{equation}
For $p = < p_1, p_2, \dots >$, we can define the monomial functions more formally by
\begin{equation}
m_p = \left( \prod_i \frac{1}{p_i!} \right) \sum_{\sigma \in S_N} x_{\sigma(1)}^{\lambda_1} x_{\sigma(2)}^{\lambda_2} \dots x_{\sigma(l(p))}^{\lambda_{l(p)}}
\label{monomial symmetric function definition}
\end{equation}
where $N \geq l(p)$ and the normalisation in front accounts for non-trivial coefficients introduced by redundancies in the 
components of $p$.

For future convenience, we also define the rescaled monomial function $M_p$ to be \eqref{monomial symmetric function definition} without the normalisation factor.
\begin{equation}
M_p = \sum_{\sigma \in S_N} x_{\sigma(1)}^{\lambda_1} x_{\sigma(2)}^{\lambda_2} \dots x_{\sigma(l(p))}^{\lambda_{l(p)}}
\label{M definition}
\end{equation}
Lowering $N$ reduces the degrees of freedom in the system, and therefore we expect the basis to change to account for this. For the monomial basis, this process is particularly simple. For $N \geq n$, which we refer to as large $N$, \eqref{M definition} forms a basis of the degree $n$ symmetric functions. For $N < n$, we set $x_{N+1} = x_{N+2} = \dots = 0$, and therefore those functions labelled by $p$ with $l(p) > N$ all vanish identically, while those with $l(p) \leq N$ form a basis for the remaining space.

\subsubsection{Multi-trace basis}

The second basis we consider is the multi-trace basis, normally called the power-sum basis in the mathematics literature. 
This is constructed from polynomials of the form
\begin{equation}
T_k = \sum_{i=1}^N x_i^k
\label{single component power-sum}
\end{equation}
For a partition $p = [\lambda_1, \lambda_2, \dots \lambda_k]$, the symmetric function is
\begin{equation}
T_p = \prod_{i=1}^{k} T_{\lambda_i}
\label{multi-component power-sum}
\end{equation}
Consider a $N \times N$ diagonal matrix $X$ with entries $x_i$, as in \eqref{X as a diagonal matrix}. Then \eqref{single component power-sum} can be written in terms of $X$ as $T_k = \tr X^k$. For the general symmetric function \eqref{multi-component power-sum} we have
\begin{equation}
T_p = \prod_i \left( \tr X^i \right)^{p_i}
\end{equation}
This is the link between symmetric functions and the multi-trace basis for the half-BPS sector of $\cN = 4$ super Yang-Mills.

When we reduce from large $N$ to $N < n$, non-trivial relationships are induced among the multi-traces of a given degree, 
which can be understood from the Cayley-Hamilton theorem. As we saw for the monomials, one can take those multi-traces 
labelled by $p$ with $l(p) \leq N$, and they will form a basis. However, in this case those multi-traces associated to $p$ 
with $l(p) > N$ do not vanish, instead they are complicated linear sums of the remaining basis elements.

\subsubsection{Schur basis}
\label{section: schur basis}

Finally, we look at the Schur basis. These are labelled by partitions $R \vdash n$, thought of as representations of the symmetric group $S_n$.
\begin{equation}
s_R = \sum_{p \vdash n} \frac{1}{z_p} \chi_R(p) T_p
\label{schur definition from traces}
\end{equation}
where $z_p$ is the order of the centralizer of a permutation in $S_n$ with cycle type $p = < p_1, p_2, \dots >$, and is given explicitly by
\begin{equation}
z_p = \prod_i i^{p_i} p_i !
\label{z_p}
\end{equation}
Since the Schur and monomial functions form a basis for the degree $n$ symmetric functions, there is a basis change matrix 
transforming between them. This is given by the Kostka numbers $K_{Rp}$
\begin{equation}
s_R = \sum_{p \vdash n} K_{R p} m_p
\label{schur definition from monomials}
\end{equation}
The Kostka have a combinatoric interpretation in terms of the number of semi-standard Young tableaux of shape $R$ and 
evaluation $p$. These Young tableaux are defined in section \ref{section: U(2) reps}.

Let $V$ be the  $N$-dimensional space on which the Yang-Mills scalar field $X$ acts, and consider $V^{\otimes n}$, the $n$-times tensor product of $V$. The permutation group $S_n$ acts on $V^{\otimes n}$ by permuting the factors, and we also have the operator $X^{\otimes n}$, which applies $X$ to each tensor factor. Take the trace over $V^{\otimes n}$ of a product between $\sigma \in S_n$, a permutation of cycle type $p \vdash n$, and $X^{\otimes n}$. This is
\begin{equation}
\tr \left( \sigma X^{\otimes n} \right) = \prod_i \left( \tr X^i \right)^{p_i} = T_p
\label{traces from permutations}
\end{equation}
Using this, we can re-write \eqref{schur definition from traces} as
\begin{equation}
s_R = \mathcal{O}_R = \frac{1}{n!} \sum_{\sigma \in S_n} \chi_R (\sigma) \tr \left( \sigma X^{\otimes n} \right)
\label{schur definition from permutations}
\end{equation}
This is the standard definition of the Schur operators in the half-BPS sector first introduced in \cite{CJR2001}.

The Schur basis behaves in the same way as monomial basis for $N < n$. Those basis elements with $l(R) > N$ vanish identically, while those with $l(R) \leq N$ form the basis for the reduced space. In the context of the $\cN = 4$ SYM theory, this behaviour with respect to finite $N$ will be called SEP-compatible.

Inverting \eqref{schur definition from traces} using character orthogonality relations gives
\begin{equation}
T_p = \sum_{R \vdash n} \chi_R (p) s_R
\end{equation} 
Combining this with \eqref{schur definition from monomials} leads to a an expression for multi-traces in terms of monomials.
\begin{equation}
T_p = \sum_{R,q \vdash n} \chi_R(p) K_{R q} m_q = \sum_{R,q \vdash n} \frac{\chi_R(p) K_{Rq}}{\prod_i q_i!} M_q
\label{traces from monomials}
\end{equation}
We will study this relationship in much greater depth in section \ref{sec:MSSP}, including the inverse relation giving monomials in terms of multi-traces.

\subsubsection{Correlators}

The physical $\mathcal{N}=4$ inner product on multi-traces of the form \eqref{traces from permutations} is given by
\begin{align}
\left\langle \tr \left( \tau X^{\otimes n} \right) | \tr \left( \sigma X^{\otimes n} \right) \right\rangle = \langle T_{p_\tau} | T_{p_\sigma} \rangle & = \sum_{\alpha,\beta \in S_n} N^{c(\beta)} \delta (\alpha \sigma \alpha^{-1} \tau^{-1} \beta ) \nonumber \\
& = \sum_{\alpha \in S_n} \delta \left( \alpha \sigma \alpha^{-1} \tau^{-1} \Omega_N \right)
\label{physical half-BPS inner product}
\end{align}
where $p_\sigma \vdash n$ is the cycle type of $\sigma$, $c(\sigma) = l(p_\sigma)$ is the number of cycles in $\sigma$, and
\begin{equation}
\Omega_N = \sum_{\sigma \in S_n} N^{c(\sigma)} \sigma
\label{Omega definition}
\end{equation}
We also use $\delta (\sigma)$, defined on $S_n$ by
\begin{equation}
\delta ( \sigma ) =
\begin{cases}
1 & \sigma = 1 \\
0 & \text{otherwise}
\end{cases}
\label{delta}
\end{equation}
and extended linearly to the group algebra $\bC(S_n)$.

Under conjugation of $\sigma$, the number of cycles $c(\sigma)$ is invariant, and therefore $\Omega_N$ commutes with any permutation in $S_n$
\begin{equation}
\alpha \Omega_N \alpha^{-1} = \sum_{\sigma \in S_n} N^{c(\sigma)} \alpha \sigma \alpha^{-1} = \sum_{\sigma \in S_n} N^{c(\alpha^{-1} \sigma \alpha)} \sigma = \sum_{\sigma \in S_n} N^{c(\sigma)} \sigma = \Omega_N 
\end{equation}
At large $N$, the leading $N$ behaviour of $\Omega_N$ is
\begin{equation}
\Omega_N = N^n \left[ 1 + O \left( \frac{1}{N} \right) \right]
\label{Omega N expansion}
\end{equation}
and consequently the inner product \eqref{physical half-BPS inner product} reduces to $N^n$ times the $S_n$ inner product
\begin{equation}
\left\langle \tr \left( \tau X^{\otimes n} \right) | \tr \left( \sigma X^{\otimes n} \right) \right\rangle_{S_n} = \langle T_{p_\tau} | T_{p_\sigma} \rangle_{S_n} := \sum_{\alpha \in S_n} \delta ( \alpha \sigma \alpha^{-1} \tau^{-1} ) = \delta_{p_\sigma p_\tau} z_{p_\sigma}
\label{Sn half-BPS inner product}
\end{equation}
where we have used \eqref{z_p} for the size of the centraliser of $\sigma$ in order to evaluate the $S_n$ inner product. 

When we take $S_n$ inner products of operators whose coefficients are $N$-independent, this is the same as the planar inner product. Therefore \eqref{Sn half-BPS inner product} shows that different multi-traces are orthogonal in the planar limit with norm $z_p$. However in section \ref{sec:Construction}  we consider operators whose coefficients depend on $N$. For these cases, \eqref{Sn half-BPS inner product} is distinct from the planar inner product, which would also take the large $N$ limit in coefficients.

Although the definition \eqref{Sn half-BPS inner product} was motivated by large $N$ considerations, it is also a valid definition at finite $N$. For $ N < n $, it is useful to introduce the definition
\begin{equation}
\delta_N ( \sigma ) = \frac{1}{n!} \sum_{\substack{R \vdash n \\ l(R) \leq N}} d_R \chi_R ( \sigma )
\label{finite N delta}
\end{equation}
where $d_R$ is the dimension of the $S_n$ irreducible representation $R$. So a finite $N$ version  of the $S_n$ inner product can be defined as 
\begin{equation}
\left\langle \tr \left( \tau X^{\otimes n} \right) | \tr \left( \sigma X^{\otimes n} \right) \right\rangle_{S_n} := \sum_{\alpha \in S_n} \delta_N ( \alpha \sigma \alpha^{-1} \tau^{-1} ) 
\label{Sn half-BPS finite N inner product}
\end{equation}
The relation between the physical and $S_n$ inner products is determined by the properties of $\Omega_N$, which we now go through. 

In a representation $R \vdash n$, $\Omega_N$ has representative
\begin{equation}
D^R \left( \Omega_N \right) = \frac{n! \operatorname{Dim}(R)}{d_R} = \prod_{b \in R} ( N + c_b ) =: f_R ( N )
\label{omega in a representation}
\end{equation}
where $\operatorname{Dim}(R)$ is the dimension of the $U(N)$ irreducible representation $R$ and for a box $b$ in $R$, $c_b$ is the \textit{contents} of $b$. If $b$ sits in the $r$th row and $c$th column of $R$, then $c_b = r - c$. For example if we take $R=[4,4,2]$, the contents of each box are
\begin{equation}
\begin{ytableau}
0 & 1 & 2 & 3 \\
-1 & 0 & 1 & 2 \\
-2 & -1
\end{ytableau}
\label{contents definition}
\end{equation}
If $l(R) > N$, then $f_R = \operatorname{Dim}(R) = 0$, and therefore $\Omega_N $ imposes the finite $N$ cut-off on Young diagrams.

Define the projector onto an irreducible representation $R$ of $S_n$ by
\begin{equation}
P_R = \frac{d_R}{n!} \sum_{\sigma \in S_n} \chi_R ( \sigma ) \sigma
\end{equation}
This projects onto the $R$ representation of $S_n$ and commutes with all permutations in $S_n$. We can write $\Omega_N $ in terms of these projectors
\begin{equation}
\Omega_N = \sum_{R \vdash n} f_R P_R
\end{equation}
Since $f_R = 0$ for $R$ with $l(R) > N$, it follows that $\Omega_N$ is invertible only in those representations $R$ with $l(R) \leq N$. Define
\begin{equation}
\Omega_N^{-1} = \sum_{\substack{R \vdash n \\l(R) \leq N}} \frac{1}{f_R} P_R
\label{omega inverse}
\end{equation}
which has representatives
\begin{equation}
D^R \left( \Omega_N^{-1} \right)= \begin{cases}
\frac{1}{f_R} & l(R) \leq N \\
0 & l(R) > N
\end{cases}
\label{omega N in a representation}
\end{equation}
This is inverse to $\Omega_N$ in all representations $R$ with $l(R) \leq N$. If $N \geq n$, it is inverse to $\Omega_N$ in all irreducible representations of $S_n$, and is therefore inverse in the full group algebra $\bC(S_n)$. 

Define maps $\cF_N$ and $\cG_N$ as  versions of $\Omega_N $ and $\Omega_N^{-1}$ on gauge invariant operators
\begin{align}
\cF_N \left[ \tr ( \sigma  X^{\otimes n} ) \right] & = \tr \left( \Omega_N \sigma X^{\otimes n} \right) 
\label{F map definition} \\
\cG_N \left[ \tr ( \sigma  X^{\otimes n} ) \right] & = \tr \left( \Omega_N^{-1} \sigma X^{\otimes n} \right) 
\label{G map definition}
\end{align}
On operators, $\cF_N$ and $\cG_N$ are inverse to each other at all $N$.

It follows from \eqref{Omega N expansion} that at large $N$, $\cF_N$ and $\cG_N$ are multiplication and division by $N^n$ respectively. At finite $N$, they are more complicated operators.

Extending the definition linearly to generic degree $n$ operators $\cO_1$ and $\cO_2$, the relation between the physical and $S_n$ inner products is
\begin{align}
\begin{aligned}\label{SnFG} 
\langle \cO_1 | \cO_2 \rangle & = \langle \cO_1 | \cF_N \cO_2 \rangle_{S_n} \\
\langle \cO_1 | \cG_N \cO_2 \rangle & = \langle \cO_1 | \cO_2 \rangle_{S_n}
\end{aligned}
\end{align}

Motivated by the interaction of $\cF_N$ and $\cG_N$ with the different inner products, we call the physical inner product the $\cF_N$-weighted inner product, and will sometimes denote it by adding a $\cF$ subscript. We introduce a similar $\cG$-weighted inner product
\begin{equation}
\langle \cO_1 | \cO_2 \rangle_\cG := \left\langle \cO_1 | \cG_N \cO_2 \right\rangle_{S_n} = \left\langle \cG_N \cO_1 | \cG_N \cO_2 \right\rangle_\cF = \sum_{\alpha \in S_n} \delta ( \alpha \sigma \alpha^{-1} \tau^{-1} \Omega_N^{-1})
\label{G-weighted inner product}
\end{equation}
The behaviour of multi-traces and monomials under $\cF_N$ and $\cG_N$ are in general quite complicated. The Schur operators \eqref{schur definition from permutations} have much simpler properties, stemming from the expressions \eqref{omega in a representation} and \eqref{omega N in a representation} for $\Omega_N$ and $\Omega_N^{-1}$ in the representation $R$.
\begin{align}
\cF ( s_R ) & = f_R s_R \\ 
\cG_N ( s_R ) & = \begin{cases} \frac{1}{f_R} s_R & l(R) \leq N \\ 0 & l(R) > N \end{cases}
\end{align}
It follows these and the orthogonality of characters that
\begin{align}
\langle s_R |  s_S \rangle_\cF & = \delta_{RS} f_R \\ 
\langle s_R |  s_S \rangle_{S_n} & = \begin{cases} \delta_{RS} & l(R) \leq N \\ 0 & l(R) > N \end{cases} \\ 
\langle s_R |  s_S \rangle_\cG & = \begin{cases} \delta_{RS} \frac{1}{f_R} & l(R) \leq N \\ 0 & l(R) > N \end{cases}
\label{Schur correlator}
\end{align}
From their action on Schur operators, it is clear that $\cF_N$ and $\cG_N$ are Hermitian with respect to all three inner products.

The maps $\cF_N$ and $\cG_N$ were first defined in \cite{CtoC}, where they were used to describe a general method of constructing weak coupling quarter-BPS operators. In section \ref{sec:Projectors} we describe how that construction is related to the approach taken in this paper.

When evaluating correlators, we will often express answers as polynomials or rational functions in $N$. It will often be convenient to think of $N$ as a formal variable for these purposes. This is valid provided $n \leq N$.

We give explicit examples of the monomial and Schur operators. Using the definitions \eqref{M definition} and \eqref{schur definition from traces} with $n=4$, we have
\begin{align}
M_{ [4] } & = m_{[4]} = \tr X^4 \\
M_{ [3,1] } & = m_{[3,1]} = \tr X \, \tr X^3 - \tr X^4  \\
M_{ [2,2] } & = 2 m_{[2,2]} = \left( \tr X^2 \right)^2 - \tr X^4 \\
M_{ [2,1,1] } & = 2 m_{[2,1,1]} = \left( \tr X \right)^2 \tr X^2 - 2 \, \tr X \, \tr X^3 - \left( \tr X^2 \right)^2 + 2 \, \tr X^4   \\
M_{ [1,1,1,1] } & = 24 m_{[1,1,1,1]} = \left( \tr X \right)^4 - 6 \left( \tr X \right)^2 \tr X^2 + 8 \, \tr X \, \tr X^3 \nonumber \\ & \qquad \qquad \qquad \qquad + 3 \left( \tr X^2 \right)^2 - 6 \, \tr X^4
\end{align}
and
\begin{align}
\ytableausetup{boxsize=5pt}
s_{[4]} & = \frac{1}{24} \left[ 
\left( \tr X \right)^4 
+ 6 \left( \tr X \right)^2 \tr X^2 
+ 8 \, \tr X \, \tr X^3 
+ 3 \left( \tr X^2 \right)^2 
+ 6 \, \tr X^4 
\right]
\label{schur [4]} \\
s_{[3,1]} & = \frac{1}{8} \left[ 
\left( \tr X \right)^4 
+ 2 \left( \tr X \right)^2 \tr X^2 
- \left( \tr X^2 \right)^2 
- 2 \, \tr X^4 
\right] 
\label{schur [3,1]} \\
s_{[2,2]} & = \frac{1}{12} \left[ 
\left( \tr X \right)^4 
- 4 \, \tr X \, \tr X^3 
+ 3 \left( \tr X^2 \right)^2 
\right] \\
s_{[2,1,1]} & = \frac{1}{8} \left[ 
\left( \tr X \right)^4 
- 2 \left( \tr X \right)^2 \tr X^2 
- \left( \tr X^2 \right)^2 
+ 2 \, \tr X^4 
\right]  \\
s_{[1,1,1,1]} & = \frac{1}{24} \left[ 
\left( \tr X \right)^4 
- 6 \left( \tr X \right)^2 \tr X^2 
+ 8 \, \tr X \, \tr X^3 
+ 3 \left( \tr X^2 \right)^2 
- 6 \, \tr X^4 
\right]
\label{schur [1,1,1,1]} 
\end{align}
\ytableausetup{boxsize=normal}

\subsection{Quarter-BPS covariant operators}
\label{section:quarter-bps}

The zero coupling quarter-BPS sector is spanned by multi-traces of two complex matrices $X$ and $Y$.
There are several natural extensions of the Schur basis \eqref{schur definition from permutations} to the 2-matrix sector \cite{Brauer,BCD0801,BCD0805}. We will use the covariant basis, introduced in \cite{BHR0711,BHR0806}, which is SEP-compatible and orthogonal to all orders in $N$.. The basis elements are also referred to as BHR operators after the authors.

\subsubsection{$U(2)$ action on traces}

Let $X_1 = X$ and $X_2 = Y$. Then there is an action of $U(2)$ on the $i$ index in $X_i$. By extension this acts on all traces and operators, so we can choose our basis to be $U(2)$ covariant. Since $U(2)$ turns $X$s into $Y$s and vice versa, this basis mixes states with different numbers of $X$s and $Y$s while keeping the total number of matrices, $n$, constant. We will say an operator has \textit{field content} $(n_1,n_2)$ if it contains $n_1$ $X$s and $n_2$ $Y$s.

The $\fu (2)$ operators on traces are given by 
\begin{align}
R^i_j = 
\begin{pmatrix}
\cX & \cJ_+  \\
\cJ_- & \cY
\end{pmatrix}
=
\begin{pmatrix}
\tr X \frac{\partial}{\partial X} & \tr X \frac{\partial}{\partial Y} \\
\tr Y \frac{\partial}{\partial X} & \tr Y \frac{\partial}{\partial Y}
\end{pmatrix}
=
\begin{pmatrix}
X^i_j \frac{\partial}{\partial X^i_j} & X^i_j \frac{\partial}{\partial Y^i_j} \\
Y^i_j \frac{\partial}{\partial X^i_j} & Y^i_j \frac{\partial}{\partial Y^i_j}
\end{pmatrix}
\label{U(2) lowering and raising operators}
\end{align}
The operator $\cX$ counts the number of $X$ matrices in a trace, similarly for $\cY$. The lowering operator $\cJ_-$ `lowers' 
a trace by turning an $X$ into a $Y$, and the raising operator $\cJ_+$ `raises' a trace by turning a $Y$ into an $X$.

Acting on the matrices $X_i$ with a $U(2)$ index
\begin{equation}
R^i_j X_k = \delta^i_k X_j
\end{equation}
Define new operators
\begin{align}
\cJ_0 &  = \cX + \cY & \cJ_3 = \cX - \cY
\end{align}
Then $\cJ_0$ counts the total number of matrices, while $\cJ_3$ counts the difference between the number of $X$s and $Y$s. As 
the notation suggests, $\cJ_3, \cJ_\pm$ form an $\fsu (2)$ subalgebra of $\fu (2)$, while $\cJ_0$ spans a $\fu (1)$ that 
commutes with the $\fsu (2)$. This split decomposes $\fu(2)$ into a sum of $\fsu(2)$ and $\fu(1)$.

The operators \eqref{U(2) lowering and raising operators} obey standard hermiticity conditions $\left( R^i_j \right)^\dagger = R^j_i$ for $R$-symmetry generators
\begin{equation}
\left( \cJ_0 \right)^\dagger = \cJ_0 \hspace{50pt} \left( \cJ_3 \right)^\dagger = \cJ_3 \hspace{50pt} \left( \cJ_+ \right)^\dagger = \cJ_-
\label{hermiticity of U(2)}
\end{equation}
It follows that operators with different $U(2)$ quantum numbers must be orthogonal.

\subsubsection{$U(2)$ representations}
\label{section: U(2) reps}

Semi-standard Young tableaux are defined to be Young tableaux in which the positive integers in the boxes increase weakly 
along the rows and strictly down the columns. For example if we take $R=[2,1]$ and allow entries of 1,2 and 3, the possible 
semi-standard tableaux are:
\begin{equation}
\begin{ytableau}
1 & 1 \\ 2
\end{ytableau}
\quad
\begin{ytableau}
1 & 1 \\ 3
\end{ytableau}
\quad
\begin{ytableau}
1 & 2 \\ 2
\end{ytableau}
\quad
\begin{ytableau}
1 & 2 \\ 3
\end{ytableau}
\quad
\begin{ytableau}
1 & 3 \\ 2
\end{ytableau}
\quad
\begin{ytableau}
1 & 3 \\ 3
\end{ytableau}
\quad
\begin{ytableau}
2 & 2 \\ 3
\end{ytableau}
\quad
\begin{ytableau}
2 & 3 \\ 3
\end{ytableau}
\label{semi-standard tableaux example}
\end{equation}
The evaluation of a semi-standard tableau $r$ is a sequence of numbers $\rho (r) = [\rho_1, \rho_2, \dots]$ where
\begin{equation}
\rho_i = (\# \text{ of occurences of the number } i \text{ in } r)
\end{equation}
So for example the evaluations of the tableaux in \eqref{semi-standard tableaux example} are respectively
\begin{equation}
[2,1,0] \hspace{15pt} [2,0,1] \hspace{15pt} [1,2,0] \hspace{15pt} [1,1,1] \hspace{15pt} [1,1,1] \hspace{15pt} [1,0,2] \hspace{15pt} [0,2,1] \hspace{15pt} [0,1,2]
\end{equation}
When the evaluation $\rho (r)$ is a partition (i.e. $\rho_1 \geq \rho_2 \geq \dots$), these tableaux contribute to the Kostka numbers $K_{R \rho}$ seen in \eqref{schur definition from monomials}.

For a representation $\Lambda \vdash n$ of $U(2)$ with $l(\Lambda) \leq 2$, the basis vectors of $\Lambda$ are labelled by the semi-standard Young tableaux of shape $\Lambda$ containing only 1s and 2s. For $\Lambda = \left[ \frac{n}{2} + j, \frac{n}{2} - j \right]$, there are $2j+1$ possible tableaux, where $j$ runs over the non-negative half-integers up to $\frac{n}{2}$. These possibilities are
\begin{equation}
\begin{gathered}
\begin{tikzpicture}
\draw [decorate,decoration={brace,amplitude=10pt},yshift=2pt,line width=0.8pt]
(0,3em) -- (6em,3em) node [midway,yshift=+0.7cm] {$\frac{n}{2} - j$} ;
\draw [decorate,decoration={brace,amplitude=10pt},yshift=2pt,line width=0.8pt]
(6em,3em) -- (12em,3em) node [midway,yshift=+0.7cm] {$k$} ;
\draw [decorate,decoration={brace,amplitude=10pt},yshift=2pt,line width=0.8pt]
(12em,3em) -- (18em,3em) node [midway,yshift=+0.7cm] {$2j - k$} ;
\draw (0,0) rectangle (3em,3em) ;
\draw (1.5em,0) -- (1.5em,3em) ;
\draw (0,1.5em) -- (3em,1.5em) ;
\node at (0.75em,2.25em) {1} ;
\node at (2.25em,2.25em) {1} ;
\node at (0.75em,0.75em) {2} ;
\node at (2.25em,0.75em) {2} ;
\node at (3.85em,0.75em) {$\dots$} ;
\node at (3.85em,2.25em) {$\dots$} ;
\draw (4.5em,0) rectangle (6em,3em) ;
\draw (4.5em,1.5em) rectangle (9em,3em) ;
\draw (7.5em,1.5em) -- (7.5em,3em) ;
\node at (5.25em,2.25em) {1} ;
\node at (6.75em,2.25em) {1} ;
\node at (8.25em,2.25em) {1} ;
\node at (5.25em,0.75em) {2} ;
\node at (9.85em,2.25em) {$\dots$} ;
\draw (10.5em,1.5em) rectangle (15em,3em) ;
\draw (12em,1.5em) rectangle (13.5em,3em) ;
\node at (11.25em,2.25em) {1} ;
\node at (12.75em,2.25em) {2} ;
\node at (14.25em,2.25em) {2} ;
\node at (15.85em,2.25em) {$\dots$} ;
\draw (16.5em,1.5em) rectangle (18em,3em) ;
\node at (17.25em,2.25em) {2} ;
\end{tikzpicture}
\end{gathered}
\end{equation}
where $0 \leq k \leq 2j$.

We can understand the representation $\Lambda = \left[ \frac{n}{2} + j, \frac{n}{2} - j \right]$ in terms of the $\fu (1)$ spanned by $\cJ_0$ and the $\fsu (2)$ spanned by $\cJ_3, \cJ_\pm$. All states in $\Lambda$ have weight $n$ under $\fu(1)$, and form a spin $j$ of $\fsu (2)$. The identification of basis vectors is
\begin{equation}
\begin{gathered}
\begin{tikzpicture}
\draw [decorate,decoration={brace,amplitude=10pt},yshift=2pt,line width=0.8pt]
(0,3em) -- (6em,3em) node [midway,yshift=+0.7cm] {$\frac{n}{2} - j$} ;
\draw [decorate,decoration={brace,amplitude=10pt},yshift=2pt,line width=0.8pt]
(6em,3em) -- (12em,3em) node [midway,yshift=+0.7cm] {$j + m_j$} ;
\draw [decorate,decoration={brace,amplitude=10pt},yshift=2pt,line width=0.8pt]
(12em,3em) -- (18em,3em) node [midway,yshift=+0.7cm] {$j - m_j$} ;
\draw (0,0) rectangle (3em,3em) ;
\draw (1.5em,0) -- (1.5em,3em) ;
\draw (0,1.5em) -- (3em,1.5em) ;
\node at (0.75em,2.25em) {1} ;
\node at (2.25em,2.25em) {1} ;
\node at (0.75em,0.75em) {2} ;
\node at (2.25em,0.75em) {2} ;
\node at (3.85em,0.75em) {$\dots$} ;
\node at (3.85em,2.25em) {$\dots$} ;
\draw (4.5em,0) rectangle (6em,3em) ;
\draw (4.5em,1.5em) rectangle (9em,3em) ;
\draw (7.5em,1.5em) -- (7.5em,3em) ;
\node at (5.25em,2.25em) {1} ;
\node at (6.75em,2.25em) {1} ;
\node at (8.25em,2.25em) {1} ;
\node at (5.25em,0.75em) {2} ;
\node at (9.85em,2.25em) {$\dots$} ;
\draw (10.5em,1.5em) rectangle (15em,3em) ;
\draw (12em,1.5em) rectangle (13.5em,3em) ;
\node at (11.25em,2.25em) {1} ;
\node at (12.75em,2.25em) {2} ;
\node at (14.25em,2.25em) {2} ;
\node at (15.85em,2.25em) {$\dots$} ;
\draw (16.5em,1.5em) rectangle (18em,3em) ;
\node at (17.25em,2.25em) {2} ;
\end{tikzpicture}
\end{gathered} = \left| j, m_j \right\rangle
\end{equation}
where $\left| j, m_j \right\rangle$ is the standard basis spanning the spin $j$ representation of $\fsu (2)$ with $-j \leq m_j \leq j$.

\subsubsection{Operator definition}

To write down the covariant basis, consider $V_2^{\otimes n}$, where $V_2$ is the fundamental of $U(2)$, and in particular the basis vector $a = e_{a_1} \otimes e_{a_2} \otimes \dots \otimes e_{a_n}$ of $V_2^{\otimes n}$ where $a_j \in \{1,2\}$ for each $j$. Then we define $\mathbb{X}_a = X_{a_1} \otimes X_{a_2} \otimes \dots \otimes X_{a_n}$. Combined with a permutation $\sigma \in S_n$, we write
\begin{equation}
\mathcal{O}_{a , \sigma} = \tr \left( \sigma \mathbb{X}_a \right)
\label{U(2) covariant operator definition}
\end{equation}
These operators are multi-traces, where each single trace factor corresponds to a cycle in $\sigma$.

The labelling set for the covariant basis is: $\Lambda \vdash n$, a partition with at most 2 rows; $M_\Lambda$, a semi-standard tableau of shape $\Lambda$ that indexes the basis vectors of the $\Lambda$ representation of $U(2)$; $R \vdash n$, a partition with at most $N$ rows; and $\tau$, a multiplicity index satisfying $1 \leq \tau \leq C(R,R,\Lambda)$, where $C(R,R,\Lambda)$ is the multiplicity of the trivial representation within $R \otimes R \otimes \Lambda$  (for $\Lambda$ as an $S_n$ representation), or equivalently the multiplicity of the $\Lambda$ representation within $R \otimes R$.

Using these labels, the BHR operators are defined by
\begin{equation}
\mathcal{O}_{\Lambda, M_{\Lambda}, R, \tau} = \frac{\sqrt{d_R}}{n!} \sum_{\sigma, a,i,j,m} S^{R \, R \, \Lambda, \, \tau}_{i 
	\ j \ m} D^R_{ij}(\sigma) C^a_{\Lambda, M_{\Lambda}, m} \mathcal{O}_{a,\sigma}
\label{U(2) basis definition}
\end{equation}
where $D^R_{ij} (\sigma)$ is the $R$ representation matrix for $\sigma$, $C^a_{\Lambda, M_\Lambda, m}$ are the Clebsch-Gordon coefficients for the Schur-Weyl decomposition
\begin{equation}
V_2^{\otimes n} = \bigoplus_{\substack{\Lambda \vdash n \\ l(\Lambda) \leq 2}} V^{U(2)}_\Lambda \otimes V^{S_n}_\Lambda
\end{equation}
and $S^{R \, R \, \Lambda \, \tau}_{i \ j \ m}$ are the Clebsch-Gordon coefficients for the $\tau$th copy of the trivial $S_n$ representation inside $R \otimes R \otimes \Lambda$.

The field content $(n_1,n_2)$ can be read off from the numbers in the tableau $M_\Lambda$. The number of 1s in $M_\Lambda$ is 
$n_1$, while the number of 2s is $n_2$.

When $\Lambda = [n]$ and $M_\Lambda$ is the tableau consisting only of 1s, the operators \eqref{U(2) basis definition} reduce to the standard Schur operators \eqref{schur definition from permutations}. Since other values of $M_\Lambda$ are related to these via $\fu (2)$ operators, which are part of the R-symmetry algebra, the entire $\Lambda = [n]$ sector is part of a half-BPS multiplet. The multiplicity index $\tau$ is trivial since $R \otimes R$ always contains a unique copy of the trivial representation
\begin{equation}
C(R,R,[n]) = 1
\label{[n] multiplicity}
\end{equation}
The $\Lambda = [n-1,1]$ sector also has special properties. In \cite{Dolan_2003} it was proved that these multiplets cannot recombine to form long non-BPS multiplets and therefore must remain quarter-BPS at all values of the coupling. In these cases, the multiplicity $\tau$ in \eqref{U(2) basis definition} runs over the number of corners of $R$ minus 1.
\begin{equation}
C(R,R,[n-1,1]) = (\# \text{ of corners in } R) - 1
\label{[n-1,1] multiplicity}
\end{equation}
This is proved most simply by comparing the covariant basis with the combinatorics of the restricted Schur basis defined in \cite{BCD0801,BCD0805}.

The covariant basis is SEP-compatible, so when $N < n$, those operators with $l(R) > N$ vanish and the reduced set forms a basis for the space.

In table \ref{table: U(2) field content 2,2 basis}, we give explicit expressions for the $U(2)$ covariant basis vectors with field content $(2,2)$. The equivalent expressions for field content $(4,0)$ were already given in \eqref{schur [4]}-\eqref{schur [1,1,1,1]}. It's simple to check that one can travel between these different sectors using the $U(2)$ operators $\cJ_\pm$ given in \eqref{U(2) lowering and raising operators}.

\begin{table}
	\ytableausetup{boxsize=7.5pt}
	\begin{center}
		\begin{tabular}{ p{2.6cm} | p{1cm} | p{1cm} | p{1cm} | p{1cm} | p{1cm} | p{0.72cm} | p{0.72cm} | p{0.72cm} | p{0.44cm} | p{0.44cm} }
			$M_{\Lambda}$ &
			\fontsize{6pt}{0} $\begin{gathered} \begin{ytableau} 1 & 1 & 2 & 2 \end{ytableau} \end{gathered}$ &
			\fontsize{6pt}{0} $\begin{gathered} \begin{ytableau} 1 & 1 & 2 & 2 \end{ytableau} \end{gathered}$ &
			\fontsize{6pt}{0} $\begin{gathered} \begin{ytableau} 1 & 1 & 2 & 2 \end{ytableau} \end{gathered}$ &
			\fontsize{6pt}{0} $\begin{gathered} \begin{ytableau} 1 & 1 & 2 & 2 \end{ytableau} \end{gathered}$ &
			\fontsize{6pt}{0} $\begin{gathered} \begin{ytableau} 1 & 1 & 2 & 2 \end{ytableau} \end{gathered}$ &
			\fontsize{6pt}{0} $\begin{gathered} \begin{ytableau} 1 & 1 & 2 \\ 2 \end{ytableau} \end{gathered}$ &
			\fontsize{6pt}{0} $\begin{gathered} \begin{ytableau} 1 & 1 & 2 \\ 2 \end{ytableau} \end{gathered}$ &
			\fontsize{6pt}{0} $\begin{gathered} \begin{ytableau} 1 & 1 \\ 2 & 2 \end{ytableau} \end{gathered}$ &
			\fontsize{6pt}{0} $\begin{gathered} \begin{ytableau} 1 & 1 \\ 2 & 2 \end{ytableau} \end{gathered}$ &
			\fontsize{6pt}{0} $\begin{gathered} \begin{ytableau} 1 & 1 \\ 2 & 2 \end{ytableau} \end{gathered}$ \\[6pt] \hline
			R &
			$\begin{gathered} \ydiagram{4} \end{gathered}$ &
			$\begin{gathered} \ydiagram{3,1} \end{gathered}$ &
			$\begin{gathered} \ydiagram{2,2} \end{gathered}$ &
			$\begin{gathered} \ydiagram{2,1,1} \end{gathered}$ &
			$\begin{gathered} \ydiagram{1,1,1,1} \end{gathered}$ &
			$\begin{gathered} \ydiagram{3,1} \end{gathered}$ &
			$\begin{gathered} \ydiagram{2,1,1} \end{gathered}$ &
			$\begin{gathered} \ydiagram{3,1} \end{gathered}$ &
			$\begin{gathered} \ydiagram{2,2} \end{gathered}$ &
			$\begin{gathered} \ydiagram{2,1,1} \end{gathered}$ \\[13pt] \hline 
			Normalisation Coefficient & 
			$\begin{gathered} \frac{\sqrt{6}}{24} \end{gathered}$ & 
			$\begin{gathered} \frac{\sqrt{6}}{24} \end{gathered}$ & 
			$\begin{gathered} \frac{\sqrt{6}}{12} \end{gathered}$ & 
			$\begin{gathered} \frac{\sqrt{6}}{24} \end{gathered}$ & 
			$\begin{gathered} \frac{\sqrt{6}}{24} \end{gathered}$ & 
			$\begin{gathered} \frac{1}{4} \end{gathered}$ & 
			$\begin{gathered} \frac{1}{4} \end{gathered}$ & 
			$\begin{gathered} \frac{\sqrt{3}}{12} \end{gathered}$ & 
			$\begin{gathered} \frac{\sqrt{6}}{12} \end{gathered}$ & 
			$\begin{gathered} \frac{\sqrt{3}}{12} \end{gathered}$ \\ \hline
			$(\tr X)^2 (\tr Y)^2$ & 1 & 3 & 1 & 3 & 1 & 0 & 0 & 0 & 0 & 0 \\ \hline
			$\tr X^2 (\tr Y)^2$ & 1 & 1 & 0 & -1 & -1 & 1 & -1 & 1 & -1 & -1 \\ \hline
			$\tr XY \tr X \tr Y$ & 4 & 4 & 0 & -4 & -4 & 0 & 0 & -2 & 2 & 2 \\ \hline
			$(\tr X)^2 \tr Y^2$ & 1 & 1 & 0 & -1 & -1 & -1 & 1 & 1 & -1 & -1\\ \hline
			$\tr X^2 Y \tr Y$ & 4 & 0 & -2 & 0 & 4 & 2 & 2 & 0 & 0 & 0 \\ \hline
			$\tr X Y^2 \tr X$ & 4 & 0 & -2 & 0 & 4 & -2 & -2 & 0 & 0 & 0 \\ \hline
			$\tr X^2 \tr Y^2$ & 1 & -1 & 1 & -1 & 1 & 0 & 0 & 2 & 0 & 2 \\ \hline
			$(\tr XY)^2$ & 2 & -2 & 2 & -2 & 2 & 0 & 0 & -2 & 0 & -2 \\ \hline
			$\tr X^2 Y^2$ & 4 & -4 & 0 & 4 & -4 & 0 & 0 & 2 & 2 & -2 \\ \hline
			$ \tr (XY)^2$ & 2 & -2 & 0 & 2 & -2 & 0 & 0 & -2 & -2 & 2
		\end{tabular}
	\end{center}
	\ytableausetup{boxsize=normal}
	\caption{The covariant basis for field content $(2,2)$ in terms of multi-traces. Each element is identified by its $M_\Lambda$ and $R$ labels. We give the overall normalisation and the coefficient of each multi-trace within the operator.}
	\label{table: U(2) field content 2,2 basis}
\end{table}

\subsubsection{Correlators}

The two-point function of generic operators of the form \eqref{U(2) covariant operator definition} is given by \cite{CtoC}.
\begin{equation}
\langle \cO_{b,\tau} | \cO_{a, \sigma} \rangle = \sum_{\alpha \in S_n} \delta_{\alpha(a), b} \delta ( \alpha \sigma \alpha^{-1} \tau^{-1} \Omega_N )
\label{physical inner product}
\end{equation}
The $\delta_{\alpha(a), b}$ factor restricts the sum over $\alpha$ to a subgroup $S_{n_1} \otimes S_{n_2}$ of $S_n$ for operators of field content $(n_1,n_2)$. Except for this restriction, the inner product looks very similar to the half-BPS case \eqref{physical half-BPS inner product}, and we proceed in an analogous manner. Define $\cF_N$ and $\cG_N$ by
\begin{align}
\cF_N \cO_{a, \alpha} = \cO_{a, \Omega_N \alpha} && \cG_N \cO_{a, \alpha} = \cO_{a, \Omega_N^{-1} \alpha}
\end{align}
These are inverse to each other on finite $N$ operators. Define the $S_n$ and $\cG$-weighted inner products on operators to be
\begin{align}
\langle \cO_{b,\tau} | \cO_{a, \sigma} \rangle_{S_n} & := \sum_{\alpha \in S_n} \delta_{\alpha(a), b} \delta_N ( \alpha \sigma \alpha^{-1} \tau^{-1} ) = \langle \cO_{b,\tau} | \cG_N \cO_{a, \sigma} \rangle
\label{quarter-bps Sn inner product} \\
\langle \cO_{b,\tau} | \cO_{a, \sigma} \rangle_{\cG} & := \sum_{\alpha \in S_n} \delta_{\alpha(a), b} \delta ( \alpha \sigma \alpha^{-1} \tau^{-1} \Omega_N^{-1} ) = \langle \cG_N \cO_{b,\tau} | \cG_N \cO_{a, \sigma} \rangle
\label{quarter-bps G inner product}
\end{align}
Then the three distinct inner products are related by multiplying operators by the appropriate $\cF_N$ or $\cG_N$.

The $U(2)$ acts only on the label $a$ in $\cO_{a,\sigma}$, while $\cF_N$ and $\cG_N$ act only on $\sigma$, and therefore the two commute. This means the same hermiticity conditions \eqref{hermiticity of U(2)} apply in each of the physical, $S_n$ and $\cG$-weighted inner products. Therefore operators with different $U(2)$ quantum numbers are orthogonal in all three inner products.

The behaviour \eqref{omega in a representation} of $\Omega_N$ in a representation $R$ of $S_n$ implies that
\begin{align}
\cF_N \cO_{\Lambda, M_{\Lambda}, R, \tau} & = f_R \cO_{\Lambda, M_{\Lambda}, R, \tau} \\ \cG_N \cO_{\Lambda, M_{\Lambda}, R, \tau} & = \begin{cases} \frac{1}{f_R} \cO_{\Lambda, M_{\Lambda}, R, \tau} & l(R) \leq N \\ 0 & l(R) > N \end{cases}
\label{quarter-BPS F and G}
\end{align}
and it follows \cite{CtoC} that the correlators of the covariant basis \eqref{U(2) basis definition} are
\begin{align}
\langle \mathcal{O}_{\Lambda, M_{\Lambda}, R, \tau} | \mathcal{O}_{\Lambda', M_{\Lambda'}, R', \tau'} \rangle & = \delta_{\Lambda, \Lambda'} \delta_{M_{\Lambda}, M_{\Lambda'}} \delta_{R, R'} \delta_{\tau, \tau'} f_R
\label{F inner product} \\
\langle \mathcal{O}_{\Lambda, M_{\Lambda}, R, \tau} | \mathcal{O}_{\Lambda', M_{\Lambda'}, R', \tau'} \rangle_{S_n} & = \begin{cases} \delta_{\Lambda, \Lambda'} \delta_{M_{\Lambda}, M_{\Lambda'}} \delta_{R, R'} \delta_{\tau, \tau'} & l(R) \leq N \\ 0 & l(R) > N \end{cases}
\label{S_n inner product} \\
\langle \mathcal{O}_{\Lambda, M_{\Lambda}, R, \tau} | \mathcal{O}_{\Lambda', M_{\Lambda'}, R', \tau'} \rangle_{\cG} & = \begin{cases} \frac{ \delta_{\Lambda, \Lambda'} \delta_{M_{\Lambda}, M_{\Lambda'}} \delta_{R, R'} \delta_{\tau, \tau'}}{ f_R } & l(R) \leq N \\ 0 & l(R) > N \end{cases}
\label{quarter-bps G-weighted inner product}
\end{align}

\section{Quarter-BPS at weak coupling: Key concepts}
\label{section: weak coupling}

In this section, we review some key concepts which form the background and motivation for the counting and construction of quarter-BPS operators at weak coupling developed in sections \ref{sec:MSSP}- \ref{sec:Projectors}. It is known that while the counting of quarter BPS operators jumps in going from zero coupling to first order in coupling, there are no further corrections to the counting in the transition to strong coupling \cite{KMMR07}.  It is further expected that some class of correlators of quarter-BPS operators are subject to non-renormalization theorems, analogous to those proved in the half-BPS sector (see \cite{EHSW0004} and references therein). Indeed one such class of correlators involving two half-BPS and one quarter BPS have been proved to be non-renormalized \cite{BBP1203}.  We therefore expect that the finite $N$ counting and construction of quarter-BPS operators at weak coupling in SYM will reflect many properties of quarter BPS states arising from geometrical constructions at strong coupling SYM, or  the dual weakly coupled gravity. These include quarter BPS giant gravitons and LLM geometries.

\subsection{Background on construction of Quarter BPS operators}
\label{sec:BCOV}

When we turn the coupling on in $\mathcal{N} = 4$ SYM, some of the short quarter-BPS multiplets at zero coupling recombine and form long non-BPS multiplets. We give two equivalent ways of characterising which 2-matrix multi-trace operators remain quarter-BPS, and which do not.

Firstly, non-BPS multi-traces of $ X , Y $ are SUSY descendants. It was explained in \cite{dHRyz0301} that these are exactly commutator traces. That is, they are multi-traces (or linear combinations thereof) where at least one of the constituent single traces contains a commutator $ [ X , Y ] $. 

Secondly, consider the one-loop dilatation operator \cite{BKPSS0208,BKS03}
\begin{equation}
\cH_2 = -  \tr \left( [X,Y] \left[ \frac{\partial}{\partial X} , \frac{\partial}{\partial Y} \right] \right)
\label{one-loop dilatation operator}
\end{equation}
Quarter-BPS operators are annihilated by $\cH_2$, and as $\cH_2$ is hermitian in the free field inner product, they are orthogonal to the image. It is clear from the definition \eqref{one-loop dilatation operator} that states in the image are commutator traces. While it is not immediately obvious that all commutator traces live in the image, our numerical calculations indicate that they are, and consistency with \cite{dHRyz0301} implies they should be.

Therefore the multi-traces that remain quarter-BPS as we move to weak coupling are those that are orthogonal to the commutator traces under the physical  free field inner product \eqref{physical inner product}. This inner product is difficult to evaluate on multi-trace operators, so \cite{CtoC} took a different approach, using $\cF_N$ and $\cG_N$ defined in \eqref{quarter-BPS F and G} to express the physical inner product in terms of the better understood $S_n$ inner product \eqref{S_n inner product}.

Let $\cO^c$ be a commutator trace, and $\cO^s$ be a \textit{pre-BPS} operator, defined to be orthogonal to commutator traces under the $S_n$ inner product. Then using the relation \eqref{quarter-bps Sn inner product} between the physical and $S_n$ inner products
\begin{equation}
\langle \cO^c | \cG_N \cO^s \rangle = \langle \cO^c | \cO^s \rangle_{S_n} = 0
\label{commutator orthogonal}
\end{equation}
So the operator $\cG_N \cO^s$ is a quarter-BPS operator at weak coupling.

The natural next step is to determine the form of the operators $\cO^s$. It was demonstrated in \cite{CtoC} that for $N \geq n$ these are symmetrised traces. For a single trace, the symmetrised version is
\begin{equation}
\tr ( X_{a_1} X_{a_2} \dots X_{a_n} ) \to \frac{1}{n!} \sum_{\sigma \in S_n} \tr ( X_{a_{\sigma(1)}} X_{a_{\sigma(2)}} \dots X_{a_{\sigma(n)}} ) =: \str ( X_{a_1} X_{a_2} \dots X_{a_n} )
\label{symmetrised trace}
\end{equation}
where $a_i \in \{1,2\}$ for each $i$ and $X_1 = X$, $X_2 = Y$. For a multi-trace, this process is applied to each of the constituent single traces. A symmetrised trace is determined by the field content of each single trace factor. For a particular total field content $(n_1,n_2)$, the possible symmetrised traces are labelled by \textit{vector partitions} $\bp$. A vector partition is a set of integer 2-vectors which sum to $(n_1,n_2)$, which we denote by $\bp \vdash (n_1,n_2)$. For a vector partition $\bp = [(k_1,l_1), \dots (k_m,l_m)]$ the associated symmetrised trace is
\begin{equation}
T_\bp = \str \left( X^{k_1} Y^{l_1} \right) \str \left( X^{k_2} Y^{l_2} \right) \dots \str \left( X^{k_m} Y^{l_m} \right)
\label{symmetrised trace operator}
\end{equation}
We conclude a generic quarter-BPS operator for $N \geq n$ can simply be written as
\begin{equation}
\cO^{BPS} = \cG_N T_\bp
\end{equation}
At finite $N$, non-trivial relations among different multi-traces reduce the dimensionality of the quarter-BPS sector, and correspondingly the pre-BPS operators as well. A finite $N$ relation among traces could have three distinct behaviours with respect to the large $N$ space of symmetrised traces
\begin{enumerate}
	
	\item It is internal to the space of symmetrised traces. In this case, under an appropriate choice of basis, a symmetrised trace reduces to the zero operator. Correspondingly, the dimension of the pre-BPS and BPS sectors reduce by 1.
	
	\item It is internal to the space of commutator traces. This does not affect the pre-BPS or BPS sectors.

	\item   It is a linear combination of  symmetrised traces and commutator traces. In this case, under an appropriate choice of basis, a symmetrised trace reduces to a commutator trace. This means it is no longer pre-BPS, as it is not $S_n$ orthogonal to descendants. Correspondingly, the dimension of the pre-BPS and BPS sectors reduce by 1.
	
\end{enumerate}
Therefore, SEP-compatibility in the pre-BPS sector has a different interpretation to the BPS equivalent. A basis for pre-BPS operators is SEP-compatible if operators with labels longer than $N$ reduce to \textit{either} the zero operator \textit{or} a commutator trace after applying finite $N$ relations. After applying $\cG_N$ to such a basis, we obtain an SEP-compatible basis for the quarter-BPS sector.

\subsection{Steps in the construction of an  SEP-compatible orthogonal BPS  basis}
\label{section: isomorphism}

The key ingredient that will allow the construction of an SEP-compatible basis for pre-BPS operators is an isomorphism proved by Vaccarino \cite{Vaccarino2007} and Domokos \cite{Domokos}, summarised nicely by Procesi in \cite{Procesi}.

From the definition \eqref{symmetrised trace}, the non-commuting matrices $X$ and $Y$ commute within a symmetrised trace, and therefore we naively expect that that symmetrised traces of non-commuting matrices correspond to ordinary multi-traces of commuting matrices via
\begin{equation}
\str ( X^{k_1} Y^{l_1} ) \dots \str ( X^{k_m} Y^{l_m} ) \longleftrightarrow \tr \left( A^{k_1} B^{l_1} \right) \dots \tr \left( A^{k_m} B^{l_m} \right)
\label{symmetrised trace isomorphism}
\end{equation}
where $A$ and $B$ are two commuting $N \times N$ matrices. The isomorphism of \cite{Vaccarino2007,Domokos,Procesi} makes this expectation rigorous.

Consider the ring $R(X,Y)$ generated by the  matrix elements of two $N \times N$ matrices $X,Y$. This ring is acted on by $U(N)$ via simultaneous conjugation of the two matrices. Given a $\cU \in U(N)$, we have
\begin{equation}
( X , Y ) \rightarrow ( \cU X \cU^{ \dagger} , \cU Y \cU^{ \dagger} ) 
\label{near isomorphism description}
\end{equation}
Then invariants of $R(X,Y)$ under this action are multi-traces of $X$ and $Y$, and correspond to the quarter-BPS sector at zero coupling. At weak coupling, we consider $R(X,Y)$ modulo the ideal $I$ generated by the commutator $[X,Y]$. We call the quotient ring $R_s(X,Y)$. Each $U(N)$ invariant of the quotient ring corresponds to an equivalence class of multi-trace operators related by addition of a commutator trace. In each class there is a unique pre-BPS representative that is orthogonal to all commutator traces (under the $S_n$ inner product).There is also a unique BPS operator orthogonal to commutator traces under the $ \cF_N$ inner product. Conversely, given a pre-BPS or BPS operator, there is a unique equivalence class to which it belongs. Therefore the invariants of $R_s(X,Y)$ give the combinatorics of the pre-BPS and BPS sectors, both at large $N$ and finite $N$.
	
Finding the pre-BPS operator in a given equivalence class is simple when $N \geq n$; as discussed in section \ref{sec:BCOV}, the representative is a symmetrised trace. When $N < n$, the multi-trace expansion of an operator is non-unique, and it is more difficult to identify the pre-BPS operator. In section \ref{sec:Construction}, we describe how to find the pre-BPS operator by orthogonalisation.

On the other side of the isomorphism are \textit{multi-symmetric functions}. Take two commuting $N \times N$ matrices $A = \Diag ( x_1, x_2 , \cdots , x_N )$ and $B = \Diag ( y_1, y_2 , \cdots , y_N )$. Then a multi-trace of $A$ and $B$ will be a polynomial in the $2N$ variables invariant under permutations
\begin{equation}
(x_i, y_i) \to \left( x_{\sigma(i)}, y_{\sigma(i)} \right)
\end{equation}
for $\sigma \in S_N$. These are called multi-symmetric functions, and generalise the symmetric functions of section \ref{section: half-bps} to two families of variables. They are discussed in detail in section \ref{sec:MSSP}.
	
The theorem in \cite{Vaccarino2007,Domokos,Procesi} tells us that the ring of invariants of $R_s(X,Y)$ is isomorphic to the ring of multi-symmetric functions in $2N$ variables.

This isomorphism is simple to give explicitly. Take a multi-trace of $X$ and $Y$ and restrict the two matrices to be diagonal. This is now a multi-symmetric function in the $2N$ eigenvalues. Clearly the commutator $[X,Y]$ vanishes for the diagonal $X$ and $Y$, and therefore any multi-traces related by a commutator trace lead to the same multi-symmetric function.

Conversely, given a multi-trace of two commuting matrices $A$ and $B$, we use the map \eqref{symmetrised trace isomorphism} to pick a representative of the isomorphic equivalence class. At large $N$, this correctly identifies the pre-BPS operator.  However, for $N < n$, this map does not associate a unique symmetrised trace with a given multi-symmetric function. Finite $N$ relations mean a multi-symmetric function can be written in multiple ways as the trace of commuting matrices. These different expressions give genuinely different operators in the gauge theory, related by commutator traces. For the multi-symmetric functions we use in this paper, we will give a defining representation as a multi-trace of commuting matrices, and then \eqref{symmetrised trace isomorphism} defines the equivalent symmetrised trace operator in an unambiguous way. 
	
We will generally use the same notation for either side of the isomorphism. For example we will use $X$ and $Y$ to refer to both the commuting matrices on the right of \eqref{symmetrised trace isomorphism} and the non-commuting matrices of the super Yang-Mills theory on the left. Similarly, both a symmetrised trace and its isomorphic multi-symmetric function will be denoted $T_\bp$. When the distinction is important, we will be clear which is under discussion.

For symmetric functions, we introduced two SEP-compatible bases, the monomial basis of section \ref{section: monomial basis} and the Schur basis of section \ref{section: schur basis}. There is no obvious analogue of the Schur basis for multi-symmetric functions, however the monomial basis does generalise, and provides a good finite $N$ description for multi-symmetric functions. We denote these monomials by $M_\bp$, where the label is a vector partition $\bp$, as already seen for symmetrised traces in \eqref{symmetrised trace operator}. The length of $\bp$ determines the finite $N$ behaviour. In section \ref{sec:MSSP} we study the basis change between $T_\bp$ and $M_\bp$, both as multi-symmetric functions and their isomorphic image as symmetrised trace operators.

Under the map  \eqref{symmetrised trace isomorphism}, the $M_\bp$ give a basis for pre-BPS operators for $N \geq n$. When $N < n$, the operators with $l(\bp) > N$ reduce to commutator traces. As discussed in section \ref{sec:BCOV}, this is a feature of an SEP-compatible basis of pre-BPS operators. However, the operators with $l(\bp) \leq N$ are not $S_n$ orthogonal to all commutator traces, and therefore are not pre-BPS. This is because the map \eqref{symmetrised trace isomorphism} did not choose the correct pre-BPS operator from the equivalence class of operators related by addition of a commutator trace. We say $M_\bp$ is SEP-compatible \textit{modulo commutators} for the pre-BPS sector, and this is a key stepping stone to an SEP-compatible basis.

In section \ref{sec:Counting} we organise the $M_\bp$ according to  representations of the $U(2)$ symmetry, replacing the label $ \bp $ with  $ ( \Lambda , M_{ \Lambda} , p , \nu  ) $.  $\Lambda $ is a $U(2)$ Young diagram with $n$ boxes, where $n$ is the total number of $ X , Y $ matrices in the operator. $M_\Lambda$ labels a basis state in the $\Lambda$ representation of $U(2)$. $p$ is an integer partition of $n$ whose components are related to the vector partition $\bp$ simply by summing each of the two-vector components of $\bp$. We call $p$ the associated partition of $\bp$. Since $l(\bp) = l(p)$, the SEP-compatibility (modulo commutators) of $M_\bp$ is transferred to the covariant basis. This restricts $l(p) \leq N$, which is the usual constraint associated with a $U(N)$ Young diagram, and we will therefore refer to $p$ as a $U(N)$ Young diagram label. The final label $\nu$ runs over a multiplicity space of dimension  $\cM_{\Lambda, p}$. Much of section \ref{sec:Counting} is devoted to calculating and understanding $\cM_{\Lambda, p}$ as it describes the finite $N$ combinatorics of the weak coupling quarter-BPS sector.

Section \ref{sec:Construction} takes the covariant monomials $M_{\Lambda, M_\Lambda, p, \nu}$ and uses orthogonalisation algorithms to produce an SEP-compatible basis of pre-BPS operators. There are three  separate  steps in producing an orthogonal SEP-compatible basis of BPS operators.

\begin{enumerate}
	
	\item  For $l(p) \leq N < n$, the covariant monomial $M_{\Lambda, M_\Lambda, p, \nu}$ differs from a pre-BPS operator by a commutator trace. Orthogonalising $M_{\Lambda, M_\Lambda, p, \nu}$ against $M_{\Lambda, M_\Lambda, q, \eta}$ with $l(q) > N$ using the $S_n$ inner product identifies  pre-BPS operators denoted $ \bar M_{ \Lambda , M_{ \Lambda} , q , \eta } $.  If $N \geq n$, this step is trivial. 
	
	\item   Applying  the operator $ \cG_N$ to the pre-BPS operators produces BPS  operators. We orthogonalize these BPS operators using the physical $ \cF_N$ inner product.  
	
	\item   We normalize these orthogonal operators using the  $S_n$ inner product. This ensures that the basis is SEP-compatible: if we apply the construction using $ \cG_N$ and matrices of size $N$, and subsequently substitute in our expressions matrices of size $ \hN$ while making substitutions $ N \rightarrow \hN$, then all operators with $ l(p) > \hN$ vanish and the non-zero operators with $ l(p) \le \hN$ are those produced by applying the 3-step construction directly at $\hN$.  
	
\end{enumerate} 
The first step is explained in detail in section \ref{sec:Projectors} in a more general context where the $2$-matrix problem is generalized to allow any number of matrices. Section \ref{sec:Construction} puts together all the steps and  proves that the outcome  $S^{BPS}_{\Lambda, M_\Lambda, p, \nu}$  indeed form an orthogonal SEP-compatible basis for BPS operators.

\begin{figure}
	\begin{center}
		\begin{tikzpicture}
		\node[draw, align=center] (T) at (0,9) {Symmetrised traces \\ $T_\bq$};
		\node[draw, align=center] (M) at (0,6) {Monomials \\ $M_\bp$};
		\node[draw, align=center] (M2) at (0,3) {Covariant monomials \\ $M_{\Lambda, M_\Lambda, p, \nu}$};
		\node[draw, align=center] (Mbar) at (0,0) {Orthogonalised \\ covariant monomials \\ $\bar{M}_{\Lambda, M_\Lambda, p, \nu}$};
		\node[draw, align=center] (S) at (0,-3) {$\cG$-orthogonal SEP-compatible \\ pre-BPS operators \\ $S_{\Lambda, M_\Lambda, p, \nu}$};
		\node[draw, align=center] (TBPS) at (7.5,9) {BPS symmetrised traces \\ $T^{BPS}_\bq$};
		\node[draw, align=center] (MBPS) at (7.5,6) { BPS monomials \\ $M^{BPS}_{\bp}$};
		\node[draw, align=center] (M2BPS) at (7.5,3) { BPS covariant monomials \\ $M^{BPS}_{\Lambda, M_\Lambda, p, \nu}$};
		\node[draw, align=center] (MbarBPS) at (7.5,0) { Orthogonalised BPS \\ covariant monomials \\ $\bar{M}^{BPS}_{\Lambda, M_\Lambda, p, \nu}$};
		\node[draw, align=center] (BPS) at (7.5,-3) {Orthogonal SEP-compatible \\ BPS operators \\ $S^{BPS}_{\Lambda, M_\Lambda, p, \nu}$};

		\node[align=center, left] (C) at (-0.25,7.5) {Multiply by \\ matrix $\tC^\bq_\bp$.};
		\node[align=center, left] (U2) at (-0.25,4.5) {Sort into $U(2)$ \\ representations.};
		\node[align=center, left] (o) at (-0.25,1.5) {$S_n$ orthogonalise.};
		\node[align=center, left] (o) at (-0.25,-1.5) {$\cG$ orthogonalise.};
		\node[align=center, right] (C2) at (7.75,7.5) {Multiply by \\ matrix $\tC^\bq_\bp$.};
		\node[align=center, right] (U22) at (7.75,4.5) {Sort into $U(2)$ \\ representations.};
		\node[align=center, right] (o2) at (7.75,1.5) {$S_n$ orthogonalise.};
		\node[align=center, right] (o2) at (7.75,-1.5) {$\cF$ orthogonalise.};
		\node[align=center, above] (G) at (3.5,9) {Apply $\cG_N$ \\ and normalise.};
		\node[align=center, above] (G) at (3.5,6) {Apply $\cG_N$ \\ and normalise.};
		\node[align=center, above] (G) at (3.5,3) {Apply $\cG_N$ \\ and normalise.};
		\node[align=center, above] (G) at (3.5,0) {Apply $\cG_N$ \\ and normalise.};
		\node[align=center, below] (G) at (3.5,-4) {Apply $\cG_N$ \\ and normalise.};
		
		\draw[->,red] (T) edge (M) (M) edge (M2) (M2) edge (Mbar) (Mbar) edge (MbarBPS) (MbarBPS) edge (BPS);
		\draw[->] (T) edge (TBPS) (TBPS) edge (MBPS) (MBPS) edge (M2BPS) (M2BPS) edge (MbarBPS);
		\draw[->] (M) edge (MBPS);
		\draw[->] (M2) edge (M2BPS);
		\draw[->] (Mbar) edge (S) (S) edge (BPS);
		\end{tikzpicture}
	\end{center}
	\caption{An outline of the algorithm presented in this paper starting with symmetrised trace operators $T_\bp$ and deriving a $U(2)$ covariant, orthogonal, SEP-compatible basis $S^{BPS}_{\Lambda, M_\Lambda, p, \nu}$ for BPS operators. The route taken in this paper is down the left side and across the bottom, shown in red. The first step is studied in detail in section \ref{sec:MSSP}, the second step in section \ref{sec:Counting} and the last three steps in section \ref{sec:Construction}.}
	\label{figure: main algorithm}
\end{figure}
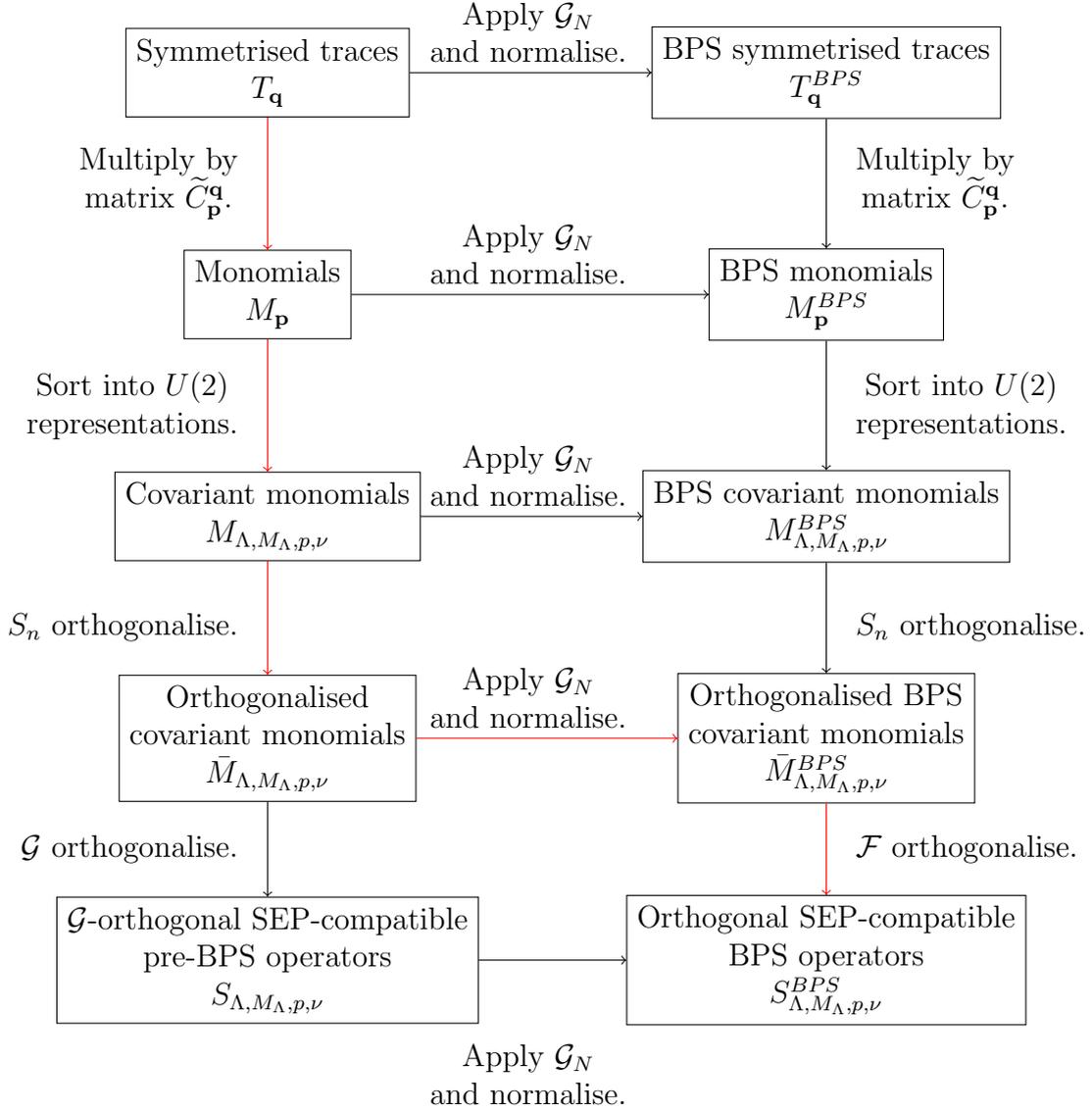

The orthogonalisation and $ \cG_N$ application processes involved in the construction are linear, so there is some flexibility in the order of the application of the different steps.  Figure  \ref{figure: main algorithm} shows the algorithm we have implemented in SAGE  to obtain the basis $S^{BPS}_{\Lambda, M_\Lambda, p, \nu}$ starting from symmetrised traces $T_\bp$. The red arrows indicate the route taken in this paper, while the other arrows indicate different routes where $\cG_N$ is applied at a different stage.

The operators $T^{BPS}_\bp$, $M^{BPS}_\bp$ and $M^{BPS}_{\Lambda, M_\Lambda, p, \nu}$, obtained by applying $\cG_N$ (and normalising) to $T_\bp$, $M_\bp$ or $M_{\Lambda, M_\Lambda, p, \nu}$ respectively, are BPS bases at $N \geq n$, but in general for $N < n$ are no longer BPS, although the latter two do capture some of the finite $N$ behaviour. Table \ref{table: BPS bases} shows the properties of the different BPS bases.

{\renewcommand{\arraystretch}{1.5}
	\begin{table}
		\begin{center}
			\begin{tabular}{c|c|c|c|c}
				& SEP-compatible & $U(2)$ covariant & BPS at $N<n$ & Orthogonal \\ \hline
				$T^{BPS}_\bp$ & $\times$ & $\times$ & $\times$ & $\times$ \\ \hline
				$M^{BPS}_\bp$ & modulo commutators & $\times$ & $\times$ & $\times$  \\ \hline
				$M^{BPS}_{\Lambda, M_\Lambda, p, \nu}$ & modulo commutators & $\checkmark$ & $\times$ & $\times$ \\ \hline
				$\bar{M}^{BPS}_{\Lambda, M_\Lambda, p, \nu}$ & modulo commutators & $\checkmark$ & $\checkmark$ & $\times$ \\ \hline
				$S^{BPS}_{\Lambda, M_\Lambda, p, \nu}$ & $\checkmark$ & $\checkmark$ & $\checkmark$ & $\checkmark$
			\end{tabular}
		\end{center}
		\caption{Properties of the different BPS bases constructed. All are BPS at $N \geq n$.}
		\label{table: BPS bases}
	\end{table}
}

In the case where $\Lambda $ is taken to be $ [n]$, the $S^{BPS}_{\Lambda, M_\Lambda, p, \nu}$ are half-BPS operators built purely from $X$.  In  section \ref{section: lambda=[n],[n-1,1]} we show that the construction for Young diagram $p$  reproduces the Schur operator \eqref{schur definition from permutations} labelled by  Young diagram $R = p$. Further, for $\Lambda = [n-1,1]$, numerical calculations suggest that $S^{BPS}_{\Lambda, M_\Lambda, p, \nu}$ match the free field quarter-BPS operators \eqref{U(2) basis definition} with $R=p$. This justifies the view that $p$ is a $U(2)$ invariant  generalization of the $R$-label of the half-BPS sector to the quarter-BPS sector. 

An important perspective on the half-BPS $R$ label comes from the analysis of the asymptotics of LLM geometries. Specifically $U(N)$ Casimirs of $R$ are measurable from the asymptotics of the supergravity fields \cite{IILoss}.  We propose that that the $U(2)$  quadratic Casimir for $ \Lambda $ as well as the  sequence of Casimirs identifying $p$ should be measurable from the multipole moments that can be read off from the long-distance expansions of the sugra fields  of LLM geometries corresponding to quarter-BPS operators at $ n \sim N^2 $. Precision holography of LLM geometries is also developed using correlators of small operators in the LLM background \cite{SkenTay} which should provide complementary insights into the labels  $\Lambda , p$. 

The $R$ Young diagram label of the half BPS sector is related to the free fermions underlying the holomorphic sector of the complex matrix model which describes half-BPS combinatorics \cite{CJR2001,Berenstein,TakTsu}. Remarkably, this free fermion description also shows up in the droplet description of half-BPS LLM geometries \cite{LLM}. The droplet description generalizes to the quarter-BPS LLM geometries, with some significant differences \cite{Donos06,CCD0704,Lunin0802} . We have colourings of regions in a four-dimensional space instead of a two-dimensional plane. The two colours are now associated with the collapse of an $S^1 \subset S^5$ or an $S^3\subset S^5$. A natural conjecture is  that the $p$-label of quarter-BPS operators is analogously associated to colourings of regions in $\mR^4$ as the $R$-label of the half-BPS operators is associated to colourings of the plane.

\section{Finite N combinatorics from many-boson states: multi-symmetric functions and set partitions}
\label{sec:MSSP}

As explained in section \ref{section: weak coupling}, the key result that will enable us to give an SEP-compatible construction of quarter-BPS operators is an isomorphism of Vaccarino and Domokos \cite{Vaccarino2007,Domokos} between multi-symmetric functions and the ring of gauge invariants of two matrices modulo commutator traces. This section focuses on the multi-symmetric function side of this isomorphism.

An important aspect of multi-symmetric functions, which plays a central role in this paper, is the  transformation between two bases for these multi-symmetric functions. The first basis will be referred to as the ``monomial multi-symmetric basis'' and the second as the ``multi-trace  basis''. The physical importance of these two bases, and their transformations, can be understood using perspectives from many-body quantum mechanics \cite{MBtextbook}. This draws on an important insight from the  AdS/CFT correspondence: the strong coupling limit of the quarter BPS sector in $ \cN=4$ SYM corresponds to a Hilbert space of $N$ bosons in a two-dimensional harmonic oscillator \cite{BGLM06,MS06,Ber0507}.   

We begin this section by developing the link between multi-symmetric functions and the Hilbert space of $N$ identical bosons in a two-dimensional harmonic oscillator. We then introduce the monomial and multi-trace bases and investigate the combinatorics of the matrix that transforms between the two. This leads to the interesting mathematics of the poset of set partitions and the associated M\"obius function.

\subsection{Multi-symmetric functions as wavefunctions of a harmonic oscillator} 

The Lagrangian for one particle in a two-dimensional harmonic oscillator is 
\begin{equation}
L = { 1 \over 2 } \left( {\dot x}^2 + { \dot y}^2 \right) - { 1 \over 2 } \left( x^2 + y^2 \right)  
\end{equation}
In terms of creation and annihilation operators, the one-particle Hamiltonian is 
\begin{equation} 
H = a_x^{ \dagger} a_x + a_y^{\dagger} a_y 
\end{equation}
Define the coherent state 
\begin{equation} 
\langle  x , y | = \langle 0 |  e^{ x a_x  + y a_y } 
\end{equation}
We have 
\begin{equation} 
\langle x , y | (a_x^{ \dagger} )^{ \lambda } ( a_y^{ \dagger} )^{ \mu} | 0 \rangle 
 = x^{ \lambda } y^{ \mu} 
\end{equation}
In this coherent state representation, the Hamiltonian acts as the degree operator for the 2-variable  polynomial 
\begin{align}
\langle x , y | H (a_x^{ \dagger} )^{ \lambda } ( a_y^{ \dagger} )^{ \mu} | 0 \rangle & = 
\left ( x { \partial \over \partial x } + y { \partial \over \partial y }  \right ) 
\langle x , y | (a_x^{ \dagger} )^{ \lambda } ( a_y^{ \dagger} )^{ \mu} | 0 \rangle  \nonumber \\
& = \left ( x { \partial \over \partial x } + y { \partial \over \partial y }  \right )  ( x^{ \lambda } y^{ \mu} ) \nonumber \\
&  = ( \lambda + \mu ) ( x^{ \lambda } y^{ \mu} )
\end{align}
For the system of  $N$-particles in the two-dimensional harmonic oscillator, we have the coherent state
\begin{equation} 
\langle  x_1 , y_1 ; x_2 , y_2 ; \dots ;  x_N , y_N |
= \langle 0 | e^{ \sum_{ i=1}^N x_i a_{i ; x}  + y_i a_{i ; y} }   
\end{equation}
The energy eigenstates of the Hamiltonian correspond to the product of one-particle wavefunctions 
\begin{align}
& \langle  x_1 , y_1 ; x_2 , y_2 ; \cdots ;  x_N , y_N |
( a_{1; x}^{\dagger} )^{ \lambda_1}  ( a_{1; y}^{\dagger}  )^{ \mu_1}
( a_{2; x}^{\dagger} )^{ \lambda_2}  ( a_{2; y}^{ \dagger}  )^{ \mu_2}
\dots 
( a_{k; x}^{ \dagger} )^{ \lambda_k}  ( a_{k; y}^{\dagger}  )^{ \mu_k} | 0 \rangle \nonumber \\ 
& \hspace{100pt} =  x_1^{ \lambda_1} y_1^{ \mu_1}  \cdots  x_k^{ \lambda_k} y_k^{ \mu_k}
\end{align}
It is useful to write 
\begin{align} 
x_1^{ \lambda_1} y_1^{ \mu_1}  \dots  x_k^{ \lambda_k} y_k^{ \mu_k}
& = \psi_{ \lambda_1 , \mu_1 } ( x_1 , y_1 )  \psi_{ \lambda_2 , \mu_2 } ( x_2 , y_2 ) \cdots 
\psi_{ \lambda_k , \mu_k } ( x_k , y_k  ) \nonumber \\
& =  \psi_{ \lambda_1 , \mu_1 } (  x_1 , y_1 ) \dots \psi_{ \lambda_k , \mu_k } ( x_k , y_k ) 
 \psi_{ 0 , 0 } ( x_{ k +1} , y_{ k +1 } ) \dots \psi_{ 0  , 0 } ( x_N , y_N ) 
\end{align}
In a system of $N$ identical bosons, we must symmetrise the product of annihilation operators using $S_N$ permutations. The product wavefunctions and their symmetrisations are a standard tool 
in many-body quantum mechanics (see. e.g. \cite{MBtextbook}). The permutations $ \sigma \in S_N$ act as 
\begin{equation}
( x_{ i } , y_i  ) \rightarrow  ( x_{ \sigma (i) } , y_{ \sigma (i) } ) 
\end{equation}
These states are polynomials, symmetric  under these simultaneous permutations of  $ x , y $ pairs, which are exactly multi-symmetric functions. In fact these form the monomial multi-symmetric functions that we will study presently. They have the nice property that finite $N$ effects are nicely encoded  in the fact that, by definition,  $ k \le N$.

A quantum state where a single particle is excited, after symmetrisation, has a coherent state representation 
\bea\label{1particleWfn}  
\phi_{ \lambda_1 , \mu_1} ( x_i , y_i )  = \sum_{ i =1 }^N  x_i^{ \lambda_1} y_i^{ \mu_1} 
\eea
When we have two particles excited, the symmetrisation of the product wavefunction is proportional to 
\bea\label{2particleWfn}  
\sum_{ i_1 \ne i_2  }^N  x_{i_1}^{ \lambda_1} y_{i_1}^{ \mu_1} x_{i_2}^{ \lambda_2} y_{i_2}^{ \mu_2}
\eea
The restriction $ i_1 \ne i_2$, when extended to $ i_1 \ne i_2 \cdots \ne i_k $,   is closely related to the finite $N $ property, but also has the consequence that the 2-particle wavefunction \ref{2particleWfn} is not equal to the product of 1-particle wavefunctions. It is rather a linear combination of the product wavefunction $\phi_{\lambda_1, \mu_1} \phi_{\lambda_2, \mu_2}$ along with a 1-particle wavefunction $\phi_{\lambda_1 + \lambda_2, \mu_1 + \mu_2}$. 

Defining  diagonal matrices $ X = \Diag ( x_1 , \cdots , x_N )$ and  $Y = \Diag ( y_1 , y_2 , \cdots , y_N )$,  we observe that the 1-particle wavefunction is  a trace 
\bea 
\phi_{ \lambda_1  , \mu_1} ( x_i , y_i ) = \tr ( X^{ \lambda_1} Y^{ \mu_1} ) 
\eea
We now draw on an idea from collective field theory, where one associates creation operators to invariant traces \cite{JS,DJ,AMOS9411} to define a map from traces and products of traces to Fock space states 
\begin{align}
\tr ( X^{ \lambda}  Y^{ \mu}  )  & \rightarrow B^{ \dagger}_{ \lambda   , \mu  } |0 \rangle  \\
\prod_{ a =1}^k  \tr ( X^{ \lambda_a } Y^{ \mu_a } )  & \rightarrow \prod_{ a =1}^k  B^{ \dagger}_{ \lambda_a  , \mu_a } |0 \rangle 
\end{align}
This map is a homomorphism between the product structure on the polynomials and the product structure on
oscillators. It can also be obtained from a coherent state construction 
\begin{align} 
& \langle X , Y | = \langle 0 | e^{ \sum_{ \lambda , \mu   } B_{ \lambda , \mu  } 
\tr (X^{ \lambda}  Y^{ \mu}  ) }  \nonumber \\ 
& \langle X , Y | \prod_{ a=1 }^k  B^{ \dagger}_{ \lambda_a , \mu_a }  | 0 \rangle 
=  \prod_{ a=1}^k \tr  ( X^{ \lambda_a } Y^{ \mu_a } ) 
\end{align} 
In section \ref{section: C combinatorics} we will be studying in detail the transformation between the monomial multi-symmetric functions and the trace wavefunctions. As a result of a triangular property of this transformation, the finite $N$ cutoff on multi-symmetric functions can also be described by restricting the number of factors in the product of traces to be less than or equal to $N$.

\subsection{Monomial and trace bases for multi-symmetric functions } 
\label{monomial-and-trace}

We now give a formal definition of multi-symmetric functions in a completely analogous way to the symmetric functions of section \ref{section: half-bps}, just with two families of variables $x_1, \dots, x_N$ and $y_1, \dots, y_N$ instead of one. They are polynomials in these $2N$ variables that are invariant under all $S_N$ permutations on the pairs $(x_i, y_i)$. Given a polynomial $f(x_1, x_2, \dots, x_N; y_1, y_2, \dots, y_N)$, $f$ is a multi-symmetric function if
\begin{equation}
f \left( x_1, x_2, \dots, x_N; y_1, y_2, \dots, y_N \right) = f \left( x_{\sigma(1)}, x_{\sigma(2)}, \dots, x_{\sigma(N)}; y_{\sigma(1)}, y_{\sigma(2)}, \dots, y_{\sigma(N)} \right)
\label{multi-sym permutation}
\end{equation}
for all $\sigma \in S_N$.

Similarly to section \ref{section: half-bps}, we can take a large $N$ limit and work with formal power series rather than polynomials. We return to the finite $N$ case by setting $x_{N+1} = y_{N+1} = x_{N+2} = y_{N+2} = \dots = 0$.

We can also define multi-symmetric functions with $M$ families of variables $x^{(k)}_i$, for $1 \leq k \leq M$, $1 \leq i \leq N$, invariant under $S_N$ permutations of the $i$ index. These would be relevant for systems of $M$ commuting matrices.

For a mathematical overview of multi-symmetric functions and their properties see \cite{multi-sym functions}.

The monomial and multi-trace (power-sum) bases for symmetric polynomials defined in \eqref{M definition} and \eqref{multi-component power-sum} have direct analogues in the multi-symmetric case. As before, they are graded bases, this time graded by both the $x$ degree $n_1$ and the $y$ degree $n_2$.

A vector partition $\bp$ of $(n_1,n_2)$ is defined to be a sequence of pairs of non-negative integers (at least one of each pair must be non-zero) summing to $(n_1, n_2)$. We use a bold $\bp$ to distinguish between vector and integer partitions, and write $\bp \vdash (n_1,n_2)$ to denote that $\bp$ sums to $(n_1, n_2)$. The basis elements at degree $(n_1, n_2)$ are labelled by $\bp \vdash (n_1,n_2)$ with length $l(\bp) \leq N$.

To construct the monomial basis, take a vector partition $\bp = [(\lambda_1, \mu_1), (\lambda_2, \mu_2), $ $ \dots, (\lambda_k, \mu_k)]$ of $(n_1, n_2)$ with $l( \bp ) = k \leq N$ and consider the un-symmetrised monomial
\begin{equation}
x_1^{\lambda_1} y_1^{\mu_1} x_2^{\lambda_2} y_2^{\mu_2} \dots x_k^{\lambda_k} y_k^{\mu_k}
\end{equation}
After adding all distinct permutations of the lower indices, one arrives at the monomial basis element. Explicitly
\begin{equation}
m_\bp = \frac{1}{Z_\bp} \sum_{\sigma \in S_N} x_{\sigma(1)}^{\lambda_1} y_{\sigma(1)}^{\mu_1} x_{\sigma(2)}^{\lambda_2} y_{\sigma(2)}^{\mu_2} \dots x_{\sigma(k)}^{\lambda_k} y_{\sigma(k)}^{\mu_k}
\end{equation}
where the factor in front removes the normalisation introduced by redundancies in the elements of $\bp$, so that the coefficient in front of each individual monomial is 1. Using multiplicity notation analogous to
\eqref{partition from multiplicities}, let $ \bp = < \bp_{(0,1)} , \bp_{(1,0)}, \dots >$. Then the normalisation is given by 
\begin{equation}
Z_\bp = \prod_{i,j} \bp_{(i,j)}!
\end{equation}
As in the symmetric case, we will use a modified version of the monomial basis, obtained by leaving out this normalisation factor
\begin{equation}
M_\bp = Z_\bp m_\bp = \sum_{\sigma \in S_N} x_{\sigma(1)}^{\lambda_1} y_{\sigma(1)}^{\mu_1} x_{\sigma(2)}^{\lambda_2} y_{\sigma(2)}^{\mu_2} \dots x_{\sigma(k)}^{\lambda_k} y_{\sigma(k)}^{\mu_k}
\label{monomial definition}
\end{equation}
As discussed below \eqref{multi-sym permutation}, we can lower $N$ to $N-1$ by setting $x_N = y_N = 0$, causing a reduction in the size of the space. Starting from $N > n_1 + n_2$ and reducing stepwise, this implies those monomial functions with $l(\bp) > N$ vanish identically, while the remaining $M_\bp$ with $l(\bp) \leq N$ form a basis for the smaller space. So the monomial basis is SEP-compatible for multi-symmetric functions.

Note that the the isomorphism, as described around \eqref{near isomorphism description}, states that multi-symmetric functions are isomorphic to invariants of matrices $X,Y$ modulo commutator traces. Therefore the isomorphic image of $M_\bp$, also referred to as $M_\bp$, is not necessarily zero if $l(\bp) > N$, but could instead be a commutator trace. This is the version of SEP-compatibility for pre-BPS operators as discussed in section \ref{section: weak coupling}.

As multi-symmetric functions, the monomial functions $M_\bp$ are SEP-compatible. Using the map \eqref{symmetrised trace isomorphism} to give the equivalent symmetrised trace operators, they form a basis for pre-BPS operators at $N \geq n$. As we decrease $N < n$, the SEP-compatibility implies any operator with $l(\bp) > N$ reduces to a commutator trace. However, the operators with $l(\bp) \leq N$ are not in general $S_n$ orthogonal to commutator traces when $N < n$, and therefore they do not form a basis for pre-BPS operators. This is due to \eqref{symmetrised trace isomorphism} not selecting the right representative of the equivalence class of operators isomorphic to the multi-symmetric function, as discussed below \eqref{near isomorphism description}. We say the Yang-Mills operators $M_\bp$ are SEP-compatible modulo commutators, and in section \ref{sec:Construction}, describe how to transform this into a genuinely SEP-compatible basis for pre-BPS operators.

The multi-trace basis for multi-symmetric functions, also called the power-sum basis in the mathematics literature, is built out of
\begin{equation}
T_{(n_1, n_2)} = \sum_{i=1}^N x_i^{n_1} y_i^{n_2}
\end{equation}
Given a vector partition $\bp = [(\lambda_1, \mu_1), (\lambda_2, \mu_2), \dots (\lambda_k, \mu_k)]$, the associated multi-symmetric function is
\begin{equation}
T_\bp = \prod_{i=1}^k T_{(\lambda_i, \mu_i)}
\label{multi-symmetric multi-trace definition}
\end{equation}
Introduce two $N \times N$ diagonal matrices $X$ and $Y$ with diagonal elements $x_1, x_2, \dots x_N$ and $y_1, y_2, \dots y_N$ respectively. Then $T_{(n_1,n_2)} = \tr X^{n_1} Y^{n_2}$, and the multi-trace multi-symmetric functions are exactly given by the multi-traces of these two matrices, justifying the name.

The isomorphism of \cite{Vaccarino2007,Domokos2005} identifies the multi-symmetric functions \eqref{multi-symmetric multi-trace definition} with the the symmetrised trace operators \eqref{symmetrised trace operator}, establishing the connection between multi-symmetric functions and the quarter-BPS sector of $\cN = 4$ super Yang-Mills at weak coupling. 

Note that while \eqref{multi-symmetric multi-trace definition} and \eqref{symmetrised trace operator} are conceptually different, the isomorphism between the two means we abuse notation slightly and use the same symbol $T_\bp$ for both.

At finite $N$, non-trivial relationships appear between the different multi-traces leading to a reduction in the dimensionality of the space of multi-symmetric functions. Those multi-traces labelled by $\bp$ with $l(\bp) \leq N$ form a basis for the reduced space. However, unlike the monomial functions, the remaining multi-traces (labelled by $\bp$ with $l(\bp) > N$) do not vanish, but become complicated linear combinations of the reduced basis.

Define a matrix $C^\bp_\bq$, indexed by vector partitions $\bp$ and $\bq$ to be the change of basis matrix from $M_\bp$ to $T_\bq$, with inverse $\widetilde{C}$
\begin{equation}
T_\bq = \sum_\bp C^\bp_\bq M_\bp  \qquad \qquad \qquad  M_\bp = \sum_\bq \widetilde{C}^\bq_\bp T_\bq 
\label{definition of C and C-tilde}
\end{equation}
At finite $N$, the $\bp$ label for monomials is SEP-compatible (modulo commutators). Therefore the second of the equations above gives the finite $N$ relations imposed on commuting matrices. On the other side of the isomorphism, this gives the linear combinations of symmetrised traces that reduce to commutator traces at finite $N$.

The $C$ and $\widetilde{C}$ matrices have very interesting combinatorial properties, which we will now investigate in depth.

\subsection{Basis change for multi-symmetric functions}
\label{section: C combinatorics}

In this section, we show that  the properties of the linear transformations $C$ and $\widetilde{C} $ are illuminated by considering set partitions. Set partitions form a partially ordered set (poset), and the M\"obius inversion formula for posets plays an important role. 

To find an expression for $C^\bp_\bq$, first expand the product in the definition \eqref{multi-symmetric multi-trace definition} for multi-trace functions. For $\bp=[(\lambda_1, \mu_1), \dots, (\lambda_k, \mu_k)]$,
\begin{align}
T_\bp & = \prod_{j=1}^{k} \left( \sum_{i=1}^N x_i^{\lambda_j} y_i^{\mu_j} \right) \\
& = \sum_{j_1, \dots, j_{k} = 1}^N x_{j_1}^{\lambda_1} y_{j_1}^{\mu_1} x_{j_2}^{\lambda_2} y_{j_2}^{\mu_2} \dots x_{j_{k}}^{\lambda_{k}} y_{j_{k}}^{\mu_{k}}
\label{multi-trace sum of monomials}
\end{align}
To further sort this sum, note the different ways the $j$s could coincide. If, for example, $k=3$, we could have
\begin{gather}
j_1 = j_2 = j_3 \qquad \qquad \qquad j_1 = j_2 \neq j_3 \\
j_1 = j_3 \neq j_2 \qquad \qquad \qquad j_1 \neq j_2 = j_3 \\
j_1 , j_2 , j_3 \text{ all distinct}
\end{gather}
These correspond to the 5 different ways of partitioning the set $\{ 1,2,3 \}$ into subsets
\begin{gather}
\pi_1 = \{\{1,2,3\}\} \qquad \qquad \qquad \pi_2 = \{ \{ 1,2 \} , \{ 3 \} \} \\
\pi_3 = \{ \{ 1,3 \} , \{ 2 \} \} \qquad \qquad \qquad \pi_4 = \{ \{ 1 \} , \{ 2,3 \} \} \\
\pi_5 = \{ \{ 1 \} , \{ 2 \} , \{ 3 \} \} 
\end{gather}
Continuing with the example, we can sort the sum \eqref{multi-trace sum of monomials} into the different partitions
\begin{align}
\begin{aligned}
T_\bp & = \qquad \sum_j x_j^{\lambda_1 + \lambda_2 + \lambda_3} y_j ^{\mu_1 + \mu_2 + \mu_3} 
+ \sum_{j_1,j_3 \text{ distinct}} x_{j_1}^{\lambda_1 + \lambda_2} y_{j_1}^{\mu_1 + \mu_2} x_{j_3}^{\lambda_3} y_{j_3}^{\mu_3} \\
& \qquad + \sum_{j_1,j_2 \text{ distinct}} x_{j_1}^{\lambda_1 + \lambda_3} y_{j_1}^{\mu_1 + \mu_3} x_{j_2}^{\lambda_2} y_{j_2}^{\mu_2}
+ \sum_{j_1,j_2 \text{ distinct}} x_{j_1}^{\lambda_1} y_{j_1}^{\mu_1} x_{j_2}^{\lambda_2 + \lambda_3} y_{j_2}^{\mu_2 + \mu_3} \\
& \qquad + \sum_{j_1, j_2, j_3 \text{ distinct}} x_{j_1}^{\lambda_1} y_{j_1}^{\mu_1} x_{j_2}^{\lambda_2} y_{j_2}^{\mu_2} x_{j_3}^{\lambda_3} y_{j_3}^{\mu_3}
\end{aligned}
\label{multi-trace expanded over set partitions}
\end{align}
The first term is just the monomial function associated to the vector partition $\pi_1 (\bp) = [(\lambda_1 + \lambda_2 + \lambda_3, \mu_1 + \mu_2 + \mu_3)]$. Similarly, the second term is related to the monomial function with vector partition $\pi_2(\bp) = [(\lambda_1 + \lambda_2, \mu_1 + \mu_2) , (\lambda_3, \mu_3)]$ via
\begin{equation}
\sum_{j_1,j_2 \text{ distinct}} x_{j_1}^{\lambda_1 + \lambda_2} y_{j_1}^{\mu_1 + \mu_2} x_{j_2}^{\lambda_3} y_{j_2}^{\mu_3} = 
\begin{cases}
2 m_{\pi_2(p)} & (\lambda_1 + \lambda_2, \mu_1 + \mu_2) = (\lambda_3, \mu_3) \\
m_{\pi_2(p)} & \text{otherwise}
\end{cases}
\end{equation}
We can simplify this expression by noting that if $(\lambda_1 + \lambda_2, \mu_1 + \mu_2) = (\lambda_3, \mu_3)$ then $Z_{\pi_2(\bp)} = 2$, and otherwise $Z_{\pi_2 (\bp)} = 1$. Therefore
\begin{equation}
\sum_{j_1,j_2 \text{ distinct}} x_{j_1}^{\lambda_1 + \lambda_2} y_{j_1}^{\mu_1 + \mu_2} x_{j_2}^{\lambda_3} y_{j_2}^{\mu_3} = Z_{\pi_2(\bp)} m_{\pi_2(\bp)} = M_{\pi_2(\bp)}
\end{equation}
Similarly the third, fourth and fifth terms of \eqref{multi-trace expanded over set partitions} are just $M_{\pi_3(\bp)}, M_{\pi_4(\bp)}$ and $M_{\pi_5(\bp)}$ respectively, where
\begin{align}
\pi_3(\bp) & = [(\lambda_1 + \lambda_3, \mu_1 + \mu_3) , (\lambda_2, \mu_2)] \\
\pi_4(\bp) & = [(\lambda_1, \mu_1), (\lambda_2 + \lambda_3, \mu_2 + \mu_3)] \\
\pi_5(\bp) & = \bp
\end{align}
Putting this together, we have
\begin{equation}
T_\bp = \sum_{i=1}^5 M_{\pi_i(\bp)}
\label{multi-trace expansion example}
\end{equation}
Repeating this analysis more generally, let the set of set partitions of $\{ 1, 2, 3, \dots k \}$ be denoted by $\Pi(k)$. Then given a set partition $\pi \in \Pi(k)$ and a vector partition $\bp = [(\lambda_1, \mu_1), \dots, (\lambda_k, \mu_k)]$ of length $l(\bp) = k$, we define the vector partition $\pi(k)$ to be that with components
\begin{equation}
\left( \sum_{i \in b} \lambda_i \quad , \quad \sum_{i \in b} \mu_i \right)
\label{set partition on vector partition}
\end{equation}
where the blocks $b \in \pi$ run over the subsets into which $\{ 1,2,3, \dots k \}$ have been partitioned. Conceptually, this should be thought of as summing up $\bp$ into a new, shorter vector partition, where the summation structure is given by $\pi$.

Given this notation, we can now write the generalisation of \eqref{multi-trace expansion example} to any $k$
\begin{equation}
T_\bp = \sum_{\pi \in \Pi(l(\bp))} M_{\pi(\bp)}
\label{multi-symmetric traces from monomials}
\end{equation}
Proving this result in the general case is just an exercise in repeating the logic that led from \eqref{multi-trace sum of monomials} to \eqref{multi-trace expansion example}.

So the coefficient of $M_\bp$ in $T_\bq$ is just the number of set partitions $\pi \in \Pi(l(\bq))$ that have $\pi(\bq)=\bp$.
\begin{equation}
\boxed{
C^\bp_\bq = \sum_{\pi \in \Pi(l(\bq))} \delta_{\bp \ \pi(\bq)}
}
\label{c}
\end{equation}
We can see that for vector partitions of a particular length $k = l(\bq)$, it is the set partitions of $\{ 1,2,3, \dots, k\}$ that control the behaviour. 

The poset (partially ordered set) structure of set of set partitions is well studied \cite{StanleyEnum}, and will help further explain the structure of the matrix $C$ and its inverse $\tC$. The partial ordering is defined by saying that one set partition, $\pi$, is less than another, $\pi'$, if every block $b \in \pi$ is contained within some block $b' \in \pi'$. We call $\pi$ a refinement of $\pi'$ or $\pi'$ a coarsening of $\pi$. 

Intuitively, if $\pi < \pi'$, then the blocks of $\pi$ are smaller in size than those in $\pi'$. However, this means that there are more blocks in $\pi$ than in $\pi'$, so confusingly $\pi < \pi'$ implies that $|\pi| > |\pi'|$.

Now instead of looking at $T_\bp$, we look at $T_{\pi(\bp)}$, for some $\pi \in \Pi(l(\bp))$. Clearly we can still use the formula \eqref{multi-symmetric traces from monomials} just by replacing $\bp$ with $\pi(\bp)$. Then summing over $\pi' \in \Pi(l(\pi(\bp)))$ with summand $M_{\pi'(\pi(\bp))}$ is equivalent to summing over all coarsenings $\pi'' \geq \pi$ with summand $M_{\pi''(\bp)}$, so we can write
\begin{equation}
T_{\pi(\bp)} = \sum_{\pi' \geq \pi} M_{\pi'(\bp)}
\end{equation}
Considering $T_\pi$ and $M_\pi$ as functions from vector partitions to multi-symmetric functions, we have
\begin{equation}
T_\pi = \sum_{\pi' \geq \pi} M_{\pi'}
\label{pre-Mobius inversion}
\end{equation}
Equations like \eqref{pre-Mobius inversion} are standard the theory of posets \cite{StanleyEnum}, and can be inverted using the M\"obius inversion formula \eqref{mobius inversion formula}. We explain this formula in more detail in section \ref{section: mobius function}, including a combinatoric interpretation that allows a simple explanation of the inversion property.

In this case, the M\"obius inversion formula implies
\begin{equation}
M_\pi = \sum_{\pi' \geq \pi} \mu (\pi, \pi') T_{\pi'}
\label{post-Mobius inversion}
\end{equation}
where the M\"obius function $\mu (\pi, \pi')$ is defined in \eqref{mobius function}.

Choosing a vector partition $\bp$ on which to act, we have
\begin{equation}
M_{\pi(\bp)} = \sum_{\pi' \geq \pi} \mu ( \pi, \pi' ) T_{\pi'(\bp)}
\label{M in terms of T}
\end{equation}
We can now use this to find an explicit expression for $\tC^\bq_\bp$. Let $k = l(\bp)$ and $\pi = \pi_k$ to be the minimal set partition in $\Pi(k)$, in which every element has its own block so that $\pi_k (\bp) = \bp$. Applying \eqref{M in terms of T} gives
\begin{equation}
M_{\bp} = \sum_{\pi \in \Pi(k)} \mu(\pi_k, \pi) T_{\pi(\bp)}
\label{M from T}
\end{equation}
and therefore
\begin{equation}
\boxed{
\widetilde{C}^\bp_\bq = \sum_{\pi \in \Pi(k)} \mu(\pi_k, \pi) \ \delta_{\bp \ \pi(\bq)}
}
\label{ctilde}
\end{equation}

\subsection{M\"obius function for the poset of set partitions and combinatoric interpretation}
\label{section: mobius function}

In this section we introduce the M\"obius function for a general poset, and give its value on the poset of set partitions. There is a combinatoric interpretation for the M\"obius function in terms of permutations on the blocks of the set partitions, and this interpretation allows us to simply see why the M\"obius inversion formula works in this case.

The M\"obius function is defined recursively for a generic poset by
\begin{equation}
\mu(\pi, \pi') = 
\begin{cases}
1 & \pi = \pi' \\
- \sum\limits_{\pi \leq \pi'' < \pi'} \mu (\pi, \pi'') & \pi < \pi' \\
0 & \text{otherwise}
\end{cases}
\label{general mobius function}
\end{equation}
The key utility of this definition is in the M\"obius inversion formula, which states that given two functions $f,g$ from a 
poset into a vector space, the following two relations are equivalent
\begin{align}
\begin{aligned}
f(\pi) & = \sum_{\pi' \geq \pi} g(\pi') \\
g(\pi) & = \sum_{\pi' \geq \pi} \mu (\pi, \pi') f(\pi')
\end{aligned}
\label{mobius inversion formula}
\end{align}
In order to give an explicit expression for the M\"obius function on set partitions, consider $\pi = \{ b_1, b_2, \dots , b_k \}$ for $k = |\pi|$. We then look at the set partitions of $\pi$ itself. For example if $k = 3$ the five possible set partitions of $\pi$ are
\begin{gather}
\begin{gathered}
\rho_1 = \{\{b_1,b_2,b_3\}\} \qquad \qquad \qquad \rho_2 = \{ \{ b_1,b_2 \} , \{ b_3 \} \} \\
\rho_3 = \{ \{ b_1,b_3 \} , \{ b_2 \} \} \qquad \qquad \qquad \rho_4 = \{ \{ b_1 \} , \{ b_2,b_3 \} \} \\
\rho_5 = \{ \{ b_1 \} , \{ b_2 \} , \{ b_3 \} \} 
\end{gathered}
\label{set partitions}
\end{gather}
The set of set partitions of $\pi$ is denoted by $\Pi(\pi)$, and there is an obvious correspondence between this and $\Pi(|\pi|)$. For any particular $\pi \in \Pi(n)$ and $\rho \in \Pi(\pi)$, we define $\rho (\pi)$ to be the following set partition in $\Pi(n)$.
\begin{equation}
\left\{ \bigcup_{b \in B} b :  B \in \rho \right\}
\end{equation}
So for the examples in \eqref{set partitions}, we have
\begin{gather}
\begin{gathered}
\rho_1 (\pi) = \{ b_1 \cup b_2 \cup b_3 \} \qquad \qquad \qquad \rho_2 (\pi) = \{ b_1 \cup b_2, b_3 \} \\
\rho_3 (\pi) = \{ b_1 \cup b_3, b_2\} \qquad \qquad \qquad \rho_4 (\pi) = \{b_1, b_2 \cup b_3 \} \\
\rho_5 (\pi) = \{ b_1, b_2, b_3 \} = \pi
\end{gathered}
\end{gather}
Given $\pi \leq \pi'$, by definition each block $b \in \pi$ is a subset of a block $b' \in \pi'$. Therefore there is a set partition $\rho \in \Pi(\pi)$ such that $\rho(\pi) = \pi'$, we call this set partition $\pi' / \pi$.

Using the definition of $\pi' / \pi$ for $\pi' \geq \pi$, we can now give an expression for $\mu(\pi, \pi')$, which is a standard result in the field of posets \cite{StanleyEnum}. Firstly, by definition $\mu (\pi, \pi')$ vanishes unless $\pi' < \pi$, so we assume $\pi' \geq \pi$. This means $\pi' / \pi$ exists, and we can write
\begin{equation}
\boxed{
\mu(\pi, \pi') = (-1)^{|\pi| - |\pi'|} \prod_{b \in \pi' / \pi} \left( |b| - 1 \right)!
} 
\label{mobius function}
\end{equation} 
There is a combinatoric interpretation for the magnitude of $\mu(\pi, \pi')$ in terms of permutations, where the sign of $\mu$ is given by the sign of these permutations. In order to describe this, consider a permutation $\sigma \in S_n$ and take an arbitrary subset $A \subseteq \{1,2,\dots,n\}$. Then $\sigma$ acts on $A$ by permuting the numbers $1$ to $n$, leading to a distinct subset $\sigma(A)$. We can then define the subgroup $G(\pi) \leq S_n$ by
\begin{equation}
G(\pi) = \left\{ \sigma : \sigma(b) = b \text{ for all blocks } b \in \pi \right\}
\label{G definition}
\end{equation}
For $\pi$ with block sizes of $[\lambda_1, \lambda_2, \dots, \lambda_k ] \vdash n$, we have
\begin{equation}
G(\pi) \cong S_{\lambda_1} \times S_{\lambda_2} \times \dots \times S_{\lambda_k}
\label{G as a product}
\end{equation}
Intuitively, the $S_{\lambda_i}$ factor permutes the elements of the corresponding block with size $\lambda_i$. The exact embedding of $S_{\lambda_1} \times \dots \times S_{\lambda_k}$ into $S_n$ depends on the set partition.

Take a $\sigma \in S_n$. The cycle structure of $\sigma$ defines a partition $\pi(\sigma) \in \Pi(n)$. Formally, the set partition $\pi(\sigma)$ is simply the set of orbits of $\{ 1,2, \dots, n \}$ under the action of $\sigma$. We also define a set of permutations associated with each $\pi \in \Pi(n)$
\begin{equation}
\text{Perms}(\pi) = \left\{ \sigma : \pi(\sigma) = \pi \right\}
\label{perms of pi}
\end{equation}
For any $\sigma \in S_n$ with $\pi(\sigma) = \pi$, $\text{Perms}(\pi)$ is just the conjugacy class of $\sigma$ under conjugation by $G(\pi)$.

Clearly $\text{Perms}(\pi)$ are disjoint for different $\pi$, and between them they cover $S_n$.
\begin{equation}
\bigsqcup_{\pi \in \Pi(n)} \text{Perms}(\pi) = S_n
\label{splitting S_n}
\end{equation}
We have a similar result for $G(\pi)$, obtained by taking the decomposition \eqref{G as a product} and applying \eqref{splitting S_n} to each factor individually.

\begin{equation}
\bigsqcup_{\pi' \leq \pi} \text{Perms}(\pi') = G(\pi)
\label{splitting G}
\end{equation}
To illustrate the above, we now give some examples. If we fix $\pi = \{ \{1,2,3\} , \{4,5\}, \{6\} \}$ then
\begin{align}
G(\pi) & = S_{ \{1,2,3\} } \times S_{ \{4,5\} } \times S_{ \{6\} } \cong S_3 \times S_2 \times S_1 \\
\text{Perms}(\pi) & = \{ \ (1,2,3)(4,5) \ , \ (1,3,2)(4,5) \ \}
\end{align}
Enumerating the elements of $G(\pi)$, we can see that it splits as specified in \eqref{splitting G}.
\begin{align}
G(\pi) & = \{ e, (1,2), (1,3), (2,3), (1,2,3), (1,3,2), \nonumber \\  
& \qquad (4,5), (1,2)(4,5), (1,3)(4,5), (2,3)(4,5), (1,2,3)(4,5), (1,3,2)(4,5) \} \nonumber \\
& = \{ e \} \sqcup \{ (1,2) \} \sqcup \{ (1,3) \} \sqcup \{ (2,3) \} \sqcup \{ (1,2,3), (1,3,2) \} \nonumber \\
& \qquad \sqcup \{ (4,5) \} \sqcup \{ (1,2)(4,5) \} \sqcup \{ (1,3)(4,5) \} \sqcup \{ (2,3)(4,5) \} \nonumber \\
& \qquad \sqcup \{ (1,2,3)(4,5), (1,3,2)(4,5) \} \nonumber \\
& = \text{Perms} \big( \{\{1\}, \{2\}, \{3\}, \{4\}, \{5\}, \{6\} \} \big) \sqcup \text{Perms} \big( \{ \{1,2\}, \{3\}, \{4\}, \{5\}, \{6\} \} \big) \nonumber \\ 
& \qquad \sqcup \text{Perms} \big( \{ \{ 1,3\} , \{2\}, \{4\}, \{5\}, \{6\} \} \big) \sqcup \text{Perms} \big( \{ \{1\}, \{2,3\}, \{4\}, \{5\}, \{6\} \} \big) \nonumber \\
& \qquad \sqcup \text{Perms} \big( \{ \{1,2,3\}, \{4\}, \{5\}, \{6\} \} \big) \sqcup \text{Perms} \big( \{ \{1\}, \{2\}, \{3\}, \{4,5\}, \{6\} \} \big) \nonumber \nonumber \\
& \qquad \sqcup \text{Perms} \big( \{ \{1,2\}, \{3\}, \{4,5\}, \{6\} \} \big) \sqcup \text{Perms} \big( \{ \{1,3\}, \{2\}, \{4,5\}, \{6\} \} \big) \nonumber \\
& \qquad \sqcup \text{Perms} \big( \{ \{1\}, \{2,3\}, \{4,5\}, \{6\} \} \big) \sqcup \text{Perms} \big( \{ \{1,2,3\}, \{4,5\}, \{6\} \} \big) \\
& = \bigsqcup_{\pi' \leq \pi} \text{Perms}(\pi')
\label{G splitting example}
\end{align}
Equations (\ref{G definition}-\ref{G splitting example}) are based on using permutations $\sigma \in S_n$ and set partitions $\pi \in \Pi(n)$. However, if we pick $\pi \in \Pi(n)$, we can use the exact same constructions for permutations of $\pi$ itself - we call this group $S_\pi$ - and set partitions $\rho \in \Pi(\pi)$. Then $\text{Perms} (\pi'/\pi)$ provides our combinatoric interpretation for $\mu$
\begin{equation}
\boxed{ 
\mu(\pi, \pi')  = \sum_{\sigma \in \text{Perms} (\pi'/\pi)} \text{sgn} (\sigma) = (-1)^{|\pi| - |\pi'|} \left| \text{Perms} (\pi'/\pi) \right|
} 
\end{equation} 
So the magnitude of $\mu$ is just the number of permutations in a certain conjugacy class, and its sign is just the sign of these permutations.

This permutation interpretation of $\mu$ allows us to easily prove the M\"obius inversion formula for set partitions. Fix $\pi$ and $\pi''$ with $\pi'' \geq \pi$ and consider the sum
\begin{equation}
\sum_{\pi'' \geq \pi' \geq \pi} \mu (\pi, \pi')
\label{sum}
\end{equation}
The simplest way to parameterise the sum over $\pi'$ is to look at the possible $\pi' / \pi \in \Pi(\pi)$. The condition $\pi'' \geq \pi'$ becomes $(\pi'' / \pi ) \geq (\pi' / \pi)$, so instead of summing over $\pi' \in \Pi(n)$, we sum over $\pi' / \pi = \rho  \in \Pi(\pi)$.
\begin{align}
\sum_{\pi'' \geq \pi' \geq \pi} \mu (\pi, \pi') 
& = \sum_{\rho \leq (\pi'' / \pi)} \ \ \sum_{\sigma \in \text{Perms} (\rho)} \text{sgn} ( \sigma ) \nonumber \\
& = \sum_{\sigma \in G(\pi''/\pi)} \text{sgn}( \sigma ) \nonumber \\
& = \begin{cases}
1 & G(\pi''/\pi) \cong S_1 \times S_1 \times \dots \times S_1 \\
0 & \text{otherwise}
\end{cases}
\end{align}
Where we have used \eqref{splitting G} to change the sum into one over $G(\pi''/\pi)$, and the final line is a simple fact from permutation group theory. Now the only case for which $G( \pi''/\pi ) \cong S_1 \times \dots \times S_1$ is when $\pi'' = \pi$, so we conclude that
\begin{equation}
\sum_{\pi'' \geq \pi' \geq \pi} \mu(\pi, \pi') = \delta_{\pi \pi''}
\label{mobius sum}
\end{equation}
Using this result and substituting \eqref{pre-Mobius inversion}, we have
\begin{align}
\sum_{\pi' \geq \pi} \mu(\pi, \pi') T_{\pi'} & = \sum_{\pi'' \geq \pi' \geq \pi} \mu(\pi, \pi')  M_{\pi''} \nonumber \\
& = M_\pi
\end{align}
This proves the M\"obius inversion formula for set partitions.

For a more thorough overview of the M\"obius function on general posets and for the poset of set partitions see \cite{StanleyEnum}.

\subsection{More general $C$ and $\tC$ matrices for $M$-matrix systems}
\label{sec:M-matrix} 

The structures explained in sections \ref{section: C combinatorics} and \ref{section: mobius function}, involving the poset of set partitions and the associated M\"obius function, can be used not just for the two matrix system, but for an $M$-matrix system. In \eqref{pre-Mobius inversion} we introduced $T_\pi$ and $M_\pi$ as functions from 2-vector partitions of length $|\pi|$ into the space of multi-symmetric functions defined on two families of variables, $x_i$ and $y_i$. However, we could equally consider them as functions from $M$-vector partitions of length $|\pi|$ into the space of multi-symmetric functions defined on $M$ families of variables, $x^{(1)}_i, x^{(2)}_i, \dots, x^{(M)}_i$. These multi-symmetric functions have monomial and multi-trace bases defined in direct analogy to the 2-vector versions in section \eqref{monomial-and-trace}. For each $M$ there are corresponding $C$ and $\widetilde{C}$ matrices, defined in a completely analogous way to \eqref{c} and \eqref{ctilde}.

To think about these possibilities in a unified way, we define a more general $C$ and $\widetilde{C}$ that transform between $T_\pi$ and $M_\pi$.
\begin{equation}
T_\pi = \sum_{\pi'} C^{\pi'}_{\pi} M_{\pi'}  \qquad \qquad \qquad  M_\pi = \sum_{\pi'} \widetilde{C}^{\pi'}_\pi T_{\pi'} 
\end{equation}
We already have expressions for these from \eqref{pre-Mobius inversion} and \eqref{post-Mobius inversion}, given by
\begin{equation}
C^{\pi'}_\pi = \zeta(\pi, \pi') = 
\begin{cases}
1  & \pi' \geq \pi \\
0 & \text{otherwise}
\end{cases}
\qquad \qquad \widetilde{C}^{\pi'}_{\pi} = \mu(\pi, \pi')
\label{general C and Ctilde}
\end{equation}
where the first equation defines $\zeta(\pi, \pi')$. The above also serves as the definition for the $\zeta$ function of a general poset, just with $\pi, \pi'$ arbitrary elements of the poset rather than set partitions. We have already seen the M\"obius function for a general poset in \eqref{general mobius function}. An equivalent way of stating the M\"obius inversion formula seen in \eqref{mobius inversion formula} is that the $\zeta$ and $\mu$ are inverses of each other when multiplied as matrices
\begin{equation}
\sum_{\pi'} \zeta(\pi, \pi') \mu (\pi', \pi'') = \delta_{\pi \pi''}
\end{equation}
The $C$ and $\widetilde{C}$ for vector partitions (both 2-vectors and $M$-vectors) can easily be obtained from these more general objects. For $\bp, \bq \vdash (n_1,n_2)$ and some $\pi \in \Pi(n)$ such that $\pi([(1,0)^{n_1}, (0,1)^{n_2}]) = \bp$, we have
\begin{equation}
C_\bp^\bq = \sum_{\pi' \in \Pi(n)} C^{\pi'}_\pi \qquad \qquad \qquad \widetilde{C}_\bp^\bq = \sum_{\pi' \in \Pi(n)} \widetilde{C}^{\pi'}_\pi
\label{flavour projection}
\end{equation}
where the sums run over $\pi'$ with $\pi' \left( [(1,0)^{n_1} , (0,1)^{n_2}] \right) = \bq$. Analogous formulae hold for $M$-vectors.

We can think of the \eqref{flavour projection} as a flavour projection from the general system of $M_{\pi}, T_{\pi}$ to the $M$-flavour system consisting of $M_\bp$ and $T_\bp$. Physically, one flavour corresponds to the half-BPS sector, two flavours to the quarter-BPS sector, and three to the eighth-BPS sector. We give an alternative viewpoint on the flavour projection using permutations in section \ref{section: flavour projection}.

As an example, consider $n=4$. There are 15 different set partitions in $\Pi(4)$, so to simplify things we only give the transformations for the five different orbits under $S_4$, corresponding to the integer partitions of 4. The $C$ matrix can be read off from the relationships showing $T_\pi$ in terms of $M_\pi$.
\begin{align}
T_{\{\{1,2,3,4\}\}} & = M_{\{\{1,2,3,4\}\}} \\
T_{\{\{1\},\{2,3,4\}\}} & = M_{\{\{1\},\{2,3,4\}\}} + M_{\{\{1,2,3,4\}\}} \\
T_{\{\{1,2\},\{3,4\}\}} & = M_{\{\{1,2\},\{3,4\}\}} + M_{\{\{1,2,3,4\}\}} \\
T_{\{\{1,2\},\{3\},\{4\}\}} & = M_{\{\{1,2\},\{3\},\{4\}\}} + M_{\{\{1,2,3\},\{4\}\}} + M_{\{\{1,2,4\},\{3\}\}} \nonumber \\
& \qquad + M_{\{\{1,2\},\{3,4\}\}} + M_{\{\{1,2,3,4\}\}} \\
T_{\{\{1\},\{2\},\{3\},\{4\}\}} & = M_{\{\{1\},\{2\},\{3\},\{4\}\}} + M_{\{\{1,2\},\{3\},\{4\}\}} + M_{\{\{1,3\},\{2\},\{4\}\}} \nonumber \\
& \qquad + M_{\{\{1,4\},\{2\},\{3\}\}} + M_{\{\{1\},\{2,3\},\{4\}\}} + M_{\{\{1\},\{2,4\},\{3\}\}} \nonumber \\
& \qquad + M_{\{\{1\},\{2\},\{3,4\}\}} + M_{\{\{1,2\},\{3,4\}\}} + M_{\{\{1,3\},\{2,4\}\}} \nonumber \\
& \qquad + M_{\{\{1,4\},\{2,3\}\}} + M_{\{\{1,2,3\},\{4\}\}} + M_{\{\{1,2,4\},\{3\}\}} \nonumber \\
& \qquad + M_{\{\{1,3,4\},\{2\}\}} + M_{\{\{1\},\{2,3,4\}\}} + M_{\{\{1,2,3,4\}\}}
\end{align}
The $\widetilde{C}$ matrix can be shown in an analogous way by writing $M_\pi$ in terms of $T_\pi$.
\begin{align}
M_{\{\{1,2,3,4\}\}} & = T_{\{\{1,2,3,4\}\}} \\
M_{\{\{1\},\{2,3,4\}\}} & = T_{\{\{1\},\{2,3,4\}\}} - T_{\{\{1,2,3,4\}\}} \\
M_{\{\{1,2\},\{3,4\}\}} & = T_{\{\{1,2\},\{3,4\}\}} - T_{\{\{1,2,3,4\}\}} \\
M_{\{\{1,2\},\{3\},\{4\}\}} & = T_{\{\{1,2\},\{3\},\{4\}\}} - T_{\{\{1,2,3\},\{4\}\}} - T_{\{\{1,2,4\},\{3\}\}} \nonumber \\
& \qquad - T_{\{\{1,2\},\{3,4\}\}} + 2 T_{\{\{1,2,3,4\}\}} \\
M_{\{\{1\},\{2\},\{3\},\{4\}\}} & = T_{\{\{1\},\{2\},\{3\},\{4\}\}} - T_{\{\{1,2\},\{3\},\{4\}\}} - T_{\{\{1,3\},\{2\},\{4\}\}} \nonumber \\
& \qquad - T_{\{\{1,4\},\{2\},\{3\}\}} - T_{\{\{1\},\{2,3\},\{4\}\}} - T_{\{\{1\},\{2,4\},\{3\}\}} \nonumber \\
& \qquad - T_{\{\{1\},\{2\},\{3,4\}\}} + T_{\{\{1,2\},\{3,4\}\}} + T_{\{\{1,3\},\{2,4\}\}} \nonumber \\
& \qquad + T_{\{\{1,4\},\{2,3\}\}} + 2 T_{\{\{1,2,3\},\{4\}\}} + 2 T_{\{\{1,2,4\},\{3\}\}} \nonumber \\
& \qquad + 2 T_{\{\{1,3,4\},\{2\}\}} + 2 T_{\{\{1\},\{2,3,4\}\}} - 6 T_{\{\{1,2,3,4\}\}}
\end{align}
We can then apply these to the vector partition $[(1,0)^2, (0,1)^2]$ to get $C$ and $\widetilde{C}$ for field content $(2,2)$. Again we choose to display them by writing out the relations between $T_\bp$ and $M_\bp$, but this time all possibilities are included. The $C$ matrix is
\begin{align}
\str X^2 Y^2 & = M_{[(2,2)]} \label{first C example} \\
\tr X^2 Y \tr Y & = M_{[(2,1),(0,1)]} + M_{[(2,2)]} \\
\tr X^2 \tr Y^2 & = M_{[(2,0),(0,2)]} + M_{[(2,2)]} \\
\tr XY^2 \tr X & = M_{[(1,2),(1,0)]} + M_{[(2,2)]} \\
\left( \tr XY \right)^2 & = M_{[(1,1),(1,1)]} + M_{[(2,2)]} \\
\tr X^2 \left( \tr Y \right)^2 & = M_{[(2,0),(0,1),(0,1)]} + M_{[(2,0),(0,2)]} + 2 M_{[(2,1),(0,1)]} + M_{[(2,2)]} \\ 
\tr XY \tr X \tr Y & = M_{[(1,1),(1,0),(0,1)]} + M_{[(1,1),(1,1)]} + M_{[(1,2),(1,0)]} \nonumber \\ & \qquad + M_{[(2,1),(0,1)]} + M_{[(2,2)]} \\ 
\left( \tr X \right)^2 \tr Y^2 & = M_{[(1,0),(1,0),(0,2)]} + 2 M_{[(1,2),(1,0)]} + M_{[(2,0),(0,2)]} + M_{[(2,2)]} \\ 
\left( \tr X \right)^2 \left( \tr Y \right)^2 & = M_{[(1,0),(1,0),(0,1),(0,1)]} + M_{[(1,0),(1,0),(0,2)]} + 4 M_{[(1,1),(1,0),(0,1)]} \nonumber \\ & \qquad + M_{[(2,0),(0,1),(0,1)]} + 2 M_{[(1,1),(1,1)]} + 2 M_{[(1,2),(1,0)]} \nonumber \\ & \qquad + M_{[(2,0),(0,2)]} + 2 M_{[(2,1),(0,1)]} + M_{[(2,2)]}
\end{align}
The $\widetilde{C}$ matrix for $(2,2)$ is
\begin{align}
M_{[(2,2)]} & = \str X^2 Y^2 
\label{monomial example 1} \\
M_{[(2,1),(0,1)]} & = \tr X^2 Y \tr Y - \str X^2 Y^2 \\
M_{[(2,0),(0,2)]} & = \tr X^2 \tr Y^2 - \str X^2 Y^2 \\
M_{[(1,2),(1,0)]} & = \tr X Y^2 \tr X - \str X^2 Y^2 \\
M_{[(1,1),(1,1)]} & = \left( \tr XY \right)^2 - \str X^2 Y^2 \\
M_{[(2,0),(0,1),(0,1)]} & = \tr X^2 \left( \tr Y \right)^2 - \tr X^2 \tr Y^2 - 2 \tr X^2 Y \tr Y + 2 \str X^2 Y^2 \\ 
M_{[(1,1),(1,0),(0,1)]} & = \tr XY \tr X \tr Y - \left( \tr XY \right)^2 - \tr XY^2 \tr X \nonumber \\ & \qquad - \tr X^2 Y \tr Y + 2 \str X^2 Y^2  \\ 
M_{[(1,0),(1,0),(0,2)]} & = \left( \tr X \right)^2 \tr Y^2 - 2 \tr XY^2 \tr X - \tr X^2 \tr Y^2 + 2 \str X^2 Y^2 
\label{monomial length 3} \\ 
M_{[(1,0),(1,0),(0,1),(0,1)]} & = \left( \tr X \right)^2 \left( \tr Y \right)^2 - \left( \tr X \right)^2 \tr Y^2 - 4 \tr XY 
\tr X \tr Y \nonumber \\ & \qquad - \tr X^2 \left( \tr Y \right)^2 + 2 \left( \tr XY \right)^2 + 4 \tr XY^2 \tr X \nonumber \\ & \qquad + \tr X^2 \tr Y^2 + 4 \tr X^2 Y \tr Y - 6 \str X^2 Y^2
\label{last monomial example}
\end{align}
Note that we have used
\begin{equation}
\str X^2 Y^2 = \frac{2}{3} \tr X^2 Y^2 + \frac{1}{3} \tr ( XY )^2
\end{equation}
rather than just $\tr X^2 Y^2$. This means the expressions (\ref{first C example}-\ref{last monomial example}) give the relations between $T_\bp$ and $M_\bp$ both as multi-symmetric functions and symmetrised trace operators in $\cN = 4$ super Yang-Mills.

\subsection{Relation to other combinatorial quantities}

Stirling numbers of the second kind, $S(n,k)$, are defined to be the number of ways of partitioning a set of $n$ objects into $k$ non-empty subsets. Combinatorically, these are a coarsened version of the 2-vector and set partition $C$ matrices. Starting with the 2-vector version, $S(n,k)$ is given by
\begin{equation}
S(n,k) = \sum_{\substack{ \bp \vdash (n_1,n_2) \\ l(\bp) = k }} C_{[(1,0)^{n_1}, (0,1)^{n_2}]}^\bp
\label{stirling numbers 1}
\end{equation}
where $n_1 + n_2 = n$.

Alternatively, consider an arbitrary $m > n$ and $\bq \vdash (m_1, m_2)$ with $m_1 + m_2 = m$ to be a vector partition with $l(\bq) = n$, then
\begin{equation}
S(n,k) = \sum_{ \substack{ \bp \vdash (m_1,m_2) \\ l(\bp) = k}} C_\bq^\bp
\end{equation}
Define $\pi_n \in \Pi(n)$ to be the unique set partition of length $n$, meaning each number has its own block. Then in terms of the more general set partition $C$
\begin{equation}
S(n,k) = \sum_{ \substack{ \pi \in \Pi(n) \\ |\pi| = k }} C_{\pi_n}^\pi
\end{equation}
Or alternatively, taking $\pi \in \Pi(m)$ to be any set partition with $|\pi| = n$, then
\begin{equation}
S(n,k) = \sum_{\substack{\pi' \in \Pi(m) \\ |\pi'| = k }} C_{\pi}^{\pi'}
\label{stirling numbers 2}
\end{equation}
Unsigned Stirling numbers of the first kind, $|s(n,k)|$, are defined to be the number of permutations in $S_n$ with $k$ cycles. The signed Stirling numbers $s(n,k)$ have the same magnitude, but are multiplied by the sign of the permutations $(-1)^{n-k}$. This is related to the 2-vector and set partition $\widetilde{C}$ matrices in the same way as $S(n,k)$ was related to $C$. Using the same notation as (\ref{stirling numbers 1}-\ref{stirling numbers 2}), we have
\begin{align}
s(n,k) & = \sum_{\substack{\bp \vdash (n_1,n_2) \\ l(\bp) = k }} \widetilde{C}_{[(1,0)^{n_1}, (0,1)^{n_2}]}^\bp &
s(n,k) & = \sum_{\substack{\bp \vdash (m_1,m_2) \\ l(\bp) = k }} \widetilde{C}_\bq^\bp \\ \ \nonumber \\
s(n,k) & = \sum_{\substack{\pi \in \Pi(n) \\ |\pi|=k }} \widetilde{C}_{\pi_n}^\pi &
s(n,k) & = \sum_{\substack{\pi' \in \Pi(m) \\ |\pi'|=k }} \widetilde{C}_{\pi}^{\pi'}
\end{align}
Bell numbers, $B_n$, count the number of set partitions of $n$ objects. In terms of $C$, these are
\begin{equation}
B_n = \sum_k S(n,k) = \sum_{\bp \vdash (n_1,n_2)} C_{[(1,0)^{n_1},(0,1)^{n_2}]}^\bp = \sum_{\bp \vdash (m_1,m_2)} C_\bq^\bp = \sum_{\pi \in \Pi(n)} C_{\pi_n}^\pi = \sum_{\pi' \in \Pi(m)} C_{\pi}^{\pi'}
\end{equation}
Finally, in \eqref{traces from monomials} we gave the 1-matrix basis change from the monomial basis to the multi-trace basis in terms of characters of $S_n$ representations and the Kostka numbers. Comparing with the (1-matrix equivalent of) definition \eqref{definition of C and C-tilde}, we see
\begin{equation}
C^p_q = \frac{1}{\prod_i p_i!} \sum_{R \vdash n} \chi_R(q) K_{Rp}
\end{equation}

\section{Counting:  $U(2) \times U(N)$ Young diagram labels and multiplicities   at weak coupling }  
\label{sec:Counting} 

The space of states spanned by symmetrised traces $T_{ \bp } $ of general matrices $X,Y$ admits a $U(2)$ action on the pair $ X , Y$ as in section \ref{section:quarter-bps}. These symmetrised traces are representatives of the elements  of  the ring of gauge invariants modulo commutators. Specialising to diagonal matrices $X = \Diag ( x_1, x_2 , \cdots , x_N ) ,Y= \Diag ( y_1, y_2 , \cdots , y_N ) $ gives the isomorphism \cite{Vaccarino2007,Procesi} to multi-symmetric polynomials in $x_i , y_i$ discussed in section \ref{section: isomorphism}. For economy of notation, we  are  generally  using $ T_{ \bp} $ also for the image $ \iota ( T_{ \bp } )$ of the isomorphism. There is an analogous $U(2)$ action on multi-symmetric functions which transforms the pairs $x_i , y_i$. Applying the isomorphism and then doing a $U(2)$ transformation is equivalent to doing a $U(2)$ transformation on symmetrised traces and then applying the isomorphism. In other words the isomorphism between gauge invariants modulo commutators and multi-symmetric polynomials is a $U(2)$ equivariant isomorphism. The $U(2)$ transformations \eqref{U(2) lowering and raising operators} on the monomial multi-symmetric functions, $ M_{ \bp } $,  are obtained either by expressing them in terms of $ T_{ \bp}$ using the $\tC$ transformation or equivalently using the $U(2)$ on the pairs $ (x_i , y_i)$. In this latter picture, the $U(2)$ generators are
\begin{align}
\begin{aligned}
\cJ_0 & =  \sum_{ i =1 }^{N} \left(  x_i \frac{\partial}{\partial x_i} + y_i \frac{\partial}{\partial y_i} \right) & \hspace{50pt} \cJ_3 & = \sum_{ i =1 }^{N}  \left(  x_i \frac{\partial}{\partial x_i} - y_i \frac{\partial}{\partial y_i} \right) \\
\cJ_+ & = \sum_{ i =1 }^{N}  x_i \frac{\partial}{\partial y_i} & \hspace{50pt} \cJ_- & =\sum_{ i =1 }^{N}   y_i \frac{\partial}{\partial x_i} 
\end{aligned}
\end{align}
A $U(2)$ covariant basis will be sorted by $U(2)$ representations $\Lambda$ and an index $M_\Lambda$ labelling the basis states. As in section \ref{section:quarter-bps}, $M_\Lambda$ runs over the semi-standard tableaux of shape $\Lambda$ and determines the field content. In order to  parameterise the space for a specific $\Lambda$, we observe that for each  vector partition $\bp \vdash (n_1,n_2)$, there is an associated integer partition $p(\bp) \vdash n_1 + n_2$ obtained by summing the pairs
\begin{equation}
\bp = [(\lambda_1, \mu_1), (\lambda _2, \mu_2), \dots, (\lambda_k, \mu_k)] \to p(\bp) = [\lambda_1 + \mu_1, \lambda_2 + \mu_2, \dots, \lambda_k + \mu_k]
\label{associated partition}
\end{equation}
Consider the action of $U(2)$ on a simple monomial $x^\lambda y^\mu$. We have
\begin{align}
\begin{aligned}
\cJ_0 \, x^\lambda y^\mu & = (\lambda + \mu) x^\lambda y^\mu & \hspace{20pt} \cJ_3 \, x^\lambda y^\mu & = (\lambda - \mu) x^\lambda y^\mu \\ 
\cJ_+ \, x^\lambda y^\mu & = \mu \, x^{\lambda+1} y^{\mu-1}  & \hspace{20pt} \cJ_- \, x^\lambda y^\mu & = \lambda \, x^{\lambda-1} y^{\mu+1}
\end{aligned}
\label{U(2) action on monomial}
\end{align}
The operators $\cJ_{\pm}$ send $\lambda \rightarrow \lambda \pm 1$, $\mu \rightarrow \mu \mp 1$ while $\cJ_0$ and $\cJ_3$ leave $\lambda, \mu$ invariant. For all $U(2)$ generators, the sum $\lambda + \mu$ is unchanged. More generally, for a monomial $x_1^{\lambda_1} y_1^{\mu_1} \dots x_k^{\lambda_k} y_k^{\mu_k}$, the sums $\lambda_i + \mu_i$ are unchanged in each monomial term arising from the action of the $U(2)$ generators.

Applying this analysis to each of the monomials in $M_\bp$, we see that $U(2)$ preserves the associated partition $p(\bp)$, and therefore $p$ serves as another label in the $U(2)$ covariant basis. We denote the multiplicity of a given pair $\Lambda, p$ in the covariant monomial basis by $\cM_{\Lambda, p}$.

For a given associated partition $p =< p_1, p_2, \dots >$ we have monomial multi-symmetric functions $M_{\bp}$ with $p(\bp) = p$. The constituent monomials in $M_\bp$ (recall the defining equation (\ref{M definition})) contain products of $p_i$ factors each with $i$ variables that can be $x$ or $y$ and are transformed between the two using $\cJ_{\pm}$. We will show that these fit into the representation
\begin{equation}
\cR_p^{U(2)} = \bigotimes_i \Sym^{p_i} \left( \Sym^i \left( V_2 \right) \right) 
\label{R_p U(2) rep}
\end{equation}
where $V_2$ is the 2-dimensional fundamental representation of $U(2)$. We can decompose $\cR_p^{U(2)}$ in terms of irreducible representations $R_\Lambda^{U(2)}$
\begin{equation}
\cR^{U(2)}_p = \bigoplus_{\substack{\Lambda \vdash n \\ l(\Lambda) \leq 2}} R^{U(2)}_\Lambda \otimes V^{mult}_{\Lambda,p}
\label{U(2) decomposition}
\end{equation}
for some multiplicity space $V^{mult}_{\Lambda,p}$. The direct sum is restricted to run only over $\Lambda \vdash n$ since $\cR_p^{U(2)}$ is a subspace of $\left( V_2 \right)^{\otimes n}$, and therefore the $U(1)$ weight of all sub-representations is $n$. The analogous representation of the global symmetry $U(3)$ in the case of eighth-BPS states is discussed in \cite{CtoC, PasRam1204}. The multiplicity of $R_\Lambda^{U(2)}$ in $\cR^{U(2)}_p$ is just the dimension of the multiplicity space $V^{mult}_{\Lambda,p}$, and is also the multiplicity of the pair $\Lambda, p$ in the covariant monomial basis
\begin{equation}
\cM_{\Lambda, p} = \Mult \left( \Lambda, \cR^{U(2)}_p \right) = \operatorname{Dim} \left( V^{mult}_{\Lambda,p} \right)
\label{multiplicity definition}
\end{equation}
To find this multiplicity we split $U(2)$ into its $U(1)$ and $SU(2)$ components as discussed in section \ref{section: U(2) reps}. As already mentioned, $\cR_p^{U(2)}$ is in the weight $n$ representation of $U(1)$, so
\begin{equation}
\cR_p^{U(2)} = R_n^{U(1)} \otimes \cR_p^{SU(2)}
\end{equation}
where $\cR_p^{SU(2)}$ is
\begin{equation}
\cR_p^{SU(2)} = \bigotimes_i \Sym^{p_i} \left( \Sym^i \left( R_{\frac{1}{2}} \right) \right) = \bigotimes_i \Sym^{p_i} \left( R_{\frac{i}{2}} \right)
\label{R_p as symmetric product}
\end{equation}
for $R_j$ the spin $j$ representation of $SU(2)$. Then the $U(2)$ decomposition \eqref{U(2) decomposition} of $\cR_p^{U(2)}$ is equivalent to the $SU(2)$ decomposition
\begin{equation}
\cR^{SU(2)}_p = \bigoplus_{j} R_j \otimes V^{mult}_{\left[ \frac{n}{2} + j, \frac{n}{2} - j \right],p}
\label{R_p multiplicity space decomposition}
\end{equation}
where we have used the correspondence, discussed in section \ref{section: U(2) reps}, between a $U(2)$ representation $\Lambda = \left[ \frac{n}{2} + j, \frac{n}{2} - j \right]$ of $U(1)$ weight $n$ and an $SU(2)$ representation of spin $j$. 
The question of calculating the dimension of the multiplicity space in \eqref{R_p multiplicity space decomposition} is called an $SU(2)$ plethysm problem and is addressed in \cite{King}. We will use a formula derived there shortly.

The monomials $M_\bp$ with $p(\bp) = p$ define states $| \bp \rangle$ in $\cR_p^{U(2)}$, whose normalisation is given by the $S_n$ inner product on $M_\bp$
\begin{equation}
\langle \bp | \bq \rangle = \left\langle M_\bp | M_\bq \right\rangle
\end{equation}
There is a change of basis to $U(2)$ orthonormal covariant states of the form
\begin{equation}
| \Lambda , M_{ \Lambda } , p , \nu \rangle
\label{KetLMP} 
\end{equation}
where $\nu$  is a multiplicity index with $1 \leq \nu \leq \cM_{ \Lambda, p}$. This change of basis is implemented using Clebsch-Gordan coefficients
\begin{equation}
| \Lambda , M_{ \Lambda } , p , \nu\rangle = \sum_{\bp \, : \, p(\bp) = p} B^{ \bp}_{ \Lambda , M_{ \Lambda }, p , \nu } | \bp \rangle
\label{CG coefficients}
\end{equation}
We define the covariant monomial operators by
\begin{equation}
M_{\Lambda, M_\Lambda, p, \nu} = \sum_{\bp \, : \, p(\bp) = p} B^{ \bp}_{ \Lambda , M_{ \Lambda }, p , \nu } M_\bp
\label{U(2) covariant monomial basis}
\end{equation}
\ytableausetup{boxsize=7pt} 
As an example, consider the multi-symmetric monomials for field content $(2,2)$, given explicitly in (\ref{monomial example 1}-\ref{last monomial example}). We only give the $M_\Lambda$ and $p$ labels, as the shape of the Young tableau specifies $\Lambda$, and the multiplicity for these operators is trivial. The covariant monomials are
\begin{align} 
M_{\, \fontsize{6pt}{0} 
	\begin{ytableau} 
	1 & 1 & 2 & 2 
	\end{ytableau} \, , \,  
	[4] 
	\fontsize{12pt}{0} } 
& = \sqrt{ \frac{3}{2} } M_{[(2,2)]}
\label{covariant monomial example 1} \\ 
M_{\, \fontsize{6pt}{0} 
	\begin{ytableau} 
	1 & 1 & 2 & 2 
	\end{ytableau} \, , \,  
	[3,1]
	\fontsize{12pt}{0} } 
& = \sqrt{ \frac{3}{14} } \left( M_{[(2,1),(0,1)]} + M_{[(1,2),(1,0)]} \right) \\ 
M_{\, \fontsize{6pt}{0} 
	\begin{ytableau} 
	1 & 1 & 2 & 2 
	\end{ytableau} \, , \,  
	[2,2]
	\fontsize{12pt}{0} } 
& = \frac{1}{3 \sqrt{2}} \left( M_{[(2,0),(0,2)]} + 2 M_{[(1,1),(1,1)]} \right) \\ 
M_{\, \fontsize{6pt}{0} 
	\begin{ytableau} 
	1 & 1 & 2 & 2 
	\end{ytableau} \, , \,  
	[2,1,1]
	\fontsize{12pt}{0} } 
& = \frac{1}{4 \sqrt{15}} \left( M_{[(2,0),(0,1),(0,1)]} + 4 M_{[(1,1),(1,0),(0,1)]} + M_{[(1,0),(1,0),(0,2)]} \right) \\ 
M_{\, \fontsize{6pt}{0} 
	\begin{ytableau} 
	1 & 1 & 2 & 2 
	\end{ytableau} \, , \,  
	[1,1,1,1]
	\fontsize{12pt}{0} } 
& = \frac{1}{4 \sqrt{6}}  M_{[(1,0),(1,0),(0,1),(0,1)]} \\ 
M_{\, \fontsize{6pt}{0} 
	\begin{ytableau} 
	1 & 1 & 2  \\ 2 
	\end{ytableau} \, , \,  
	[3,1]
	\fontsize{12pt}{0} } 
& = \frac{1}{\sqrt{2}} \left( M_{[(2,1),(0,1)]} - M_{[(1,2),(1,0)]} \right) \\ 
M_{\, \fontsize{6pt}{0} 
	\begin{ytableau} 
	1 & 1 & 2  \\ 2 
	\end{ytableau} \, , \,  
	[2,1,1]
	\fontsize{12pt}{0} } 
& = \frac{1}{4} \left( M_{[(2,0),(0,1),(0,1)]} - M_{[(1,0),(1,0),(0,2)]} \right) \\ 
M_{\, \fontsize{6pt}{0} 
	\begin{ytableau} 
	1 & 1  \\ 2 & 2 
	\end{ytableau} \, , \,  
	[2,2]
	\fontsize{12pt}{0} } 
& = \frac{1}{\sqrt{6}} \left( M_{[(2,0),(0,2)]} - M_{[(1,1),(1,1)]} \right) \\ 
M_{\, \fontsize{6pt}{0} 
	\begin{ytableau} 
	1 & 1  \\ 2 & 2 
	\end{ytableau} \, , \,  
	[2,1,1]
	\fontsize{12pt}{0} } 
& = \frac{1}{6} \left( M_{[(2,0),(0,1),(0,1)]} - 2 M_{[(1,1),(1,0),(0,1)]} + M_{[(1,0),(1,0),(0,2)]} \right)
\label{covariant monomial last example}
\end{align}
The associated partition has length $l(p(\bp)) = l(\bp)$, and therefore the SEP compatibility (modulo commutators) of the $M_\bp$ basis is transferred to the new basis. 

If $p$ has length $l(p) > N$ then the multi-symmetric function $M_{\Lambda, M_\Lambda, p, \nu}$ vanishes identically, while on the other side of the isomorphism, the operator $M_{\Lambda, M_\Lambda, p, \nu}$ reduces to a commutator trace and therefore is no longer pre-BPS. Operators with $l(p) \leq N$ are in general not pre-BPS, but differ from such an operator by a commutator trace. In section \ref{sec:Construction} we show how to remove this commutator trace component to derive a pre-BPS basis. For now, we note that the multiplicity $\cM_{\Lambda, p}$ determines the finite $N$ combinatorics of the quarter-BPS sector.

The half-BPS operators $\cO_R$ defined in \eqref{schur definition from permutations} are dual to giant gravitons. There are two types of giant gravitons: those that have an extended $ S^3 \subset S^5$ as part of the world-volume, and those that have an extended $S^3 \subset AdS_5$. We will refer to these as sphere giants and AdS giants respectively: they are also sometimes distinguished  in the AdS/CFT literature as giants versus dual-giants respectively. $U(2)$ rotations of these half-BPS giants produces giant graviton states in the $ \Lambda = [ n ] $ representation, where $n$ is the number of boxes in the Young diagram $R$. The nature of the Young diagram $R$ is related to the type of giant graviton system. As we deform $ \Lambda $ to $ \Lambda = [ n - m , m ] $ we move away from the half-BPS sector. The deformations of sphere giant states are described in terms of moduli spaces of polynomials in three complex variables \cite{Mikhailov} while deformations of AdS giant states are described in terms of a family of solutions with $S^3 \subset AdS_5$ world-volumes orbiting great circles on the $S^5$ \cite{MS06}.

In section \ref{sec:Construction} we will produce a basis $S^{BPS}_{\Lambda, M_\Lambda, p, \nu}$ for the quarter-BPS sector with the same labels as \eqref{U(2) covariant monomial basis}. For $\Lambda = [n]$, this basis agrees with the half-BPS Schur basis \eqref{schur definition from permutations} by identifying $p$ with $R$. This matching between the Young diagrams $p$ labelling the quarter-BPS sector and $R$ labelling the half-BPS states suggests that for a particular diagram $p$, we can follow  the half-BPS sector states  into the quarter-BPS sector by considering $\Lambda = [n-m,m]$ and slowly increasing the length of the second row, $m$. We expect that if we keep $p$ fixed in this half to quarter transition, we qualitatively preserve the physical nature of the giant graviton: Young diagrams with a few long rows of lengths order $N$ correspond to $AdS$-giants while diagrams with a few long columns of lengths order $N$ correspond to sphere giants.  It is  reasonable to think of Young diagrams $p$  (for more general $ \Lambda$) with $k$  rows of length comparable to $N$ as an AdS-giant system formed as some form of composite of $k$ giants. Likewise, in the following discussion, we will think of a Young diagram $p$ with $k$ rows of length order $N$ as some composite involving $k$ sphere giants. There will be interesting differences between sphere giants and AdS giants, so the precise meaning of ``composite system of $k$ giants'' is something which should be explored through future comparisons between bulk physics and CFT correlators.

The multiplicities \eqref{multiplicity definition} interpolate from half-BPS in the case $ \Lambda=[n]$ 
to more general quarter-BPS for $ \Lambda = [ n - \Lambda_2  , \Lambda_2 ] $, with small $ \Lambda_2$ being close to half-BPS.  These multiplicities should be reproducible from the stringy physics of $D3$-branes in $AdS_5 \times S^5$. In each part of this section, we discuss the giant graviton interpretation of the multiplicity results.

\ytableausetup{boxsize=normal}

\subsection{$\Lambda, p$ multiplicities and plethysms of $SU(2)$ characters  }
\label{section: lambda, p multiplicities}

We consider the space of multi-symmetric functions $M_\bp$ with a given associated partition $p$, and how this can be split into $U(2)$ representations.

As discussed in section \ref{section:quarter-bps}, $U(2)$ can be split into a product of $U(1)$ and $SU(2)$. The $U(1)$ weight of a given $p$ is just $n = |p|$, so to derive the $U(2)$ representation we first study the $SU(2)$ part, then recombine with the $U(1)$ piece at the end.

For the sake of simplicity, we will primarily work with non-symmetrised monomials, since such a choice determines the associated multi-symmetric function by adding all permuted monomials. The construction of the multi-symmetric function from the non-symmetrised monomial can affect the $SU(2)$ structure, and we will describe this in more detail as it occurs.

Start by considering $p = [n]$. This allows $\bp = [(\lambda,\mu)]$ for $\lambda + \mu = n$. The non-symmetrised monomials corresponding to these are just $x_1^\lambda y_1^\mu$, whose action under $U(2)$ we gave in \eqref{U(2) action on monomial}. From the action of $\cJ_\pm, \cJ_3$, they lie in the spin $\frac{n}{2}$ representation of $SU(2)$. Symmetrising the monomials does not change the $SU(2)$ structure, so $p = [n]$ produces the $R_{\frac{n}{2}}$ representation of $SU(2)$. 

Next consider $p = [k_1,k_2]$. This allows $\bp = [(\lambda_1,\mu_1), (\lambda_2, \mu_2)]$, with corresponding non-symmetrised monomials $x_1^{\lambda_1} y_1^{\mu_1} x_2^{\lambda_2} y_2^{\mu_2}$ subject to $\lambda_i + \mu_i = k_i$ for $i=1,2$. There are $(k+1)(l+1)$ different states, living in the tensor product representation $R_{\frac{k_1}{2}} \otimes R_{\frac{k_2}{2}}$. When $k_1 \neq k_2$, this is the correct $SU(2)$ representation for the symmetrised version as well. However, if $k_1=k_2$, then the states $x_1^{\lambda_1} y_1^{\mu_1} x_2^{\lambda_2} y_2^{\mu_2}$ and $x_1^{\lambda_2} y_1^{\mu_2} x_2^{\lambda_1} y_2^{\mu_1}$ both lead to the same multi-symmetric function and should be identified with each other. The correct representation here is the symmetric part of the tensor product, written as $\Sym^2 \left( R_{\frac{k_1}{2}} \right)$.

As our final example, take $p = [k_1,k_2,k_3]$, allowing $\bp = [(\lambda_1,\mu_1), (\lambda_2, \mu_2), (\lambda_3, \mu_3)]$. By the same considerations as the previous two examples, the non-symmetrised monomials $x_1^{\lambda_1} y_1^{\mu_1} x_2^{\lambda_2} y_2^{\mu_2} x_3^{\lambda_3} y_3^{\lambda_3}$ fit into the $R_{\frac{k_1}{2}} \otimes R_{\frac{k_2}{2}} \otimes R_{\frac{k_3}{2}}$ representation of $SU(2)$. If all three of the $k$s are distinct, this is the correct representation for the symmetrised monomials. If two of the $k$s coincide and the third is distinct, e.g. $k_1 = k_2 \neq k_3$, then the $M_\bp$ live in $\text{Sym}^2 \left( R_{\frac{k_1}{2}} \right) \otimes R_{\frac{k_3}{2}}$. Finally, if $k_1 = k_2 = k_3$, then there are 6 permutations of the basic monomial that lead to the same multi-symmetric function and should be identified. These are
\begin{align}
x_1^{\lambda_1} y_1^{\mu_1} x_2^{\lambda_2} y_2^{\mu_2} x_3^{\lambda_3} y_3^{\lambda_3} & & x_1^{\lambda_2} y_1^{\mu_2} x_2^{\lambda_3} y_2^{\mu_3} x_3^{\lambda_1} y_3^{\lambda_1} & & x_1^{\lambda_3} y_1^{\mu_3} x_2^{\lambda_1} y_2^{\mu_1} x_3^{\lambda_2} y_3^{\lambda_2} \nonumber \\
x_1^{\lambda_1} y_1^{\mu_1} x_2^{\lambda_3} y_2^{\mu_3} x_3^{\lambda_2} y_3^{\lambda_2} & & x_1^{\lambda_3} y_1^{\mu_3} x_2^{\lambda_2} y_2^{\mu_2} x_3^{\lambda_1} y_3^{\lambda_1} & & x_1^{\lambda_2} y_1^{\mu_2} x_2^{\lambda_1} y_2^{\mu_1} x_3^{\lambda_3} y_3^{\lambda_3}
\end{align}
In an analogous way to the single coincidence, this leads to us using the completely symmetric part of the triple tensor product, written $\text{Sym}^3 \left( R_{\frac{k_1}{2}} \right)$. This is the part of $R_{\frac{k_1}{2}}^{\otimes 3}$ that is invariant under all $S_3$ permutations.

From the principles established in these three examples, we can generalise to a generic integer partition $p$. The multi-symmetric functions with associated partition $p = < p_1, p_2, \dots >$ fit into the representation of $SU(2)$ given by
\begin{equation}
\cR_{p}^{SU(2)} = R_p^{SU(2)} = \bigotimes_i \text{Sym}^{p_i} \left( R_{\frac{i}{2}} \right)
\label{plethysm}
\end{equation}
Restoring the $U(1)$ weight, as a $U(2)$ representation this is
\begin{align}
\cR_p^{U(2)} & = R_n^{U(1)} \otimes \cR^{SU(2)}_p \nonumber \\
& =  \bigotimes_i R_{i p_i}^{U(1)} \otimes \Sym^{p_i} \left( \Sym^i \left( R_{\frac{1}{2}} \right) \right) \nonumber \\
& = \bigotimes_{i} \Sym^{p_i} \left( R_i^{U(1)} \otimes \Sym^i \left( R_{\frac{1}{2}} \right) \right) \nonumber \\
& = \bigotimes_i \Sym^{p_i} \left(  \Sym^i \left( R_1^{U(1)} \otimes R_{\frac{1}{2}} \right) \right) \nonumber \\
& = \bigotimes_i \Sym^{p_i} \left(  \Sym^i \left( V_2 \right) \right)
\end{align}
where we have used $R_{\frac{i}{2}} = \Sym^i \left( R_{\frac{1}{2}} \right)$ for $SU(2)$ representations and the fundamental representation of $U(2)$ is $V_2 = R_1^{U(1)} \otimes R_{\frac{1}{2}}$.

So the problem of finding $\cM_{\Lambda, p}$ reduces to a $U(2)$ representation theory problem of finding the multiplicity of $R_\Lambda^{U(2)}$ within the representation $\cR_p^{U(2)}$, or equivalently the $SU(2)$ representation theory problem of finding the multiplicity of $R_j$ within $\cR_p^{SU(2)}$ and using the correspondence $j \leftrightarrow \Lambda = \left[ \frac{n}{2} + j, \frac{n}{2} - j \right]$ for $U(2)$ representations of $U(1)$ weight $n$.

We will solve the $SU(2)$ problem. In order to do this, we calculate the character of the representation \eqref{plethysm} and compare it to the known characters of the spin representations. From standard $SU(2)$ representation theory we know that
\begin{equation}
\chi_{R_j} \left( q^{J_3} \right) = q^j + q^{ j-1} + \cdots + q^{ - j} =   q^{ - j} \frac{  ( 1- q^{ 2j+1} ) }{ ( 1 - q) }  
\end{equation}
So the multiplicity of $R_j$ inside a direct sum representation $R$ is given by
\begin{equation}
\Mult \left( R_j , R \right) = \cof \left[ q^{-j}, (1-q) \chi_R \left( q^{J_3} \right)  \right]
\end{equation}
Taking a single factor of \eqref{plethysm}, the character of $\text{Sym}^{p_i} \left( R_{\frac{i}{2}}\right)$ was calculated in \cite{King} and is given by
\begin{align}
\chi_{\text{Sym}^{p_i} \left( R_{\frac{i}{2}} \right)} \left( q^{J_3} \right) & = q^{ - \frac{i p_i}{2} } { ( 1 - q^{ p_i +1} ) \over ( 1 - q ) }   { ( 1 - q^{ p_i +2} ) \over ( 1 - q^2) } \cdots { (  1 - q^{ p_i + i } ) \over ( 1 - q^i ) } \nonumber \\
& = q^{ - \frac{i p_i}{2} } F_{i, p_i}(q) 
\end{align}
where 
\begin{equation}
F_{i,p_i} = { ( 1 - q^{ p_i +1} ) \over ( 1 - q ) }   { ( 1 - q^{ p_i +2} ) \over ( 1 - q^2) } \cdots { (  1 - q^{ p_i + i } ) \over ( 1 - q^i ) }
\label{F definition}
\end{equation}
So the multiplicity of $R_j$ inside $\cR_p^{SU(2)}$ is
\begin{align}
\Mult \left( R_j, \cR_p^{SU(2)} \right) & = \cof \left( q^{-j}, (1-q) \prod_i q^{ - \frac{i p_i}{2} } F_{i, p_i}(q) \right) \nonumber \\ 
& = \cof \left( q^{-j}, (1-q) q^{ - \frac{n}{2} }  \prod_i F_{i, p_i}(q) \right)
\end{align}
Since the $\Lambda = [n-m,m]$ representation of $U(2)$ corresponds to spin $j = \frac{n}{2} - m$, this means
\begin{align}
\Mult \Big( [n-m, m], \cR^{U(2)}_p \Big) & = \cof \left( q^{m-\frac{n}{2}}, (1-q) q^{ - \frac{n}{2} }  \prod_i F_{i, p_i}(q) \right) \nonumber \\
& = \cof \left( q^m, (1-q) \prod_i F_{i, p_i}(q) \right)
\end{align}
Writing 
\begin{equation}
F_p (q) = \prod_i F_{i, p_i}(q)
\label{F_p}
\end{equation}
we can give a simple formula for the multiplicity in terms of the coefficients of $F_p$
\begin{empheq}[box=\fbox]{align}
\hspace{5pt} \cM_{[n-m,m],p} = \cof \left( q^m, F_p \right) - \cof \left( q^{m-1}, F_p \right) \hspace{5pt} \vphantom{\begin{gathered} \ydiagram{1,1} \end{gathered}}
\label{lambda p multiplicity}
\end{empheq}
We now take two distinct approaches to studying $F_p$. Firstly we derive a generic formula for $\cM_{\Lambda, p}$ that allows simple computational calculations of the multiplicity for any $\Lambda, p$ of reasonable size. Secondly, we study sets of $p$ which have identical multiplicities for all $\Lambda$ and give explicit results of $\cM_{\Lambda, p}$ for the simplest such sets.

\subsection{Covariant trace bases}
\label{section: covariant trace bases}

In the previous section we argued from the vector partition structure of the monomial multi-symmetric functions that the $M_\bp$ fit in to the representation $\cR^{U(2)}_p$ of $U(2)$, where $p$ is the integer partition associated to $\bp$. Performing a similar process on the multi-trace multi-symmetric functions $T_\bp$ (or equivalently symmetrised trace operators), the $U(2)$ action not only preserves $p(\bp)$, it has exactly the same form as the action on monomials $M_\bp$. That is, given $\cU \in U(2)$ with action
\begin{equation}
\cU M_\bp = \sum_{\bq} a^\bq_\bp M_\bq
\end{equation}
for some coefficients $a^\bq_\bp$, then the action of $\cU$ on symmetrised traces is
\begin{equation}
\cU T_\bp = \sum_{\bq} a^\bq_\bp T_\bq
\label{U(2) action on symmetrised traces}
\end{equation}
Therefore sorting $M_\bp$ into a $U(2)$ covariant basis is mathematically identical to sorting $T_\bp$ into a $U(2)$ covariant basis. It follows that the linear maps $C , \tC$  relating  $M_{\bp }$ and $ T_{ \bp }$ are $U(2)$ equivariant, and we can define a $U(2)$ covariant symmetrised trace basis
\begin{equation}
T_{\Lambda, M_\Lambda, p, \nu} = \sum_{\bp \, : \, p(\bp) = p} B^{ \bp}_{ \Lambda , M_{ \Lambda }, p , \nu } T_\bp
\label{covariant symmetrised trace}
\end{equation}
In \cite{CtoC}, the authors proved that the multiplicity of $\Lambda, p$ in the symmetrised trace covariant basis is
\begin{equation}
\cM_{\Lambda, p} = \chi_\Lambda ( \bP_p )
\label{permutation multiplicity formula}
\end{equation} 
where $\chi_\Lambda$ is the $S_n$ character of $\Lambda$ and $\bP_p$ is an element of $\bC (S_n)$ that projects onto symmetrised traces with cycle type $p$. We discuss this projector in section \ref{sec: multiplicity space basis}. 

The formulae \eqref{lambda p multiplicity} and \eqref{permutation multiplicity formula} give $\cM_{\Lambda, p}$ from $U(2)$ and $S_n$ representation theory respectively. The former is more amenable to explicit calculations.

As discussed above \eqref{definition of C and C-tilde}, the symmetrised trace operators $T_\bp$ with $l(\bp) \leq N$ form a basis for symmetrised traces (but not pre-BPS operators) at finite $N$. Since the $p$ label in \eqref{covariant symmetrised trace} has the same length as $\bp$, this property also holds for $T_{\Lambda, M_\Lambda, p, \nu}$.

In addition to the symmetrised trace covariant basis, there is a corresponding $U(2)$ covariant basis for commutator traces. It follows from the definitions \eqref{U(2) lowering and raising operators} that the $U(2)$ generators act on a simple commutator as
\begin{equation}
R^i_j [X,Y] = \delta^i_j [X,Y]
\label{U(2) action on commutator}
\end{equation}
Any commutator trace, generically containing a more complicated commutator than $[X,Y]$, can be written as a linear combination of traces containing $[X,Y]$. So \eqref{U(2) action on commutator} shows that the space of commutator traces forms a $U(2)$ representation. By similar considerations to $M_\bp$ and $T_\bp$, these can be further sorted by an integer partition $p \vdash n$ that describes the factorisation of a commutator multi-trace into single traces. 

In \cite{dHRyz0301}, the authors used superspace techniques in the $SU(N)$ gauge theory to develop candidate quarter-BPS operators and SUSY descendent operators. These are exactly the covariant symmetrised trace and commutator trace bases respectively, though they did not include partitions with components of size 1, since in the $SU(N)$ theory, traces of individual matrices vanish.

The covariant bases for symmetrised and commutator traces are used in appendices \ref{appendix: lambda = [3,2]}, \ref{appendix: lambda = [4,2]} and \ref{appendix: lambda = [3,3]} to describe the final BPS operators at $n=5,6$. In this section we will focus on the covariant monomials and not comment further on the covariant symmetrised or commutator traces.

\subsection{General multiplicity formula}
\label{section: general multiplicity formula}

We now find an expression for 
\begin{equation}
\cof \left( q^m, F_p (q) \right)
\end{equation}
This is done explicitly for $m = 0,1,2,3$, from which we extrapolate the general result.

The relevant parts of $p$ to describe the coefficients in \eqref{F_p} are
\begin{equation}
c_{j,k} = \left| \left\{ i : i > j, p_i \geq k \right\} \right|  \qquad \qquad \qquad  j \geq 0, k \geq 1
\label{c_j,k}
\end{equation}
Let $Y_j(p)$, $j \geq 0$ be the Young diagram of $p$ with the first $j$ columns removed. Then intuitively, $c_{j,k}$ is the number of vertical edges of length $k$ or greater in $Y_j(p)$. It follows that $c_{j,1}$ is the number of corners in $Y_j(p)$, and $c_{0,1}$ is the number of corners in the full Young diagram $Y(p)$. Figure \ref{figure: c_jk examples} shows some examples to illustrate this. The full set of $c_{j,k}$ completely determines the partition $p$.

It will also be useful to define
\begin{equation}
s_l = \sum_{j+k=l} c_{j,k}
\label{s_l}
\end{equation}
We have included examples of the $s_l$ in figure \ref{figure: c_jk examples}. In  contrast to the $c_{j,k}$, the $s_l$ do not define the partition $p$. For example, $p = [2]$ and $p = [1,1]$ both have $s_1 =1$, $s_2 = 1$ and all others zero. The sets of partitions which have identical $s_l$ for all $l$ are studied in section \ref{section: orbits of identical multiplicities}.

\begin{figure}
	\begin{center}
		\begin{tabular}{c|c|c|c}
			$p$ & Young diagram & Table of $c_{j,k}$ & Table of $s_l$ \\ \hline
			$[4,3,2,1]$ & $\begin{gathered}\ \\ \ydiagram{4,3,2,1} \\ \ \end{gathered}$ &	
			$\begin{array}{c|c}
			c_{j,k} & k = 1 \\ \hline
			j=0 & 4 \\
			j=1 & 3 \\
			j=2 & 2 \\
			j=3 & 1
			\end{array}$ &
			$\begin{array}{c}
			s_1 = 4 \\ s_2 = 3 \\ s_3 = 2 \\ s_4 = 1
			\end{array}$ \\ \hline
			$[3,3,2,1,1]$ & $ \begin{gathered}\ \\ \ydiagram{3,3,2,1,1} \\ \ \end{gathered}$ &
			$\begin{array}{c|cc}
			c_{j,k} & k = 1 & k=2 \\ \hline
			j=0 & 3 & 2 \\
			j=1 & 2 & 1 \\
			j=2 & 1 & 1
			\end{array}$ &
			$\begin{array}{c}
			s_1 = 3 \\ s_2 = 4 \\ s_3 = 2 \\ s_4 = 1
			\end{array}$ \\ \hline
			$[3,3,3]$ & $\begin{gathered} \ \\ \ydiagram{3,3,3} \\ \ \end{gathered}$ &
			$\begin{array}{c|ccc}
			c_{j,k} & k = 1 & k=2 & k=3 \\ \hline
			j=0 & 1 & 1 & 1 \\
			j=1 & 1 & 1 & 1 \\
			j=2 & 1 & 1 & 1
			\end{array}$ &
			$\begin{array}{c}
			s_1 = 1 \\ s_2 = 2 \\ s_3 = 3 \\ s_4 = 2 \\ s_5 = 1
			\end{array}$
		\end{tabular}
		
	\end{center}
	\caption{Examples of the non-zero $c_{j,k}$ and $s_l$ for various integer partitions $p$.}
	\label{figure: c_jk examples}
\end{figure}

To find the coefficients of $q^{0,1,2,3}$ in $F_p (q)$, we look at the low order terms from the definitions  \eqref{F definition} and \eqref{F_p}. For all $ i > 0$, $F_{i, p_i} $ contains the factor 
\begin{equation}
{  ( 1 - q^{ p_i +1 }   ) \over ( 1 - q)   } = 1 + q + q^2 + \cdots + q^{ p_i }
\end{equation}
For all $ i > 1$, $ F_{i , p_i}$ contains, in addition to the above, the factor 
\begin{align}
\frac{  ( 1 - q^{ p_i +2 }   ) }{ ( 1 - q^2 )   } & = ( 1 + q^2   + q^4 + \cdots ) ( 1- q^{ p_i +2} ) \\ 
& = 1 + q^2 - q^{p_i + 2} + O(q^4)
\end{align}
For all $ i > 2$, the factor $ F_{i , p_i} $ contains, in addition to the above, 
\begin{align}
\frac{  ( 1 - q^{ p_i + 3 }   ) }{ ( 1 - q^3 )   } & = ( 1 + q^3 + \cdots ) ( 1 - q^{ p_i +3} ) \\
& = 1 + q^3 - q^{p_i + 3} + O(q^4)
\end{align}
All other factors in the definition \eqref{F definition} of $F_{i,p_i}$ are of the form $1 + O(q^4)$ so we can ignore them for our purposes, giving
\begin{align}
F_{p} & =  f_1  f_2 f_3 + O(q^4)
\label{prodFexp}  
\end{align}
where
\begin{align}
f_1 & = \prod_{ i > 0} ( 1 + q + q^2 + \cdots + q^{ p_i  } )  \\
f_2 & = \prod_{ i > 1}  ( 1  + q^2 - q^{p_i + 2} + \cdots ) \\
f_3 & = \prod_{ i > 2 } ( 1 + q^3 - q^{p_i + 3} + \cdots ) 
\end{align}
From this we can read off
\begin{equation}
\cof \left( q^0, F_p \right) = 1
\label{0th coefficient}
\end{equation}
All the $q$s in the expansion of $F_p$ come from $f_1$, with the coefficient given by the number of $p_i \geq 1$. From the definitions \eqref{c_j,k} and \eqref{s_l}, we can express this as
\begin{equation}
\cof \left( q, F_p \right) = c_{0,1} = s_1
\label{1st coefficient}
\end{equation}
There are three ways to arrive at a $q^2$ from the product \eqref{prodFexp}. 
\begin{enumerate}
	
	\item We can take a $q^2$ from a factor of $f_2$ and 1 from every other factor. Within $f_2$, this happens whenever $p_i \geq 1$ for $i > 1$, so there are $c_{1,1}$ different ways of doing this. Therefore this route contributes $c_{1,1}$ to the coefficient of $q^2$.
	
	\item We can take a $q^2$ from a factor of $f_1$. Each factor contains a $q^2$ term only if $p_i \geq 2$, so the number of different ways of doing this is $c_{0,2}$.
	
	\item We can take a $q$ from a pair of the $f_1$ factors. There are $\binom{c_{0,1}}{2}$ different ways of doing this.
	
\end{enumerate}
So we arrive at the expression
\begin{equation}
\cof \left( q^2, F_p \right) = c_{0,2} + c_{1,1} + \binom{c_{0,1}}{2} = s_2 + \frac{s_1(s_1 - 1)}{2}
\label{2nd coefficient}
\end{equation}
Looking at $q^3$, there are six distinct ways to arrive at a $q^3$ from the product \eqref{prodFexp}.
\begin{enumerate}
	\item We can take a $q^3$ from a factor of $f_3$. There are $ c_{2,1}$ different ways of doing this.
	\item We can take a $q^3$ from a factor of $f_2$. This can only be done if $p_i = 1$, as it comes from the term $q^{p_i + 1}$. The number of factors with $p_i = 1$ is given by $c_{1,1} - c_{1,2}$. Noting that any $q^3$ obtained in this manner comes with a minus sign, this contributes $c_{1,2} - c_{1,1}$ to the coefficient.
	\item We can take a $q^3$ from a factor of $f_1$. There are $c_{0,3}$ ways of doing this.
	\item We can take a $q^2$ from a factor of $f_2$ and a $q$ from a factor of $f_1$. There are $c_{1,1} c_{0,1}$ ways of doing this.
	\item We can take a $q^2$ from a factor of $f_1$ and a $q$ from a different factor of $f_1$. There are $c_{0,2} (c_{0,1} - 1)$ different ways of doing this.
	\item We can take a $q$ from three different factors of $f_1$. There are $\binom{c_{0,1}}{3}$ different ways of doing this.
\end{enumerate}
Collecting everything, we have
\begin{align}
\cof \left( q^3 , F_p \right) & = c_{2,1} + c_{1,2} - c_{1,1} + c_{0,3} + c_{1,1} c_{0,1} +  c_{0,2} ( c_{0,1} - 1 ) + \binom{c_{0,1}}{3} \nonumber \\ 
& =  c_{2,1} + c_{1,2} + c_{0,3} +  ( c_{1,1} + c_{0,2} ) ( c_{0,1} - 1 ) + \binom{c_{0,1}}{3} \nonumber \\
& = s_3 + s_2 (s_1 - 1) + \frac{s_1 ( s_1 - 1 ) ( s_1 - 2 )}{6}
\label{3rd coefficient}
\end{align}
A similar process for the coefficient of $q^4$ leads to
\begin{align}
\cof \left( q^4 , F_p \right) & = c_{0,4} + c_{1,3} + c_{2,2} + c_{3,1} + (c_{0,3} + c_{1,2} + c_{2,1})(c_{0,1} -1)  \nonumber \\ & \qquad + \binom{c_{0,2} + c_{1,1}}{2} + (c_{0,2} + c_{1,1}) \binom{c_{0,1} - 1}{2} + \binom{c_{0,1}}{4} \nonumber \\
& = s_4 + s_3 (s_1 - 1) + \frac{s_2 (s_2 - 1)}{2} + \frac{s_2 (s_1 - 1) (s_1 - 2)}{2} \nonumber \\ & \qquad + \frac{s_1 (s_1 - 1) (s_1 - 2) (s_1 - 3)}{24}
\label{4th coefficient}
\end{align}
In \eqref{0th coefficient}, \eqref{1st coefficient}, \eqref{2nd coefficient}, \eqref{3rd coefficient} and \eqref{4th coefficient} we have expressed the first 5 coefficients in the expansion of $F_p$ in terms of the $s_l$. The terms in these sums correspond to the partitions of the exponent of $q$. For example in \eqref{4th coefficient}, the terms correspond respectively to the partitions $[4], [3,1], [2,2], [2,1,1]$ and $[1,1,1,1]$. This leads us to suggest the general formula
\begin{empheq}[box=\fbox]{align} 
\hspace{5pt}
\cof \left( q^m , F_p \right) = \sum_{\lambda \vdash m} \left[ \prod_{k} \left( s_{\lambda_k} - k + 1 \right) \right] \left[ \prod_i \frac{1}{\mu_i !} \right]
 \hspace{5pt} \vphantom{\begin{gathered} \ydiagram{1,1,1} \end{gathered}}
\label{general F coefficient} 
\end{empheq}
where we have used both the component notation $\lambda = [ \lambda_1, \lambda_2, ...]$ and the multiplicity notation $\lambda = < \mu_1, \mu_2, \dots >$ for $\lambda$.

In our work for this paper, we have algebraically proved this formula for $m \leq 6$, and have numerically checked it up to $m = 20$. A proof for general $m$ and $p$ is a problem for future work.

It is interesting to note that since $F_p$ is a palindromic polynomial (arising from the $q \rightarrow q^{-1}$ invariance of $SU(2)$ characters), these coefficients form a palindromic sequence. Explicitly,
\begin{equation}
\cof \left( q^m, F_p \right) = \cof \left( q^{n-m}, F_p \right)
\label{palindromic F}
\end{equation}
As the sums over $\lambda$ in \eqref{general F coefficient} get extremely complicated for large $m$, this is quite surprising, and leads us to suspect there is more hidden structure in the sum \eqref{general F coefficient}.

Combining \eqref{lambda p multiplicity} with \eqref{general F coefficient} gives us an explicit formula for $\cM_{\Lambda, p}$
\begin{empheq}[box=\fbox]{align}
\hspace{5pt} \cM_{[n-m,m], p } & = \sum_{\lambda \vdash m} \left[ \prod_{k} \left( s_{\lambda_k} - k + 1 \right) \right] \left[ \prod_i \frac{1}{\mu_i(\lambda) !} \right] \nonumber \\ & \hspace{70pt} - \sum_{\lambda \vdash m-1} \left[ \prod_{k} \left( s_{\lambda_k} - k + 1 \right) \right] \left[ \prod_i \frac{1}{\mu_i(\lambda) !} \right] \hspace{5pt}
\label{general multiplicity formula}
\end{empheq}
Applying this to $m=0$ to $4$, the formulae are
\begin{align}
\cM_{[n], p} & = 1 
\label{lambda=[n] counting} \\
\cM_{[n-1,1], p} & = s_1 - 1 
\label{lambda=[n-1,1] counting} \\
\cM_{[n-2,2], p } & = s_2 + \frac{s_1 (s_1 - 3)}{2} \\
\cM_{ [n-3,3], p } & = s_3 + s_2(s_1 - 2) + \frac{s_1(s_1 - 1)(s_1 - 5)}{6} \\
\cM_{ [n-4,4], p } & = s_4 + s_3(s_1 - 2) + s_2(s_2 - 1) + \frac{s_2 ( s_2 - 1 )}{2} \nonumber \\ & \qquad + \frac{s_2 (s_1 - 1)(s_1 - 4)}{2} + \frac{s_1 (s_1 - 1)(s_1 - 2)(s_1 - 7)}{24} 
\label{lambda=[n-4,4] counting}
\end{align}
These formulae are independent of $N$, so to get finite $N$ multiplicities we impose the finite $N$ cut-off on $p$. Including this, the general multiplicity formula is
\begin{equation}
\cM_{ [n-m,m], p} = \begin{cases}
\eqref{general multiplicity formula} & l(p) \leq N \\
0 & l(p) > N
\end{cases}
\end{equation}
We can also look at the total $\Lambda$ multiplicity $\cM_\Lambda$ by summing over all $p \vdash n$
\begin{align}
\cM_{ [n-m,m] } & = \sum_{p \vdash n, l(p) \leq N} \cM_{ [n-m,m], p} \nonumber \\
& = \sum_{p \vdash n, l(p) \leq N} \left( \sum_{\lambda \vdash m} \left[ \prod_{k} \left( s_{\lambda_k} - k + 1 \right) \right] \left[ \prod_i \frac{1}{\mu_i(\lambda) !} \right] \right. \nonumber \\
& \hspace{120pt} - \left. \sum_{\lambda \vdash m-1} \left[ \prod_{k} \left( s_{\lambda_k} - k + 1 \right) \right] \left[ \prod_i \frac{1}{\mu_i(\lambda) !} \right] \right)
\label{total lambda multiplicity}
\end{align}
From representation theory considerations \cite{Dolan_2003}, the sectors with $\Lambda = [n]$ and $[n-1,1]$ do not undergo a step-change as we turn on the coupling constant. Therefore the weak coupling combinatorics of these sectors should match the free field combinatorics of section \ref{section:quarter-bps}. A priori, the combinatorics should agree when considering the entire $\Lambda = [n]$ or $[n-1,1]$ sector. We find a stronger result: the combinatorics of the Young diagram label $R$ in \eqref{U(2) basis definition} matches the partition $p$ of this section. 

From \eqref{lambda=[n-4,4] counting}, for $\Lambda = [n]$ the multiplicity of any given $p$ is 1, while for $\Lambda = [n-1,1]$ recall that $s_1 = c_{0,1}$ is the number of corners of $p$, so the multiplicity of $p$ is simply the number of corners subtract 1. As expected, these match \eqref{[n] multiplicity} and \eqref{[n-1,1] multiplicity} respectively and therefore
\begin{equation}
C(R,R,\Lambda) = \cM_{ \Lambda, R}
\label{multiplicity matching}
\end{equation}
for $\Lambda = [n]$ and $[n-1,1]$.

\subsection{Hermite reciprocity and $p$-orbits of fixed  $ \cM_{\Lambda , p } $ }
\label{section: orbits of identical multiplicities}

There are   collections of $p$ which lead to the same multiplicities for all $\Lambda$. To understand these, we look at the definition \eqref{F definition} of $F_{i, p_i}$. If $i > p_i$, the numerator and denominator start cancelling, and we end up with
\begin{equation}
F_{i, p_i} = \frac{ \left( 1 - q^{ i+1} \right) }{ ( 1 - q ) }   \frac{ ( 1 - q^{ i+2} ) }{ ( 1 - q^2) } \cdots
\frac{ (  1 - q^{ i + p_i } ) }{ \left( 1 - q^{p_i}  \right) } = F_{p_i, i}
\end{equation}
We can rewrite this to be explicitly symmetric in $i \leftrightarrow p_i$
\begin{equation}
F_{i , p_i} = F_{p_i, i} = \frac{ \left( 1 - q^{\max(i,p_i)+1} \right) }{ ( 1 - q ) }   \frac{ ( 1 - q^{\max(i,p_i)+2} ) }{ ( 1 - q^2) } \cdots
\frac{ (  1 - q^{ i + p_i } ) }{ \left( 1 - q^{\min(i,p_i)}  \right) }
\end{equation}
This symmetry is  known as  Hermite reciprocity \cite{Goodman_Wallach} and can be viewed as 
a property of  $ SU(2)$ characters.

We can use this $i \leftrightarrow p_i$ symmetry to do transformations on partitions that keep the product $F_p$ the same, and by extension all the associated $\Lambda$ multiplicities.

As our first example, take $p$ to be rectangular, so $p = [i^{p_i}]$ for some particular choice of $i$. Then the conjugate partition $p^c = [(p_i)^i]$ has the same $F$, leading to the same multiplicities for all $\Lambda$. Note that $p^c = p$ if $i = p_i$.

Now suppose $p = [i^{p_i}, j^{p_j}]$ for $i < j$. Then there are three candidates for partitions with the same $F$, namely
\begin{align}
p^{(1)} & = \left[ (p_i)^i, j^{p_j} \right] \\
p^{(2)} & = \left[ i^{p_i}, (p_j)^j \right] \\
p^{(3)} & = \left[ (p_i)^i, (p_j)^j \right]
\end{align}
The partition given by $p^{(1)}$ will only produce the same $F$ if $j \neq p_i$. If $j = p_i$, then $p^{(1)}$ should be written as $[j^{i + p_j}]$ and the $F$s no longer match. Similarly $p^{(2)}$ will only match if $i \neq p_j$ and $p^{(3)}$ if $p_i \neq p_j$.

To visualise the transformations taking $p$ to $p^{(1,2,3)}$, split $p$ into two rectangles stacked on top of each other. Then $p^{(1)}$ is obtained by rotating the $i$ rectangle through 90 degrees, reordering the two rectangles if appropriate, and re-stacking them. In the same manner, $p^{(2)}$ is obtained by rotating the $j$ rectangle, and $p^{(3)}$ by rotating both. We take $p = [4, 3,3]$ as an example
\begin{align}
\ytableausetup{boxsize=12pt}
\begin{aligned}
p & = \begin{gathered} \ydiagram{4,3,3} \end{gathered} 
&  p^{(1)} & = \begin{gathered} \ydiagram{4,2,2,2} \end{gathered}
& p^{(2)} & = \begin{gathered} \ydiagram{3,3,1,1,1,1} \end{gathered}
&  p^{(3)} & = \begin{gathered} \ydiagram{2,2,2,1,1,1,1} \end{gathered}
\end{aligned}
\label{4 distinct dimension}
\end{align}
When one of the dimensions of the first rectangle coincides with one of the dimensions of the second rectangle, one or more of these four options will reduce from two distinct rectangles into one larger rectangle, and hence to a different $F$. If there is one coincidence, for example $p = [3,2]$ where we have $p_2 = p_3 = 1$, we only have three partitions with the same $F$
\begin{align}
p & = \begin{gathered}\ydiagram{3,2} \end{gathered}
& p^{(1)} & = \begin{gathered} \ydiagram{3,1,1} \end{gathered}
& p^{(2)} & = \begin{gathered} \ydiagram{2,1,1,1} \end{gathered}
\end{align}
If there are two coincidences, then the partition is not related to any other via these transformations. There are three distinct ways for these two coincidences to occur. Firstly, three of the dimensions could be the same, while the fourth is different, for example $p = [2,2,1,1]$. Secondly, both rectangles are squares, with distinct sizes, for example $p = [2,2,1]$. Finally, the two rectangles are identical, but are non-square, for example $p = [2,1,1]$. These three partitions are shown below
\begin{equation}
\begin{gathered} \ydiagram{2,2,1,1} \end{gathered} \hspace{100pt} \begin{gathered} \ydiagram{2,2,1} \end{gathered} \hspace{100pt} \begin{gathered} \ydiagram{2,1,1}
\end{gathered}
\end{equation}
The generalisation to more rectangles is straightforward. A partition made from $k$ rectangles can be related to as many as $2^k$ others by rotating a subset of the $k$ rectangles. These rotations are only valid if the widths of all the rotated rectangles are distinct. As an example, consider all partitions of 5. These fall into 4 orbits under these rotations
\begin{align}
\ytableausetup{boxsize = 5pt}
\begin{aligned}
o_1 & = \left\{ \begin{gathered} \ydiagram{5} \end{gathered} \ , \ \begin{gathered} \ydiagram{1,1,1,1,1} \end{gathered} \right\} & \qquad
o_2 & = \left\{ \begin{gathered} \ydiagram{4,1} \end{gathered} \right\} \\
o_3 & = \left\{ \begin{gathered} \ydiagram{3,2} \end{gathered} \ , \ \begin{gathered} \ydiagram{3,1,1} \end{gathered} \ , \ \begin{gathered} \ydiagram{2,1,1,1} \end{gathered} \right\} & \qquad
o_4 & = \left\{ \begin{gathered} \ydiagram{2,2,1} \end{gathered} \right\}
\end{aligned}
\end{align}
The equivalent classification for partitions of 6 is
\begin{align}
\begin{aligned}
o_1 & = \left\{ \begin{gathered} \ydiagram{6} \end{gathered} \ , \ \begin{gathered} \ydiagram{1,1,1,1,1,1} \end{gathered} \right\} &
o_2 & = \left\{ \begin{gathered} \ydiagram{5,1} \end{gathered} \right\} &
o_3 & = \left\{ \begin{gathered} \ydiagram{4,2} \end{gathered} \ , \ \begin{gathered} \ydiagram{4,1,1} \end{gathered} \ , \ \begin{gathered} \ydiagram{2,1,1,1,1} \end{gathered} \right\} \\
o_4 & = \left\{ \begin{gathered} \ydiagram{3,3} \end{gathered} \ , \ \begin{gathered} \ydiagram{2,2,2} \end{gathered} \right\} &
o_5 & = \left\{ \begin{gathered} \ydiagram{3,2,1} \end{gathered} \right\} &
o_6 & = \left\{ \begin{gathered} \ydiagram{3,1,1,1} \end{gathered} \right\} \\
&& o_7 & = \left\{ \begin{gathered} \ydiagram{2,2,1,1} \end{gathered} \right\}
\end{aligned}
\end{align}
In appendices \ref{appendix: lambda = [3,2]}, \ref{appendix: lambda = [4,2]} and \ref{appendix: lambda = [3,3]}, we give explicit formulae for the $n=5,6$ basis elements. As expected, the families of partitions above have the same multiplicities for all $\Lambda$.

The first orbit constructed from 2 rectangles to have all 4 dimensions distinct, and therefore achieve the maximum size of $4 = 2^2$ is found at $n=10$ and is shown in \eqref{4 distinct dimension}. The first orbit constructed from 3 rectangles to have all 6 dimensions distinct and have maximum size $8 = 2^3$ is found at $n=28$, and consists of
\begin{gather}
\left\{
\begin{gathered}
\begin{gathered} \ydiagram{6,5,5,4,4,4} \end{gathered} \  , \ 
\begin{gathered} \ydiagram{6,5,5,3,3,3,3} \end{gathered} \ , \
\begin{gathered} \ydiagram{6,4,4,4,2,2,2,2,2} \end{gathered} \ , \
\begin{gathered} \ydiagram{6,3,3,3,3,2,2,2,2,2} \end{gathered} \ , \
\begin{gathered} \ydiagram{5,5,4,4,4,1,1,1,1,1,1} \end{gathered} \ , \
\begin{gathered} \ydiagram{5,5,3,3,3,3,1,1,1,1,1,1} \end{gathered} \ , \ \begin{gathered} \ydiagram{4,4,4,2,2,2,2,2,1,1,1,1,1,1} \end{gathered} \ , \ \begin{gathered} \ydiagram{3,3,3,3,2,2,2,2,2,1,1,1,1,1,1} \end{gathered}
\end{gathered}
\right\}
\end{gather}
In the half-BPS sector, the Young diagram label of the Schur basis \eqref{schur definition from permutations} gives us the properties of the corresponding giant graviton in the AdS/CFT correspondence. A partition with $k \sim 1$ rows of length $L \sim N$ describes $k$ giant gravitons (or a single giant graviton wrapped $k$ times) extended in $AdS_5$ of angular momentum $L$. Similarly, a partition with $k$ columns of length of length $L$ describes $k$ giant gravitons extended in $S^5$ of angular momentum $L$.

These orbits of partitions with identical multiplicities at all $\Lambda$ allow us to identify families of different giant graviton states that behave the same under quarter-BPS deformations.

Taking the simplest example of a single rectangle with $p = [k^L]$, this rectangle rotation symmetry means $k$ coincident giant gravitons rotating with angular momentum $L$ in $AdS_5$ behave the same way as $k$ coincident giant gravitons rotating with angular momentum $L$ in $S^5$. 

A system of two $AdS$ giant gravitons with different angular momenta has $p = [k_1,k_2]$ with $k_1, k_2 \sim N$ and $k_1 \neq k_2$. Then we can rotate each of the rows individually to get $p = [k_1, 1^{k_2}]$ or $p = [k_2, 1^{k_1}]$, however we cannot rotate both at the same time. Therefore the behaviour of two non-coincident sphere giants under quarter-BPS deformations is different to that of two non-coincident $AdS$ giants. This is studied further in the next section.

\ytableausetup{boxsize = normal}

\subsection{Calculation of multiplicities for simplest orbits}
\label{section: simplest orbits}

For some of the simplest orbits of partitions under rectangle rotation, we can describe the $\Lambda$ multiplicities explicitly. These are dual to one or two giant gravitons wrapped around the $AdS_5$ or $S^5$ factors of $AdS_5 \times S^5$. 

Recall from \eqref{lambda p multiplicity} that $(1-q) F_p$ is the generating function for the $\Lambda$ multiplicities. More specifically, the coefficient of $q^m$ is the multiplicity of $\Lambda = [n-m,m]$ for $m \leq \left\lfloor \frac{n}{2} \right\rfloor$.

We start with $p = [1^n]$ or equivalently $p = [n]$.
\begin{equation}
(1-q) F_{[n]} = (1-q) F_{1,n} = 1 - q^{n+1}
\end{equation}
So $\Lambda = [n]$ appears with multiplicity 1, and all other $\Lambda$ have multiplicity 0. In the dual string theory, this means a single half-BPS giant graviton cannot deform into the quarter-BPS sector.

We next consider rectangles with side lengths 2 and $k \geq 2$, so $p = [2^k]$ or $[k,k]$.
\begin{align}
(1-q) F_{[k,k]} & = (1-q) F_{2, k} =  \frac{(1-q^{k+1})(1-q^{k+2})}{1-q^2} \nonumber \\
& = ( 1 + q^2 + q^4 + \dots) (1 - q^{k+1} - q^{k+2} + q^{2k + 3} ) \nonumber \\
& = 1 + q^2 + q^4 + \dots + q^{\lfloor k \rfloor_2} + O(q^{k+1})
\label{lambda = [k,k] multiplicity}
\end{align}
where $\left\lfloor k \right\rfloor_2 = 2 \left\lfloor \frac{k}{2} \right\rfloor$ is $k$ rounded down to the nearest multiple of 2. Since we are only interested in the terms with exponent $\leq k = \frac{n}{2}$, we can ignore the $O(q^{k+1})$ parts of the expression. Then $\Lambda = [n-m,m]$ appears with multiplicity 1 if $m$ is even, and 0 otherwise.

The dual interpretation of $p = [k,k]$ is two coincident $AdS$ giants, while $p = [2^k]$ is two coincident sphere giants. Then \eqref{lambda = [k,k] multiplicity} states that these states can be deformed deep into the quarter-BPS sector. In some sense, the quarter-BPS state `furthest' from half-BPS is $\Lambda = [\frac{n}{2}, \frac{n}{2}]$, and this arrangement of giants can be deformed right up to that limit if $\frac{n}{2}$ is even (and only one away if $\frac{n}{2}$ odd). However, not all quarter-BPS deformations are available. In particular the `smallest' deformation $\Lambda = [n-1,1]$ does not exist, and we must deform by `twice' as much for each step into the quarter-BPS.

Now look at a combination of two rectangles, both with one dimension of length 1. Let the other dimensions be $k \geq l$. If $k = l$, then the orbit has size 1, namely $p = [k, 1^k]$. Otherwise, the orbit consists of three partitions, $p = [k,l]$, $p = [k, 1^l]$, $p = [l, 1^k]$. The considerations of the orbit size do not affect the calculation of multiplicities. This calculation is
\begin{align}
(1-q) F_{[k,l]} & = (1-q) F_{1,k} F_{1,l} = \frac{(1 - q^{k+1})(1-q^{l+1})}{1-q} \nonumber \\
& = (1 + q + q^2 + \dots )( 1 - q^{l+1} - q^{k+1} + q^{k+l+2}) \nonumber \\
& = 1 + q + q^2 + \dots + q^l + O(q^{k+1})
\label{lambda = [k,l] mult}
\end{align}
So $\Lambda = [n-m,m]$ appears with multiplicity 1 if $m \leq l$ and 0 otherwise.

For $k > l$, based on the argument that keeping $Y(p)$ fixed and deforming $ \Lambda = [ n ] $ to $ \Lambda = [ n - n_2 , n_2 ] $ preserves the qualitative physics  of the giant states, we expect the partition $p = [k,l]$ corresponds to two non-coincident $AdS$ giants when $ k ,  l $ are of order $N$. The multiplicity of $ \Lambda $ we are getting above is precisely the multiplicity of $U(2)$ reps in $\left( \Lambda = [ k] \right) \otimes \left( \Lambda = [ l] \right)$. Indeed the $U(2)$ representation for the quantum states constructed from  multi-symmetric functions $ M_{ {\bf p}}$ with associated partition     $ p = [ k , l ] $  is  $ Sym^k ( V_2 ) \otimes Sym^l ( V_2) =   \cR^{ U(2)}_{ p = [k] }  \otimes  \cR^{ U(2)}_{ p = [ l] } $. The case $p = [ k , k ] $ corresponds, by the same argument, to bound state of two AdS giants  of angular momentum $k$. In this case the multi-symmetric construction gives the $U(2)$ representation $ Sym^2 ( Sym^k ( V_2)) = Sym^2 ( \Lambda = [k] ) $, which is the symmetric subspace of $ \cR^{ U(2)}_{ p = [k] } \otimes   \cR^{ U(2)}_{ p = [k] } $. The projection to the symmetric part accounts for the missing powers of $q$ in \eqref{lambda = [k,k] multiplicity} compared to the $ p = [ k , l ] $ case \eqref{lambda = [k,l] mult}.

To look at two non-coincident sphere giants, we consider $p = [l,l,k-l]$ for $k > l$. After rotating both rectangles, this is equivalent to $p = [2^l, 1^{k-l}]$, corresponding to two sphere giants of momenta $k$ and $l$ respectively. This has
\begin{align}
(1-q) F_{[l,l,k-l]} & = (1-q) F_{2,l} F_{1,k-l} = \frac{(1-q^{l+1}) (1-q^{l+2}) (1-q^{k-l+1})}{(1-q)(1-q^2)} \nonumber \\
& = \big( 1 - q^{k-l+1} - q^{l+1} - q^{l+2} + q^{2l+3} + O(q^{k})  \big) \nonumber \\ & \hspace{100pt} (1 + q + 2 q^2 + 2 q^3 + 3 q^4 + 3 q^5 + \dots)
\label{two sphere giant mult}
\end{align}
Where we can ignore terms of order $k$ and higher as these exponents are greater than $\frac{n}{2} = \frac{k+l}{2}$.

Let $a_m = 0$ for $m < 0$ and $a_m = \left\lfloor \frac{m}{2} \right\rfloor + 1$ for $m \geq 0$, the coefficient of $q^m$ in the second factor of \eqref{two sphere giant mult}. Then the coefficient of $q^m$ in \eqref{two sphere giant mult} is
\begin{align}
\cM_{[n-m,m], [l,l,k-l] } & = a_m - a_{m-k+l-1} - a_{m-l-1} - a_{m-l-2} + a_{m-2l-3}
\label{two sphere giant general mult}
\end{align}
The exact formulae for the multiplicities depend on the relative sizes of $k$ and $l$. If $k \leq 2l$, then
\begin{align}
\cM_{[n-m,m], [l,l,k-l]}
& = \begin{cases}
\vspace{5pt}
\left\lfloor \frac{m}{2} \right\rfloor + 1 & \ \ 0 \leq m \leq k-l \\ \vspace{5pt}
\left\lfloor \frac{m}{2} \right\rfloor - \left\lfloor \frac{m-k+l-1}{2} \right\rfloor & \ \ k-l+1 \leq m \leq l \\
\left\lfloor \frac{m}{2} \right\rfloor - \left\lfloor \frac{m-k+l-1}{2} \right\rfloor - m + l & \ \ l+1 \leq m \leq \frac{k+l}{2}
\end{cases}
\label{two sphere giant mult 1}
\end{align}
where we have used
\begin{align}
\left\lfloor \frac{c-1}{2} \right\rfloor + \left\lfloor \frac{c-2}{2} \right\rfloor & = \frac{c-1}{2} + \frac{c-2}{2} - \frac{1}{2} = c-2 
\end{align}
for $c = m - l$.

If $2l \leq k \leq 3l$, then
\begin{align}
\cM_{[n-m,m], [l,l,k-l]}
& = \begin{cases}
\vspace{5pt}
\left\lfloor \frac{m}{2} \right\rfloor + 1 & \ \ 0 \leq m \leq l \\ \vspace{5pt}
\left\lfloor \frac{m}{2} \right\rfloor - m + l + 1 & \ \ l+1 \leq m \leq k-l \\
\left\lfloor \frac{m}{2} \right\rfloor - \left\lfloor \frac{m-k+l-1}{2} \right\rfloor - m + l & \ \ k-l+1 \leq m \leq \frac{k+l}{2}
\end{cases}
\label{two sphere giant mult 2}
\end{align}
Finally, if $k \geq 3l$
\begin{align}
\cM_{[n-m,m], [l,l,k-l]}
& = \begin{cases}
\vspace{5pt}
\left\lfloor \frac{m}{2} \right\rfloor + 1 & \ \ 0 \leq m \leq l \\ \vspace{5pt}
\left\lfloor \frac{m}{2} \right\rfloor - m + l + 1 & \ \ l+1 \leq m \leq 2l \\
0 & \ \ 2l+1 \leq m \leq \frac{k+l}{2}
\end{cases}
\label{two sphere giant mult 3}
\end{align}
For two sphere giants of momenta $k,l \lessapprox N$, the multiplicities fall into category \eqref{two sphere giant mult 1}. Roughly speaking, the multiplicity of $\Lambda = [n-m,m]$ increases as $\frac{m}{2}$ until reaching $\frac{k-l}{2}$. It then stays constant until $m$ reaches $l$ before turning around and decreasing for $m \geq l$, reaching 0 at $m = \frac{n}{2}$.

From the construction based on multi-symmetric functions,  the states for $ p  = [2^l , 1^{k-l} ] $ form, in all the cases, the $U(2)$ representation 
\begin{align} 
\Sym^{ l } ( \Sym^2 ( V_2) ) \otimes \Sym^{ k-l} ( V_2 ) & = \Sym^{ 2 } ( \Sym^l ( V_2 ) ) \otimes \Sym^{ k  -l} ( V_2 ) \nonumber \\
& = \cR^{U(2)}_{ p = [ 2^l ] } \otimes \cR^{U(2)}_{ p = [ 1^{ k-l} ] } \nonumber \\
& =  \cR^{U(2)}_{ p = [ l , l ] }  \otimes \cR^{U(2)}_{ p = [k-l]} 
\end{align}
So the construction  implies that the $2$-sphere-giant system for $ p = [ 2^k , 1^{k-l} ]$ 
have the same multiplicities as a composite consisting of the 2-sphere-giant bound state $p=[2^l]$ 
along with a 1-sphere giant system $ [1^{(k-l)} ]$, while Hermite reciprocity further implies that these multiplicities are also the same as those of an AdS 2-giant bound state of angular momentum $l$ composed with a single AdS giant of angular momentum $k-l$.

We can see a marked difference between the behaviour of two non-coincident sphere giants compared to two non-coincident $AdS$ giants. In \eqref{lambda = [k,l] mult} the multiplicity of each $\Lambda$ was at most 1, so there was a unique way of deforming the arrangement of $AdS$ giants at each stage on their way into the quarter-BPS sector. Furthermore, the furthest possible deformation was $m=l$, the lesser of the two momenta of the gravitons. With \eqref{two sphere giant mult 1} the multiplicities can be larger than 1, and are non-zero right up to $m = \frac{n}{2}$. So there are a multitude of ways of deforming sphere giants, and they can be deformed all the way into the quarter-BPS. Interestingly, when the two momenta are more uneven, and $m$ can get as high as $m=2l$, there is a cut-off on the possible deformations. This is twice the equivalent cut-off for non-coincident $AdS$ giants.

We have interpreted $p = [2^l, 1^{k-l}]$ as corresponding to two non-coincident sphere giants in order to compare with the equivalent system of $AdS$ giants. However, when $l, k-l \sim N$, the rotation $p = [l,l,k-l]$ is exactly the system of two coincident $AdS$ giants of momenta $l$ and a third giant of momenta $k-l$. So two separated sphere giants have the same behaviour as a system of three $AdS$ giants.

It is worth remarking that there are important differences in how the same Hilbert spaces of $N$ free bosons in a harmonic oscillator are arrived at in the two problems of quantizing moduli spaces of sphere giants \cite{BGLM06} and the moduli space of AdS giants \cite{MS06}. In \cite{BGLM06} quarter-BPS multi-giant systems are described by Fock space oscillators associated with  higher order polynomials  in $ x , y $.  In \cite{MS06} there is a relatively simpler phase space of classical $AdS$ solutions which is $ \mC^2$ (and $ \mC^3$ in the more general eighth-BPS case)  and the  full Hilbert space is obtained  by considering an $N$ particle boson system based on this 1-particle system. This serves to explain why the gauge  theory  construction of BPS operators we are giving here, which is intimately tied to a weak-coupling gauge theory realization of the multi-free boson Hilbert space, leads to simpler compositeness structures for the AdS giants as discussed above.

\subsection{Partitions with one dominant row or column  }
\label{section: partitions with a dominant row}

There is another family of partitions that have nice properties. Consider $p \vdash n$ in which the first row dominates the partition, i.e. $p = [ \lambda_1, \hat{p}]$, where $\lambda_1 \geq \frac{n}{2}$ and $\hat{p} \vdash \hat{n} = n - \lambda_1$.

With one exception, when $\lambda_1 = \frac{n}{2}$ and $\hat{p} = [\lambda_1]$ (this case has already been considered in \eqref{lambda = [k,k] multiplicity}), this leads to
\begin{equation}
F_p =   F_{ \lambda_1 , 1 }   F_{\hat{p}} = \frac{1 - q^{\lambda_1 + 1}}{1-q} F_{\hat{p}}
\end{equation}
and therefore
\begin{equation}
(1-q) F_p = ( 1 - q^{\lambda_1 + 1} ) F_{\hat{p}}
\end{equation}
Using the second equation in \eqref{lambda p multiplicity}, we have 
\begin{align}
\cM_{[n-m, m], [\lambda_1, \hat{p}]} & = \cof \left[ q^m, (1-q) \prod_i F_{i, p_i}(q) \right] \nonumber \\
& = \cof \left[ q^m , ( 1 - q^{\lambda_1 + 1} ) F_{\hat{p}}  \right]
\label{previous equation}
\end{align}
Since  $ m \leq { n \over 2 } $ for $ \Lambda $ to be valid Young diagram and $\lambda_1 + 1 > \frac{n}{2}$ by the dominant first row property, it follows that 
\begin{equation}
\cof \left( q^m , q^{ \lambda_1 + 1 } F_{ \hat p } \right) = 0 
\end{equation}
and we can simplify \eqref{previous equation} to
\begin{equation}
\cM_{[n-m, m], [\lambda_1, \hat{p}]} = \cof \left( q^m , F_{\hat{p}} \right ) 
\end{equation}
Thus, the generating function for the $\Lambda$ multiplicities is just $F_{\hat{p}}$ and does not  depend on $\lambda_1$. We can now use our study of the coefficients of $F$ from section \ref{section: general multiplicity formula} to give the $\Lambda$ multiplicities. Note that the dominant first row condition has allowed us to obtain a simple formula for $ ( 1 - q ) F_p$ in terms of $ F_{ \hat p } $. As a result the multiplicities are being obtained simply from the coefficients of a known generating function ($ F_{ \hat p } $). This is simpler than the procedure of section \ref{section: general multiplicity formula} where the $\Lambda$ multiplicities were obtained from the difference of two consecutive coefficients in $ F_p$. 

Using the formulae \eqref{lambda=[n] counting}-\eqref{lambda=[n-4,4] counting} we can write
\begin{align}
F_{\hat{p}} ( q ) = \sum_{m } q^m \cM_{[ n-m , m ] , [ \lambda_1 , \hat p ]}  & = 1 + s_1 q + \left( s_2 + \frac{s_1 (s_1 - 1)}{2} \right) q^2 \nonumber \\ & \hspace{30pt} + \left( s_3 + s_2 (s_1 - 1) + \frac{s_1 (s_1 - 1)(s_1 - 2)}{6} \right) q^3 + \dots
\label{F_phat}
\end{align}
Where the $s_i$ refer to $s_i(\hat{p})$, not $s_i(p)$. 

As previously observed in \eqref{palindromic F}, the coefficients of $F_{\hat{p}}$ form a palindromic sequence, starting and ending with 1 at $q^0$ and $q^{\hat{n}}$. Adding this to \eqref{F_phat}, we have
\begin{align}
\sum_{m } q^m \cM_{[ n-m , m ] , [ \lambda_1 , \hat p ]}  & = 1 + s_1 q + \left( s_2 + \frac{s_1 (s_1 - 1)}{2} \right) q^2 \nonumber \\ 
& \hspace{30pt} + \left( s_3 + s_2 (s_1 - 1) + \frac{s_1 (s_1 - 1)(s_1 - 2)}{6} \right) q^3 \nonumber \\
& \hspace{30pt} + \dots \nonumber \\
& \hspace{30pt} + \left( s_3 + s_2 (s_1 - 1) + \frac{s_1 (s_1 - 1)(s_1 - 2)}{6} \right) q^{\hat{n} - 3} \nonumber \\
& \hspace{50pt} + \left( s_2 + \frac{s_1 (s_1 - 1)}{2} \right) q^{\hat{n}-2} + s_1 q^{\hat{n} - 1} + q^{\hat{n}}
\label{single dominant row multiplicities}
\end{align}
In summary, for $p$ of the form $p = [\lambda_1, \hat{p}]$, where $\hat{p} \vdash \hat{n}$ and $ \lambda_1 \geq { n \over  2 }$, the multiplicities of $\Lambda = [n]$ and $\Lambda=[n-\hat{n}, \hat{n}]$ are exactly 1, the multiplicities of $\Lambda = [n-1,1]$ and $\Lambda = [n-\hat{n}+1,\hat{n}-1]$ are the number of corners in $\hat{p}$, and the general multiplicity of $\Lambda = [n-m,m]$ can be read from \eqref{general F coefficient} with $p = \hat{p}$ if $m \leq \hat{n}$ or is 0 if $m > \hat{n}$. Furthermore, when $m \leq \hat{n}$, sending $m \to \hat{n} - m$ does not affect the multiplicity : 
\begin{equation}
\cM_{[ n-m , m ] , [ \lambda_1 , \hat p ]} =  \cM_{[ n - \hat{n} + m , \hat{n} - m ] , [ \lambda_1 , \hat p ]}
\end{equation}
These properties have interesting implications. For a given $n$, $\hat{n} \leq \frac{n}{2}$, there are a large class of partitions with a dominant single row of length $n - \hat{n}$ for which the combinatorics of the deep quarter-BPS sector are determined by the combinatorics of the near half-BPS sector. For $\Lambda = [n-\hat{n},\hat{n}]$, there is a multiplicity of exactly 1 for any of the partitions in this class, which is the same combinatorics as the half-BPS $\Lambda = [n]$. For $\Lambda = [n - \hat{n} + 1, \hat{n} - 1]$, the multiplicity is the same as the next to half-BPS $\Lambda = [n-1,1]$. For $\Lambda = [n]$ and $[n-1,1]$, there is no change in spectrum as we turn on the coupling constant. Therefore the combinatorics of the $\Lambda = [n-\hat{n},\hat{n}], [n-\hat{n}+1,\hat{n}-1]$ sectors (for this class of partitions) at weak coupling are determined by free field considerations.

It would be interesting to find out whether this unreasonable effectiveness of the free theory has any connections to arguments in \cite{KMMR07,BerensteinSBH} that important features of black hole physics in $AdS_5 \times S^5$ are captured by the free theory.

More generally, if $m \ll \hat{n} \lesssim \frac{n}{2}$, then $\Lambda = [n-m,m]$ is a small deviation from half-BPS while $\Lambda = [n - \hat{n} + m, \hat{n} - m]$ is a long way into the quarter-BPS sector, and yet their combinatorics are identical for this class of partitions. An  interesting question  is whether for these states with dominant first row in $p$ (or dominant first column) have a well-defined semi-classical brane or  space-time interpretation which can explain the coincidence of multiplicities between near-half-BPS and far-into-quarter-BPS regimes.  Near half-BPS states have been studied in the context of the BMN limit of AdS/CFT \cite{BMN}. In the context of giant gravitons, the physics of perturbations, in some sense small, of well-separated multi-giants has been understood \cite{GGO,double coset ansatz,DN1710}.

Using the rectangle rotation described in section \ref{section: orbits of identical multiplicities}, similar properties hold for a single large column. Consider $p$ with a first column of length $\mu_1$ and a partition $\bar{p}$ attached to the right. This is denoted by $p = [1^{\mu_1}] + \bar{p}$. In terms of rectangles we can use for the rotation symmetry, this is a partition $[1^{l(\bar{p})}] + \bar{p}$ with a single column below it of length $\mu_1 - l(\bar{p})$. So setting $\lambda_1 = \mu_1 - l(\bar{p})$ and $\hat{p} = [1^{l(\bar{p})}] + \bar{p}$, $p$ is in the same rotation orbit as $[\lambda_1, \hat{p}]$, and we can apply the logic of this section directly to $p$.

As an example, consider $\mu_1 = 8$ and $\bar{p} = [2,1]$, with corresponding $\lambda_1 = 6$, $\hat{p} = [3,2]$. This is easiest to see visually
\ytableausetup{boxsize=10pt}
\begin{align}
p = 
\begin{gathered} \ydiagram{1,1,1,1,1,1,1,1} \end{gathered}
+
\begin{gathered} \ydiagram{2,1} \end{gathered}
=
\begin{gathered} \ydiagram{3,2,1,1,1,1,1,1} \end{gathered}
\overset{\substack{rectangle \\ rotation}}{\longrightarrow} 
\begin{gathered} \ydiagram{6,3,2} \end{gathered}
\label{dominant column example}
\end{align}
It is clear that the conditions on a single dominant column are more difficult to work with than those for a single dominant row. Let $\mu_1$ and $\mu_2$ be the length of the first and second columns respectively, then to use the analysis of this section, we require $\mu_1 - \mu_2 \geq \frac{n}{2}$. This is a far smaller class of diagrams than given by the analogous condition $\lambda_1 \geq \frac{n}{2}$ for a diagram with a single dominant row.

\ytableausetup{boxsize=normal}

\subsection{Identifying a multiplicity space basis}
\label{sec: multiplicity space basis}

In discussing the decomposition  \eqref{R_p multiplicity space decomposition} we have not specified a choice of basis for $V_{\Lambda, p}^{mult}$, instead introducing a multiplicity index $\nu$ in the state \eqref{KetLMP}. In this section we outline an algebraic approach to choosing a basis and characterising $\nu$. 

As seen in \eqref{R_p U(2) rep}, $R^{U(2)}_p$ is a subspace of $\left( V_2 \right)^{\otimes n}$. There is an $S_n$ action on $\left( V_2 \right)^{\otimes n}$ by permutation of the tensor factors, and $R^{U(2)}_p$ is the subspace invariant under the subgroup $G_p$ of $S_n$. For $p = <p_1, p_2, \dots> \vdash n$, this subgroup, which is discussed in \cite{CtoC}, is
\begin{equation}
G_p = \bigtimes_{i} S_{p_i} \left[ S_i \right]
\label{G_p definition}
\end{equation}
where $S_{p_i} \left[ S_i \right]$ is the wreath product of $S_{p_i}$ with $S_i$. This is defined as the semi-direct product of $S_{p_i}$ with $\left( S_i \right)^{p_i}$, where $S_{p_i}$ acts on $\left( S_i \right)^{p_i}$ by permutation of factors.

$G_p$ contains as a subgroup the group $G(\pi)$ given in \eqref{G as a product}, where $\pi \in \Pi(n)$ is a set partition with block size sizes given by $p$. $G_p$ consists of $G(\pi)$ with the addition of the $S_{p_i}$ factors.

The projector onto the $G_p$-invariant space is
\begin{equation}
\bP_p = \frac{1}{\left| G_p \right|} \sum_{\sigma \in G_p} \sigma
\label{symmetrised p projector}
\end{equation}
which acts on a permutation $\tau \in S_n$ via the adjoint action
\begin{equation}
\bP_p (\tau) = \frac{1}{|G_p|} \sum_{\sigma \in G_p} \sigma \tau \sigma^{-1}
\end{equation}
This projector was used in \cite{CtoC} to derive the formula \eqref{permutation multiplicity formula} for $\cM_{\Lambda, p}$.

On the full space $\left( V_2 \right)^{\otimes n}$, the $U(2)$ and $S_n$ actions commute, and therefore they still commute on the $R^{U(2)}_p$ subspace. Since $R^{U(2)}_p$ is the $G_p$-invariant subspace, we should consider the action of the permutation subalgebra invariant under $G_p$-conjugation, rather than the full group algebra $\mathbb{C} ( S_n )$. This algebra is
\begin{equation}
\mathcal{A}_p = \bP_p \left[ \bC ( S_n ) \right] = \left\{ \alpha \in \mathbb{C} (S_n) \, | \, \sigma \alpha \sigma^{-1} = \alpha \, , \,  \forall \sigma \in G_p \right\} 
\end{equation}
Now $\mathcal{A}_p$ acts on $R^{U(2)}_p$, but commutes with $U(2)$, which means in the decomposition \eqref{R_p multiplicity space decomposition} it acts only on the multiplicity space components. So to choose a basis for $V^{mult}_{\Lambda, p}$, we can choose a maximally commuting set of operators in $\mathcal{A}_p$ and label the multiplicity space basis by the eigenvalues of these operators.

The algebras $\cA_p$ are in general quite complicated, and finding a maximally commuting set of operators within them is an involved computational problem that we do not attempt to find a general solution for. They are a generalisation of the algebras studied in \cite{EHS,PCA}.

\section{Construction of orthogonal $ U(2) \times U(N)$ Young-diagram-labelled basis}
\label{sec:Construction} 

In Section \ref{sec:Projectors}  we show that for any $n,N$, we can construct BPS operators 
by applying $ \cG_N $ to the subspace $ \bar \cM^{\le }_{ N }   \subset \mC ( S_n ) $, and using 
the map (\ref{U(2) covariant operator definition}) from permutations to gauge invariant operators built from $N \times N$ matrices.  The physical inner product on such operators  obtained from 2-point functions uses the element $ \cF_N $: $ \delta ( ( \cG_N \sigma_1)  \cF_N ( \cG_N \sigma_2 ) ) = \delta ( \sigma_1 \cG_N \sigma_2 )$.  An orthogonal basis is obtained by choosing an ordering (section \ref{section: Sn orthogonalisation}) on the labels of the basis elements of $ \bar \cM^{ \le }_N$ and Gram-Schmidt orthogonalising. The orthogonal basis elements $S^{ BPS}_{ \Lambda , M_{ \Lambda } , p , \nu } $ are normalised in the $S_n$ inner product given in \eqref{quarter-bps Sn inner product} (in section \ref{sec:Projectors} this is the $g_{n,N} $ inner product).  This construction algorithm  gives a basis of BPS operators which is not only orthogonal but also SEP-compatible. 

In this section we explain the construction of this orthogonal SEP-compatible basis of BPS operators from the covariant monomials $M_{\Lambda, M_\Lambda, p, \nu}$. We work with $N \times N$ matrices $X$ and $Y$, of which there are $n$ in total, where we can consider $N \geq n$ or $N < n$.

The final output will be a basis of BPS operators of the form
\begin{align}
S^{BPS}_{\Lambda, M_\Lambda, p, \nu} & = \sum_{\substack{R,\tau \\ l(R) \leq N}} s^{R,\tau}_{p,\nu}(\Lambda;N) \cO_{\Lambda, M_\Lambda, R, \tau}
\label{final output}
\end{align}
where $\cO_{\Lambda, M_\Lambda, R, \tau}$ are the free field operators defined in \eqref{U(2) basis definition} and the expansion coefficients $s^{R,\tau}_{p,\nu}(\Lambda;N)$ are functions of $N$. These will in general consist of a polynomial numerator and a denominator that is the square root of a polynomial.

Let us give a precise statement of SEP-compatibility for these operators. Take some $\hN \leq N$, and evaluate these operators on matrices $X$ and $Y$ of size $\hN \times \hN$ instead of size $N \times N$. This means the free field operators with $l(R) > \hN$ vanish, and the coefficients are evaluated at $\hN$ rather than $N$. Then the operators with $l(p) > \hN$ will vanish and the operators with $l(p) \leq \hN$ will form a basis for the reduced BPS sector. Moreover, these are exactly the operators that would be produced by applying the construction algorithm directly with matrices of size $\hN \times \hN$.

Sections \ref{section: Sn orthogonalisation} through \ref{section: F orthogonalisation} describe how to construct $S^{BPS}_{\Lambda, M_\Lambda, p, \nu}$ and prove that this basis is indeed BPS and SEP-compatible. In section \ref{section: quicker route} we give an equivalent, shorter construction that is represented in figure \ref{figure: main algorithm}. The remaining sections investigate various properties of the bases.

\subsection{Orthogonalisation and SEP compatibility }
\label{section: Sn orthogonalisation}

In order to construct the basis \eqref{final output}, we use results from section \ref{sec:Projectors}. Define
\begin{align}
\cM^>_{\widehat{N}}   ( N )    & = \Span \left\{ M_{\Lambda, M_\Lambda, p, \nu}  ( N )   : l(p) > \widehat{N} \right\}
\label{M> space} \\
\cM^\leq_{\widehat{N}}  ( N )   & = \Span \left\{ M_{\Lambda, M_\Lambda, p, \nu}  ( N )    : l(p) \leq \widehat{N} \right\}
\label{M< space}
\end{align}
for some $\hN \leq N$.  The operators $ M_{\Lambda, M_\Lambda, p, \nu} ( N )$ are constructed by employing the permutation to operator map in (\ref{U(2) covariant operator definition}) with matrices $X , Y $ of size $N$. Orthogonalise $\cM^\leq_{\hN}$ against $\cM^>_{\hN}$ using the $S_n$ inner product (the $g_{ n , \hN }$ inner product in section \ref{sec:Projectors}), and denote the orthogonalised space by $\bar{\cM}^\leq_{\hN}$. Note here the distinction between $N$ and $\hN$. The operators are defined using matrices of size $N \times N$, while $\hN$ is used to separate the operators into two classes depending on the length of $p$, the partition label for operators.

The result \eqref{orthogonalisation proof} and the discussion below it prove several useful facts.

\begin{enumerate}
	
	\item Setting $\hN = N$, $\bar{\cM}^\leq_{N}$ is the entire pre-BPS sector. 
	
	\item The subspace  $\bar{\cM}^\leq_{\hN} (N) $  is within the span of free field operators with $l(R) \leq \hN$. In particular, operators within $\bar{\cM}^\leq_{\hN}$ do not receive any contribution from free field operators with $l(R) > \hN$. To see this, note that $\bar{\cM}^\leq_{\hN}  = Im ( \cP ) \cap Im ( \cF_{\hN}) $. The general gauge invariant operators for matrices of size $N$ are constructed using permutation group algebra elements cut-off by $l(R ) \le N$. The definition of $ \bar{\cM}^\leq_{\hN} (N)  $ involves the stronger restriction $ l(R) \le \hN$.
	
	\item  $\bar{\cM}^\leq_{\hN} (N) $  gives a subspace of pre-BPS operators for matrices of size $N$. This subspace is such that, when we reduce $N$ to $\hN$ by lowering the size of the matrices $X$ and $Y$ to $\hN \times \hN$, these operators remain pre-BPS, and in fact form the entire pre-BPS sector.
	
\end{enumerate}
The first of these results tells us the minimum work necessary to create BPS operators. Take an operator $M_{\Lambda, M_\Lambda, p, \nu}(N)$ with $l(p) \leq N$, and orthogonalise it against $\cM^>_N$ to give a new operator $\bar{M}_{\Lambda, M_\Lambda, p, \nu}(N)$. These form a basis for the quarter-BPS sector. If $N \geq n$, then $\cM^>_N$ is empty, and no orthogonalisation is necessary.

However, for $\hN < N$, operators $\bar{M}_{\Lambda, M_\Lambda, p, \nu}(N)$ with $l(p) \leq \hN$ are not necessarily orthogonal to $\cM^>_{\hN}$, and therefore upon lowering $N$ to $\hN < N$, these are no longer pre-BPS. In other words, this is not an SEP-compatible basis, and more work is needed to find one. From the second and third points above, we have a sequence of pre-BPS spaces
\begin{equation}
\bar{\cM}^\leq_1 ( N ) \subset \dots \subset \bar{\cM}^\leq_{N-1} ( N )  \subset \bar{\cM}^\leq_N ( N ) 
\label{sequence of spaces}
\end{equation}
such that for any $\hN \leq N$, the corresponding subspace $\bar{\cM}^\leq_{\hN}$ is the entire pre-BPS sector when we lower $N$ to $\hN$.

It is now clear how we construct an SEP-compatible basis. Each operator $M_{\Lambda, M_\Lambda, p, \nu} (N) $ must be  orthogonalised in the $S_n$ inner product  against all operators $M_{\Lambda, M_\Lambda, q, \eta} (N) $ with $l(q) > l(p)$. Then for any $\hN$, the orthogonalised operators with $l(p) \leq \hN$ form a basis for $\bar{\cM}_{\hN}^\leq$. Note that operators in different $\Lambda$ sectors are already orthogonal from the hermiticity properties of $U(2)$, so we do not need to consider these. In the subsequent discussion we will describe in more detail the steps involved  in the construction of SEP-compatible orthogonal BPS operators starting from $M_{\Lambda, M_\Lambda, p, \nu} (N) $. We will henceforth drop the label $N$ and simply write  $M_{\Lambda, M_\Lambda, p, \nu} $, with the fact that we are describing the construction for matrices of size $N$ being  understood.  The parameter $\hN < N $ will come up in discussion of SEP compatibility of the construction.

In section \ref{section: isomorphism}, we explained that if $N < n$, then $M_{\Lambda, M_\Lambda, p, \nu}$ picked an operator that differed from a pre-BPS operator by addition of a commutator trace. Intuitively, the orthogonalisation is removing the commutator trace part to leave only the pre-BPS operator.

Before implementing the orthogonalisation, recall from section \ref{sec:BCOV} that for $N \geq n$, applying $\cG_N$ to any basis of symmetrised traces gives BPS operators, without any complicated orthogonalisation procedure. From the above, we can give a weaker bound on $N$ within a specific $\Lambda$ sector.

Let $p^*_\Lambda$ be the longest (largest in the ordering \eqref{partition ordering}) partition with $\cM_{\Lambda, p^*_\Lambda} > 0$. In section \ref{section: longest lambda} we prove that for $\Lambda = [\Lambda_1, \Lambda_2]$, we have
\begin{equation}
l(p^*_\Lambda) = n - \left\lceil \frac{\Lambda_2}{2} \right\rceil
\label{lambda constraint}
\end{equation}
Then if $N \geq l(p^*_\Lambda)$, then $\cM_N^>$ has no operators transforming on the $\Lambda $ representation of $U(2)$ and the operators $M_{\Lambda, M_\Lambda, p, \nu}$ do not need to be orthogonalised before applying $\cG_N$ to get BPS operators. 

Returning to the construction, to carry out the procedure we will use Gram-Schmidt orthogonalisation, which requires choosing an ordering on partitions. To ensure the correct properties when we lower $N$ to some $\hN \leq N$, we must begin with the longest partition, and proceed to the shortest. For those with the same length, any ordering would suffice to create an SEP-compatible basis. A natural choice is to compare the length of the second column and start with the longer. If this is also the same length, the comparisons proceed along the columns until one is longer. If the partitions are the same and the operators occupy a multiplicity space, we do not specify an ordering. Any will suffice.

More formally, this ordering on partitions is the conjugate of the standard lexicographic ordering of partitions, which compares partitions based on the length of the first row, followed by the second row, etc. Given $p^c = [\lambda_1, \lambda_2, \dots], q^c = [\mu_1, \mu_2, \dots]$, then
\begin{align}
p > q \iff  \text{ there is a } k \geq 1 \text{ such that }
\begin{cases}
\lambda_i = \mu_i & i < k \\
\lambda_k > \mu_k
\end{cases}
\label{partition ordering}
\end{align}
The operators obtained by performing the orthogonalisation are denoted by $\bar{Q}_{\Lambda, M_\Lambda, p, \mu}$.

In section \ref{section: longest lambda} we prove that $\cM_{\Lambda, p^*_\Lambda} = 1$, so we drop the multiplicity. For this partition we have
\begin{equation}
\bar{Q}_{\Lambda, M_\Lambda, p^*_\Lambda} = M_{\Lambda, M_\Lambda, p^*_\Lambda}
\label{longest partition operator}
\end{equation}
If we are performing the algorithm with $N < l\left( p^*_\Lambda \right)$, then the associated operator $M_{\Lambda, M_\Lambda, p^*_\Lambda}$ will reduce to a commutator trace or vanish. In the former case, there is no difference to the algorithm, while in the latter, we instead start the orthogonalisation with the largest $p$ such that $\cM_{\Lambda, p} > 0$ and the associated operator does not vanish. This partition will not necessarily have multiplicity 1.

For the remaining $p, \nu$ the orthogonalised operators are defined inductively
\begin{align}
\bar{Q}_{\Lambda, M_\Lambda, p,\nu} 
& = M_{\Lambda, M_\Lambda, p,\nu} - 
\sum_{(q,\eta) > (p, \nu)} \frac{ \left\langle M_{\Lambda, M_\Lambda, p,\nu} , \bar{Q}_{\Lambda, M_\Lambda, q, \eta} \right\rangle_{S_n} }
{ \left\langle \bar{Q}_{\Lambda, M_\Lambda, q, \eta}, \bar{Q}_{\Lambda, M_\Lambda, q, \eta} \right\rangle_{S_n} }
\bar{Q}_{\Lambda, 	M_\Lambda, q, \eta}
\label{orthogonalisation of multi-symmetric functions}
\end{align}
where by $(q,\eta) > (p,\nu)$ we mean either $q > p$ in the ordering \eqref{partition ordering} or $q = p$ and $\eta > \nu$.

Similarly to the first step, it may occur that $\bar{Q}_{\Lambda, M_\Lambda, p,\nu} = 0$ for some operators with $l(p) > N$. Such operators are excluded from the rest of the orthogonalisation algorithm. It is implicit that the sum in \eqref{orthogonalisation of multi-symmetric functions} does not run over these values of $q,\eta$.

In order to compare the different $\bar{Q}_{\Lambda, M_\Lambda, p, \nu}$, we normalise to have $S_n$ norm 1
\begin{equation}
Q_{\Lambda, M_\Lambda, p, \nu} = \frac{ \bar{Q}_{\Lambda, M_\Lambda, p, \nu} }{\sqrt{ \left\langle \bar{Q}_{\Lambda, M_\Lambda, p, \nu} | \bar{Q}_{\Lambda, M_\Lambda, p, \nu} \right\rangle_{S_n} }}
\label{Q normalisation}
\end{equation}
The new operators $Q_{\Lambda, M_\Lambda, p \nu}$ are an SEP-compatible basis for pre-BPS operators. They are orthonormal under the $S_n$ inner product, and form a stepping stone on the way to producing \eqref{final output}.

From the arguments above, we know that when expanding $Q_{\Lambda, M_\Lambda, p, \nu}$ in terms of the free field BPS operators \eqref{U(2) basis definition}, only those operators with $l(R) \leq l(p)$ contribute.

\subsection{An example: field content $(2,2)$}
\label{section: lambda = [2,2]}

We now give an explicit example of this construction for the field content $(2,2)$ sector at $N \geq n = 4$. While doing so, we will observe various features that generalise to higher orders. These are discussed in later subsections.

We begin with the operators $M_{\Lambda, M_\Lambda, p, \nu}$ given in (\ref{covariant monomial example 1}-\ref{covariant monomial last example}). The ordering \eqref{partition ordering} for partitions of $n=4$ is
\begin{equation}
[1,1,1,1] > [2,1,1] > [2,2] > [3,1,1] > [4]
\end{equation}

\subsubsection{$\Lambda = [4]$ and $[3,1]$ sectors}
\label{section: [4] and [3,1] orthogonalisation}

Orthogonalising in the $\Lambda = [4]$ sector, and normalising with respect to the $S_n$ inner product, we obtain
\ytableausetup{boxsize=7pt}
\begin{align}
Q_{
	\fontsize{6pt}{0} \begin{ytableau}
	1 & 1 & 2 & 2 
	\end{ytableau} \fontsize{12pt}{0} \, , \,
	[1, 1, 1, 1] 
} & =
\frac{1}{4\sqrt{6}}  \left[
- 4 \tr X^2 Y^2 
- 2 \tr \left( X Y \right)^2 
+ 4 \tr X^2 Y \tr Y 
+ 4 \tr X \tr X Y^2 \right. \nonumber \\
&  \hspace{50pt} + 2 \left( \tr X Y \right)^2 
+ \tr X^2 \tr Y^2 
- \tr X^2 \left( \tr Y \right)^2 \nonumber \\
& \left. \hspace{50pt} - 4 \tr X \tr X Y \tr Y 
- \left( \tr X \right)^2 \tr Y^2 
+ \left( \tr X \right)^2 \left( \tr Y \right)^2 
\right] 
\label{symmetrised trace operator 1} \\
Q_{
	\fontsize{6pt}{0} \begin{ytableau}
	1 & 1 & 2 & 2 
	\end{ytableau} \fontsize{12pt}{0} \, , \,
	[2, 1, 1]
} & =
\frac{1}{4\sqrt{6}} \left[
4 \tr X^2 Y^2 
+ 2 \tr \left( X Y \right)^2 
- 2 \left( \tr X Y \right)^2 
- \tr X^2 \tr Y^2 \right. \nonumber \\
&  \hspace{50pt} - \tr X^2 \left( \tr Y \right)^2 
- 4 \tr X \tr X Y \tr Y 
- \left( \tr X \right)^2 \tr Y^2 \nonumber \\
& \left. \hspace{50pt} + 3 \left( \tr X \right)^2 \left( \tr Y \right)^2 
\right] \\
Q_{
	\fontsize{6pt}{0} \begin{ytableau}
	1 & 1 & 2 & 2 
	\end{ytableau} \fontsize{12pt}{0} \, , \,
	[2, 2]
} & =
\frac{1}{2\sqrt{6}} \left[
2 \left( \tr X Y \right)^2 
+ \tr X^2 \tr Y^2 
- 2 \tr X^2 Y \tr Y 
- 2 \tr X \tr X Y^2 \right. \nonumber \\
& \left. \hspace{50pt} + \left( \tr X \right)^2 \left( \tr Y \right)^2 
\right] \\
Q_{
	\fontsize{6pt}{0} \begin{ytableau}
	1 & 1 & 2 & 2 
	\end{ytableau} \fontsize{12pt}{0} \, , \,
	[3, 1]
} & =
\frac{1}{4 \sqrt{6}} \left[
- 4 \tr X^2 Y^2 
- 2 \tr \left( X Y \right)^2 
- 2 \left( \tr X Y \right)^2 
- \tr X^2 \tr Y^2 \right. \nonumber \\
& \hspace{50pt} + \tr X^2 \left( \tr Y \right)^2 
+ 4 \tr X \tr X Y \tr Y 
+ \left( \tr X \right)^2 \tr Y^2 \nonumber \\
& \left. \hspace{50pt} + 3 \left( \tr X \right)^2 \left( \tr Y \right)^2 
\right] \\
Q_{
	\fontsize{6pt}{0} \begin{ytableau}
	1 & 1 & 2 & 2 
	\end{ytableau} \fontsize{12pt}{0} \, , \,
	[4]
} & =
\frac{1}{4\sqrt{6}} \left[
4 \tr X^2 Y^2 
+ 2 \tr \left( X Y \right)^2 
+ 4 \tr X^2 Y \tr Y 
+ 4 \tr X \tr X Y^2 \right. \nonumber \\
& \hspace{50pt} + 2 \left( \tr X Y \right)^2 
+ \tr X^2 \tr Y^2 
+ \left( \tr X \right)^2 \left( \tr Y \right)^2 \nonumber \\
& \left. \hspace{50pt} + \tr X^2 \left( \tr Y \right)^2 
+ 4 \tr X \tr X Y \tr Y 
+ \left( \tr X \right)^2 \tr Y^2 
\right]
\end{align}
where we have suppressed the trivial multiplicity indices, and omitted $\Lambda$ as this is determined by the shape of the semi-standard tableau $M_\Lambda$.

For $\Lambda = [3,1]$, the orthogonalisation process produces
\begin{align}
Q_{
	\fontsize{6pt}{0} \begin{ytableau}
	1 & 1 & 2 \\ 2 
	\end{ytableau} \fontsize{12pt}{0} \, , \,
	[2, 1, 1]
} & =
\frac{1}{4} \left[
- 2 \tr X^2 Y \tr Y 
+ 2 \tr X \tr X Y^2 
+ \tr X^2 \left( \tr Y \right)^2 - \left( \tr X \right)^2 \tr Y^2 
\right] \\
Q_{
	\fontsize{6pt}{0} \begin{ytableau}
	1 & 1 & 2 \\ 2 
	\end{ytableau} \fontsize{12pt}{0} \, , \,
	[3, 1]
} & =
\frac{1}{4} \left[
2 \tr X^2 Y \tr Y 
- 2 \tr X \tr X Y^2 
+ \tr X^2 \left( \tr Y \right)^2 
- \left( \tr X \right)^2 \tr Y^2 
\right]
\label{symmetrised trace operator 7}
\end{align}
For both these sectors, the operators obtained are identical to the free field BPS operators, given for field content $(2,2)$ in table \ref{table: U(2) field content 2,2 basis}, where the weak coupling label $p$ matches the zero coupling label $R$. At $n=4$, this is largely pre-determined by SEP-compatibility of the two bases. 

Since the $\Lambda = [4]$ and $[3,1]$ sectors remain unchanged as we turn on interactions, the free field BPS operators are also the weak coupling BPS operators. As they are also eigen-operators of $\cF_N$, the space of free field BPS operators is the same as the space of pre-BPS operators. Moreover, in these sectors there are no commutator traces, and therefore SEP-compatibility for the pre-BPS operators is the same as that for BPS operators, an operator with $l(p) > N$ vanishes identically.

Therefore for $\Lambda = [4]$ and $[3,1]$, $Q_{\Lambda, M_\Lambda, p, \nu}$ and $\cO_{\Lambda, M_\Lambda, R, \tau}$ are both SEP-compatible bases for the same space. This means the spaces spanned by partitions of a given length are the same. These spaces are all one-dimensional with the exception of $\Lambda = [4]$, $p=[2,2]$ and $p=[3,1]$. Therefore the matching of all but these two is trivial. A priori, the matching between $p$ and $R$ for $p = [3,1]$ and $[2,2]$ is surprising.

More generally, we find the same behaviour for all $n \leq 7$, which are within reach of numerical calculations. The $p$ label in $Q_{\Lambda, M_\Lambda, p, \nu}$ matches the $R$ label in $\cO_{\Lambda, M_\Lambda, R, \tau}$ for $\Lambda = [n]$ and $[n-1,1]$. Since neither basis specifies exactly how the multiplicities $\nu$ and $\tau$ are chosen, these will not necessarily match, but they do span the same space. We go into this in more detail in section \ref{section: lambda=[n],[n-1,1]}. For $\Lambda = [n]$, the matching follows from the fact the Kostka numbers converting the monomial basis to the Schur basis \eqref{schur definition from monomials} are upper diagonal in partition indices. For $\Lambda = [n-1,1]$ we leave the matching at general $n$ as a conjecture.

\subsubsection{$\Lambda = [2,2]$ sector}

The orthogonalised basis of pre-BPS operators for $\Lambda = [2,2]$ is
\begin{align}
Q_{
	\fontsize{6pt}{0} \begin{ytableau}
	1 & 1 \\ 2 & 2 
	\end{ytableau} \fontsize{12pt}{0} \, , \,
	[2, 1, 1]
} & =
\frac{1}{6} \left( - 2 t_{[2,2]}  + t_{[2,1,1]} \right)
\label{lambda = [2,2], p = [2,1,1]} \\
Q_{
	\fontsize{6pt}{0} \begin{ytableau}
	1 & 1 \\ 2 & 2 
	\end{ytableau} \fontsize{12pt}{0} \, , \,
	[2, 2]
} & =
\frac{1}{3\sqrt{2}} \left( t_{[2,2]}+ t_{[2,1,1]} \right)
\label{lambda = [2,2], p = [2,2]}
\end{align}
where the trace combinations are
\begin{align}
t_{[2,1,1]} & = \tr X^2 \left( \tr Y \right)^2  - 2 \tr X \tr X Y \tr Y + \left( \tr X \right)^2 \tr Y^2 \\
t_{[2,2]} & = \tr X^2 \tr Y^2 - \left( \tr X Y \right)^2
\end{align}
These trace combinations are the $\Lambda = [2,2]$, $M_\Lambda = 
\fontsize{6pt}{0} \begin{ytableau}
1 & 1  \\ 2 & 2
\end{ytableau} \fontsize{12pt}{0}$ parts of the symmetrised trace covariant basis defined in \eqref{covariant symmetrised trace}.

In order to check the SEP-compatibility of \eqref{lambda = [2,2], p = [2,1,1]} and \eqref{lambda = [2,2], p = [2,2]}, we express them as a linear combination of the zero coupling basis given in table \ref{table: U(2) field content 2,2 basis}. We have
\begin{align}
Q_{
	\fontsize{6pt}{0} \begin{ytableau}
	1 & 1 \\ 2 & 2 
	\end{ytableau} \fontsize{12pt}{0} \, , \,
	[2, 1, 1]
} & = - \frac{1}{2 \sqrt{3}} \left( \begin{gathered}
\mathcal{O}_{
	\fontsize{6pt}{0} \begin{ytableau}
	1 & 1 \\ 2 & 2 
	\end{ytableau} \fontsize{12pt}{0}\, , \,
	\ydiagram{3,1}
} + \sqrt{2}
\mathcal{O}_{
	\fontsize{6pt}{0} \begin{ytableau}
	1 & 1 \\ 2 & 2 
	\end{ytableau} \fontsize{12pt}{0}\, , \,
	\ydiagram{2,2}
} + 3 
\mathcal{O}_{
	\fontsize{6pt}{0} \begin{ytableau}
	1 & 1 \\ 2 & 2 
	\end{ytableau} \fontsize{12pt}{0}\, , \,
	\ydiagram{2,1,1}
} \end{gathered}
\right)
\label{random ref 3} \\
Q_{
	\fontsize{6pt}{0} \begin{ytableau}
	1 & 1 \\ 2 & 2 
	\end{ytableau} \fontsize{12pt}{0} \, , \,
	[2, 2]
} & = \frac{1}{\sqrt{3}} \left( \begin{gathered} \sqrt{2} 
\mathcal{O}_{
	\fontsize{6pt}{0} \begin{ytableau}
	1 & 1 \\ 2 & 2 
	\end{ytableau} \fontsize{12pt}{0}\, , \,
	\ydiagram{3,1}
} - \mathcal{O}_{
	\fontsize{6pt}{0} \begin{ytableau}
	1 & 1 \\ 2 & 2 
	\end{ytableau} \fontsize{12pt}{0}\, , \,
	\ydiagram{2,2}
} \end{gathered} \right)
\label{random ref 1}
\end{align}
The only commutator trace at field content $(2,2)$ is
\begin{align}
\tr \, X[X,Y]Y & = \tr X^2 Y^2 - \tr (XY)^2 \nonumber \\
& = \frac{\sqrt{3}}{2} \left( \begin{gathered}
\mathcal{O}_{
	\fontsize{6pt}{0} \begin{ytableau}
	1 & 1 \\ 2 & 2 
	\end{ytableau} \fontsize{12pt}{0}\, , \,
	\ydiagram{3,1}
} + \sqrt{2}
\mathcal{O}_{
	\fontsize{6pt}{0} \begin{ytableau}
	1 & 1 \\ 2 & 2 
	\end{ytableau} \fontsize{12pt}{0}\, , \,
	\ydiagram{2,2}
}  -
\mathcal{O}_{
	\fontsize{6pt}{0} \begin{ytableau}
	1 & 1 \\ 2 & 2 
	\end{ytableau} \fontsize{12pt}{0}\, , \,
	\ydiagram{2,1,1} 
} \end{gathered} \right)
\label{commutator}
\end{align}
It is simple to check that for $N \geq 3$
\begin{equation}
\left\langle \left. \begin{gathered} Q_{
	\fontsize{6pt}{0} \begin{ytableau}
	1 & 1 \\ 2 & 2 
	\end{ytableau} \fontsize{12pt}{0} \, , \,
	[2, 1, 1]
} \end{gathered} \right| \tr \, X[X,Y]Y \right\rangle_{S_n} 
= \left\langle \left. \begin{gathered} Q_{
\fontsize{6pt}{0} \begin{ytableau}
1 & 1 \\ 2 & 2 
\end{ytableau} \fontsize{12pt}{0} \, , \,
[2, 2]
} \end{gathered} \right| \tr \, X[X,Y]Y \right\rangle_{S_n} = 0
\end{equation}
At $N=2$, comparing \eqref{random ref 3} with \eqref{commutator} and recalling that $\mathcal{O}_{
	\fontsize{6pt}{0} \begin{ytableau}
	1 & 1 \\ 2 & 2 
	\end{ytableau} \fontsize{12pt}{0}\, , \,
	\ydiagram{2,1,1} 
}$ vanishes, we have
\begin{equation}
Q_{
	\fontsize{6pt}{0} \begin{ytableau}
	1 & 1 \\ 2 & 2 
	\end{ytableau} \fontsize{12pt}{0} \, , \,
	[2, 1, 1]
} = - \frac{1}{3} \tr X [X,Y] Y
\label{symmetrised trace = commutator trace}
\end{equation}
Therefore \eqref{random ref 3} is no longer pre-BPS. The other operator \eqref{random ref 1} is still orthogonal to the commutator.

At $N=1$, both \eqref{random ref 3} and \eqref{random ref 1} vanish identically. Combined with the behaviour at $N=2$, this demonstrates that these two operators form an SEP-compatible basis for pre-BPS operators in the $\Lambda = [2,2]$ sector.

At $N=2$, the finite $N$ relations mean that \eqref{random ref 3} can be written as both a symmetrised trace and a commutator trace. This is discussed in more generality in section \ref{sec:Projectors}, where we develop the finite $N$ vector space geometry responsible for transforming between the two types of traces.

\subsection{Normalisation conventions for BPS operators}
\label{section: normalisation of BPS operators}

The final step to obtain an SEP-compatible basis for weakly coupled BPS operators is to apply $\cG_N$ to $Q_{\Lambda, M_\Lambda, p, \nu}$, where $l(p) \leq N$. However, the operators $\cG_N Q_{\Lambda, M_\Lambda, p, \nu}$ contain denominators of the form $(N-i)$ for $i \leq l(p)$ which make it difficult to see how they should behave when we lower $N$ to $\hN = i$.

In this section we prove that by normalising $\cG_N Q_{\Lambda, M_\Lambda, p, \nu}$ under the $S_n$ inner product, we remove these denominators and obtain an SEP-compatible basis of BPS operators. We start by continuing the example of $\Lambda = [2,2]$ from the previous subsection to show some of the behaviour that occurs. Working at $N \geq 3$, we have
\begin{align}
\cG_N \, Q_{
	\fontsize{6pt}{0} \begin{ytableau}
	1 & 1 \\ 2 & 2 
	\end{ytableau} \fontsize{12pt}{0} \, , \,
	[2, 1, 1]
} & = - \frac{1}{2 \sqrt{3} N (N-1)(N+1)} \left( \begin{gathered}
\frac{1}{N+2}
\mathcal{O}_{
	\fontsize{6pt}{0} \begin{ytableau}
	1 & 1 \\ 2 & 2 
	\end{ytableau} \fontsize{12pt}{0}\, , \,
	\ydiagram{3,1}
} + \frac{\sqrt{2}}{N}
\mathcal{O}_{
	\fontsize{6pt}{0} \begin{ytableau}
	1 & 1 \\ 2 & 2 
	\end{ytableau} \fontsize{12pt}{0}\, , \,
	\ydiagram{2,2}
} \end{gathered} \right.  \nonumber \\
& \hspace{180pt} \left. \begin{gathered}
+ \frac{3}{N-2}
\mathcal{O}_{
	\fontsize{6pt}{0} \begin{ytableau}
	1 & 1 \\ 2 & 2 
	\end{ytableau} \fontsize{12pt}{0}\, , \,
	\ydiagram{2,1,1}
} \end{gathered}
\right)
\label{GQ1} \\
\cG_N \, Q_{
	\fontsize{6pt}{0} \begin{ytableau}
	1 & 1 \\ 2 & 2 
	\end{ytableau} \fontsize{12pt}{0} \, , \,
	[2, 2]
} & = \frac{1}{\sqrt{3}N (N-1) (N+1)} \left( \begin{gathered} \frac{\sqrt{2}}{N+2}
\mathcal{O}_{
	\fontsize{6pt}{0} \begin{ytableau}
	1 & 1 \\ 2 & 2 
	\end{ytableau} \fontsize{12pt}{0}\, , \,
	\ydiagram{3,1}
} - \frac{1}{N} \mathcal{O}_{
	\fontsize{6pt}{0} \begin{ytableau}
	1 & 1 \\ 2 & 2 
	\end{ytableau} \fontsize{12pt}{0}\, , \,
	\ydiagram{2,2}
} \end{gathered} \right)
\end{align}
Now consider lower $N$ to $\hN=2$ and imposing the finite $\hN$ cut-off. It is unclear how we should treat \eqref{GQ1}, since the operator  $\mathcal{O}_{
	\fontsize{6pt}{0} \begin{ytableau}
	1 & 1 \\ 2 & 2 
	\end{ytableau} \fontsize{12pt}{0}\, , \,
	\ydiagram{2,1,1} 
}$ vanishes, yet we also have a division by zero. Let us resolve the ambiguity by declaring that this term should indeed vanish. Then at $\hN=2$ we have
\begin{equation}
\cG_N \, Q_{
	\fontsize{6pt}{0} \begin{ytableau}
	1 & 1 \\ 2 & 2 
	\end{ytableau} \fontsize{12pt}{0} \, , \,
	[2, 1, 1]
} = - \frac{1}{12 \sqrt{3}} \left( \begin{gathered}
\frac{1}{4}
\mathcal{O}_{
	\fontsize{6pt}{0} \begin{ytableau}
	1 & 1 \\ 2 & 2 
	\end{ytableau} \fontsize{12pt}{0}\, , \,
	\ydiagram{3,1}
} + \frac{1}{\sqrt{2}}
\mathcal{O}_{
	\fontsize{6pt}{0} \begin{ytableau}
	1 & 1 \\ 2 & 2 
	\end{ytableau} \fontsize{12pt}{0}\, , \,
	\ydiagram{2,2}
} \end{gathered} \right)
\label{broken SEP-compatibility}
\end{equation}
This is a perfectly well defined operator, yet $\cG_N Q_{\Lambda, M_\Lambda, p, \nu}$ was meant to be an SEP-compatible basis. For BPS operators, this means \eqref{GQ1} should vanish after reducing $N$ to $\hN=2$.

The resolution of this problem is to normalise $\cG_N Q_{\Lambda, M_\Lambda, p, \nu}$ in the $S_n$ inner product. Define
\begin{equation}
Q^{BPS}_{\Lambda, M_\Lambda, p, \nu} = \frac{ \cG_N \, Q_{\Lambda, M_\Lambda, p, \nu} }{\sqrt{ \left\langle \cG_N \, Q_{\Lambda, M_\Lambda, p, \nu} | \cG_N \, Q_{\Lambda, M_\Lambda, p, \nu} \right\rangle_{S_n} }}
\label{BPS covariant basis}
\end{equation}
For $\Lambda = [2,2]$ we have
\begin{align}
Q^{BPS}_{
	\fontsize{6pt}{0} \begin{ytableau}
	1 & 1 \\ 2 & 2 
	\end{ytableau} \fontsize{12pt}{0} \, , \,
	[2, 1, 1]
} & = - \frac{1}{2 \sqrt{P_1(N)}} \left( \begin{gathered}
N(N-2)
\mathcal{O}_{
	\fontsize{6pt}{0} \begin{ytableau}
	1 & 1 \\ 2 & 2 
	\end{ytableau} \fontsize{12pt}{0}\, , \,
	\ydiagram{3,1}
} + \sqrt{2}(N+2)(N-2)
\mathcal{O}_{
	\fontsize{6pt}{0} \begin{ytableau}
	1 & 1 \\ 2 & 2 
	\end{ytableau} \fontsize{12pt}{0}\, , \,
	\ydiagram{2,2}
} \end{gathered} \right.  \nonumber \\
& \hspace{180pt} \left. \begin{gathered}
+ 3 N (N+2)
\mathcal{O}_{
	\fontsize{6pt}{0} \begin{ytableau}
	1 & 1 \\ 2 & 2 
	\end{ytableau} \fontsize{12pt}{0}\, , \,
	\ydiagram{2,1,1}
} \end{gathered}
\right)
\label{QBPS1} \\
Q^{BPS}_{
	\fontsize{6pt}{0} \begin{ytableau}
	1 & 1 \\ 2 & 2 
	\end{ytableau} \fontsize{12pt}{0} \, , \,
	[2, 2]
} & = \frac{1}{\sqrt{P_2(N)}} \left( \begin{gathered} \sqrt{2} N
\mathcal{O}_{
	\fontsize{6pt}{0} \begin{ytableau}
	1 & 1 \\ 2 & 2 
	\end{ytableau} \fontsize{12pt}{0}\, , \,
	\ydiagram{3,1}
} - (N+2)\mathcal{O}_{
	\fontsize{6pt}{0} \begin{ytableau}
	1 & 1 \\ 2 & 2 
	\end{ytableau} \fontsize{12pt}{0}\, , \,
	\ydiagram{2,2}
} \end{gathered} \right)
\label{QBPS2}
\end{align}
where the normalisation polynomials are
\begin{align}
P_1 (N) & = 3 N^4 + 8 N^3 + 6 N^2 + 8 \\
P_2 (N) & = 3 N^2 + 4N + 4
\end{align}
There is now no ambiguity in the definitions of the operators after lowering $N$ to $\hN=1,2$, and they vanish identically for $\hN < l(p)$, thereby forming an SEP-compatible basis for BPS operators.

We now generalise to arbitrary $\Lambda$. Take some operator $Q_{\Lambda, M_\Lambda, p, \nu}$ with $l(p) \leq N$ and expand it in terms of free field operators
\begin{equation}
Q_{\Lambda, M_\Lambda, p, \nu} = \sum_{\substack{R, \tau \\ l(R) \leq l(p)}} q^{R, \tau}_{p,\nu} \cO_{\Lambda, M_\Lambda, R, \tau}
\end{equation}
for some coefficients $q^{R,\tau}_{p,\nu}$. The limit $l(R) \leq l(p)$ on the sum was discussed below \eqref{Q normalisation}. Define the set
\begin{equation}
Y_{p,\nu} = \left\{ R : q^{R,\tau}_{p,\nu} \neq 0 \text{ for some } \tau \right\}
\end{equation}
Intuitively, $Y_{p,\nu}$ is the set of Young diagrams $R$ that contribute to $Q_{\Lambda, M_\lambda, p, \nu}$. Define $R^{max}_{p,\nu}$ to be the minimal Young diagram that contains every $R$ in $Y_{p, \nu}$ and $R^{min}_{p,\nu}$ to be the maximal Young diagram that is contained within every $R$ in $Y_{p,\nu}$. Define
\begin{equation}
f^{max}_{p,\nu} (N) = f_{R^{max}_{p,\nu}} (N) \hspace{50pt} f^{min}_{p,\nu} (N) = f_{R^{min}_{p,\nu}} (N)
\label{fmax}
\end{equation}
where $f_R(N)$ is a polynomial in $N$ depending on the shape of $R$. It is defined in \eqref{omega in a representation} as a product of linear factors. Intuitively, $f^{max}_{p,\nu}(N)$ is the lowest common multiple of the different $f_R$ for $R \in Y_{p, \nu}$, while $f^{min}_{p,\nu}(N)$ is the highest common factor.

As an example, consider the previous example with $\Lambda = [2,2]$ and $p=[2,1,1]$. We have
\begin{equation}
Y_{[2,1,1]} = \left\{ \begin{gathered} \ydiagram{3,1} \, , \, \ydiagram{2,2} \, , \, \ydiagram{2,1,1} \end{gathered} \right\}
\end{equation}
and
\begin{align}
R^{max}_{[2,1,1]} & = \ydiagram{3,2,1}  & \qquad  f^{max}_{[2,1,1]}(N) & = (N+2)(N+1)N^2(N-1)(N-2) \\
R^{min}_{[2,1,1]} & = \ydiagram{2,1}  & \qquad f^{min}_{[2,1,1]}(N) & = (N+1)N(N-1)
\end{align}
Using $f^{max}_{p, \nu}(N)$, we can factor out the denominators in $\cG_N Q_{\Lambda, M_\Lambda, p, \nu}$. We have
\begin{align}
\cG_N Q_{\Lambda, M_\Lambda, p, \nu} = \sum_{\substack{R, \tau \\ l(R) \leq l(p)}} \frac{q^{R, \tau}_{p,\nu}}{f_R(N)} \cO_{\Lambda, M_\Lambda, R, \tau} = \frac{1}{f^{max}_{p, \nu}(N)} \sum_{\substack{R, \tau \\ l(R) \leq l(p)}} q^{R, \tau}_{p,\nu} \frac{f^{max}_{p,\nu}(N)}{f_R(N)} \cO_{\Lambda, M_\Lambda, R, \tau}
\end{align}
where the coefficients $\frac{f^{max}_{p,\nu}(N)}{f_R(N)}$ are simple polynomials in $N$ made up of products of linear factors. Therefore the $S_n$ norm of $\cG_N Q_{\Lambda, M_\Lambda, p, \nu}$ is
\begin{equation}
\left| \cG_N \, Q_{\Lambda, M_\Lambda, p, \nu} \right|^2_{S_n} = \left\langle \cG_N \, Q_{\Lambda, M_\Lambda, p, \nu} | \cG_N \, Q_{\Lambda, M_\Lambda, p, \nu} \right\rangle_{S_n} = \frac{P_{p,\nu} (N)}{\left[ f^{max}_{p,\nu}(N) \right]^2}
\end{equation}
where the numerator polynomial is
\begin{equation}
P_{p,\nu} (N) = \sum_{\substack{R, \tau \\ l(R) \leq l(p)}} \left( q^{R, \tau}_{p,\nu}  \frac{f^{max}_{p,\nu}(N)}{f_R(N)} \right)^2
\label{norm poly}
\end{equation}
This polynomial is a sum of squares, and therefore can only vanish if all terms are zero. Since $\frac{f^{max}_{p,\nu}(N)}{f_R(N)}$ is a product of simple linear factors in $N$, this in turn can only occur if there is linear factor common to every term. From the definition of $f^{max}_{p,\nu}$ as the lowest common multiple of the $f_R$ that appear in the sum, this does not happen. Therefore $P_{p,\nu} (N)$ is positive for any $N$. In particular, it remains positive when we evaluate it on $\hN \leq N$. We have
\begin{equation}
Q^{BPS}_{\Lambda, M_\Lambda, p, \nu} = \frac{1}{\sqrt{P_{p,\nu}(N)}} \sum_{\substack{R, \tau \\ l(R) \leq l(p)}} q^{R, \tau}_{p,\nu} \frac{f^{max}_{p,\nu}(N)}{f_R(N)} \cO_{\Lambda, M_\Lambda, R, \tau}
\label{QBPS expansion}
\end{equation}
We now prove this vanishes when we lower $N$ to some $\hN < l(p)$. The properties of $P_{p,\nu}(N)$ mean there are no divisions by zero to concern us, and we focus on the coefficients $\frac{f^{max}_{p,\nu}(N)}{f_R(N)}$.

By construction, $Q_{\Lambda, M_\Lambda, p, \nu}$ reduces to a commutator trace when we lower $N$ to $\hN < l(p)$. Therefore it must contain at least one $\cO_{\Lambda, M_\Lambda, R, \tau}$ with $l(R) = l(p)$, otherwise it would remain $S_n$ orthogonal to commutator traces if we lowered $N$ to $\hN = l(p)-1$. Therefore
\begin{equation}
f^{max}_{p,\nu} (N) \supset \prod_{i=0}^{l(p)-1} (N - i )
\label{fmax factors}
\end{equation}
where $\supset$ means $f^{max}_{p,\nu}$ contains these as factors. By definition
\begin{equation}
f_R (N) \supset \prod_{i=0}^{l(R)-1} (N-i)
\end{equation}
it follows that if $R \in Y_{p, \nu}$, the coefficient in front of $\cO_{\Lambda, M_\Lambda, R, \tau}$ contains
\begin{equation}
\frac{f^{max}_{p,\nu}(N)}{f_R(N)}  \supset \prod_{i=l(R)}^{l(p)-1} (N-i)
\label{factors}
\end{equation}
This ensures that if $l(p) > \hN$, all terms in the expansion \eqref{QBPS expansion} vanish when we lower $N$ to $\hN$. If $l(R) > \hN$, $\cO_{\Lambda, M_\Lambda, R, \tau}$ vanishes by definition, while if $l(R) \leq \hN$, the factors in \eqref{factors} set it to zero. 

Therefore $Q^{BPS}_{\Lambda, M_\Lambda, p, \nu}$ vanishes identically when we lower $N$ to $\hN < l(p)$, and hence this is an SEP-compatible basis for weakly coupled BPS operators. This justifies the statement made in section \ref{sec:BCOV} that applying $\cG_N$ to an SEP-compatible basis of pre-BPS operators gives an SEP-compatible basis of BPS operators.

An alternative viewpoint is to look at the physical $\cF$-weighted inner product. We have
\begin{equation}
\left| Q^{BPS}_{\Lambda, M_\Lambda, p, \nu} \right|^2_{\cF} = \left\langle Q_{\Lambda, M_\Lambda, p, \nu} | Q_{\Lambda, M_\Lambda, p, \nu} \right\rangle_{\cF} = \frac{f^{max}_{p,\nu}(N) P^{(F)}_{p,\nu}(N)}{P_{p, \nu}(N)}
\label{QFnorm}
\end{equation}
where the new polynomial in the numerator is
\begin{equation}
P^{(F)}_{p,\nu}(N) = \sum_{\substack{R, \tau \\ l(R) \leq l(p)}} \left( q^{R,\tau}_{p,\nu} \right)^2 \frac{f^{max}_{p,\nu}(N)}{f_R(N)}
\end{equation}
The overall factor of $f^{max}_{p,\nu}(N)$ in \eqref{QFnorm} ensures that it vanishes when we lower $N$ to some $\hN < l(p)$. As a consistency check, we also prove that it is non-vanishing when $\hN \geq l(p)$. 

This relies on noticing that the linear factors in $\frac{f^{max}_{p,\nu}(N)}{f_R(N)}$ are all of the form $(N-i)$ for $-n \leq i \leq l(p) -1$. Therefore $\frac{f^{max}_{p,\nu}(\hN)}{f_R(\hN)} > 0$ when $\hN \geq l(p)$. It follows that $P^{(F)}_{p,\nu}(\hN) > 0$, and the result follows.

In the planar ($N \to \infty$) limit, applying $\cG_N$ reduces to division by $N^n$, and therefore after $S_n$-normalising we have
\begin{equation}
\left. Q^{BPS}_{\Lambda, M_\Lambda, p, \nu} \right|_{N \to \infty} = Q_{\Lambda, M_\Lambda, p, \nu}
\label{Q planar limit}
\end{equation}
Since $Q_{\Lambda, M_\Lambda, p, \nu}$ is a symmetrised trace at large $N$, this means commutator traces are sub-leading in the large $N$ multi-trace expansion of $Q^{BPS}_{\Lambda, M_\Lambda, p, \nu}$.

Having constructed an SEP-compatible basis of BPS operators, the natural next question to ask is whether we can find a formula for their correlators. This uses the physical $\cF$-weighted inner product. From the hermiticity of $U(2)$, the $Q^{BPS}_{\Lambda, M_\Lambda, p, \nu}$ are $\cF$-orthogonal in the $\Lambda, M_\Lambda$ labels, but in general are not in $p, \nu$. Therefore studying correlators involves calculating a matrix of inner products. In the next section we $\cF$-orthogonalise the $Q^{BPS}_{\Lambda, M_\Lambda, p, \nu}$ in order to produce an $\cF$-orthogonal SEP-compatible basis in which it is easier to study properties of correlators.

In this section we have normalised operators using the $S_n$ inner product. This is in some sense un-natural, as the $Q^{BPS}_{\Lambda, M_\Lambda, p, \nu}$ are not orthogonal in this inner product. There is another, alternative normalisation we could consider. In \eqref{QBPS expansion} we have a complicated normalisation factor of $\left( P_{p,\nu} (N) \right)^{-1/2}$, which if removed, would mean the coefficients of the free field operators (and multi-traces) would be expressible purely as polynomials in $N$. This is a natural normalisation to consider, and is given simply by
\begin{equation}
f^{max}_{p, \nu} \cG_N \, Q_{\Lambda, M_\Lambda, p, \nu}
\end{equation}
However, this is more difficult than the $S_n$ normalisation, since it involves knowing which free field operators appear in the expansion of $Q_{\Lambda, M_\Lambda, p, \nu}$, and the free field operators are computationally expensive to construct. In contrast, the $S_n$ normalisation can be deduced purely from an expression in terms of multi-traces, which is obtainable explicitly from the construction of $Q_{\Lambda, M_\Lambda, p, \nu}$.

In section \ref{section: lambda=[n],[n-1,1]} we prove that for $\Lambda = [n]$, the $Q_{\Lambda, M_\Lambda, p, \nu}$ exactly reproduce the free field basis $\cO_{\Lambda, M_\Lambda, R, \tau}$ up to a choice of multiplicity basis, and conjecture that the same happens for $\Lambda = [n-1,1]$. If this is true, then for these $\Lambda$, applying $\cG$ and $S_n$ normalising leaves the operators unchanged and we have
\begin{equation}
Q^{BPS}_{\Lambda, M_\Lambda, p, \nu} = Q_{\Lambda, M_\Lambda, p, \nu} = \cO_{\Lambda, M_\Lambda, R=p, \tau=\nu}
\end{equation}

\subsection{$\cF$-orthogonalisation}
\label{section: F orthogonalisation}

To define an orthogonal SEP-compatible basis, we Gram-Schmidt orthogonalise the $Q^{BPS}_{\Lambda, M_\Lambda, p, \nu}$ operators using the physical $\cF$-weighted inner product and the same ordering defined in \eqref{partition ordering}. We denote the orthogonalised operators by $\bar{S}^{BPS}_{\Lambda, M_\Lambda, p, \nu}$, and after normalising them to have $S_n$ norm 1, $S^{BPS}_{\Lambda, M_\Lambda, p, \nu}$.

Let $p^*_{\Lambda;N}$ be the largest partition with $\cM_{\Lambda, p^*_{\Lambda;N}} > 0$ and $l(p^*_{\Lambda;N}) \leq N$. Then in analogy to \eqref{longest partition operator}, \eqref{orthogonalisation of multi-symmetric functions} and \eqref{Q normalisation}, we have
\begin{align}
\bar{S}^{BPS}_{\Lambda, M_\Lambda, p^*_{\Lambda;N}} & = Q^{BPS}_{\Lambda, M_\Lambda, p^*_{\Lambda;N}}
\label{first Sbar} \\
\bar{S}^{BPS}_{\Lambda, M_\Lambda, p,\nu} & = Q^{BPS}_{\Lambda, M_\Lambda, p,\nu} - 
\sum_{\substack{(q,\eta) > (p, \nu) \\ l(q) \leq N}} \frac{ \left\langle Q^{BPS}_{\Lambda, M_\Lambda, p,\nu} , \bar{S}^{BPS}_{\Lambda, M_\Lambda, q, \eta} \right\rangle_{\cF} }
{ \left\langle \bar{S}^{BPS}_{\Lambda, M_\Lambda, q, \eta}, \bar{S}^{BPS}_{\Lambda, M_\Lambda, q, \eta} \right\rangle_{\cF} }
\bar{S}^{BPS}_{\Lambda, M_\Lambda, q, \eta} \\
S^{BPS}_{\Lambda, M_\Lambda, p, \nu} & = \frac{ \bar{S}^{BPS}_{\Lambda, M_\Lambda, p, \nu} }{\sqrt{ \left\langle \bar{S}^{BPS}_{\Lambda, M_\Lambda, p, \nu} | \bar{S}^{BPS}_{\Lambda, M_\Lambda, p, \nu} \right\rangle_{S_n} }}
\label{S normalisation}
\end{align}
where $l(p) \leq N$.

Note the difference in the starting point of the orthogonalisation compared to \eqref{longest partition operator}. When $S_n$ orthogonalising the pre-BPS operators, we began with $p^*_{\Lambda} = p^*_{\Lambda;\infty}$ even if $l(p^*_\Lambda) > N$, whereas this time we apply the finite $N$ cut-off to the partitions being orthogonalised. 

From the construction it follows that $Q^{BPS}_{\Lambda, M_\Lambda, q, \eta}$ only contributes to $S^{BPS}_{\Lambda, M_\Lambda, p, \nu}$ if $q \geq p$. Upon lowering $N$ to $\hN$, since the $Q^{BPS}$ operators with $l(q) > \hN$ vanish identically, the $S^{BPS}$ will also vanish for $l(p) > \hN$, and therefore this is an SEP-compatible basis. 

Note this relies on $Q^{BPS}_{\Lambda, M_\Lambda, q, \eta}$ not appearing with a coefficient of $\frac{1}{N-i}$ for $i \leq l(q)$, as this would upset the SEP-compatibility in a way similar to that described in \eqref{broken SEP-compatibility}. As in the previous section, the $S_n$ normalisation ensures this does not occur.

In the planar ($N \to \infty$) limit, the physical $\cF$-weighted inner product reduces to $N^n$ times the $S_n$ inner product, therefore $\cF$-orthogonalising is equivalent to $S_n$-orthogonalising. From \eqref{Q planar limit}, the $Q^{BPS}_{\Lambda, M_\Lambda, p, \nu}$ operators reduce to the pre-BPS operators $Q_{\Lambda, M_\Lambda, p, \nu}$ in the planar limit, which are already $S_n$ orthonormal. Therefore
\begin{equation}
\left. S^{BPS}_{\Lambda, M_\Lambda, p, \nu} \right|_{N \to \infty} = Q_{\Lambda, M_\Lambda, p, \nu}
\label{S planar limit}
\end{equation}
Below \eqref{Q normalisation} we explained that only those free field operators with $l(R) \leq l(p)$ contribute to $Q_{\Lambda, M_\Lambda, p, \nu}$. After orthogonalisation, the $S^{BPS}_{\Lambda, M_\Lambda, p, \nu}$ can admit operators with $l(R) > l(p)$, but \eqref{S planar limit} proves that these are sub-leading at large $N$. 

Furthermore, just as discussed for $Q^{BPS}_{\Lambda, M_\Lambda, p, \nu}$ below \eqref{Q planar limit}, commutator traces are sub-leading in the large $N$ multi-trace expansion of $S^{BPS}_{\Lambda, M_\Lambda, p, \nu}$.

\subsubsection{The $\Lambda = [2,2]$ example}
\label{section: [2,2] S operators}

We begin with the operators \eqref{QBPS1} and \eqref{QBPS2}. After $\cF$-orthogonalising and normalising with respect to the $S_n$ inner product, we obtain
\begin{align}
S^{BPS}_{
	\fontsize{6pt}{0} \begin{ytableau}
	1 & 1 \\ 2 & 2 
	\end{ytableau} \fontsize{12pt}{0} \, , \,
	[2,1,1]
}
& = - \frac{1}{2 \sqrt{P_1 (N)}} \left( \begin{gathered} N(N-2) 
\mathcal{O}_{
	\fontsize{6pt}{0} \begin{ytableau}
	1 & 1 \\ 2 & 2 
	\end{ytableau} \fontsize{12pt}{0}\, , \,
	\ydiagram{3,1}
} + \sqrt{2} (N-2)(N+2)
\mathcal{O}_{
	\fontsize{6pt}{0} \begin{ytableau}
	1 & 1 \\ 2 & 2 
	\end{ytableau} \fontsize{12pt}{0}\, , \,
	\ydiagram{2,2}
} \end{gathered} \right. \nonumber \\ & \hspace{210pt} \left. \begin{gathered} + 3 N(N+2)
\mathcal{O}_{
	\fontsize{6pt}{0} \begin{ytableau}
	1 & 1 \\ 2 & 2 
	\end{ytableau} \fontsize{12pt}{0}\, , \,
	\ydiagram{2,1,1} 
} \end{gathered} \right)  
\label{[2,2],[2,1,1] bps} \\
S^{BPS}_{
	\fontsize{6pt}{0} \begin{ytableau}
	1 & 1 \\ 2 & 2 
	\end{ytableau} \fontsize{12pt}{0} \, , \,
	[2,2]
} & = \frac{1}{\sqrt{2P_2 (N)}} \left( \begin{gathered}  (2N-1)
\mathcal{O}_{
	\fontsize{6pt}{0} \begin{ytableau}
	1 & 1 \\ 2 & 2 
	\end{ytableau} \fontsize{12pt}{0}\, , \,
	\ydiagram{3,1}
} - \sqrt{2} (N+1)
\mathcal{O}_{
	\fontsize{6pt}{0} \begin{ytableau}
	1 & 1 \\ 2 & 2 
	\end{ytableau} \fontsize{12pt}{0}\, , \,
	\ydiagram{2,2}
} \end{gathered} \right. \nonumber \\ & \hspace{210pt} \left. \begin{gathered}  +
\mathcal{O}_{
	\fontsize{6pt}{0} \begin{ytableau}
	1 & 1 \\ 2 & 2 
	\end{ytableau} \fontsize{12pt}{0}\, , \,
	\ydiagram{2,1,1} 
} \end{gathered} \right)
\label{[2,2],[2,2] bps}
\end{align}
where the normalisation polynomials $P_1$ and $P_2$ are given by
\begin{align}
P_1 (N) & = 3 \, N^{4} + 8 \, N^{3} + 6 \, N^{2} + 8  \\
P_2 (N) & = 3 N^2 + 2
\end{align}
Written in terms of traces, these operators are
\begin{align}
S^{BPS}_{
	\fontsize{6pt}{0} \begin{ytableau}
	1 & 1 \\ 2 & 2 
	\end{ytableau} \fontsize{12pt}{0} \, , \,
	[2,1,1]
}
& = \frac{1}{2 \sqrt{3 P_1 (N)}} \bigg[  \left( N^{2} + 2 \, N - 2 \right) t_{[2,1,1]} - 2 N \left(N + 1\right)  t_{[2,2]} \nonumber \\ & \hspace{100pt} + 4 \left( N + 1 \right) \tr X [X,Y] Y  \bigg]
\label{bps from traces 1} \\
S^{BPS}_{
	\fontsize{6pt}{0} \begin{ytableau}
	1 & 1 \\ 2 & 2 
	\end{ytableau} \fontsize{12pt}{0} \, , \,
	[2,2]
}
& = \frac{1}{\sqrt{6 P_2 (N)}} \bigg[ N \big( t_{[2,1,1]} + t_{[2,2]} \big) - 2 \tr X [X,Y] Y \bigg]
\label{bps from traces 2}
\end{align}

\subsection{A shorter algorithm}
\label{section: quicker route} 

In the previous sections we have given a method to derive an orthogonal SEP-compatible basis $S^{BPS}_{\Lambda, M_\Lambda, p, \nu}$. This method goes through two orthogonalisation procedures, applying $\cG_N$ inbetween. The first step used the $S_n$ inner product and involved all the partitions, the second used the $\cF$-weighted inner product and only included those partitions with $l(p) \leq N$. This means the partitions with $l(p) \leq N$ are orthogonalised among each other twice. We now prove that one may simplify the first orthogonalisation procedure to only orthogonalise against those $p$ with $l(p) > N$ and still obtain the same final output. This simplifies the computational requirements, and is also conceptually simpler, for reasons that will be outlined below.

To prove this streamlined procedure produces the same BPS operators, we first recall some useful facts.

\subsubsection{General properties of Gram-Schmidt orthogonalisation}

Consider two bases $\{v_i\}$ and $\{e_i\}$ of a vector space $V$, where the second is orthonormal. Then orthogonalising $v_i$ results in the basis $e_i$ (up to normalisation constants) if and only if the matrix connecting the two is lower diagonal.
\begin{equation}
\{v_i\} \overset{\text{GS}}{\longrightarrow} \{e_i\} \quad \iff \quad e_i = \sum_{j \geq i} A^j_i v_j
\label{upper diagonal orthogonalisation}
\end{equation}
Note that the orthogonalisation process here is the opposite way round to the standard Gram-Schmidt orthogonalisation one would find in a textbook. Ordinarily, one starts with minimal $i$ and proceeds to larger $i$, and therefore obtains an upper diagonal $A$ matrix. In \eqref{upper diagonal orthogonalisation}, we start with maximal $i$ and decrease, since this is the approach taken in \eqref{orthogonalisation of multi-symmetric functions}.

Now introduce another basis $\{u_i\}$ for $V$, not necessarily orthogonal. This is related to $\{v_i\}$ by a a matrix $B^j_i$
\begin{equation}
u_i = \sum_i B^j_i v_j
\label{same orthogonalisation}
\end{equation}
Then it follows from \eqref{upper diagonal orthogonalisation} that if $B^j_i$ is lower diagonal, $\{u_i\}$ and $\{v_i\}$ orthogonalise to the same orthonormal basis $\{e_i\}$ (up to normalisation constants).

\subsubsection{Back to SEP-compatible bases}

As explained below \eqref{M< space}, the $S_n$ orthogonalisation of $\cM^\leq_N$ against $\cM^>_N$ gives us pre-BPS operators. The continued $S_n$ orthogonalisation among partitions with $l(p) \leq N$ gives us SEP-compatibility, but is not required for the operators to be pre-BPS.

To split these two steps us, define the pre-BPS basis $\bar{M}_{\Lambda, M_\Lambda, p, \nu}$ by
\begin{align}
\bar{M}_{\Lambda, M_\Lambda, p^*_\Lambda} & = M_{\Lambda, M_\Lambda, p^*_\Lambda} \\
\bar{M}_{\Lambda, M_\Lambda, p,\nu} & = M_{\Lambda, M_\Lambda, p,\nu} - 
\sum_{\substack{(q,\eta) > (p, \nu) \\ l(q) > N}} \frac{ \left\langle M_{\Lambda, M_\Lambda, p,\nu} , \bar{M}_{\Lambda, M_\Lambda, q, \eta} \right\rangle_{S_n} }
{ \left\langle \bar{M}_{\Lambda, M_\Lambda, q, \eta}, \bar{M}_{\Lambda, M_\Lambda, q, \eta} \right\rangle_{S_n} }
\bar{M}_{\Lambda, 	M_\Lambda, q, \eta}
\label{Mbar}
\end{align}
where for $p$ with $l(p) > N$, we have the same caveats as mentioned around \eqref{orthogonalisation of multi-symmetric functions} regarding vanishing of operators. The $\bar{M}_{\Lambda, M_\Lambda, p, \nu}$ operators were briefly mentioned above \eqref{sequence of spaces}. The corresponding BPS operators, normalised in the $S_n$ inner product, are
\begin{equation}
\bar{M}^{BPS}_{\Lambda, M_\Lambda, p, \nu} = \frac{ \cG_N \bar{M}_{\Lambda, M_\Lambda, p, \nu}}{\left| \cG_N \bar{M}_{\Lambda, M_\Lambda, p, \nu} \right|_{S_n} }
\end{equation}
From their construction, orthogonalising $\bar{M}_{\Lambda, M_\Lambda, p, \nu}$ all the way down the partitions using the $S_n$ inner product would result in the $S_n$ orthogonal basis $Q_{\Lambda, M_\Lambda, p, \nu}$ for pre-BPS operators. From \eqref{upper diagonal orthogonalisation}, these two bases are related by a lower diagonal matrix. Applying $\cG_N$ and $S_n$ normalising both bases, the equivalent BPS operators are also related by a (rescaled) lower diagonal matrix. Then from the discussion below \eqref{same orthogonalisation}, it follows that orthogonalising the two bases in the physical $\cF$-weighted inner product will result in the same final basis $S^{BPS}_{\Lambda, M_{\Lambda}, p, \nu}$.

Therefore one may obtain the $S^{BPS}_{\Lambda, M_\Lambda, p, \nu}$ basis in a simpler manner by $\cF$-orthogonalising the $\bar{M}_{\Lambda, M_\Lambda, p, \nu}$ basis, much as we did in (\ref{first Sbar}-\ref{S normalisation})
\begin{align}\label{SBPSGFconst} 
\bar{S}^{BPS}_{\Lambda, M_\Lambda, p^*_{\Lambda;N}} & = \bar{M}^{BPS}_{\Lambda, M_\Lambda, p^*_{\Lambda;N}} \\
\bar{S}^{BPS}_{\Lambda, M_\Lambda, p,\nu} & = \bar{M}^{BPS}_{\Lambda, M_\Lambda, p,\nu} - 
\sum_{\substack{(q,\eta) > (p, \nu) \\ l(q) \leq N}} \frac{ \left\langle \bar{M}^{BPS}_{\Lambda, M_\Lambda, p,\nu} , \bar{S}^{BPS}_{\Lambda, M_\Lambda, q, \eta} \right\rangle_{\cF} }
{ \left\langle \bar{S}^{BPS}_{\Lambda, M_\Lambda, q, \eta}, \bar{S}^{BPS}_{\Lambda, M_\Lambda, q, \eta} \right\rangle_{\cF} }
\bar{S}^{BPS}_{\Lambda, M_\Lambda, q, \eta} \\
S^{BPS}_{\Lambda, M_\Lambda, p, \nu} & = \frac{ \bar{S}^{BPS}_{\Lambda, M_\Lambda, p, \nu} }{\sqrt{ \left\langle \bar{S}^{BPS}_{\Lambda, M_\Lambda, p, \nu} | \bar{S}^{BPS}_{\Lambda, M_\Lambda, p, \nu} \right\rangle_{S_n} }}
\end{align}
where $l(p) \leq N$.

This approach to producing the $S^{BPS}_{\Lambda, M_\Lambda, p, \nu}$ operators still involves two orthogonalisation steps, but the first is now computationally less demanding, and can be completely skipped if $N \geq l \left( p^*_\Lambda \right)$. Moreover, there is a clearer conceptual separation between the two steps. The first one obtains pre-BPS operators, while the second one finds an SEP-compatible basis.

This process skips the $Q_{\Lambda, M_\Lambda, p, \nu}$ operators, but they still have physical relevance as the planar limit of $S^{BPS}_{\Lambda, M_\Lambda, p, \nu}$, and were mathematically useful in proving the SEP-compatibility of these operators.

\subsection{Choice of SEP-compatible basis}

In the section \ref{section: [2,2] S operators} we derived an orthogonal SEP-compatible basis of operators for the $\Lambda = [2,2]$ sector. This sector was also investigated in appendix C of \cite{CtoC}, and a different orthogonal SEP-compatible basis was found. One can check that the two bases span the same two dimensional space for any $N \geq 3$, and when $N=2$ we find exact agreement of operators. 

The SEP-compatibility determines the behaviour of such a basis for $N=1,2$, but for higher $N$ there is a large degree of freedom. Define the $\cF$-normalised operators $\widehat{S}^{BPS}_{\Lambda, M_\Lambda, p, \nu}$ for a generic $\Lambda$ by
\begin{equation}
\widehat{S}^{BPS}_{\Lambda, M_\Lambda, p, \nu} = \frac{ S^{BPS}_{\Lambda, M_\Lambda, p, \nu} }{ \left\langle S^{BPS}_{\Lambda, M_\Lambda, p, \nu} | S^{BPS}_{\Lambda, M_\Lambda, p, \nu} \right\rangle_\cF }
\label{normalised by F inner product}
\end{equation}
Let $c(N)$ and $s(N)$ be two functions of $N$ satisfying
\begin{equation}
c(2) = 1 \hspace{50pt} s(2) = 0 \hspace{50pt} c(N)^2 + s(N)^2 = 1
\end{equation}
We can use $c$ and $s$ to rotate the $\Lambda = [2,2]$ $\cF$-normalised basis operators to a new configuration
\begin{align}
\cO_1 & = c(N) \widehat{S}^{BPS}_{
	\fontsize{6pt}{0} \begin{ytableau}
	1 & 1 \\ 2 & 2 
	\end{ytableau} \fontsize{12pt}{0} \, , \,
	[2, 1, 1]
} + s(N) \widehat{S}_{
	\fontsize{6pt}{0} \begin{ytableau}
	1 & 1 \\ 2 & 2 
	\end{ytableau} \fontsize{12pt}{0} \, , \,
	[2, 2]
} \\
\cO_2 & = - s(N) \widehat{S}^{BPS}_{
	\fontsize{6pt}{0} \begin{ytableau}
	1 & 1 \\ 2 & 2 
	\end{ytableau} \fontsize{12pt}{0} \, , \,
	[2, 1, 1]
} + c(N) \widehat{S}_{
	\fontsize{6pt}{0} \begin{ytableau}
	1 & 1 \\ 2 & 2 
	\end{ytableau} \fontsize{12pt}{0} \, , \,
	[2, 2]
}
\end{align}
To avoid problems with vanishing denominators at $N=2$, we normalise $\cO_1, \cO_2$ to have norm 1 in the $S_n$ inner product. These then define an alternative orthogonal, SEP-compatible basis for weak coupling quarter-BPS operators in the $\Lambda = [2,2]$ sector.

As $s(N)$ is determined by $c(N)$, there is effectively a function's worth of freedom in defining an orthogonal SEP-compatible basis. Clearly the vast majority of these will have definitions with far more complicated coefficients than those in \eqref{[2,2],[2,1,1] bps} and \eqref{[2,2],[2,2] bps} (or equivalently \eqref{bps from traces 1} and \eqref{bps from traces 2}). An interesting question is whether we can uniquely characterise a basis by having the `nicest' coefficients. For example, the coefficients in the basis of \cite{CtoC} are of the form
\begin{equation}
N+1 \pm \sqrt{2N^2 + 1}
\end{equation}
These involve a sum of polynomial and surd terms, whereas the basis \eqref{[2,2],[2,1,1] bps} and \eqref{[2,2],[2,2] bps} has coefficients that are polynomial in $N$ up to an overall normalisation. One possible criterion would be to demand a basis with polynomial coefficients (up to overall normalisation) whose polynomials have minimal degree. If unique, these operators would in some sense be the `simplest' orthogonal, SEP-compatible basis. It is reasonable to conjecture that $S^{BPS}_{\Lambda, M_\Lambda, p, \nu}$ form this basis.

For more general $\Lambda$, take an $N$-dependent orthogonal rotation matrix $R(\Lambda;N)$ of size $\cM_\Lambda \times \cM_\Lambda$, where
\begin{equation}
\cM_\Lambda = \sum_{\substack{p \vdash n \\ l(p) \leq N}} \cM_{\Lambda, p}
\end{equation}
When evaluated at $\hN \leq N$, the matrix should split into diagonal blocks
\begin{equation}
R(\Lambda;\hN) = \begin{pmatrix}
R^\leq(\Lambda;\hN) & 0 \\
0 & R^> (\Lambda;\hN)
\end{pmatrix}
\end{equation}
where $R^>(\Lambda;\hN)$ rotates those partitions with length $l(p) > \hN$ among themselves and $R^\leq (\Lambda;\hN)$ rotates those partitions with length $l(p) \leq \hN$ among themselves.

Then consider the $S_n$ normalised versions of the operators
\begin{equation}
\sum_{\substack{q,\eta\\ l(q) \leq N}} R(\Lambda;N)^{p,\nu}_{q,\eta} \, \widehat{S}^{BPS}_{\Lambda, M_\Lambda, p, \nu}
\end{equation}
These form an alternative orthogonal, SEP-compatible basis for the weakly coupled quarter-BPS sector.

\subsection{Physical norms of BPS operators}
\label{section: norms of BPS operators}

For the free field operators \eqref{U(2) basis definition}, the physical norm $f_R$ is a polynomial in $N$ that is closely related to the corresponding Young diagram $R$. It has mathematical significance as the numerator of the Weyl dimension formula for $U(N)$ representations. We now investigate the physical norms of the weak coupling BPS operators $S^{BPS}_{\Lambda, M_\Lambda, p, \nu}$, beginning with an example at $\Lambda = [2,2]$. These operators are given in \eqref{[2,2],[2,1,1] bps} and \eqref{[2,2],[2,2] bps}, and have physical $\cF$-weighted norms
\begin{align}
\left| S^{BPS}_{
	\fontsize{6pt}{0} \begin{ytableau}
	1 & 1 \\ 2 & 2 
	\end{ytableau} \fontsize{12pt}{0} \, , \,
	[2,1,1]
} \right|^2 & =
\frac{{\left(3 \, N^{2} + 4 \, N - 2\right)} {\left(N + 2\right)} {\left(N + 1\right)} N^2 {\left(N - 1\right)} {\left(N - 2\right)} }{P_1}
\label{[2,2],[2,1,1] norm} \\
\left| S^{BPS}_{
	\fontsize{6pt}{0} \begin{ytableau}
	1 & 1 \\ 2 & 2 
	\end{ytableau} \fontsize{12pt}{0} \, , \,
	[2,2]
} \right|^2 & = 
\frac{{\left(3 \, N^{2} + 4 \, N - 2\right)} {\left(N + 1\right)} N^2 {\left(N - 1\right)}}{P_2}
\label{[2,2],[2,2] norm}
\end{align}
This has two key features that will generalise. Firstly, the liner factors in the norms reflect the SEP-compatibility, enforcing that the first operator vanish for when we lower $N$ to $\hN = 1,2$ and the second operator vanish when we lower $N$ to $\hN = 1$. However, both norms have more linear factors than just those required by SEP-compatibility. For these two $p$, the numerators contain $f_p$ as a factor. This does not generalise to all $p$; in \eqref{p=[2,2,2] norm} and \eqref{p=[3,3] norm} we see that the numerators in the norms of $p=[2,2,2]$ and $[3,3]$ operators are one linear factor short of containing $f_p$. It is unclear whether these are exceptions, or whether at large $n$, very few operators contain $f_p$ in the numerator of the norm. It would be interesting to understand the linear factors that appear in the numerator and whether these have a physical interpretation.

Secondly, the numerators share a factor of $(3 N^2 + 2N - 2)$. In appendices \ref{appendix: lambda = [3,2]}, \ref{appendix: lambda = [4,2]} and \ref{appendix: lambda = [3,3]} we see that consecutive partitions (in the ordering \eqref{partition ordering}) share a complicated polynomial factor in the numerators. We believe this generalises to larger $n$, though it may be an artefact of the orthogonalisation process. This is discussed further in appendix \ref{section: [4,2] norms}.

While the numerators of BPS norms have interesting properties, we have not found any structure in the denominators. They arise by dividing through by the square root of the $S_n$ norm, and from our numerical calculations do not seem to factorise into smaller units.

\subsubsection{Norms of operators with multiplicity}

The norms \eqref{[2,2],[2,1,1] norm} and \eqref{[2,2],[2,2] norm} can be considered as characteristic functions of $\Lambda$ along with the partitions $[2,1,1]$ and $[2,2]$ respectively, just as the norms of the free field basis are characteristic polynomials of the Young diagrams.

For the free field covariant basis \eqref{U(2) basis definition}, the physical norms depended only on $R$ and not the $U(2)$ Young diagram $\Lambda$ or the multiplicity $\tau$. For the weak coupling basis, the norms can now depend on $\Lambda, p$ and $\nu$. The dependence on $\Lambda$ and $p$ is completely determined by the construction, while the dependence on $\nu$ is dictated by the choice of multiplicity space basis. We now outline a way of extracting functions of $N$ that do not depend on this choice, and are therefore associated to the pair $p, \Lambda$.

In \eqref{normalised by F inner product}, a rescaled BPS basis $\widehat{S}^{BPS}_{\Lambda, M_\Lambda, p, \nu}$ was defined, orthonormal in the physical $\cF$-weighted inner product. A different choice of multiplicity basis would result in an orthogonal rotation of these operators, and any trace over the multiplicity index is therefore independent of the this choice. 

In particular, consider the matrix of $S_n$ inner products. For a trivial multiplicity space, this would be a $1 \times 1$ matrix containing the reciprocal of the norm $| S^{BPS}_{\Lambda, M_\Lambda, p, \nu} |^2$. The appropriate generalisation to non-trivial multiplicity should therefore be to take the reciprocal of the trace, and we should also divide by the dimension of the multiplicity space. So the invariant function is
\begin{equation}
f_{\Lambda, p} =  \left( \frac{1}{\cM_{\Lambda, p}} \sum_{\nu = 1}^{\cM_{\Lambda, p}} \left\langle \widehat{S}^{BPS}_{\Lambda, M_\Lambda,  p, \nu} | \widehat{S}^{BPS}_{\Lambda, M_\Lambda, p, \nu} \right\rangle \right)^{-1}
\end{equation}
Note that the the hermiticity properties of $U(2)$ imply that we can choose any semi-standard tableau $M_\Lambda$ and it will not affect the calculation.

We can also use the square/cube/... of the $S_n$ inner product matrix to extract further basis-invariant functions of $N$. Let $A$ be the $S_n$ inner product matrix. Then we have
\begin{align}
f^{(k)}_{\Lambda, p} =  \left( \frac{1}{\cM_{\Lambda, p}} \tr A^k \right)^{-\frac{1}{k}}
\end{align}
Where we have taken the $k$th root in order to have functions of the same degree in $N$. This stack of powers only goes up to $k \leq \cM_{\Lambda, p}$ before the invariants are no longer independent.

In appendix \ref{appendix: lambda = [4,2]}, we see two examples of non-trivial multiplicity spaces in the $\Lambda = [4,2]$ sector, both of dimension two. In section \ref{section: p=[2,2,1,1] and [3,2,1] norms} we show the calculation for $p = [2,2,1,1]$ in some detail, while we are more schematic for $p=[3,2,1]$.

In both examples, the numerator of $f^{(2)}_{\Lambda, p}$ is the same as $f^{(1)}_{\Lambda, p}$, though the denominator is not. As discussed below \eqref{[2,2],[2,2] norm}, this is further evidence that we should only look at the numerators of the BPS norms, giving a single characteristic function for a given $\Lambda, p$.

\subsection{Longest $p$ for a given $\Lambda$ and explicit quarter-BPS operators}
\label{section: longest lambda}

Due to the computational nature of the orthogonalisation process to derive SEP-compatible BPS operators, it is difficult to give explicit formulae for many of the $S^{BPS}_{\Lambda, M_\Lambda, p, \nu}$ operators. The exception is for $p^*_\Lambda,$ the longest partition with $\cM_{\Lambda, p} > 0$. Shortly, we will prove that this has $\cM_{\Lambda, p^*_\Lambda} = 1$, so we drop the multiplicity index. Provided $N \geq l \left( p^*_\Lambda \right)$, the operators with partition $p^*_\Lambda$ do not get orthogonalised, so we have the formula
\begin{empheq}[box=\fbox]{align}
\hspace{5pt} S^{BPS}_{\Lambda, M_\Lambda, p^*_\Lambda} = \frac{ \cG_N M_{\Lambda, M_\Lambda, p^*_\Lambda} }{\left| \cG_N M_{\Lambda, M_\Lambda, p^*_\Lambda} \right|_{S_n}} \hspace{5pt}
\end{empheq}
where 
\begin{align}
M_{\Lambda, M_\Lambda, p^*_\Lambda} & = \sum_{\bp \vdash (n_1,n_2) \, : \, p(\bp) = p^*_\Lambda} B^\bp_{\Lambda, M_\Lambda, p^*_\Lambda} M_\bp \\
& = \sum_{\substack{\bp \vdash (n_1,n_2) \, : \, p(\bp) = p^*_\Lambda \\ \bq \vdash (n_1,n_2)}} B^\bp_{\Lambda, M_\Lambda, p^*_\Lambda} \tC_\bp^\bq T_\bq
\end{align}
and we have used \eqref{U(2) covariant monomial basis} to express $M_{\Lambda, M_\Lambda, p^*_\Lambda}$ in terms of $M_\bp$, \eqref{definition of C and C-tilde} to write $M_\bp$ in terms of $T_\bq$, and $T_\bq$ is the symmetrised trace operator \eqref{symmetrised trace operator}.

Explicit formulae for other $S^{BPS}_{\Lambda, M_\Lambda, p, \nu}$ operators in each $\Lambda$ sector are much more difficult to write down as they involve first orthogonalising down the partitions. Of course one may find non-orthogonal BPS operators by applying $\cG_N$ to the covariant monomials $M_{\Lambda, M_\Lambda, p, \nu}$ (provided $N \geq l\left( p^*_\Lambda \right)$).

We can use the results of sections \ref{section: simplest orbits} and \ref{section: partitions with a dominant row} to find $p^*_\Lambda$ explicitly. As discussed at the end of section \ref{section: partitions with a dominant row}, we consider a partition $p = [1^{\mu_1}] + \bar{p}$ with a single dominant column of length $\mu_1$ attached to a smaller partition $\bar{p} \vdash \bar{n} = n - \mu_1$. By rectangle rotations, this has the same multiplicities as a single dominant row partition with first row of length $\lambda_1 = \mu_1 - l(\bar{p})$ above a smaller partition $\hat{p} = [1^{l(\bar{p})}] + \bar{p}$, where $\hat{p} \vdash \hat{n} = l(\bar{p}) + \bar{n}$ and we are using the notation of section \ref{section: partitions with a dominant row}. We give an example of these relations between single dominant column and single dominant row partitions in \eqref{dominant column example}.

Applying the general formula \eqref{single dominant row multiplicities} for a partition with a single dominant row, we see that $\Lambda = [\Lambda_1, \Lambda_2]$ only has non-trivial multiplicity with $p$ if
\begin{equation}
\Lambda_2 \leq n - \mu_1 + l(\bar{p})
\end{equation}
Rearranging to constrain $\mu_1$ in terms of $\Lambda$ and $\bar{p}$
\begin{equation}
\mu_1 \leq \Lambda_1 + l(\bar{p})
\label{longest partition constraint}
\end{equation}
Not only does \eqref{single dominant row multiplicities} give this constraint on $\mu_1$, it also tells us that the multiplicity is 1 if the inequality is saturated. If it is not quite saturated, and instead $\mu_1 = \Lambda_1 + l(\bar{p}) - 1$, then the multiplicity is the number of corners in $p$ minus 1. Since $p$ has a dominant first column, this is just the number of corners in $\bar{p}$.

Therefore the maximum possible $\mu_1$ is obtained when $l(\bar{p})$ is at its largest. This occurs when $\bar{p} = [1^{\bar{n}}]$ and $l(\bar{p}) = \bar{n} = n - \mu_1$. Plugging this in, we have
\begin{equation}
\mu_1 \leq n - \frac{\Lambda_2}{2}
\end{equation}
Therefore the maximal $\mu_1$ is $n - \left\lceil \frac{\Lambda_2}{2} \right\rceil$, with associated $\bar{p} = \left[ 1^{\left\lceil \frac{\Lambda_2}{2} \right\rceil} \right]$. If $\Lambda _2$ is even then \eqref{longest partition constraint} is saturated and the multiplicity is 1. If $\Lambda_2$ is odd, then the multiplicity is the number of corners in $\bar{p}$, which is also 1. These multiplicities agree with the explicit calculation for two column partitions in \eqref{two sphere giant general mult}.

Stated fully, for $\Lambda = [\Lambda_1, \Lambda_2] \vdash n$, the longest $p$ with non-trivial multiplicity is
\begin{equation}
p^*_\Lambda = \left[ n - \left\lceil \frac{\Lambda_2}{2} \right\rceil , \left\lceil \frac{\Lambda_2}{2} \right\rceil \right]^c = \left[ 2^{\left\lceil \frac{\Lambda_2}{2} \right\rceil}, 1^{n - 2 \left\lceil \frac{\Lambda_2}{2} \right\rceil} \right]
\end{equation}
and this multiplicity is 1.

\subsection{Orthogonalisation at $\Lambda = [n]$ and $[n-1,1]$}
\label{section: lambda=[n],[n-1,1]}

In section \ref{section: [4] and [3,1] orthogonalisation}, we observed that Gram-Schmidt orthogonalising the $M_{\Lambda, M_\Lambda, p, \nu}$ with $\Lambda = [4]$ or $[3,1]$ in the $S_n$ inner product led to the free field operators $\cO_{\Lambda, M_\Lambda, R, \tau}$. We now prove that this behaviour is general for $\Lambda = [n]$, and motivate a conjecture that this also happens for $\Lambda = [n-1,1]$.

For this subsection, when we use $\Lambda$, we will be referring specifically to $\Lambda = [n]$ or $[n-1,1]$, and stated results will apply only to those $\Lambda$.

Recall that for these $\Lambda$, a free field BPS operator is also a weak coupling pre-BPS operators. Therefore we have expansions of the form
\begin{equation}
\cO_{\Lambda, M_\Lambda, R, \tau} = \sum_{p, \nu} b^{p,\nu}_{R, \tau} M_{\Lambda, M_\Lambda, p, \nu} \hspace{50pt} M_{\Lambda, M_\Lambda, p, \nu} = \sum_{R,\tau} \left(b^{-1}\right)_{p,\nu}^{R, \tau} \cO_{\Lambda, M_\Lambda, R, \tau}
\label{free field from covariant monomials}
\end{equation}
Next, recall from (\ref{F inner product}-\ref{quarter-bps G-weighted inner product}) that the free field operators are orthogonal in the $S_n$ inner product. Therefore from the property \eqref{upper diagonal orthogonalisation} of Gram-Schmidt orthogonalisation, the $M_{\Lambda, M_\Lambda, p, \nu}$ will orthogonalise to $\cO_{\Lambda, M_\Lambda, R, \tau}$ if and only if $b^{p,\nu}_{R,\tau}$ is lower diagonal. That is, if the coefficients satisfy
\begin{align}
b^{p,\nu}_{R, \tau}  = 0 \quad \text{if} \quad (p,\nu) < (R,\tau)
\label{lower diagonal coefficients 1}
\end{align}
where the comparison between multiplicities makes sense since the size of the multiplicity space for a given $p=R$ is the same for the two bases. This is proved in \eqref{multiplicity matching}. 

More generally, since neither basis specifies a choice of multiplicity space basis, it is sufficient to prove that
\begin{align}
b^{p,\nu}_{R, \tau}  = 0 \quad \text{if} \quad p < R
\label{lower diagonal coefficients 2}
\end{align}
then after choosing the multiplicity space bases appropriately, \eqref{lower diagonal coefficients 1} and the result will follow.

\subsubsection{$\Lambda = [n]$}

In the $\Lambda = [n]$ sector at field content $(n,0)$, the covariant monomials $M_{\Lambda, M_\Lambda, p, \nu}$ reduce to the monomial symmetric functions $M_p$ defined in \eqref{M definition}, while the free field operators $\cO_{\Lambda, M_\Lambda, R, \tau}$ reduce to the Schur operators $s_R$ defined in terms of monomials in \eqref{schur definition from monomials}. From these definitions, the two bases are related by the (rescaled) Kostka numbers 
\begin{equation}
s_R = \sum_{p \vdash n} \frac{ K_{Rp} }{ \prod_i p_i ! } M_p
\label{schur to monomial}
\end{equation}
The Kostka numbers are the number of semi-standard Young tableaux of shape $R$ and evaluation $p$, where these terms are defined in section \ref{section: U(2) reps}. To prove \eqref{lower diagonal coefficients 2} in this case we need
\begin{equation}
K_{Rp} = 0  \qquad \text{for} \qquad R > p 
\label{Kostka upper diagonal}
\end{equation}
Consider $R > p$ with column lengths $R^c = [\lambda_1, \lambda_2, \dots]$ and $p^c = [\rho_1, \rho_2, \dots]$. By definition there is some $l$ for which $\lambda_i = \rho_i$ for all $i < l$ and $\lambda_l > \rho_l$. Now take a semi-standard Young tableaux of shape $R$ and evaluation $p$. The entries in the first column must strictly increase, so the entry at the bottom of the first column is $\geq \lambda_1$. Since the evaluation is $p$, the available numbers to use are $1,2, \dots, \rho_1 = \lambda_1$, so we must fill this column with exactly the numbers $1$ to $\lambda_1$. Similarly the second column must be filled with the numbers $1$ to $\lambda_2$ and so on until we reach the $l$th column. At this point, the entry at the bottom of the $l$th column must be $\geq \lambda_l$, while the maximum available number to use is $\rho_l < \lambda_l$. So the Young tableaux cannot have evaluation $p$, and therefore $K_{Rp} = 0$.

This proves \eqref{lower diagonal coefficients 2} for $\Lambda = [n]$ and the highest weight state $M_\Lambda$. Applying the $U(2)$ lowering operator $\cJ_-$, the same will happen for any $M_\Lambda$ within the $\Lambda = [n]$ sector, and this gives the result.

\subsubsection{$\Lambda = [n-1,1]$}

For $\Lambda = [n-1,1]$, there are two principal reasons we might expect $Q_{\Lambda, M_{\Lambda}, p, \nu}$ to agree with $\mathcal{O}_{\Lambda, M_\Lambda, R, \tau}$. Firstly, we know both bases are SEP-compatible. Since the space of operators that vanish when $N \to N-1$ is well-defined, it follows that for a fixed length $k = l(p) = l(R)$, the two bases must have the same span
\begin{equation}
\Span \left\{ S_{\Lambda, M_\Lambda, p, \nu} : l(p) = k  \right\} = \Span \left\{ \mathcal{O}_{\Lambda, M_\Lambda, R, \tau} : l(R) = k  \right\}
\label{same spans for given length}
\end{equation} 
Secondly, from \eqref{lower diagonal coefficients 1} the multiplicity for a given $p$ matches the multiplicity for $R = p$. Mathematically
\begin{align}
& \text{Dim} \big( \Span \left\{ S_{\Lambda, M_\Lambda, p, \nu} : 1 \leq \nu \leq \cM_{\Lambda,p}  \right\} \big) \nonumber \\
& \hspace{130pt} = \text{Dim} \big( \Span \left\{ \mathcal{O}_{\Lambda, M_\Lambda, R, \tau} : 1 \leq \tau \leq C(R,R,\Lambda)  \right\} \big)
\end{align}
A rigorous proof that the $Q_{\Lambda, M_\Lambda, p, \nu}$ and $\mathcal{O}_{\Lambda, M_\Lambda, R, \tau}$ operators match is more difficult. Numerical calculations for this paper indicate it holds true up to at least $n=7$, and we leave the general case as a conjecture.

\subsection{Alternative algorithm}
\label{sec: alternative algorithm}

\begin{figure}
	\begin{center}
		\begin{tikzpicture}
		\node[draw, align=center] (0) at (0,9) {Free field operators \\ $\cO_{\Lambda, M_\Lambda, R, \tau}$};
		\node[draw, align=center] (symm) at (0,6) {Symmetrised operators \\ $\cO^{symm}_{\Lambda, M_\Lambda, R, \tau}$};
		\node[draw, align=center] (orth) at (0,3) {$S_n$-orthogonal pre-BPS operators \\ $\cO^{orth}_{\Lambda, M_\Lambda, R, \rho}$};
		\node[draw, align=center] (BPS) at (0,0) {Orthogonal BPS operators \\ $\cO^{BPS}_{\Lambda, M_\Lambda, R, \rho}$};
		
		\node[align=center, right] (C) at (0.5,7.5) {Remove commutator trace components.};
		\node[align=center, right] (GS) at (0.5,4.5) {Gram-Schmidt orthogonalise using the $S_n$  \\ inner producs. Multiplicity space is reduced.};
		\node[align=center, right] (G) at (0.5,1.5) {Apply $\cG$, $\cF$-orttognalise and $S_n$-normalise.};
		
		\draw[->] (0) edge (symm) (symm) edge (orth) (orth) edge (BPS);
		\end{tikzpicture}
	\end{center}
	\caption{Outline of the alternative algorithm of this section. Our numerical calculations suggest that $\cO^{BPS}_{\Lambda, M_\Lambda, R, \rho}$ agrees (up to a choice of multiplicity basis) with the operators $S^{BPS}_{\Lambda, M_\Lambda, p, \nu}$ derived from the algorithm in figure \ref{figure: main algorithm}.}
	\label{figure: alternative algorithm}
\end{figure}
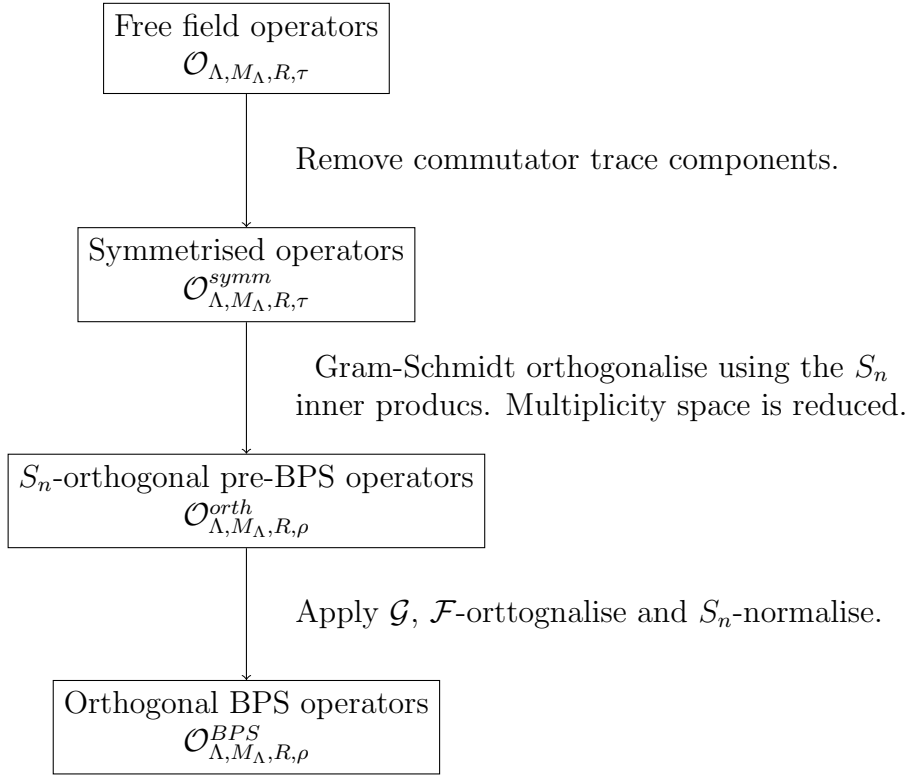

There is an alternative approach to capturing the finite $N$ behaviour of the pre-BPS sector starting from the free field operators. Following a similar process to that given in sections \ref{section: Sn orthogonalisation}-\ref{section: quicker route}, one may use this to derive an orthogonal SEP-compatible basis of BPS operators. This alternative algorithm is outlined in figure \ref{figure: alternative algorithm}. At first glance, there is no reason to expect agreement between this and the $S^{BPS}_{\Lambda, M_\Lambda, p, \nu}$ basis defined in \eqref{SBPSGFconst}. However our numerical calculations show that they do agree up to $n=6$, and we conjecture that this is a general result.

Start by considering the free field basis $\cO_{\Lambda, M_\Lambda, R, \tau}$. This has a symmetrised trace component and a commutator trace component. As discussed around \eqref{U(2) action on commutator}, there is a $U(2)$-covariant basis for commutator traces that we will denote by $c_{\Lambda, M_\Lambda, p, \xi}$, where $p \vdash n$ is a partition that describes the trace structure of the commutator trace and $\xi$ is a multiplicity index. For an example of these operators see (\ref{covariant commutators 1}-\ref{covariant commutators end}), where we give the highest weight states in the $\Lambda = [4,2]$ covariant commutator trace basis. 

Then $\cO_{\Lambda, M_\Lambda, R, \tau}$ can be written
\begin{equation}
\cO_{\Lambda, M_\Lambda, R, \tau} = \sum_{p, \nu} b_{R,\tau}^{p,\nu} M_{\Lambda, M_\Lambda, p, \nu} + \sum_{p, \xi} d_{R,\tau}^{p,\xi} c_{\Lambda, M_\Lambda, p, \xi}
\label{symm and comm trace expansion}
\end{equation}
The coefficients $b_{R,\tau}^{p,\nu}$ are a generalisation of those seen in \eqref{free field from covariant monomials} to generic $\Lambda$.

The expansion coefficients in \eqref{symm and comm trace expansion} are only defined uniquely at $N \geq n$. For $N < n$, finite $N$ relations make the choice non-unique. We will choose to use the large $N$ coefficient, even when working at $N < n$.

After removing the commutator trace component, we are left with
\begin{equation}
\cO^{symm}_{\Lambda, M_\Lambda, R, \tau} = \cO_{\Lambda, M_\Lambda, R, \tau} - \sum_{p, \xi} d_{R,\tau}^{p,\xi} c_{\Lambda, M_\Lambda, p, \xi} = \sum_{p, \nu} b_{R,\tau}^{p,\nu} M_{\Lambda, M_\Lambda, p, \nu}
\label{symmetrised zero coupling basis}
\end{equation}
which is a redundant spanning set for symmetrised traces. These operators were considered in \cite{CtoC}, where they were referred to as $\cO^S_{\Lambda, M_\Lambda, R, \tau}$.

Since $\cO^{symm}_{\Lambda, M_\Lambda, R, \tau}$ are symmetrised traces, they form pre-BPS operators for $N \geq n$. We also know from the construction that if $N < l(R)$, the operator $\cO^{symm}_{\Lambda, M_\Lambda, R, \tau}$ reduces to a commutator trace. However, if $l(R) \leq N < n$, it is not necessarily true that $\cO^{symm}$ is orthogonal to all commutator traces. This is the same situation as the monomial basis, discussed in section \ref{section: isomorphism}.

We may therefore use the same processes described in sections \ref{section: Sn orthogonalisation}-\ref{section: quicker route} in order to find an orthogonal SEP-compatible basis for BPS operators. In this section, we sill use the route given in sections \ref{section: Sn orthogonalisation}-\ref{section: F orthogonalisation} rather than the shorter one from section \ref{section: quicker route}.

In particular, we produce a basis $\cO^{orth}_{\Lambda, M_\Lambda, R, \rho}$ by following the $S_n$-orthogonalisation procedure (\ref{longest partition operator}-\ref{Q normalisation}). We then apply $\cG_N$, $\cF$-orthogonalise the resulting operators and $S_n$-normalise. The final basis is then denoted by $\cO^{BPS}_{\Lambda, M_\Lambda, R, \rho}$. This algorithm is outlined in figure \ref{figure: alternative algorithm}.

As the operators $\cO^{symm}_{\Lambda, M_\Lambda, R, \tau}$ are linearly dependent, some of the them will vanish during the orthogonalisation process. Unlike the orthogonalisation of monomials, this can occur for both $l(R) > N$ and $l(R) \leq N$. At such a point, remove that operator and continue with the orthogonalisation. This means the multiplicities for each pair $\Lambda, R$ with $l(R) \leq N$ could reduce, and in some cases will reduce to zero. Denote the reduced multiplicity for a pair by $\cM^{orth}_{\Lambda,R}$. To indicate this reduction, we use a multiplicity index $\rho$ for the orthogonalised operators rather than $\tau$.

The $\cO^{orth}_{\Lambda, M_\Lambda, R, \rho}$ with $l(R) \leq N$ form a $S_n$-orthogonal SEP-compatible basis for pre-BPS operators. The $\Lambda, M_\Lambda$ labels match the equivalents in $Q_{\Lambda, M_\Lambda, p, \nu}$, and by similar reasoning to \eqref{same spans for given length}, the length of $R$ must match the length of $p$. However, a priori, there is no reason to suspect that the $R$ label should match the $p$ label, or even that the multiplicities should be the same for any given partition.

From our numerical calculations up to $n=6$, we find that for each pair $\Lambda, R$ the multiplicities match, and the span of the operators with those labels is the same. Mathematically,
\begin{align}
\cM^{orth}_{\Lambda, R} & = \cM_{\Lambda,p=R}
\label{conjecture statement 1} \\
\Span \left\{ \cO^{orth}_{\Lambda, M_\Lambda, R, \rho} : 1 \leq \rho \leq \cM^{orth}_{\Lambda, R} \right\} & = \Span \left\{ Q_{\Lambda, M_\Lambda, p=R, \nu} : 1 \leq \nu \leq \cM_{\Lambda, p} \right\} 
\label{conjecture statement 2}
\end{align}
We conjecture that this is a general result for all $\Lambda, R$.

From this, it clearly follows that the BPS bases $\cO^{BPS}_{\Lambda, M_\Lambda, R, \rho}$ and $S^{BPS}_{\Lambda, M_\Lambda, p, \nu}$ also match.

We can consider \eqref{conjecture statement 1} and \eqref{conjecture statement 2} as a generalisation to all $\Lambda$ of the orthogonalisation results discussed in section \ref{section: lambda=[n],[n-1,1]} for $\Lambda = [n]$ and $[n-1,1]$. In that case, we showed that to prove the results, it was sufficient for the coefficients in \eqref{free field from covariant monomials} to be lower diagonal in partition indices. We now prove a proposition that generalises this lower diagonality to arbitrary $\Lambda$.

The proposition involves the coefficients $b^{p,\nu}_{R,\tau}$ in \eqref{symmetrised zero coupling basis}. Consider the sub-matrix of $b^{p,\nu}_{R,\tau}$ with $p,R \geq q$ for some partition $q$ and denote this by $\left( b_q \right)^{p,\nu}_{R,\tau}$. Then $b_q$ is an $m_{q;1} \times m_{q;2}$ matrix where
\begin{equation}
m_{q;1} = \sum_{p \geq q} \cM_{\Lambda, p} \hspace{50pt} m_{q;2} = \sum_{R \geq q} C(R,R,\Lambda)
\end{equation}
Denote the rank of $b_q$ by $r_q$. This satisfies $r_q \leq m_{q;1}$. \\
\ \\
\textbf{Proposition} \\
\ \\
Suppose the coefficients $b^{p,\nu}_{R,\tau}$ are lower diagonal in the partition indices, so that
\begin{equation}
\cO^{symm}_{\Lambda, M_\Lambda, R, \tau} = \sum_{\substack{p, \nu \\ p \geq R}} b_{R,\tau}^{p,\nu} M_{\Lambda, M_\Lambda, p, \nu}
\label{symmetrised upper diagonality}
\end{equation}
In addition, suppose that $b_q$ has maximal rank $r_q = m_{q;1}$ for each $q$. Then \eqref{conjecture statement 1} and \eqref{conjecture statement 2} hold. \\
\ \\
\textbf{Proof} \\
\ \\
Define
\begin{align}
V_R = \operatorname{Span} \{\cO^{symm}_{\Lambda, M_\Lambda, S, \tau} : S \geq R, 1 \leq \tau \leq C(S,S,\Lambda) \} \\
\widetilde{V}_R = \operatorname{Span} \{ M_{\Lambda, M_\Lambda, p, \nu} : p \geq R , 1 \leq \nu \leq \cM_{\Lambda, p} \}
\end{align}
It follows from \eqref{symmetrised upper diagonality} and the maximal rank condition for $b_R$ that $V_R = \widetilde{V}_R$. 

Denote the partition immediately higher than $R$ by $R+1$. By construction, for $R$ with $l(R) \leq N$, the additional monomials included in $\widetilde{V}_R$ by lowering $R+1$ to $R$ are linearly independent, and therefore the dimension increases by $\cM_{\Lambda, p=R}$. 

For $V_R$, we have $C(R,R,\Lambda)$ new operators included by lowering $R+1$ to $R$, but the dimension only increases by $\cM_{\Lambda, R}^{orth}$. Then since $V_R = \widetilde{V}_R$, we have
\begin{equation}
\cM_{\Lambda, R}^{orth} = \cM_{\Lambda, p=R}
\end{equation}
Therefore for $R$ with $l(R) \leq N$ we may choose the multiplicity space basis such that $\cO^{symm}_{\Lambda, M_\Lambda, R, \tau}$ is linearly independent of $V_{R+1}$ if $1 \leq \tau \leq \cM_{\Lambda, p=R}$, and is linearly dependent on $V_{R+1}$ if $\tau >  \cM_{\Lambda, p=R}$. Under the orthogonalisation procedure, those operators with $\tau > \cM_{\Lambda, p=R}$ will vanish. We can therefore equivalently start the procedure with a reduced set of operators $\cO^{symm}_{\Lambda, M_\Lambda, R, \rho}$ where we only consider $1 \leq \rho \leq \cM_{\Lambda, p=R}$. The coefficients relating these with $M_{\Lambda, M_\Lambda, p, \nu}$ are still lower diagonal in partition indices.

Now split the orthogonalisation into two steps, first orthogonalising against the $p$ or $R$ with $l(p), l(R) > N$ using the $S_n$ inner product. For the monomials, this results in the $\bar{M}{\Lambda, M_\Lambda, p, \nu}$ operators given in \eqref{Mbar}. Let the equivalent operators for the $\cO^{symm}$ orthogonalisation be $\bar{O}^{symm}_{\Lambda, M_\Lambda, R, \rho}$. These are both basis for the set of pre-BPS operators, and are still related by coefficients that are lower diagonal in partition indices. Therefore, as discussed below \eqref{same orthogonalisation}, they must orthogonalise to the same basis, in particular
\begin{equation}
\cO^{orth}_{\Lambda, M_\Lambda, R, \rho} = Q_{\Lambda, M_\Lambda, p=R, \nu=\rho}
\end{equation}
$\square$\\
\ \\
We have proved that \eqref{symmetrised upper diagonality}, along with the maximal rank condition, is sufficient for \eqref{conjecture statement 1} and \eqref{conjecture statement 2}. By a similar argument, it is also a necessary condition, though we will not prove this here.

The maximal rank condition $r_q = m_{q;1}$ is equivalent to saying that we can choose bases for the free field and covariant monomial multiplicity spaces such that
\begin{align}
b^{p,\nu}_{R=p,\tau} & = 0 & \text{if } \tau > \nu \\
b^{p, \nu}_{R=p, \tau} & \neq 0 & \text{if } \tau = \nu
\end{align}
where $1 \leq \nu \leq \cM_{\Lambda, p}$ and $1 \leq \tau \leq C(p,p,\Lambda)$. Intuitively, this says that $b$ is lower diagonal with non-zero elements on the diagonal. 

The coefficients $b^{p,\nu}_{R,\tau}$ are in some sense a covariant generalisation of the Kostka numbers, which have a nice combinatoric interpretation. It would be interesting to investigate whether there is a choice of normalisation for $\cO^{symm}_{\Lambda, M_\Lambda, p, \tau}$ and $M_{\Lambda, M_\Lambda, p, \nu}$ such that these coefficients are integers, and whether they have any combinatoric interpretation.

\section{Vector space Geometry in $ \mC ( S_n)$: BPS states from Projectors for the intersection of finite $N$ and symmetrisation constraints in  symmetric group algebras}
\label{sec:Projectors}

The construction algorithm for quarter BPS states in section \ref{sec:Construction} involves  a $U(2)$ global symmetry which provides labels for the states constructed. Alongside the $U(2)$ state labels,  there is a $U(N)$ Young diagram $Y  (p)$ which emerges  from the combinatorics of multi-symmetric functions and their relation to the  space of gauge invariant 2-matrix operators modulo commutators $ [ X , Y ] $. We have observed  in Section \ref{sec:M-matrix} that the combinatorics of multi-symmetric functions admits a generalization to the multi-matrix case where we have $M$ different matrices $ X^1 , X^2 , \cdots , X^M $. In this section we take a different viewpoint on the $M$-matrix system using permutations in analogy to the 1 and 2-matrix constructions in section \ref{section: zero coupling}.

This is used to investigate the vector space geometry in $\bC (S_n)$ that lies behind the constructions of BPS bases in the previous sections, involving the interplay between a projector $\cP_H$ for the $U(M)$ flavour symmetry, a projector for the symmetrisation of traces $\cP$ and an operator $\cF_N$ whose kernel implements finite $N$ constraints. Restricting to the image of $ \cF_N$,  there is a well-defined inverse $ \cG_N$. These operators are $M$-matrix analogues of $\cF_N$ and $\cG_N$ discussed in section \ref{section: zero coupling}. It was proved in \cite{CtoC} that BPS states are in 
\bea\label{BPSGP} 
\im ( \cG_N \cP_N ) 
\eea
where $ \cP_{N}  $ is an orthogonal (with respect to the $S_n$ inner product) projector acting on $ \mC ( S_n )$ with
\bea
\im \cP_N \equiv  \im \cP \cap \im \cF_N
\label{defcPN}  
\eea
The isomorphism between multi-symmetric functions and the ring of gauge invariants modulo commutators and the associated combinatorics of set partitions explained in section \ref{sec:MSSP} allows us to give a general explicit construction of $ \cP_N$. This general discussion also serves to explain why the construction algorithm in section \ref{sec:Construction} is able to handle the finite $N$ constraints on BPS operators systematically. The flavour projection $\cP_H$, for any chosen flavour group $H$, commutes with $\cP , \cF_N$ and can be done at the end.

\subsection{$n$-matrix model from permutations}

Let $V$ be the carrier space for the $N$-dimensional fundamental representation of $U(N)$, and consider the $n$-fold tensor product $V^{\otimes n}$. $S_n$ acts on this space by permutation of the factors. Let
\begin{equation}
\bZ = Z_1 \otimes Z_2 \otimes \dots \otimes Z_n
\end{equation}
where the $Z_a$, $a= 1, \dots, n$ are $n$ different $U(N)$ matrices. Conjugating $\bZ$ by $\sigma$ leads to
\begin{align}
\sigma \mathbb{Z}  \sigma^{-1} & = \sigma \left( Z_1 \otimes Z_2 \otimes \dots \otimes Z_n \right) \sigma^{-1} \nonumber \\
& = Z_{\sigma(1)} \otimes Z_{\sigma(2)} \otimes \dots \otimes Z_{\sigma(n)} \nonumber \\
& = \sigma \left( \mathbb{Z} \right) 
\label{permutation conjugation of n matrices}
\end{align}
where the last line defines $\sigma \left( \mathbb{Z} \right)$.

Taking the trace over $V^{\otimes n}$ of $\sigma \bZ$ gives a multi-trace featuring each of the $n$ matrices exactly once. A single cycle $(a_1, a_2, \ldots, a_k)$ in $\sigma$ leads to the single trace $\tr Z_{a_1} Z_{a_2} \ldots Z_{a_k}$, while a permutation with several cycles leads to a multi-trace. Conversely, given a multi-trace of $Z_1, Z_2, \dots, Z_n$ in which each matrix only appears once, we can identify the permutation $\sigma \in S_n$ which produces this trace. Therefore we have an invertible map from $\bC (S_n)$ into the space of $n$-matrix multi-traces of degree $(1,1,\dots, 1)$
\begin{equation}
\sigma \leftrightarrow \tr ( \sigma \bZ )
\label{permutation isomorphism}
\end{equation}
Provided $N \geq n$, this is a vector space isomorphism between $\bC (S_n)$ and the degree $(1,1,\dots,1)$ subspace of the ring of $U(N)$ invariants of an $n$-matrix system.

There is a natural inner product on $\bC(S_n)$. On permutations $\sigma, \tau \in S_n$ it is defined by
\begin{equation}
g_n ( \sigma, \tau ) = \delta \left( \sigma \tau^{-1} \right)
\label{inner product}
\end{equation}
where $\delta$ is 1 on the identity and 0 on all other permutations. The inner product extends linearly to the rest of $\bC(S_n)$. Using the isomorphism to $n$-matrix traces, this is the $n$-matrix equivalent of the $S_n$ inner product (at large $N$) defined for 1 and 2-matrix systems in \eqref{Sn half-BPS inner product} and \eqref{quarter-bps Sn inner product}.

We now look at three specialisations of this isomorphism. Firstly, to include the finite $N$ relations among traces. Secondly, to the degree $(n_1, n_2, \dots, n_M)$ subspace of an $M$-matrix system. Thirdly, to symmetrised traces of the $n$-matrices.

\subsubsection{Finite $N$ relations}

If $N < n$, there are finite $N$ relations among the traces to consider. These can be captured using the Fourier basis for $\bC (S_n)$
\begin{equation}
Q^R_{IJ} = {d_R \over n !  }  \sum_{ \sigma \in S_n } D^R_{IJ} ( \sigma ) \sigma 
\end{equation}
where $R \vdash n$ labels a representation of $S_n$ and $I,J$ are basis indices within $R$. The finite $N$ cut-off on traces corresponds to removing those $Q^R_{IJ}$ with $l(R) > N$.

We impose the finite $N$ cut-off using an operator $\cF_N$
\begin{equation}
\cF_N \sigma = \Omega_N \sigma
\end{equation}
where $\Omega_N$ is an $N$ dependent element in the centre of $\bC(S_n)$ defined in \eqref{Omega definition}. On the Fourier basis it acts as
\begin{equation}
\cF_N Q^R_{IJ} = \Omega_N  Q^R_{IJ} = f_R (N) Q^R_{IJ}
\end{equation}
where the polynomial $f_R(N)$ is given in \eqref{omega in a representation}. Since $f_R = 0$ when $l(R) > N$, this imposes the finite $N$ cut-off on the permutation space
\begin{align}
\Ker \cF_N & = \Span \{ Q^{R}_{IJ} : l(R)  > N \} \\
\im \cF_N & = \Span \{ Q^{R}_{IJ} : l(R) \le N \} 
\end{align}
For $\alpha \in \im \cF_N$, the inverse operator to $\cF_N$ is
\begin{equation}
\cG_N \alpha = \Omega_N^{-1} \alpha
\end{equation}
where $\Omega_N^{-1}$ is defined in \eqref{omega inverse}. On the Fourier basis it acts as
\begin{equation}
\cG_N Q^R_{IJ} = \Omega_N^{-1} Q^R_{IJ} =
\begin{cases}  
\frac{1}{f_R} Q^R_{IJ} & l(R) \leq N \\
0 & l(R) > N
\end{cases}
\end{equation}
Using standard character orthogonality relations, the Fourier basis for $\bC(S_n)$ is orthogonal under the $g_n$ inner product
\begin{equation}
g_n \left( Q^R_{IJ} , Q^S_{KL} \right) = \frac{d_R}{n!} \delta^{RS} \delta_{IK} \delta_{JL}
\end{equation}
therefore $\im \cF_N$ and $\Ker \cF_N$ are orthogonal subspaces of $\bC (S_n)$
\begin{equation}
( \im  \cF_N )^{ \perp } = \Ker  \cF_N 
\label{finite N hermitian}
\end{equation}
where $(.)^\perp$ denotes the orthogonal complement of a vector space.

For $N < n$, define a second inner product $g_{n,N}$ on $\bC(S_n)$ by
\begin{equation}\label{gnNdef} 
g_{n,N} (\sigma, \tau) = \delta_N ( \sigma \tau^{-1})
\end{equation}
where $\delta_N$ is given in \eqref{finite N delta}. Taken across to operators using the map \eqref{permutation isomorphism}, this is the $S_n$ inner product for $N < n$. It vanishes on $\Ker \cF_N$, and is identical to $g_n$ on $\im \cF_N$. In particular, for an element $\alpha \in \im \cF_N$, we have
\begin{equation}
g_{n,N} (\alpha, \tau) = g_n (\alpha, \tau)
\end{equation}
for any $\tau \in S_n$.

There are two ways of consider the map \eqref{permutation isomorphism} as an isomorphism in the finite $N$ setting. Firstly, we could consider the quotient space. For $\alpha \in \Ker \cF_N$, the corresponding trace $\tr \alpha \bX$ vanishes, and therefore any two elements of $\bC(S_n)$ related by $\Ker \cF_N$ produce the same multi-trace. Therefore this is a map from the quotient space
\begin{equation}
\bC(S_n) / \Ker \cF_N
\label{two forms of isomorphism}
\end{equation}
into the space of multi-traces of $n$ $N \times N$ matrices of degree $(1,1,\dots,1)$, where now we can allow $N < n$.

Alternatively, we can take a representative of each equivalence class in the quotient space. The most natural way to do this is to take the element orthogonal to $\Ker \cF_N$, which by \eqref{finite N hermitian} is in $\im \cF_N$. Therefore \eqref{permutation isomorphism} is a vector space isomorphism between $\im \cF_N$ and the appropriate space of multi-traces.

\subsubsection{Flavour projection}
\label{section: flavour projection}

To study an $M$-matrix system $X_1, X_2, \dots, X_M$, we set each of the $Z_i$ to be equal to one of the $X_j$. To study the degree $(n_1, n_2, \dots, n_M)$ subspace of the matrix system, where $\sum_i n_i = n$, we set $Z_1, Z_2, \dots, Z_{n_1}$ equal to $X_1$; $Z_{n_1 + 1}, \dots, Z_{n_1 + n_2}$ equal to $X_2$ and so on. Then $\bZ$ becomes
\begin{equation}
\bX = X_1^{\otimes n_1} \otimes X_2^{\otimes n_2} \otimes \dots \otimes X_M^{\otimes n_M}
\end{equation}
Tracing $\sigma \bX$ over $V^{\otimes n}$ leads to multi-traces of degree $(n_1, n_2, \dots, n_M)$.

Let $H = S_{n_1} \times S_{n_2} \times \dots \times S_{n_M}$. From \eqref{permutation conjugation of n matrices} we see that for any $\tau \in H$ we have $\tau (\bX) = \bX$. Therefore
\begin{equation}
\tr ( \sigma \bX ) = \tr ( \tau \sigma \tau^{-1} \bX )
\label{trace invariance}
\end{equation}
Define a projector
\begin{equation}
\cP_H (\sigma) = \frac{1}{|H|} \sum_{\tau \in H} \tau \sigma \tau^{-1}
\label{flavour projector}
\end{equation}
To justify calling $\cP_H$ a projector, we prove that it has the two properties
\begin{equation}
\left( \cP_H \right)^2 = \cP_H \hspace{50pt} \left( \cP_H \right)^\dagger = \cP_H
\end{equation}
where the Hermitian conjugate is with respect to the $g_n$ inner product. These are both simple consequences of the definition. We have
\begin{align}
\left( \cP_H \right)^2 ( \alpha ) & = \cP_H ( \cP_H ( \alpha  ) )  \nonumber \\
& = { 1 \over |H|^2 }  \sum_{ \sigma , \tau \in H }  \sigma \tau \alpha \tau^{-1} \sigma^{-1}  \nonumber \\
& = { 1 \over |H|^2 } \sum_{ \sigma, \tau  \in H }   \sigma \alpha \sigma^{-1} \nonumber \\ 
& = { 1 \over |H| } \sum_{ \mu \in H }  \sigma \alpha \sigma^{-1} \nonumber \\
& = \cP_H ( \alpha ) 
\end{align}
and
\begin{align}
g ( \alpha  , \cP_H ( \beta ) ) & = { 1 \over |H| } \sum_{ \sigma \in H } \delta ( \alpha \sigma \beta^{-1} \sigma^{-1} ) \nonumber \\
& = { 1 \over |H| } \sum_{ \sigma \in H } \delta ( \sigma^{-1} \alpha \sigma \beta^{-1} ) \nonumber \\
& = { 1 \over |H| } \sum_{ \sigma \in H } \delta ( \sigma \alpha \sigma^{-1} \beta^{-1} ) \nonumber \\ 
& = g ( \cP_H ( \alpha ) , \beta ) 
\end{align}
The image of $\cP_H$ is the sub-algebra of $\bC (S_n)$ invariant under conjugation by $\cH$, which we denote by $\cA_H$. It follows from \eqref{trace invariance} that
\begin{equation}
\tr ( \sigma \bX ) = \tr \big( \cP_H (\sigma) \bX \big)
\end{equation}
Therefore $\cA_H$ is the algebra of permutations responsible for constructing multi-traces of degree $(n_1, \dots, n_M)$. We call $\cP_H$ the flavour projector, as it projects from the $n$-matrix system to one with $M$ flavours.

As discussed for the finite $N$ relations around \eqref{two forms of isomorphism}, the map \eqref{permutation isomorphism} gives an isomorphism between $\cA_H$ or $\bC(S_n)/\Ker\cP_H$ and the space of $M$-matrix multi-traces of degree $(n_1,\dots, n_M)$.

In section \ref{section: zero coupling} we discussed the half-BPS sector, corresponding to $M=1$, and the free field quarter-BPS sector, corresponding to $M=2$. In the latter case we have $H = S_{ n_1} \times S_{ n_2}$, and the matching permutation centraliser algebra $\cA_H$ was studied in \cite{PCA}. The free-field eighth BPS sector is larger than just multi-traces of three matrices, since it also includes fermion contractions. However, the sub-sector consisting only of the scalar fields can be found by setting $M=3$. This subsector is dual to the class of eighth-BPS giant gravitons considered in \cite{Mikhailov}.

\subsubsection{Symmetrised traces}
\label{symmetrised traces}

A symmetrised trace of $Z_1, Z_2, \dots, Z_n$ is defined in a completely analogous manner to the 2-matrix version in \eqref{symmetrised trace}, allowing $a_i \in \{1,2,\dots,n\}$ instead of $\{1,2\}$. Degree $(1,\dots,1)$ symmetrised traces are labelled by set partitions $\pi \in \Pi(n)$. These naturally correspond to $n$-vector partitions of weight $(1,\dots,1)$.

Take $b \subseteq \{1,2,\dots,n\}$. Then there is an associated symmetrised single trace
\begin{equation}
T_b = \str \left( \prod_{i \in b} Z_i \right)
\end{equation}
where the symmetrisation implicit in $\str$ means the ordering of the product is irrelevant. For a set partition $\pi \in \Pi(n)$ we have
\begin{equation}
T_\pi = \prod_{b \in \pi} T_b
\label{symmetrised trace set partition}
\end{equation}
where $b$ runs over the blocks of $\pi$.

The equivalent permutation picture comes from the set of permutations $\text{Perms}(\pi)$, defined in \eqref{perms of pi}. We have
\begin{equation}
T_\pi = \frac{1}{|\text{Perms}(\pi)|} \sum_{\sigma \in \text{Perms}(\pi)} \sigma
\label{T pi permutation}
\end{equation}
where we use the same notation $T_\pi$ for both the sum over permutations and the associated symmetrised trace operator. For the remainder of the section we only work with the permutation sum, so this ambiguity will not be an issue.

More generally, one can define a symmetrisation projector $\cP$ which projects a permutation onto the space isomorphic to symmetrised traces. This is
\begin{equation}
\cP ( \sigma )  = \frac{1}{|G(\pi(\sigma))|} \sum_{\tau \in G(\pi(\sigma))} \tau \sigma \tau^{-1}
\end{equation}
where the set partition $\pi(\sigma)$ is defined naturally from the cycle structure of $\sigma$ and is discussed above \eqref{perms of pi}. The subgroup $G(\pi)$ for a given set partition is defined in \eqref{G definition} and permutes each block of $\pi$ within itself but does not mix the different blocks. As mentioned below \eqref{perms of pi}, the set $\text{Perms}(\pi(\sigma))$ is just the conjugacy class of $\sigma$ under $G(\pi(\sigma))$, and therefore
\begin{equation}
\cP (\sigma) = T_{\pi(\sigma)}
\label{T in imP}
\end{equation}
As with the flavour projection, we now prove
\begin{equation}
\cP^2 = \cP \hspace{50pt} \cP^\dagger = \cP
\label{projector properties}
\end{equation}
These follow immediately from the definition. We have
\begin{align}
\cP ( \cP ( \sigma ) ) & = { 1 \over | G ( \pi ( \sigma ))|^2   } \sum_{ \tau \in G ( \pi ( \sigma ) ) } \sum_{ \mu \in G ( \pi ( \tau \sigma \tau^{ -1}  ) )  } \mu \tau  \sigma \tau^{-1}  \mu^{-1}    \nonumber \\
& =  { 1 \over | G ( \pi ( \sigma ))|^2   } \sum_{ \tau \in G ( \pi ( \sigma ) ) } \sum_{ \mu \in G ( \pi ( \tau \sigma \tau^{ -1}  ) )  } \tau \tau^{-1} \mu \tau  \sigma \tau^{-1}  \mu^{-1} \tau \tau^{ -1}     \nonumber \\
& = { 1 \over | G ( \pi ( \sigma ))|^2   } \sum_{ \tau, \tilde{\mu} \in G ( \pi ( \sigma ) ) } \tau \tilde{\mu} \sigma \tilde \mu^{-1} \tau^{-1} \nonumber \\
& =  { 1 \over   | G ( \pi ( \sigma ))|^2     } \sum_{ \tau \in G ( \pi ( \sigma )) } ~~ | G  ( \pi ( \sigma ) ) |    \tau \alpha \tau^{-1} \nonumber \\
& = \cP ( \alpha ) 
\end{align}
where in the third line, we have defined $ \tilde \mu = \tau^{-1} \mu \tau$. The conjugation by $\tau$ takes $\mu \in G(\pi(\tau \sigma \tau^{-1}))$ to $\tilde{\mu} \in G(\pi(\sigma))$. To prove $\cP$ is Hermitian, we note that 
\begin{align}
g ( \sigma , \cP ( \tau ) )  & = { 1 \over | G ( \pi ( \tau ) )| } \sum_{ \mu \in  G ( \pi ( \tau )) }  \delta ( \sigma \mu \tau^{-1} \mu^{-1} )  
\end{align}
is only non-zero if $\sigma, \tau$ belong to the same $\text{Perms}(\pi)$. In particular, they have $\pi(\sigma) = \pi(\tau)$. Therefore
\begin{align}
g ( \sigma , \cP ( \tau ) ) & = { 1 \over | G ( \pi ( \sigma )) | }  \sum_{ \mu \in G ( \pi ( \sigma ) ) } \delta ( \mu^{-1} \sigma \mu \tau^{-1} ) \nonumber \\ 
& =  { 1 \over | G ( \pi ( \sigma )) | }  \sum_{ \mu \in G ( \pi ( \sigma ) ) } \delta ( \mu \sigma \mu^{-1} \tau^{-1} ) \nonumber \\ 
& =  g ( \cP(\sigma)  , \tau  ) 
\end{align}
Therefore the map \eqref{permutation isomorphism} gives an isomorphism between the symmetrised traces of $n$ matrices with degree $(1,\dots,1)$ and either $\im \cP$ or $\bC(S_n) / \Ker \cP$. This is true when $N \geq n$. To deal with $N < n$ we have to include the finite $N$ relations as well, which is discussed later.

As $\cP$ is satisfies \eqref{projector properties}, it is expressible in the standard projector form
\begin{equation}
\mathcal{P} = \sum_i | i \rangle \langle i | 
\label{projector general form}
\end{equation}
for orthonormal basis states $| i \rangle$ for $\im(\mathcal{P})$. These states $| i \rangle = \alpha_i$ belong to $\mathbb{C}(S_n)$, so to avoid doubling of notation, we will write this as
\begin{equation}
\mathcal{P} = \sum_i \alpha_i \otimes \alpha_i
\end{equation}
which acts on $\sigma \in S_n$ via
\begin{equation}
\mathcal{P}(\sigma) = \sum_i g(\alpha_i, \sigma) \alpha_i
\end{equation}
It is clear from \eqref{T in imP} that $T_\pi$ spans $\im \cP$. For $\pi \neq \pi'$, $\text{Perms}(\pi)$ is disjoint from $\text{Perms}(\pi')$, and therefore
\begin{align}
g \left( T_{\pi}, T_{\pi'} \right) & = \frac{1}{|\text{Perms}(\pi)| |\text{Perms}(\pi')|} \sum_{\substack{ \sigma \in \text{Perms} ( \pi ) \\ \tau \in \text{Perms} ( \pi') }} \delta ( \sigma \tau^{-1} )  \nonumber \\
& = \frac{1}{|\text{Perms}(\pi)|} \delta_{ \pi  , \pi' } 
\end{align}
So an orthonormal basis for $\im \cP$ is given by
\begin{equation}
\alpha_\pi = \sqrt{|\text{Perms}(\pi)|} \, T_\pi
\end{equation}
and the corresponding expression for $\cP$ is
\begin{equation}
\cP = \sum_{\pi \in \Pi(n)} |\text{Perms}(\pi)| \ T_\pi \otimes T_\pi
\label{P from T pi}
\end{equation}
In section \ref{section: C combinatorics} we defined another $T_\pi$ as a map from 2-vector partitions to multi-symmetric functions. Composing these with the map \eqref{symmetrised trace isomorphism} between multi-symmetric functions and symmetrised trace operators, we can identify $T_\pi \in \bC(S_n)$ with these using the flavour projector $\cP_H$ with $H = S_{n_1} \times S_{n_2}$. 

Let $\bp \vdash (n_1,n_2)$ be a vector partition, and $\pi$ a set partition such that $\pi ([(1,0)^{n_1}, (0,1)^{n_2}]) = \bp$, where the action of a set partition on a vector partition was given in \eqref{set partition on vector partition}. Then
\begin{equation}
T_\bp = \tr \left[ \cP_H \left( T_\pi \right) \bX \right]
\end{equation}
where $T_\bp$ is the 2-matrix symmetrised trace operator given in \eqref{symmetrised trace operator}. 

Intuitively, the flavour projection and symmetrisation projectors should commute, since symmetrising a trace and renaming matrices from $Z_i$ to $X_j$ are commuting operations. Indeed
\begin{align}
\cP P_H ( \sigma ) & = { 1 \over | G ( \pi ( \tau \sigma \tau^{-1}  ) ) | } { 1 \over |H| }  \sum_{\substack{ \tau \in H \\ \mu \in G ( \pi ( \tau \sigma \tau^{-1}  )  ) }} \mu \tau \sigma \tau^{-1} \mu^{-1} \nonumber \\
& = \frac{1}{|G(\pi(\sigma))| |H|} \sum_{\substack{\tau \in H \\ \mu \in G ( \pi ( \tau \sigma \tau^{-1}  )  ) } } \tau \left( \tau^{-1} \mu \tau \right) \sigma (\tau^{-1} \mu \tau )^{-1} \tau^{-1} \nonumber \\
& = \frac{1}{|G(\pi(\sigma))| |H|} \sum_{\substack{\tau \in H \\ \tilde{\mu} \in G ( \pi ( \sigma  )  ) } } \tau \tilde{\mu} \sigma \tilde{\mu}^{-1} \tau^{-1} \nonumber \\
& = P_{ H } \cP ( \sigma )
\end{align}
$\cP$ was first considered in \cite{CtoC}, though a slightly different group $G(\pi)$ was used in the definition. This involves wreath products and is given in section \ref{sec: multiplicity space basis} for an integer partition. The difference in the defining group does not affect the action of the projector.

\subsection{Multi-symmetric function isomorphism for $n$ matrices}

In section \ref{section: isomorphism} we described the isomorphism of \cite{Vaccarino2007,Domokos} between $U(N)$ gauge invariant of 2 complex matrices $X_1$ and $X_2$, modulo the ideal generated by commutators, and the ring of multi-symmetric functions in 2 families of variables. This was then used to construct a basis for 2-matrix symmetrised traces that respected the finite $N$ behaviour. We now generalise this to the $n$-matrix case, and use the previous section to identify the space of multi-symmetric functions with sub-algebras of $\bC(S_n)$. This will in turn allow a construction of the projector $\cP_N$ that describes the interaction of the finite $N$ cut-off with the symmetrisation projector $\cP$.

Consider the $M$ matrix variables $ X_1 , X_2, \cdots , X_M$. For each $ a \in \{ 1, \cdots , M \}$, we have $N^2$ variables 
\bea 
(X_a ) _{ ij} 
\eea
where $ i,  j \in \{ 1, \cdots , N \}$. Consider the ring of polynomials in these $M N^2$ variables. In this ring, there is an ideal generated by the elements of the commutators  
\bea
\left[ X_a, X_b \right]_{ik} = \sum_j  ( X_a)_{ i j } ( X_b )_{ j k } - ( X_b)_{ i j } ( X_a )_{ j k }
\eea
where $a \neq b \in \{ 1, 2, \cdots , M \}$. We can form a quotient ring from this ideal. The ring of polynomial functions in the $M$ matrix variables admits an action by $\cU \in U(N)$ (or $GL(N, \mC)$): 
\bea 
X_a \rightarrow \cU X_a \cU^{-1} 
\eea
The ideal generated by the commutators is invariant under  the action of $U(N)$, so there is a quotient ring of $U(N)$ invariant polynomials. This is the  ring of gauge invariants modulo commutator traces. This quotient ring of gauge invariants consists of multi-traces where any two traces differing by commutator traces define the same element of the ring. This is denoted by $ A_P^G$ in Theorem 3 of \cite{Vaccarino2007}. 

There is a polynomial ring $D$ generated by $x^a_{ i } $ for $ a \in \{ 1 , \cdots , M \} $ and $ i \in \{ 1 , \cdots , N \}$.  These polynomials have an $S_N$ action given by 
\bea 
x^{ a}_i  \rightarrow x^{ a }_{ \sigma(i ) } 
\eea 
The $S_N$ invariant polynomials form multi-symmetric functions in $M$ families of variables, and the ring of these functions is denoted $D^{ S_N }$. Theorem 3 of \cite{Vaccarino2007} states that these two rings $D^{ S_N } $ and $ A^{ G}_{ P} $ are isomorphic. 

To summarise, we have an isomorphism between gauge invariant polynomial functions  of $M$ matrices, modulo commutator traces, and permutation invariant polynomial functions of the $M$ diagonal matrices. The map from the ring of $U(N)$ gauge invariant polynomial functions of matrices, modulo the commutator trace, to the space of $S_N$ invariant polynomials is obtained by evaluating the gauge invariant functions on diagonal matrices.  This map, denoted by  $ \iota $, is proved to be an isomorphism in \cite{Vaccarino2007,Domokos}.

In the following, we will use a special case of this isomorphism where we have $M =n$ matrices and we consider gauge invariants containing exactly one of each matrix.

The space of matrix invariants appearing in this special case is important for the construction of BPS states. As discussed in section \ref{section: weak coupling}, the construction of quarter-BPS states is based on finding the orthogonal complement to the operators which are expressible as commutator traces at finite $N$. This orthogonalisation admits a generalization to the present case of $M = n$ matrices and gauge invariants containing one matrix of each type. Using the permutation description of $n$-matrix traces given in the previous section, it can be expressed as a problem in $ \mC ( S_n )$ or constructing the  orthogonal complement of $ \Ker \cP + \Ker \cF_N $. 

To see this, recall that permutations in $\Ker \cP$ correspond to commutator traces via \eqref{permutation isomorphism}, while those in $\Ker \cF_N$ correspond to the zero operator. Therefore any permutation in $\Ker \cP + \Ker \cF_N$ is a commutator trace. It follows that

\vskip.2cm 

\noindent 
{\bf Lemma 1 } 
\begin{align}
& \Big( \mC [ X_1 , X_2 , \cdots , X_n ] \ / \ \langle \{ [ X_a , X_b ]  : 1 \le a < b \le n  \}  \rangle   \Big)^{ U(N)} \Bigg\vert_{ ( 1, 1, \cdots , 1 ) } \nonumber \\
\ \nonumber \\
& \hspace{50pt} =  \mC ( S_n ) / \big( \Ker \cP + \Ker \cF_N \big) 
\label{L1}
\end{align}
Composing these two isomorphisms gives an identification between multi-symmetric functions and the quotient space of permutations.

The ring of multi-symmetric functions in $n$ families of variables is spanned by multi-traces of $n$ commuting matrices or monomials functions, denoted by $T_\pi$ and $M_{ \pi }$ respectively. As discussed in section \ref{symmetrised traces}, these are labelled by set partitions $\pi \in \Pi(n)$ when the degree of each family of variables is 1, since this is equivalent to a $n$-vector partition of $(1,\dots,1)$. The size $N$ of each family of variables limit the number of subsets in the $\pi$ to be less than $N$. This is denoted by $|\pi| \le N$ and follows immediately from the definition of the $ M_{ \pi}$, given for 2 families of variables in \eqref{monomial definition}.

We use the same notation $T_\pi$ and $M_\pi$ for the multi-symmetric functions and the equivalent permutations. For $T_\pi$, this is given in \eqref{T pi permutation}. The $M_\pi $ and $T_\pi$ are related by
\begin{equation}
M_\pi = \sum_{\pi'} \tC^{\pi'}_\pi T_{\pi'} \hspace{50pt} T_\pi = \sum_{\pi'} C^{\pi'}_\pi M_{\pi'}
\label{M_pi}
\end{equation}
where the $C$ and $\tC$ matrices are described in section \ref{sec:M-matrix}.

We now investigate the decomposition of $\bC (S_n)$ in terms of the images and kernels of $\cP$ and $\cF_N$.

\vskip.2cm 

\noindent 
\textbf{Lemma 2}
\vskip.2cm 
\noindent Consider two subspaces $ S_1 , S_2$ of a vector space, equipped with an inner product. 
Let $ S_1^{ \perp} , S_2^{ \perp} $ be the orthogonal complements to $S_1, S_2$ respectively.
Let $ S_1 + S_2$ be the set of vectors of the form $  v_1 + v_2$, where $ v_1 \in S_1, v_2 \in S_2$. It is a  standard result that 
\bea 
(S_1 + S_2)^{ \perp } = S_1^{\perp } \cap S_2^{ \perp } 
\eea
which is stated as ``The orthogonal complement of a sum of vector spaces is the intersection of orthogonal complements''. 
\vskip.2cm
\noindent 
{\bf Proof }
\vskip.2cm 
\noindent Suppose $ w \in S_1^{\perp } \cap S_2^{ \perp } $, then 
\bea 
v_1 \cdot w = v_2 \cdot w =  0 
\eea
for all $ v_1 \in S_1 , v_2 \in S_2$. It follows that $ w \cdot ( v_1 + v_2 ) = w \cdot v_1 + w \cdot v_2 = 0 $. So we conclude that $ w \in (S_1 + S_2)^{ \perp }  $. 

Conversely, suppose $ w \notin S_1^{\perp } \cap S_2^{ \perp }  $, then $ w \notin S_i^{ \perp }$ for $i = 1$ or $2$. This means there is some  $v \in S_i$, such that $ w \cdot v \ne 0$. But $ v \in S_1 + S_2$, so $ w \notin ( S_1 + S_2)^{ \perp }$.  $\square$

\ \\

\noindent Taking $S_1 = \im \cP$ and $S_2 = \im \cF_N$, we have an orthogonal decomposition for $\bC (S_n)$ with respect to $g_n$
\vskip.2cm
\noindent 
{\bf Lemma  2 } 
\begin{align} 
\mC ( S_n ) & = ( \im \cP \cap \im  \cF_N )  \oplus_{ g_n }   ( \Ker \cP + \Ker \cF_N )
\end{align}
Using the fact that the monomial multi-symmetric functions form a basis, we have 
\bea 
\im \cP = \cM =  \Span \{ M_{ \pi}  : \pi \in \Pi(n)  \} 
\eea
We will also define  
\begin{align}
\cM^{ \le }_{ N } & = \Span \{ M_{ \pi} :  \pi \in \Pi(n)  ,  | \pi | \le N \} \cr 
\cM^{ > }_{ N } & = \Span \{ M_{ \pi} : \pi \in \Pi(n)  ,  |\pi | > N \} 
\end{align}
For $n < N $, we have to consider both operators  $\cF_N$ and $ \cP$. They are both hermitian operators wrt the $g_n$ inner product, but they do not commute. The space $\Ker \cP +  \Ker\cF_N$, spanned by sums of vectors in $ \Ker \cP$ and $ \Ker\cF_N$  is in general bigger than  $ \Ker \cP$.  There is non-trivial intersection 
\bea 
\im \cP \cap (  \Ker \cP +  \Ker \cF_N ) 
\eea
The non-triviality of this intersection is reflected in the fact that some symmetrised traces can also be written as as a symmetrised trace at finite $N$. An example of this is given in \eqref{symmetrised trace = commutator trace}.

Since $\cP$ is a Hermitian projector, we have the orthogonal decomposition
\begin{equation}
\bC(S_n) = \im \cP \oplus_{ g_n } \Ker \cP
\end{equation}
It follows that we have an orthogonal decomposition of $\Ker \cP + \Ker \cF_N$ 
\vskip.2cm
\noindent 
{\bf Lemma 3 } 
\begin{align} 
 \Ker \cP + \Ker \cF_N
=   \big( ( \Ker \cP + \Ker \cF_N  )  \cap  \im \cP \big) 
\oplus_{ g_n }  \Ker \cP \hspace{.3cm} 
\end{align} 
\noindent 
{\bf Lemma 4 } 
\begin{empheq}[box=\fbox]{align} 
& \cr  
& \hspace{.3cm} (  \Ker \cP +  \Ker\cF_N ) \cap  \im \cP  = \cM^{ > }_N  
= \Span \{ M_{ \pi } , |\pi | > N \} \hspace{.3cm}  \cr 
& \nonumber
\end{empheq} 

\noindent{\bf Proof } 
\vskip.2cm
\noindent$ \cM^{ > }_N $ is  exactly the subspace of 
$ \cM  $ which is not  in the image of the isomorphism $\iota $. Therefore
\bea 
\cM^{ >  }_N  \subset \Ker \cP  + \Ker \cF_N   \subset \mC ( S_n ) 
\eea
Additionally $\cM^{ >  }_N \subset \im \cP  = (\Ker \cP )^{ \perp }  $. Then using Lemma 4, 
the result follows. $\square$

\ \\
\noindent Using Lemmas  2, 3 and 4 we have the result
\vskip.2cm
\noindent 
{\bf Theorem } 
\begin{empheq}[box=\fbox]{align} 
& \cr  
& \hspace{.5cm} 
 \mC ( S_n )  = ( \im P \cap \im \cF_N ) \oplus_{ g_n} \cM_N^{ > } \oplus_{ g_n } \Ker \cP
 \hspace{.5cm} 
\label{orthogonalisation proof} \\
& \nonumber
\end{empheq} 
Using the definition of $ \cP_N$ in \eqref{defcPN} we can also write this 
as 
\bea 
\mC ( S_n )  =  \im \cP_N \oplus_{ g_n} \cM_N^{ > } \oplus_{ g_n } \Ker \cP
\eea
This gives a procedure, based on the combinatorics of multi-symmetric functions, for constructing  the projector  $ \cP_N$. This projector is built by constructing the  projector for the susbspace of $\cM$ orthogonal,  with respect to  the inner product $g_n$ on $ \mC ( S_n )$, to $ \cM^{>}_N$.  Since $ \cM^{ \le }_{ N } + \cM^{ > }_{N } =  \im \cP$, this can be built  by taking the vectors in $ \cM^{ \le }_N $ and subtracting vectors in $ \cM^{ > }_N$ to ensure the orthogonality. The construction of $\im \cP \cap \im \cF_N$ uses the  following elements: 
\begin{itemize} 
	
	\item  Vectors $M_{ \pi}$ in  $ \mC ( S_n)$ labelled by set partitions, spanning $ \im \cP$. 
	
	\item Finite $N$ cut-off  implemented using the set partition labels: the condition $ | \pi | \le N$ which defines $ \cM^{ \le }_N$. 
	
	\item Orthogonalization of $ \cM^{ \le }_N$ to $ \cM_N^{ > }$ with respect to 	the inner product $g_n$. 

\end{itemize} 
This procedure is used in section \ref{sec:Construction} to construct quarter-BPS bases.

The result of this procedure is a vector subspace of $\im \cF_N$, and therefore the orthogonalisation can equivalently be done using the inner product $g_{n,\hN}$ for any $\hN \geq N$, as on these permutations the $g_n$ and $g_{n,\hN}$ inner products are the same.

We now give a construction of the projector $\cP_N$ that captures this process. The formula for this is given in \eqref{finite N projector}.

\subsection{Finite $N$ symmetrisation operator on $ \mC ( S_n )$ }
\label{sec:projectorformulae}

We now construct the projector $\cP_N$ onto $\im \cP \cap \im \cF_N$ and prove it has the projector properties
\begin{equation}
\left(\cP_N \right)^2 = \cP_N \hspace{50pt} \left( \cP_N \right)^\dagger = \cP_N
\label{projector properties PN}
\end{equation}
and commutes with the flavour projector
\begin{equation}
\cP_N \cP_H = \cP_H \cP_N
\label{PN PH commute}
\end{equation}
Before producing $\cP_N$, we give an alternative formula for $\cP$, the large $N$ symmetrisation projector. Using \eqref{P from T pi} and substituting using \eqref{M_pi}, we have
\begin{equation}
\cP = \sum_{\pi, \pi', \pi'' \in \Pi(n)} |\text{Perms}(\pi)| C^{\pi'}_\pi C^{\pi''}_\pi M_{\pi'} \otimes M_{\pi''} = \sum_{\pi,\pi' \in \Pi(n)} \left( C D C^T \right)^{\pi}_{\pi'} M_\pi \otimes M_{\pi'}
\end{equation}
where $D$ is the diagonal matrix
\begin{equation}
D^\pi_{\pi'} = |\text{Perms}(\pi)| \delta_{\pi, \pi'}
\end{equation}
To understand the appearance of $CDC^T$, note that this is the inverse metric on the subspace $\im(\mathcal{P})$ of $\mathbb{C}(S_n)$. 

We know $g$ is a positive definite inner product on the entirety of $\mathbb{C}(S_n)$. It is therefore also a positive definite inner product on the subspace $\im \mathcal{P}$. Hence there is an inverse metric on this subspace, which we call $G$. Using \eqref{M_pi} we have
\begin{align}
g ( M_{ \pi } , M_{ \pi' } ) & = \sum_{ \pi_1 , \pi_2 \in  \Pi(n)  } {\widetilde C}^{ \pi_1}_{\pi } {\widetilde C}^{\pi_2}_{\pi' } g ( T_{ \pi_1} , T_{ \pi_2} ) \nonumber \\ 
& = \sum_{ \pi_1  \in  \Pi(n)  }  \frac{{\widetilde C }^{\pi_1}_{\pi } {\widetilde C}^{\pi_1}_{\pi' }}{|\text{Perms}(\pi_1)|} \nonumber \\
& = \left(  {\widetilde C }^T D^{-1} \widetilde C  \right)^{\pi}_{\pi' } 
\label{g from C}  
\end{align}
Since $C$ and $\tC$ are inverses of each other, this implies
\begin{align}
G(M_\pi, M_{\pi'}) = (C D C^T)^\pi_{\pi'}
\label{G from C}
\end{align}
We can therefore write
\begin{equation}
\mathcal{P} = \sum_{\pi, \pi' \in \Pi(n)} (C D C^T)^\pi_{\pi'} \ M_\pi \otimes M_{\pi'} = \sum_{\pi, \pi' \in \Pi(n)} G(M_\pi, M_{\pi'}) M_\pi \otimes M_{\pi'}
\label{orthogonal projection}
\end{equation}
This form for a projector is a generalisation of \eqref{projector general form} to a basis of the image that is not orthonormal. We now find a basis for $\im \cP_N$, and can use the form above to write down $\cP_N$.

At finite $N$, we want to project onto the orthogonal complement of $\cM_N^>$ within $\im \mathcal{P}$. The $M_\pi$ with $|\pi| \leq N$ do not suffice for this as they are not orthogonal to $M_\pi$ with $|\pi| > N$; we need to orthogonalise them first.

As already noted, $g$ is an inner product on any subspace of $\mathbb{C}(S_n)$. This time the relevant subspace is $\cM_N^>$. This means that the matrix of inner products $g( M_\pi , M_{\pi'} )$ for $|\pi|, |\pi'| > N$ is invertible and has an inverse metric that we call $G^>$. Note that $G^>$ is distinct to $G$, which is the inverse inner product on $\im \cP = \cM$. Practically, the difference is
\begin{align}
\sum_{\pi' \in \Pi(n)} g(M_\pi, M_{\pi'}) G(M_{\pi'}, M_{\pi''}) & = \delta_{\pi \pi''} & \pi, \pi'' \text{ unrestricted}
\label{inverse property G}  \\
\sum_{\substack{\pi' \in \Pi(n) \\ |\pi'| > N}} g(M_\pi, M_{\pi'}) G^> (M_{\pi'}, M_{\pi''}) & = \delta_{\pi \pi''} & |\pi|, |\pi''| > N 
\label{inverse property G>}
\end{align}
We can use $G^>$ to construct a basis for $\cM_N^\leq$, labelled by those set partitions with $|\pi| \leq N$.
\begin{align}
\bar{M}_\pi = M_\pi - \sum_{\substack{\pi_1, \pi_2 \in \Pi(n) \\ |\pi_1|, |\pi_2| > N} } G^> (M_{\pi_1}, M_{\pi_2}) g(M_\pi, M_{\pi_1})  M_{\pi_2}
\label{Mbar definition}
\end{align}
The simplest way of looking at this is to notice that the second term is using a projector of the form \eqref{orthogonal projection} applied to $M_\pi$. This is the projector
\begin{align}
\mathcal{P}^> = \sum_{\substack{\pi_1, \pi_2 \in \Pi(n) \\ |\pi_1|, |\pi_2| > N} } G^> (M_{\pi_1}, M_{\pi_2}) ( M_{\pi_1} \otimes M_{\pi_2} )
\end{align}
that orthogonally projects onto $\cM_N^>$. The construction of $\bar{M}_\pi$ just applies $1 - \mathcal{P}^>$ to $M_\pi$ to produce something orthogonal to $\cM_N^>$ while remaining in $\im \mathcal{P}$. We can now use the $\bar{M}_\pi$ to define the finite $N$ symmetrisation projector. Again, we need to produce a new inverse metric $G^\leq$ on the space spanned by $\bar{M}_\pi$. This satisfies
\begin{align}
\sum_{\substack{\pi' \in \Pi(n) \\ |\pi'| \leq N}} g(\bar{M}_\pi, \bar{M}_{\pi'}) G^\leq (\bar{M}_{\pi'}, \bar{M}_{\pi''}) & = \delta_{\pi \pi''} & |\pi|, |\pi''| \leq N 
\label{inverse metric leq N}
\end{align}
Using this, we construct the finite $N$ symmetrisation projector
\begin{equation}
\boxed{ 
\hspace{.5cm } \cP_{ N } ~~~  = \sum_{ \substack { \pi , \pi' \in \Pi(n) \\ | \pi |, |\pi'| \leq N } }    G^\leq ( \bar{M}_{\pi} , \bar{M}_{ \pi'} ) ( \bar{M}_{ \pi } \otimes \bar{M}_{ \pi'} )
\hspace{.4cm} 
}
\label{finite N projector}
\end{equation}
We now prove the properties \eqref{projector properties PN} and \eqref{PN PH commute} for $\cP_N$.
\vskip.2cm
\noindent {\bf $\cP_N$ is a projector } 
\vskip.2cm
\noindent To prove this, we act with the square of the projector 
\begin{align}
\cP_{ N } \cP_{ N } ( \alpha ) 
& = \cP_N  \sum_{ \substack  { \pi_1 ,\pi_2  \in \Pi(n) \\  |\pi_1|, |\pi_2 |  \le N } }
G^\leq ( \bar{M}_{ \pi_1 } , \bar{M}_{ \pi_2 } ) \bar{M}_{ \pi_1 } \ g ( \bar{M}_{ \pi_2} , \alpha ) \nonumber \\
&  =  \sum_{ \substack  { \pi_1, \pi_2, \pi_3 ,\pi_4   \in \Pi(n) \\ |\pi_1|, |\pi_2|, |\pi_3|, |\pi_4 |  \le N } }
G^\leq ( \bar{M}_{ \pi_3 } , \bar{M}_{ \pi_4 } )  G^\leq ( \bar{M}_{ \pi_1 } , \bar{M}_{ \pi_2 } )
g ( \bar{M}_{ \pi_4} , \bar{M}_{ \pi_1} ) g ( \bar{M}_{ \pi_2} , \alpha ) \bar{M}_{\pi_3} \nonumber \\
&  = \sum_{ \substack  { \pi_1 ,\pi_2, \pi_3   \in \Pi(n) \\  |\pi_1|, |\pi_2 |, |\pi_3|  \le N } }  
\delta ( \pi_1 , \pi_3 )   G^\leq ( \bar{M}_{ \pi_1 } , \bar{M}_{ \pi_2 } )
g ( \bar{M}_{ \pi_2} , \alpha ) \bar{M}_{\pi_3}  \nonumber \\
& =  \sum_{ \substack  { \pi_1 ,\pi_2  \in \Pi(n) \\  |\pi_1|, |\pi_2 |  \le N } }   
G^\leq ( \bar{M}_{ \pi_1 } , \bar{M}_{ \pi_2 } ) \bar{M}_{\pi_1}
g ( \bar{M}_{ \pi_2} , \alpha )  \nonumber \\
& = \cP_{ N  } ( \alpha ) 
\end{align}
where we have used \eqref{inverse metric leq N} to get from the second to third line.
\vskip.2cm
\noindent {\bf $\cP_N$ is hermitian }
\vskip.2cm
\noindent This follows from the symmetry between $\pi$ and $\pi'$ in \eqref{finite N projector}
\begin{align}
g ( \alpha , \cP_{ N } ( \beta ) ) & = \sum_{ \substack  { \pi_1 ,\pi_2  \in \Pi(n) \\  |\pi_1|, |\pi_2 |  \le N } } G^\leq ( \bar{M}_{\pi_1}, \bar{M}_{\pi_2} ) g( \bar{M}_{\pi_2}, \beta) g(\alpha, \bar{M}_{\pi_1}) \nonumber \\
&  = g ( \cP_{ N } ( \alpha ) , \beta )  
\end{align}
{\bf $ \cP_N$ commutes with $\cP_H$} 
\vskip.2cm
\noindent This relies on some smaller results. We start with
\bea 
\sigma^{-1} T_{ \pi } \sigma = T_{ \sigma (\pi) } 
\label{permutation of T}
\eea
where we define $ \sigma (\pi)$ as the set partition obtained by substituting $ i \rightarrow \sigma ( i ) $ in the set partition $ \pi$. It is useful to recall the fact that $ \sigma^{-1} \mu \sigma $ is the permutation obtained by the substitution $ i \rightarrow \sigma ( i ) $ in the cycle decomposition of $\mu$, and therefore 
\begin{equation}
\text{Perms}(\sigma(\pi)) = \sigma^{-1} \text{Perms}(\pi) \sigma = \left\{ \sigma^{-1} \mu \sigma : \mu \in \text{Perms}(\pi) \right\}
\end{equation}
It follows that
\begin{align}
\sigma^{-1} T_{ \pi } \sigma & = { 1 \over   |\text{Perms} ( \pi ) |   } \sum_{ \mu \in \text{Perms} ( \pi ) } \sigma^{-1} \mu \sigma \nonumber \\ 
& = { 1 \over  | \text{Perms} ( \sigma (\pi)  ) | }   \sum_{ \tilde \mu \in \text{Perms} ( \sigma (\pi) ) }  \tilde \mu  \nonumber \\ 
&  = T_{ \sigma (\pi) } 
\end{align}
We also observe that
\bea 
C^{\pi_1}_{\pi_2} = C^{\sigma (\pi_1)}_{\sigma(\pi_2)} \hspace{100pt} \widetilde C^{\pi_1}_{\pi_2} = \widetilde C^{\sigma(\pi_1)}_{\sigma(\pi_2)} 
\label{permutation of C}
\eea
This is because the incidence relations of the poset of set partitions are unchanged when we go from 
set partitions of $ \{ 1, 2, \cdots , n \}$ to set partitions of $ \{ \sigma (1) , \cdots, \sigma ( n ) \} $.

It follows from \eqref{permutation of T} and \eqref{permutation of C} that
\begin{align}
\sigma^{-1} M_\pi \sigma & = \sum_{\pi' \in \Pi(n)} \widetilde{C}^{\pi'}_{\pi} \sigma^{-1} T_{\pi'} \sigma \nonumber \\
& = \sum_{\pi' \in \Pi(n)} \widetilde{C}^{\sigma(\pi')}_{\sigma(\pi)} T_{\sigma(\pi')} \nonumber \\
& = \sum_{\pi' \in \Pi(n)} \widetilde{C}^{\pi'}_{\sigma(\pi)} T_{\pi'} \nonumber \\
& = M_{\sigma(\pi)}
\label{conjugation of M_pi}
\end{align}
where in going from the 2nd to 3rd line we have reparameterised the sum by $\pi' \rightarrow \sigma(\pi')$, which clearly just permutes the set partitions in $\Pi(n)$ among each other.

It is immediate from \eqref{inner product} that
\begin{equation}
g(\sigma \alpha \sigma^{-1}, \sigma \beta \sigma^{-1}) = g(\alpha, \beta)
\label{permutation invariance for g}
\end{equation}
Applying this to $\alpha = M_\pi, \beta = M_{\pi'}$ and using \eqref{conjugation of M_pi}, we have
\begin{equation}
g(M_{\sigma(\pi)}, M_{\sigma(\pi')}) = g(M_\pi, M_{\pi'})
\end{equation}
We would like to show that $G$ also has this property. To see this, note that $G$ is defined by the property \eqref{inverse property G}, so we need to show that the matrix $G(M_{\sigma(\pi)}, M_{\sigma(\pi')})$ satisfies the same relation.
\begin{align}
\sum_{\pi' \in \Pi(n)} g(M_\pi, M_{\pi'}) G(M_{\sigma(\pi')}, M_{\sigma(\pi'')}) 
& = \sum_{\pi' \in \Pi(n)} g(M_\pi, M_{\sigma^{-1}(\pi')}) G(M_{\pi'}, M_{\sigma(\pi'')}) \nonumber \\
& = \sum_{\pi' \in \Pi(n)} g(M_{\sigma(\pi)}, M_{\pi'}) G(M_{\pi'}, M_{\sigma(\pi'')}) \nonumber \\
& = \delta_{\sigma(\pi) \sigma(\pi'')} \nonumber \\
& = \delta_{\pi \pi''}
\label{permutation invariance logic}
\end{align}
Therefore 
\begin{equation}
G(M_{\sigma(\pi)}, M_{\sigma(\pi')}) = G(M_\pi, M_{\pi'})
\label{permutation invariance for G}
\end{equation}
Next note that $|\pi| = |\sigma(\pi)|$, so when changing variables from $\pi$ to $\sigma(\pi)$, the restrictions $|\pi| > N$ or $|\pi| \leq N$ are maintained. This means we can repeat the steps in \eqref{permutation invariance logic} but using $G^>$ or $G^\leq$ instead. Hence
\begin{equation}
G^>(M_{\sigma(\pi)}, M_{\sigma(\pi')}) = G^>(M_\pi, M_{\pi'}) \hspace{50pt} G^\leq(M_{\sigma(\pi)}, M_{\sigma(\pi')}) = G^\leq(M_\pi, M_{\pi'})
\label{permutation invariance for G>, G<}
\end{equation}
where $\pi, \pi'$ satisfy the appropriate constraints on their length for the two operations.

Using the definition \eqref{Mbar definition}, as well as \eqref{conjugation of M_pi}, \eqref{permutation invariance for g} and \eqref{permutation invariance for G>, G<}
\begin{align}
\sigma^{-1} \bar{M}_\pi \sigma & = \sigma^{-1} M_\pi \sigma - \sum_{\substack{\pi_1, \pi_2 \in \Pi(n) \\ |\pi_1|, |\pi_2| > N}} G^> (M_{\pi_1}, M_{\pi_2}) g(M_\pi, M_{\pi_1}) \sigma^{-1} M_{\pi_2} \sigma \nonumber \\
& = M_{\sigma(\pi)} - \sum_{\substack{\pi_1, \pi_2 \in \Pi(n) \\ |\pi_1|, |\pi_2| > N}} G^> (M_{\pi_1}, M_{\pi_2}) g(M_\pi, M_{\pi_1}) M_{\sigma(\pi_2)} \nonumber \\
& = M_{\sigma(\pi)} - \sum_{\substack{\pi_1, \pi_2 \in \Pi(n) \\ |\pi_1|, |\pi_2| > N}} G^> (M_{\pi_1}, M_{\sigma^{-1}(\pi_2)}) g(M_\pi, M_{\pi_1}) M_{\pi_2} \nonumber \\
& = M_{\sigma(\pi)} - \sum_{\substack{\pi_1, \pi_2 \in \Pi(n) \\ |\pi_1|, |\pi_2| > N}} G^> (M_{\sigma(\pi_1)}, M_{\pi_2}) g(M_\pi, M_{\pi_1}) M_{\pi_2} \nonumber \\
& = M_{\sigma(\pi)} - \sum_{\substack{\pi_1, \pi_2 \in \Pi(n) \\ |\pi_1|, |\pi_2| > N}} G^> (M_{\pi_1}, M_{\pi_2}) g(M_\pi, M_{\sigma^{-1}(\pi_1)}) M_{\pi_2} \nonumber \\
& = M_{\sigma(\pi)} - \sum_{\substack{\pi_1, \pi_2 \in \Pi(n) \\ |\pi_1|, |\pi_2| > N}} G^> (M_{\pi_1}, M_{\pi_2}) g(M_{\sigma(\pi)}, M_{\pi_1}) M_{\pi_2} \nonumber \\
& = \bar{M}_{\sigma(\pi)}
\end{align}
We can now prove that $\cP_N$ and $\cP_H$ commute
\begin{align}
\cP_{ N} P_H ( \alpha ) 
& = { 1 \over |H| } \sum_{ \sigma \in H } 
\sum_{ \substack{ \pi_1 , \pi_2 \in \Pi(n) \\  |\pi_1|, |\pi_2| \le N } }
G^\leq ( \bar{M}_{ \pi_2} , \bar{M}_{ \pi_1} )  \bar{M}_{ \pi_2} g ( \bar{M}_{ \pi_1 } , \sigma \alpha \sigma^{-1} ) \nonumber \\
& = 
{ 1 \over |H| } 
\sum_{ \sigma \in H } \sum_{ \substack { \pi_1,  \pi_2 \in \Pi(n) \\  |\pi_1|, |\pi_2 | \le N }}  G^\leq ( \bar{M}_{ \pi_2} , \bar{M}_{ \pi_1} )  \bar{M}_{ \pi_2}   g ( \sigma^{-1}   \bar{M}_{ \pi_1} \sigma  , \alpha  ) \nonumber \\   
& =  { 1 \over |H| } 
\sum_{ \sigma \in H } \sum_{ \substack { 
		\pi_1 , \pi_2 \in \Pi(n) \\  |\pi_1|, |\pi_2 | \le N } } G^\leq ( \bar{M}_{ \pi_2} , \bar{M}_{ \pi_1} )
\bar{M}_{ \pi_2} g ( \bar{M}_{\sigma(\pi_1) } , \alpha )  \nonumber \\
&   = 
{ 1 \over |H| } 
\sum_{ \sigma \in H } \sum_{ \substack { 
		\pi_1 , \pi_2 \in \Pi(n) \\ |\pi_1|, |\pi_2 | \le N } }
G^\leq ( \bar{M}_{ \pi_2}  , \bar{M}_{ \sigma^{-1}(\pi_1)}  ) 
\bar{M}_{ \pi_2} g ( \bar{M}_{ \pi_1 } , \alpha ) \nonumber \\
&  =  { 1 \over |H| } 
\sum_{ \sigma \in H } \sum_{ \substack { 
		\pi_1 , \pi_2 \in \Pi(n) \\ |\pi_1|, |\pi_2 | \le N } } G^\leq (  \bar{M}_{ \sigma(\pi_2)}   , \bar{M}_{ \pi_1}  ) 
\bar{M}_{ \pi_2}  g ( \bar{M}_{ \pi_1 } , \alpha ) 
\\
&  =  { 1 \over |H| } 
\sum_{ \sigma \in H } \sum_{ \substack { 
		\pi_1 , \pi_2 \in \Pi(n) \\ |\pi_1|, |\pi_2 | \le N } } G^\leq (  \bar{M}_{ \pi_2}   , \bar{M}_{ \pi_1}  ) 
\bar{M}_{ \sigma^{-1}(\pi_2)}  g ( \bar{M}_{ \pi_1 } , \alpha ) \nonumber \\
&  =  { 1 \over |H| } 
\sum_{ \sigma \in H } \sum_{ \substack { 
		\pi_1 , \pi_2 \in \Pi(n) \\ |\pi_1|, |\pi_2 | \le N } } G^\leq (  \bar{M}_{ \pi_2}   , \bar{M}_{ \pi_1}  ) 
\sigma \bar{M}_{ \pi_2} \sigma^{-1}  g ( \bar{M}_{ \pi_1 } , \alpha ) \nonumber \\
& = P_H \cP_{N} ( \alpha )
\label{ComPNPH}  
\end{align}
We can interpret this in words as follows. Recall that permutations $ \sigma $ generate gauge invariant operators via \eqref{permutation isomorphism}. Imagine we start with the $n$-flavour gauge invariant operator generated by $ \sigma $, and then symmetrise the traces, and map that to symmetrised permutations. This means applying $ \cP$ then setting
\begin{gather}
\begin{gathered}
Z_1 , Z_2 , \dots , Z_{ n_1} \rightarrow X \\
Z_{ n_1 +1 } , Z_{ n_1 +2} , \dots , Z_n \rightarrow Y 
\end{gathered}
\end{gather}
On the other hand, we could specialise the $n$-flavour gauge invariants to $2$-flavour gauge invariants before projecting to symmetrised traces. Intuitively, thinking about traces, we don't see any reason for a difference between the two orders of arriving at symmetrised traces of two matrices. So we expect the two projectors to commute. Indeed they do as shown above.

\section{Hidden 2D topology: Permutation TFT2 for the counting and correlators at weak coupling }\label{sec:TFT2countCorr}

The connection between delta functions on symmetric group algebras and TFT2 is explained in \cite{CtoC}. 
We will give the delta function formulae and explain the TFT2 defects. 
\vskip.2cm
\noindent{\bf Lemma \ref{sec:TFT2countCorr}.1 } 
\vskip.2cm
\noindent In the problem of gauge invariants of $n$ matrices, each occuring once, the counting of symmetrised traces at large $N$ is given by 
\bea
\sum_{ \alpha \in S_n } \delta \big( \cP ( \alpha^{-1}  ) \cP ( \alpha )  \big) 
\eea
{\bf Proof}
\vskip.2cm
\noindent The symmetrised traces form the image of the hermitian projector $ \cP$. So the dimension of 
the space of symmetrised traces is calculated as 
\bea 
\operatorname{Dim}\left( \im \cP \right) & = &  \sum_{ \alpha \in S_n  } g( \alpha , \cP ( \alpha ) ) \cr
& =  & \sum_{ \alpha \in S_n } g( \alpha , \cP ( \cP ( \alpha ) ) ) \cr 
&  =  & \sum_{ \alpha \in S_n } g( \cP ( \alpha  ) , \cP ( \alpha ) ) \cr 
&  =  & \sum_{ \alpha \in S_n } \delta ( \cP ( \alpha^{-1}  ) \cP ( \alpha )  ) 
\eea
$\square$
\vskip.2cm
\noindent{\bf Proposition \ref{sec:TFT2countCorr}.2}
\vskip.2cm
\noindent The counting of quarter-BPS operators in the large $N$ limit in the free theory is given by 
\bea 
 \sum_{ \alpha \in S_{ n} } \delta \big( \cP_H( \alpha ) \alpha^{-1} \big) 
\eea
where $\cP_H$ is the flavour projector onto two flavours with $H = S_{n_1} \times S_{n_2}$ as described in \eqref{flavour projector}.
\vskip.2cm
\noindent{\bf Proof} 
\vskip.2cm
\noindent  We know that permutations can be used to construct 2-matrix gauge invariants and there is an equivalence up to conjugation by $H$, given in \eqref{trace invariance}. Using Burnside's Lemma to count the free field operators, we have 
\bea 
{ 1 \over |H| } \sum_{ \gamma \in H  } \sum_{ \alpha \in S_{ n} } 
\delta ( \gamma \alpha \gamma^{-1} \alpha^{-1} ) =  \sum_{ \alpha \in S_{ n} } \delta \big( \cP_H( \alpha ) \alpha^{-1} \big) 
\eea
This is the free field counting of 2-matrix operators \cite{CtoC}. $\square$
\vskip.2cm
\noindent{\bf Proposition \ref{sec:TFT2countCorr}.3} 
\vskip.2cm
\noindent The counting of 2-matrix symmetrised operators in the $ ( n_1 , n_2)$ sector is 
\bea 
\sum_{ \alpha \in S_{ n} } \delta ( \cP_H \cP ( \alpha )  \cP ( \alpha^{-1} )  ) 
\eea
{\bf Proof}
\vskip.2cm
\noindent Both $ \cP , \cP_H $ are hermitian with respect to the standard inner product on $ \mC S_n$ and they commute, so they can be simultaneously diagonalised. The dimension of the intersection of their images is equal to the trace of their product 
\begin{align}
\sum_{ \alpha \in S_n } g ( \alpha , \cP \cP_{H} ( \alpha ) ) & = \sum_{ \alpha \in S_n } g ( \alpha , \cP^2 \cP_{H} ( \alpha ) ) \nonumber \\
& = \sum_{ \alpha \in S_n } g ( \cP (\alpha) , \cP \cP_{H} ( \alpha ) ) \nonumber \\
& = \sum_{ \alpha \in S_n } \delta ( \cP ( \alpha^{-1}) \cP P_H ( \alpha ) ) \nonumber \\
& = \sum_{ \alpha \in S_{ n} } \delta ( \cP_H \cP ( \alpha )  \cP ( \alpha^{-1} )  ) 
\end{align}
$\square$
\vskip.2cm
\noindent{\bf Proposition \ref{sec:TFT2countCorr}.4}
\vskip.2cm
\noindent The counting formula for the finite $N$ quarter-BPS operators is
\bea 
\sum_{ \alpha \in S_n } \delta ( \cP_H \cP_{ N } ( \alpha )  \cP_{ N } ( \alpha^{-1}  ) )
\eea
{\bf Proof } 
\vskip.2cm
\noindent Given that we have proved the projector, hermiticity, and commutativity properties of $ \cP_N$ and $\cP_H$, we can calculate the dimension of the image of $ \cP_{ N } \cP_{ H}$ by repeating the steps we had for $ \cP$ and $\cP_H$
\begin{align}
\operatorname{Dim} \big( \im (\cP_N \cP_H) \big) & = \sum_{ \alpha \in S_n } g ( \alpha , \cP_N \cP_{H} ( \alpha ) ) \nonumber \\
& = \sum_{ \alpha \in S_n } g ( \alpha , \cP_N^2 \cP_{H} ( \alpha ) ) \nonumber \\
& = \sum_{ \alpha \in S_n } g ( \cP_N (\alpha) , \cP_N \cP_{H} ( \alpha ) ) \nonumber \\
& = \sum_{ \alpha \in S_n } \delta ( \cP_N ( \alpha^{-1}) \cP_N \cP_H ( \alpha ) ) \nonumber \\
& = \sum_{ \alpha \in S_{ n} } \delta ( \cP_H \cP_N ( \alpha )  \cP_N ( \alpha^{-1} )  )
\end{align}
$\square$
\vskip.2cm
\noindent{\bf Proposition \ref{sec:TFT2countCorr}.5} 
\vskip.2cm
\noindent The finite $N$ two-point function for BPS states can be written as
\begin{align}
& \left\langle \cG_N \cP_N \tr \left( \alpha_1 X^{\otimes n_1} Y^{\otimes n_2} \right) \ , \ \cG_N \cP_N \tr \left( \alpha_2 X^{\otimes n_1} Y^{\otimes n_2} \right) \right\rangle \nonumber \\ & \hspace{200pt} = \delta \Big( \cP_H \cP_{ N } ( \alpha_1 ) \cP_{ N } ( \alpha_2^{-1} )   \Omega_{ N}^{-1} \Big) 
\end{align}

This follows as in \cite{CtoC}. $ \Omega_N^{-1} \cP_N ( \alpha ) $  span the BPS states as $ \alpha $ runs over $ \mC ( S_n )$. The free field inner product is $ g_{FF} ( \alpha , \beta ) = g ( \alpha , \Omega_N \beta )$. The step forward in this paper is that we have an explicit construction of $ \cP_N$ 
using set partitions. 

\begin{figure}
	\centering
\begingroup%
  \makeatletter%
  \providecommand\color[2][]{%
    \errmessage{(Inkscape) Color is used for the text in Inkscape, but the package 'color.sty' is not loaded}%
    \renewcommand\color[2][]{}%
  }%
  \providecommand\transparent[1]{%
    \errmessage{(Inkscape) Transparency is used (non-zero) for the text in Inkscape, but the package 'transparent.sty' is not loaded}%
    \renewcommand\transparent[1]{}%
  }%
  \providecommand\rotatebox[2]{#2}%
  \newcommand*\fsize{\dimexpr\f@size pt\relax}%
  \newcommand*\lineheight[1]{\fontsize{\fsize}{#1\fsize}\selectfont}%
  \ifx\svgwidth\undefined%
    \setlength{\unitlength}{382.02094953bp}%
    \ifx\svgscale\undefined%
      \relax%
    \else%
      \setlength{\unitlength}{\unitlength * \real{\svgscale}}%
    \fi%
  \else%
    \setlength{\unitlength}{\svgwidth}%
  \fi%
  \global\let\svgwidth\undefined%
  \global\let\svgscale\undefined%
  \makeatother%
  \begin{picture}(1,0.34692653)%
    \lineheight{1}%
    \setlength\tabcolsep{0pt}%
    \put(0,0){\includegraphics[width=\unitlength,page=1]{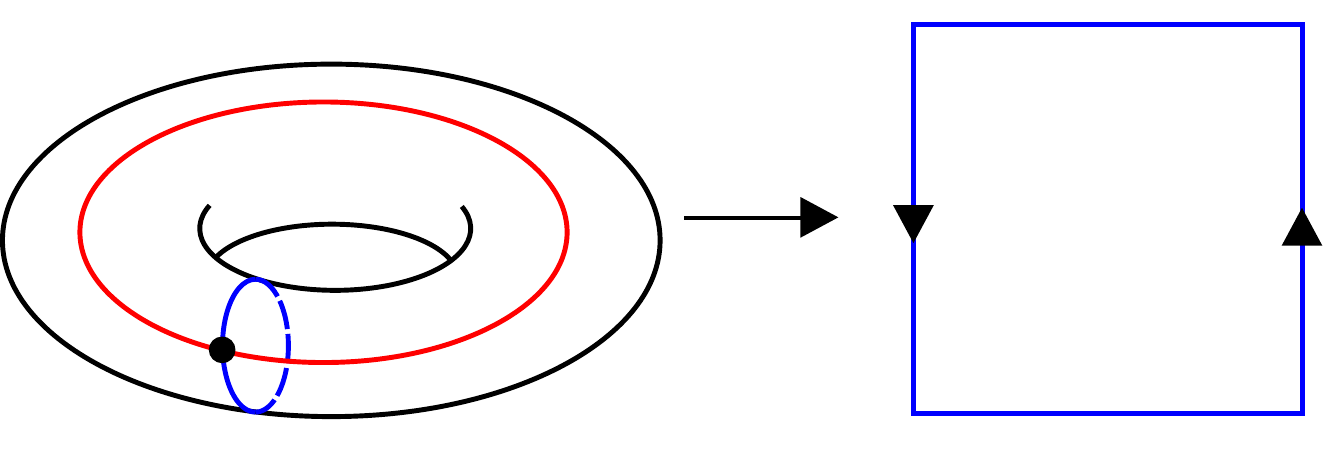}}%
    \put(0.43163977,0.18431995){\color[rgb]{0,0,0}\makebox(0,0)[lt]{\lineheight{1.25}\smash{\begin{tabular}[t]{l}$\cP_N$\end{tabular}}}}%
    \put(0.15624154,0.00347255){\color[rgb]{0,0,0}\makebox(0,0)[lt]{\lineheight{1.25}\smash{\begin{tabular}[t]{l}$H$\end{tabular}}}}%
    \put(0.79760154,0.288261){\color[rgb]{0,0,0}\makebox(0,0)[lt]{\lineheight{1.25}\smash{\begin{tabular}[t]{l}$\cP_N ( \alpha )$\end{tabular}}}}%
    \put(0.79184355,0.06086507){\color[rgb]{0,0,0}\makebox(0,0)[lt]{\lineheight{1.25}\smash{\begin{tabular}[t]{l}$\cP_N ( \alpha^{-1} )$\end{tabular}}}}%
    \put(0.70778384,0.16783337){\color[rgb]{0,0,0}\makebox(0,0)[lt]{\lineheight{1.25}\smash{\begin{tabular}[t]{l}$\gamma^{-1}$\end{tabular}}}}%
    \put(0.94025495,0.16783336){\color[rgb]{0,0,0}\makebox(0,0)[lt]{\lineheight{1.25}\smash{\begin{tabular}[t]{l}$\gamma$\end{tabular}}}}%
    \put(0,0){\includegraphics[width=\unitlength,page=2]{TFT2_partition_function.pdf}}%
  \end{picture}%
\endgroup%

	\caption{TFT2 partition function for finite $N$ weak coupling BPS counting}
	\label{fig:BPSwkCount}
\end{figure}

\begin{figure}
	\centering
	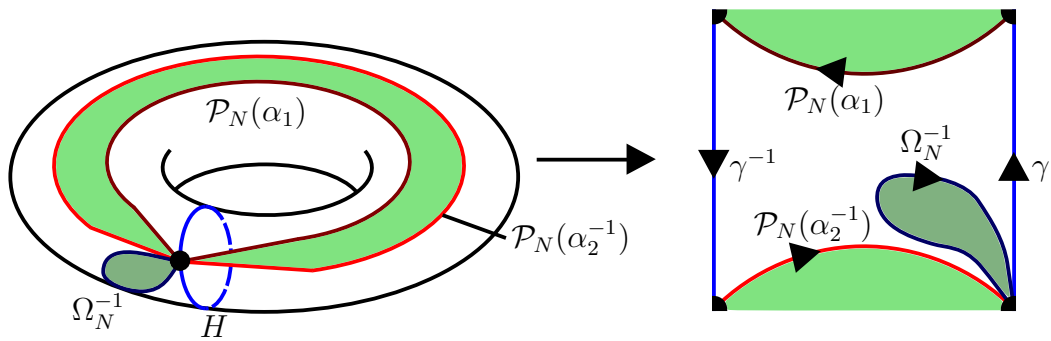
	\caption{TFT2 partition function for finite $N$ BPS 2-point function}
	\label{fig:BPSwk2pt}
\end{figure}

Now we will draw the TFT2 pictures corresponding to these delta function formulae. 
Figure \ref{fig:BPSwkCount} gives us the counting of weak coupling BPS operators. 
Figure \ref{fig:BPSwk2pt} gives the TFT2 formulation for the 2-point function of 
quarter BPS operators at weak coupling.
The one new ingredient in these TFT2 constructions is the $\cP_N$-defect which can be associated to a circle. The defect is defined by declaring that it modifies the permutation $ \alpha $ associated to that circle in the TFT2 to $ \cP_N ( \alpha )$. 

\clearpage

\section{Summary and Outlook }\label{sec:Conclusions} 

We have given a construction of quarter BPS operators in $ \cN =4 $ SYM with $U(N)$ gauge group, built from two matrices $X , Y$ and annihilated by  the 1-loop dilatation operator of  the $SU(2)$ sector. The construction depends on parameters $n , N$ which are arbitrary, with $n$ being the number of $X,Y$ matrices in the operator. The construction, given in sections \ref{sec:Construction} and \ref{sec:Projectors}, produces an orthogonal basis of operators which obeys an SEP-compatibility condition. The labels for the basis operators include a $U(2)$ Young diagram  $ \Lambda  $ and a $U(N)$ Young diagram $ p$, alongside multiplicity labels. The SEP-compatibility means that finite $N$ effects are captured simply by restricting the length of $p$ (defined as the length of the first column of the Young diagram) to be less than $N$. In section \ref{sec:Counting} we have given detailed formulae for the dimensions of the multiplicity spaces as a function of $ \Lambda , p$.

The understanding of holographic map between the quarter-BPS sector between  $ \cN =4$ SYM and  $ AdS_5 \times S^5$ is far less well-developed than the half-BPS sector.  The Young diagram labels for half-BPS states have provided valuable tools for precision mapping of states between SYM and the dual space-time. In the quarter BPS sector, there is a rich combinatoric structure involving $ \Lambda , p $ and  the plethysm problem underlying the multiplicities $ \cM ( \Lambda , p )$, which  control the structure of states. If will be fascinating to uncover the role of these structures in the dual space-time. Conceretely, reproducing the refined multiplicity formulae for specified $ \Lambda , p $ from the the weakly coupled gravitational dual, is an interesting problem. It has been shown \cite{BBP1203} that three-point functions involving quarter-BPS operators, alongside half-BPS operators are not renormalized, while the question of non-renormalization for more general quarter-BPS operators remains open. It will be interesting to compute correlators involving the operators  constructed here and compare these with calculations involving giant gravitons and LLM geometries, generalising the successful comparisons in the half-BPS giants/geometries and fluctuations thereof,  such as in  \cite{BKYZ1103,Lin2012,KSY1507,CDZ1208,DKZ}.  
A natural generalization of the work presented here is to develop the analogous discussion for $SO(N)/SP(N)$ gauge groups, building on earlier work at zero and weak coupling \cite{AABF0211,CDD1301,CDD1303,Kemp,Kemp1406,KRS1608,KRS1608,LR1804}.

\vskip.5cm 

\begin{center} 
{ \bf Acknowledgements} 
\end{center} 
SR is supported by the STFC consolidated grant ST/P000754/1 `` String Theory, Gauge Theory \& Duality” and  a Visiting Professorship at the University of the Witwatersrand, funded by a Simons Foundation grant (509116)  awarded to the Mandelstam Institute for Theoretical Physics. We are grateful for conversations on the subject of this paper to Matt Buican, Robert de Mello Koch, Joan Simon. 

\vskip.5cm 

\appendix

\section{$\Lambda = [3,2]$ sector}
\label{appendix: lambda = [3,2]}

In this and the following appendices we give explicit examples of quarter-BPS operators constructed using the algorithm presented in this paper. This appendix gives the operators in the $\Lambda = [3,2]$ sector with $M_\Lambda$ the highest weight state with field content $(3,2)$. Other states in the $U(2)$ representation can be reached by applying the lowering operator $\cJ_-$.

Throughout this section we will work with $\Lambda = [3,2]$ and $M_\Lambda = 
\fontsize{6pt}{0} \begin{ytableau}
1 & 1 & 1 \\ 2 & 2
\end{ytableau} \fontsize{12pt}{0}$, so we will suppress this index in operator labels.

For each BPS operator, we will first present it as a sum over the free field basis \eqref{U(2) basis definition} and then as a sum over symmetrised traces and commutator traces, for which we use the covariant bases discussed in section \ref{section: covariant trace bases}. The covariant symmetrised trace basis is
\begin{align}
t_{[3,2]} & = \tr X^3 \tr Y^2 - 2 \tr X^2 Y \tr X Y + \tr X^2 \tr X Y^2
\label{trace combination 1} \\
t_{[3,1,1]} & = \tr X^3  \left( \tr Y \right)^2 - 2 \tr X \tr X^2 Y \tr Y + \left( \tr X \right)^2 \tr X Y^2 \\
t_{[2,2,1]} & = \tr X \tr X^2 \tr Y^2  - \tr X \left( \tr X Y \right)^2 \\
t_{[2,1,1,1]} & = \tr X \tr X^2 \left( \tr Y \right)^2 - 2 \left( \tr X \right)^2 \tr X Y \tr Y + \left( \tr X \right)^3 \tr Y^2
\label{trace combination 4}
\end{align}
and the covariant commutator trace basis is 
\begin{align}
c_{[5]} & =  \tr X^3 Y^2 - \tr X^2 Y X Y = \tr X^2 [X,Y] Y \\
c_{[4,1]} & = \tr X \tr X^2 Y^2 - \tr X \tr \left( X Y \right)^2 = \tr X \tr X[X,Y]Y
\end{align}
For these two bases, the partition label describes the cycle structure of the multi-traces.

The free field operators can be written in terms of symmetrised and commutator traces
\ytableausetup{boxsize=3pt}
\begin{align}
\mathcal{O}_{
	\ydiagram{4,1}
} & = \frac{1}{6 \sqrt{10}} \left( 3 t_{[3,2]} + t_{[3,1,1]} + 4 t_{[2,2,1]} + t_{[2,1,1,1]} + 6 c_{[5]} + 4 c_{[4,1]} \right) \\
\mathcal{O}_{
	\ydiagram{3,2}
} & = \frac{1}{6 \sqrt{2}} \left( t_{[3,1,1]} + t_{[2,2,1]} + t_{[2,1,1,1]} - 3 c_{[5]} - 2 c_{[4,1]} \right) \\
\mathcal{O}_{
	\ydiagram{3,1,1} \, , \, \text{odd}
} & = \frac{1}{2 \sqrt{30}}
\left( 2 t_{[3,2]} - t_{[2,1,1,1]} - 4 c_{[4,1]} \right) \\
\mathcal{O}_{
	\ydiagram{3,1,1} \, , \, \text{even}
} & = \frac{1}{2 \sqrt{5}}
\left( t_{[3,1,1]} - t_{[2,2,1]} + c_{[5]} \right) \\
\mathcal{O}_{
	\ydiagram{2,2,1}
} & = \frac{1}{6 \sqrt{2}} \left( t_{[3,1,1]} + t_{[2,2,1]} - t_{[2,1,1,1]} - 3 c_{[5]} + 2 c_{[4,1]} \right) \\
\mathcal{O}_{
	\ydiagram{2,1,1,1}
} & = \frac{1}{6 \sqrt{10}} \left( - 3 t_{[3,2]} + t_{[3,1,1]} + 4 t_{[2,2,1]} - t_{[2,1,1,1]} + 6 c_{[5]} - 4 c_{[4,1]} \right)
\end{align}
The odd/even labels for the $R=[3,1,1]$ multiplicity come from the odd/even permutations used to produce the respective traces. All other zero coupling operators are defined uniquely by $\Lambda$ and $R$.

\subsection{BPS operators}

Following the algorithm, the BPS operators in the $\Lambda = [3,2]$ sector are
\begin{align}
S^{BPS}_{
	[2,1,1,1]
} & = \frac{1}{2 \sqrt{ 15 P_1}} \left( {\left(N - 2\right)} {\left(N - 3\right)} \left[
2  N 
\mathcal{O}_{
	\ydiagram{4,1} 
} - \sqrt{5}  {\left(N + 3\right)} 
\mathcal{O}_{
	\ydiagram{3,2} 
} \right] \right. \nonumber \\
& \hspace{50pt} 
+ N {\left(N + 3\right)} {\left(N - 3\right)} \left[
4 \sqrt{3} 
\mathcal{O}_{
	\ydiagram{3,1,1} \, , \, \text{odd}
}
+ 3 \sqrt{2}
\mathcal{O}_{
	\ydiagram{3,1,1} \, , \, \text{even}
} \right] \nonumber \\
& \hspace{50pt} \left. - 5 {\left(N + 3\right)} {\left(N + 2\right)} \left[
\sqrt{5}   {\left(N - 3\right)}
\mathcal{O}_{
	\ydiagram{2,2,1} 
} 
+ 4 N 
\mathcal{O}_{
	\ydiagram{2,1,1,1} 
} \right]
\right) \\
& = \frac{1}{2 \sqrt{6 P_1}} \Big[
{\left(N^{3} + 5 \, N^{2} + 2 \, N - 18\right)} t_{[2,1,1,1]}
- 4 {\left(N^{2} + 3 \, N - 3\right)} {\left(N + 1\right)} t_{[2,2,1]} \nonumber \\
& \hspace{50pt} -  {\left(N^{2} + 3 \, N - 6\right)} {\left(N + 2\right)} t_{[3,1,1]}
+ 3 N {\left(N + 2\right)} {\left(N + 1\right)} t_{[3,2]} \nonumber \\
& \hspace{50pt} + 4 {\left(2 \, N + 9\right)} {\left(N + 1\right)} c_{[4,1]}
- 18 {\left(N + 2\right)} {\left(N + 1\right)} c_{[5]}
\Big]
\end{align}
\begin{align}
S^{BPS}_{
	[2,2,1]
} & = \frac{1}{2 \sqrt{3 P_2}} \left( - \sqrt{5}  {\left(N - 1\right)} {\left(N - 2\right)} \left[
2  N
\mathcal{O}_{
	\ydiagram{4,1} 
} - \sqrt{5} {\left(N + 3\right)}
\mathcal{O}_{
	\ydiagram{3,2} 
} \right] \right. \nonumber \\
& \hspace{50pt}
- N {\left(N + 3\right)} {\left(N - 1\right)}  \sqrt{5} \left[ 
4 \sqrt{5} 
\mathcal{O}_{
	\ydiagram{3,1,1} \, , \, \text{odd}
}
+ 3 \sqrt{2}
\mathcal{O}_{
	\ydiagram{3,1,1} \, , \, \text{even}
} \right] \nonumber \\
& \hspace{50pt} \left.
- {\left(15 \, N^{3} + 48 \, N^{2} + 19 \, N + 6\right)}
\mathcal{O}_{
	\ydiagram{2,2,1} 
} 
+ 2 \sqrt{5} N {\left(3 \, N + 8\right)} 
\mathcal{O}_{
	\ydiagram{2,1,1,1} 
}
\right) \\
& = \frac{1}{2 \sqrt{6 P_2}} \Big[
{\left(5 \, N^{3} + 12 \, N^{2} - 12 \, N + 6\right)} t_{[2,1,1,1]}
+ 2 {\left(3 \, N - 2\right)} {\left(N - 1\right)} t_{[2,2,1]} \nonumber \\
& \hspace{50pt} - {\left(5 \, N^{3} + 12 \, N^{2} - 2 \, N - 4\right)} t_{[3,1,1]}
- N \left(5 \, N^{2} + 8 \, N - 2\right) t_{[3,2]} \nonumber \\
& \hspace{50pt} - 4 {\left(8 \, N + 3\right)} c_{[4,1]}
+ 6 {\left(5 \, N^{2} + 8 \, N - 2\right)} c_{[5]}
\Big]
\end{align}
\begin{align}
S^{BPS}_{
	[3,1,1]
} & = \frac{1}{\sqrt{15 P_3}} \left(
- {\left(2 \, N - 1\right)} {\left(N - 2\right)} 
\left[
2  N
\mathcal{O}_{
	\ydiagram{4,1} 
} - \sqrt{5} {\left(N + 3\right)}
\mathcal{O}_{
	\ydiagram{3,2} 
} \right] \right. \nonumber \\
& \hspace{50pt}
- \sqrt{3}  {\left(3 \, N^{3} + 9 \, N^{2} - 5 \, N - 2\right)}
\mathcal{O}_{
	\ydiagram{3,1,1} \, , \, \text{odd}
} \nonumber \\
& \hspace{50pt}
+ 3 \sqrt{2} {\left(3 \, N^{3} + 6 \, N^{2} - 4 \, N + 2\right)}
\mathcal{O}_{
	\ydiagram{3,1,1} \, , \, \text{even}
} \nonumber \\
& \hspace{50pt} \left. - {\left(N + 2\right)} {\left(N + 1\right)} \left[
\sqrt{5}
\mathcal{O}_{
	\ydiagram{2,2,1} 
} 
- 2
\mathcal{O}_{
	\ydiagram{2,1,1,1} 
} \right]
\right) \\
& = \frac{1}{2 \sqrt{6 P_3}} \Big[
{\left(N^{3} + 3 \, N^{2} - 5 \, N + 2\right)} t_{[2,1,1,1]}
- 4 N^{2} {\left(N + 1\right)}  t_{[2,2,1]} \nonumber \\
& \hspace{50pt} + 2 {\left(2 \, N^{3} + 4 \, N^{2} - 5 \, N + 2\right)} t_{[3,1,1]}
- 2 N^{2} {\left(N + 1\right)}  t_{[3,2]} \nonumber \\
& \hspace{50pt} + 8 {\left(N + 1\right)} {\left(N - 1\right)} c_{[4,1]}
+ 12 N {\left(N + 1\right)}  c_{[5]}
\Big]
\end{align}
\begin{align}
S^{BPS}_{
	[3,2]
} & = \frac{1}{2 \sqrt{15 P_4}} \left(
2 {\left(5 \, N^{2} - 5 \, N + 2\right)}
\mathcal{O}_{
	\ydiagram{4,1} 
} 
+ \sqrt{5}  {\left(4 \, N^{2} + 5 \, N - 2\right)}
\mathcal{O}_{
	\ydiagram{3,2} 
} \right. \nonumber \\
& \hspace{50pt}
+ 4 \sqrt{3} {\left(N - 1\right)}
\mathcal{O}_{
	\ydiagram{3,1,1} \, , \, \text{odd}
} 
- 3 \sqrt{2}  N
\mathcal{O}_{
	\ydiagram{3,1,1} \, , \, \text{even}
} \nonumber \\
& \hspace{50pt} \left. + ( N + 2 ) \left[
\sqrt{5}
\mathcal{O}_{
	\ydiagram{2,2,1} 
} 
- 2 
\mathcal{O}_{
	\ydiagram{2,1,1,1} 
} \right]
\right) \\
& = \frac{\sqrt{6}}{12 \sqrt{P_4}} \Big[
N^2 ( t_{[2,1,1,1]} + 2 t_{[2,2,1]} + t_{[3,1,1]} + t_{[3,2]} ) - 4 (N-1) c_{[4,1]} - 6 N c_{[5]}
\Big]
\end{align}
where the normalisation polynomials are
\begin{align}
P_1 & = 10 \, N^{6} + 74 \, N^{5} + 199 \, N^{4} + 252 \, N^{3} + 351 \, N^{2} + 648 \, N + 702 \\
P_2 & = 50 \, N^{6} + 220 \, N^{5} + 192 \, N^{4} - 78 \, N^{3} + 541 \, N^{2} - 156 \, N + 78 \\
P_3 & = 15 \, N^{6} + 50 \, N^{5} + 17 \, N^{4} - 66 \, N^{3} + 115 \, N^{2} - 60 \, N + 20 \\
P_4 & = 3 \, N^{4} + 5 \, N^{2} - 4 \, N + 2
\end{align}
In \cite{dHRyz0301}, these operators were studied, though in the $SU(N)$ gauge theory rather than the $U(N)$ theory. This means all traces whose cycle structure $p \vdash n$ contained one or more 1s do not contribute. In the $\Lambda = [3,2]$ sector, they found the single operator
\begin{equation}
\cO = N t_{[3,2]} - 6 c_{[5]}
\end{equation}
One can check that in each of the expansions above, $t_{[3,2]}$ and $c_{[5]}$ only appear in this ratio. We have found that by expanding the gauge group to $U(N)$ and allowing traces of a single matrix, there are three additional quarter-BPS operators.

\subsection{Norms of BPS operators}
\label{section: [3,2] norms}

The physical $\cF$-weighted norms of the BPS operators are
\begin{align}
\left| S^{BPS}_{
	[2,1,1,1]
} \right|^2 
& = \frac{
	{\left(N + 3\right)} {\left(N + 2\right)} {\left(N + 1\right)} N^2 {\left(N - 1\right)} {\left(N - 2\right)} {\left(N - 3\right)} Q_1
}{P_1} \\
\left| S^{BPS}_{
	[2,2,1]
} \right|^2 
& = \frac{
	{\left(N + 1\right)} N^2 {\left(N - 1\right)} {\left(N - 2\right)} Q_1 Q_2
}{P_2} \\
\left| S^{BPS}_{
	[3,1,1]
} \right|^2 
& = \frac{
	{\left(N + 2\right)} {\left(N + 1\right)} N {\left(N - 1\right)} {\left(N - 2\right)} Q_2 Q_3
}{P_3} \\
\left| S^{BPS}_{
	[3,2]
} \right|^2 
& = \frac{{\left(N + 2\right)} {\left(N + 1\right)} N^3 {\left(N - 1\right)} Q_3 }{P_4}
\end{align}
Where the polynomials in the numerators are
\begin{align}
Q_1 & = 10 \, N^{3} + 37 \, N^{2} + 11 \, N - 36 \\
Q_2 & = 5 \, N^{3} + 11 \, N^{2} - 7 \, N + 2 \\
Q_3 & = 3 \, N^{3} + 5 \, N^{2} - 5 \, N + 2
\end{align}
We discuss the combination of linear factors and $Q$ polynomials in the numerators in section \ref{section: [4,2] norms}.

\section{$\Lambda = [4,2]$ sector}
\label{appendix: lambda = [4,2]}
	
\ytableausetup{boxsize=7pt}
	
We give the BPS basis for the $\Lambda = [4,2]$ sector with $M_\Lambda$ the highest weight state corresponding to field content $(4,2)$.
	
Throughout this section we will work with $\Lambda = [4,2]$ and $M_\Lambda = 
	\fontsize{6pt}{0} \begin{ytableau}
	1 & 1 & 1 & 1 \\ 2 & 2
	\end{ytableau} \fontsize{12pt}{0}$, so we will suppress these in operator labels.
	
\subsection{Free field covariant basis from traces}
	
When writing our operators as sums over the free field covariant basis \eqref{U(2) basis definition}, we have made a choice about how to span the free field multiplicity space for $R = [4,2], [4,1,1], [3,2,1],$ $[3,1,1,1], [2,2,1,1]$. These choices are:
\ytableausetup{boxsize=2pt}
\fontsize{8pt}{0}
\begin{align}
	\mathcal{O}_{
		\ydiagram{4, 2},
		1
	} & = \frac{\sqrt{10}}{120} \,  \big(
	- t_{[4, 2] } - t_{[4, 1, 1] } + 6 t_{[3, 3] } + 6 t_{[3, 2, 1] , 1 } +
	6 t_{[3, 2, 1] , 2 } - 3 t_{[2, 2, 2] } - 3 t_{[2, 2, 1, 1] , 1 } \nonumber \\ & \hspace{50pt} + 6
	c_{[6] , 2 } + 6 c_{[5, 1] } - 4 c_{[4, 2] } + 2 c_{[4, 1, 1] } \big)
	\label{free field start} \\
	\mathcal{O}_{
		\ydiagram{4, 2},
		2
	} & = \frac{\sqrt{10}}{240} \,  \big(
	- 2 t_{[4, 1, 1] } - 6 t_{[3, 2, 1] , 1 } - 6 t_{[3, 1, 1, 1] } - 3
	t_{[2, 2, 1, 1] , 1 } + 9 t_{[2, 2, 1, 1] , 2 } - 3 t_{[2, 1, 1, 1, 1] } \nonumber \\ & \hspace{50pt}
	+ 24 c_{[6] , 1 } + 12 c_{[5, 1] } + 12 c_{[4, 2] } + 4 c_{[4, 1, 1] } \big) \\
	\mathcal{O}_{
		\ydiagram{4, 1, 1},
		1
	} & = \frac{\sqrt{3}}{108} \,  \big(
	- t_{[4, 2] } + 3 \, t_{[4, 1, 1] } + 6 \, t_{[3, 3] } + 6 \, t_{[3, 2,
		1] , 2 } + 6 \, t_{[3, 1, 1, 1] } - 3 \, t_{[2, 2, 2] }  + 3 \, t_{[2, 2,
		1, 1] , 1 } \nonumber \\ & \hspace{50pt}  + 6 \, t_{[2, 2, 1, 1] , 2 } 
	+ 12 \, c_{[6] , 1 } + 6 \, c_{[6] , 2 } + 6 \, c_{[5, 1] } + 2 \, c_{[4, 2] } \big) \\
	\mathcal{O}_{
		\ydiagram{4, 1, 1},
		2
	} & = \frac{\sqrt{15}}{1080} \,  \big(
	- 10 \, t_{[4, 2] } - 48 \, t_{[3, 3] } - 18 \, t_{[3, 2, 1] , 1 } - 12
	\, t_{[3, 2, 1] , 2 } + 6 \, t_{[3, 1, 1, 1] } - 12 \, t_{[2, 2, 2] } \nonumber \\ & \hspace{50pt}  + 3 \, t_{[2, 2, 1, 1] , 1 } - 21 \, t_{[2, 2, 1, 1] , 2 } + 9 \, t_{[2, 1, 1, 1, 1] } \nonumber \\ & \hspace{50pt}  - 24 \, c_{[6] , 1 } - 48 \, c_{[6] , 2 } + 60 \, c_{[5,
		1] } + 20 \, c_{[4, 2] } + 60 \, c_{[4, 1, 1] } \big) \\
	\mathcal{O}_{
		\ydiagram{3, 2, 1},
		1
	} & = \frac{\sqrt{10}}{90} \, \big(
	- t_{[4, 2] } + 6 \, t_{[3, 3] } - 3 \, t_{[3, 1, 1, 1] } + 3 \, t_{[2,
		2, 1, 1] , 1 } - 3 \, t_{[2, 2, 1, 1] , 2 }  - 3 \, c_{[5, 1] } + 2 \,
	c_{[4, 2] } \big) 
	\label{free field [3,2,1] start} \\
	\mathcal{O}_{
		\ydiagram{3, 2, 1},
		2
	} & = \frac{\sqrt{5}}{15} \,  \big(
	t_{[3, 2, 1] , 2 } + t_{[2, 2, 2] } - c_{[6] , 1 } + c_{[6] , 2 } +
	c_{[4, 1, 1] } \big) \\
	\mathcal{O}_{
		\ydiagram{3, 2, 1},
		3
	} & = \frac{\sqrt{5}}{90} \,  \big(
	- t_{[4, 1, 1] } - 6 \, t_{[3, 2, 1] , 1 } - 3 \, t_{[3, 2, 1] , 2 } +
	3 \, t_{[2, 1, 1, 1, 1] } + 12 \, c_{[6] , 1 } + 12 \, c_{[6] , 2 } - 4
	\, c_{[4, 1, 1] } \big)
	\label{free field [3,2,1] end} \\
	\mathcal{O}_{
		\ydiagram{3, 1, 1, 1},
		1
	} & = \frac{\sqrt{3}}{108} \,  \big(
	- t_{[4, 2] } - 3 \, t_{[4, 1, 1] } + 6 \, t_{[3, 3] } - 6 \, t_{[3, 2,
		1] , 2 } + 6 \, t_{[3, 1, 1, 1] } + 3 \, t_{[2, 2, 2] }  + 3 \, t_{[2, 2,
		1, 1] , 1 }  \nonumber \\ & \hspace{50pt} + 6 \, t_{[2, 2, 1, 1] , 2 } - 12 \, c_{[6] , 1 } - 6 \,
	c_{[6] , 2 } + 6 \, c_{[5, 1] } + 2 \, c_{[4, 2] } \big) \\
	\mathcal{O}_{
		\ydiagram{3, 1, 1, 1},
		2
	} & = \frac{\sqrt{15}}{1080} \, \big(
	- 10 \, t_{[4, 2] } - 48 \, t_{[3, 3] } + 18 \, t_{[3, 2, 1] , 1 } + 12
	\, t_{[3, 2, 1] , 2 } + 6 \, t_{[3, 1, 1, 1] } + 12 \, t_{[2, 2, 2] } \nonumber \\ & \hspace{50pt} +
	3 \, t_{[2, 2, 1, 1] , 1 } - 21 \, t_{[2, 2, 1, 1] , 2 } - 9 \, t_{[2,
		1, 1, 1, 1] } \nonumber \\ & \hspace{50pt} + 24 \, c_{[6] , 1 } + 48 \, c_{[6] , 2 } + 60 \, c_{[5,
		1] } + 20 \, c_{[4, 2] } - 60 \, c_{[4, 1, 1] } \big) \\
	\mathcal{O}_{
		\ydiagram{2, 2, 1, 1},
		1
	} & = \frac{\sqrt{10}}{120} \, \big(
	- t_{[4, 2] } + t_{[4, 1, 1] } + 6 \, t_{[3, 3] } - 6 \, t_{[3, 2, 1] ,
		1 } - 6 \, t_{[3, 2, 1] , 2 } + 3 \, t_{[2, 2, 2] }  - 3 \, t_{[2, 2, 1,
		1] , 1 } \nonumber \\ & \hspace{50pt} - 6 \, c_{[6] , 2 } + 6 \, c_{[5, 1] } - 4 \, c_{[4, 2] } - 2
	\, c_{[4, 1, 1] } \big)
	\label{free field [2,2,1,1] start} \\
	\mathcal{O}_{
		\ydiagram{2, 2, 1, 1},
		2
	} & = \frac{\sqrt{10}}{240} \, \big(
	2 \, t_{[4, 1, 1] } + 6 \, t_{[3, 2, 1] , 1 } - 6 \, t_{[3, 1, 1, 1]
	} - 3 \, t_{[2, 2, 1, 1] , 1 } + 9 \, t_{[2, 2, 1, 1] , 2 }  + 3 \,
	t_{[2, 1, 1, 1, 1] } \nonumber \\ & \hspace{50pt} - 24 \, c_{[6] , 1 } + 12 \, c_{[5, 1] } + 12 \,
	c_{[4, 2] } - 4 \, c_{[4, 1, 1] } \big)
	\label{free field [2,2,1,1] end}
\end{align}
\fontsize{12pt}{0}
The zero coupling operators with $R = [5,1],[3,3],[2,2,2],[2,1^4]$ are defined uniquely (up to a minus sign) by $\Lambda$ and $R$. We use
\fontsize{8pt}{0}
\begin{align}
	\mathcal{O}_{
		\ydiagram{5, 1}
	} & = - \frac{\sqrt{10}}{720} \left( 
	8 t_{[4,2]} 
	+ 2 t_{[4,1,1]}
	+ 24 t_{[3,3]} 
	+ 30 t_{[3,2,1],1} 
	+ 6 t_{[3,1,1,1]} 
	+ 12 t_{[2,2,2]} 
	+ 3 t_{[2,2,1,1],1} 
	- 21 t_{[2,2,1,1],2}
	\right. \nonumber \\ & \hspace{50pt} \left.
	+ 3 t_{[2,1,1,1,1]} 
	+ 72 c_{[6],1} 
	+ 24 c_{[6],2} 
	+ 60 c_{[5,1]} 
	+ 20 c_{[4,2]} 
	+ 20 c_{[4,1,1]}
	\right) \\
	\mathcal{O}_{
		\ydiagram{3,3}
	} & = \frac{\sqrt{10}}{360} \left(
	t_{[4,2]} 
	+ t_{[4,1,1]}
	- 6 t_{[3,3]} 
	- 12 t_{[3,2,1],1} 
	- 18 t_{[3,2,1],2} 
	- 6 t_{[3,1,1,1]} 
	- 3 t_{[2,2,2]} 
	+ 6 t_{[2,2,1,1],1} 
	+ 3 t_{[2,2,1,1],2} 
	\right. \nonumber \\ & \hspace{50pt} \left.
	- 3 t_{[2,1,1,1,1]} 
	+ 30 c_{[6],2} 
	+ 30 c_{[5,1]} 
	- 20 c_{[4,2]} 
	+ 10 c_{[4,1,1]}
	\right) \\
	\mathcal{O}_{
		\ydiagram{2,2,2}
	} & = \frac{\sqrt{10}}{360} \left(
	t_{[4,2]} 
	- t_{[4,1,1]}
	- 6 t_{[3,3]} 
	+ 12 t_{[3,2,1],1} 
	+ 18 t_{[3,2,1],2} 
	- 6 t_{[3,1,1,1]} 
	+ 3 t_{[2,2,2]} 
	+ 6 t_{[2,2,1,1],1} 
	+ 3 t_{[2,2,1,1],2} 
	\right. \nonumber \\ & \hspace{50pt} \left.
	+ 3 t_{[2,1,1,1,1]} 
	- 30 c_{[6],2} 
	+ 30 c_{[5,1]} 
	- 20 c_{[4,2]} 
	- 10 c_{[4,1,1]}
	\right) \\
	\mathcal{O}_{
		\ydiagram{2,1,1,1,1}
	} & = \frac{\sqrt{10}}{720} \left( 
	- 8 t_{[4,2]} 
	+ 2 t_{[4,1,1]}
	- 24 t_{[3,3]} 
	+ 30 t_{[3,2,1],1} 
	- 6 t_{[3,1,1,1]} 
	+ 12 t_{[2,2,2]} 
	- 3 t_{[2,2,1,1],1} 
	+ 21 t_{[2,2,1,1],2}
	\right. \nonumber \\ & \hspace{50pt} \left.
	+ 3 t_{[2,1,1,1,1]} 
	+ 72 c_{[6],1} 
	+ 24 c_{[6],2} 
	- 60 c_{[5,1]} 
	- 20 c_{[4,2]} 
	+ 20 c_{[4,1,1]}
	\right)
	\label{free field end}
\end{align}
\fontsize{12pt}{0}
where the symmetrised trace combinations we use are defined by
\fontsize{8pt}{0}
\begin{align}
	t_{[4,2]} & = 3 \tr X^4 \tr Y^2 - 6 \tr X^3 Y \tr X Y + 2 \tr X^2 \tr X^2 Y^2 + \tr X^2 \tr \left( X Y \right)^2 \\
	t_{[4,1,1]} & = 3 \tr X^4 \left( \tr Y \right)^2 - 6 \tr X \tr X^3 Y \tr Y  + 2 \left( \tr X \right)^2 \tr X^2 Y^2 + \left( \tr X \right)^2 \tr \left( X Y \right)^2  \\
	t_{[3,3]} & = \tr X^3 \tr X Y^2 - \left( \tr X^2 Y \right)^2  \\
	t_{[3,2,1],1} & = \tr X \tr X^3 \tr Y^2 - 2 \tr X \tr X^2 Y \tr X Y + \tr X \tr X^2 \tr X Y^2  \\
	t_{[3,2,1],2} & =  \tr X^3 \tr X Y \tr Y - \tr X \tr X^3 \tr Y^2 - \tr X^2 \tr X^2 Y \tr Y + \tr X \tr X^2 Y \tr X Y \\
	t_{[3,1,1,1]} & = \tr X \tr X^3 \left( \tr Y \right)^2 - 2 \left( \tr X \right)^2 \tr X^2 Y \tr Y + \left( \tr X \right)^3 \tr X Y^2   \\
	t_{[2,2,2]} & = \left( \tr X^2 \right)^2 \tr Y^2 - \tr X^2 \left( \tr X Y \right)^2 \\
	t_{[2,2,1,1],1} & = \left( \tr X^2  \tr Y \right)^2 - 2 \tr X \tr X^2 \tr X Y \tr Y + \left( \tr X \tr X Y \right)^2 \\
	t_{[2,2,1,1],2} & = \left( \tr X \tr X Y \right)^2 - \left( \tr X \right)^2 \tr X^2 \tr Y^2 \\
	t_{[2,1,1,1,1]} & = \left( \tr X \right)^2 \tr X^2 \left( \tr Y \right)^2 - 2 \left( \tr X \right)^3 \tr X Y \tr Y + \left( \tr X \right)^4 \tr Y^2 
\end{align}
\fontsize{12pt}{0}
along with the commutators
\fontsize{8pt}{0}
\begin{align}
	c_{[6],1} & = \tr X^4 Y^2 - \tr X^3 Y X Y = \tr X^3 [X,Y] Y 
	\label{covariant commutators 1} \\
	c_{[6],2} & = \tr X^3 Y X Y - \tr \left( X^2 Y \right)^2 = \tr X^2 [X,Y] Y^2 \\
	c_{[5,1]} & = \tr X \tr X^3 Y^2 - \tr X \tr X^2 Y X Y = \tr X \tr X^2 [X,Y] Y \\
	c_{[4,2]} & = \tr X^2 \tr X^2 Y^2 - \tr X^2 \tr \left( X Y \right)^2 = \tr X^2 \tr X [X,Y] Y \\
	c_{[4,1,1]} & = \left( \tr X \right)^2 \tr X^2 Y^2 - \left( \tr X \right)^2 \tr \left( X Y \right)^2 = \left( \tr X \right)^2 \tr X [X,Y] Y
	\label{covariant commutators end}
\end{align} \fontsize{12pt}{0}
These are respectively the covariant symmetrised trace and commutator trace bases for the $\Lambda = [4,2]$ sector with $M_\Lambda$ the highest weight state, as discussed in section \ref{section: covariant trace bases}.

\subsection{Quarter-BPS basis}

We now give the end result of the construction algorithm for quarter-BPS operators in the $\Lambda = [4,2]$ sector. The operators in this section are very lengthy to write out, so in the interests of brevity we only express them as a sum of free field operators. An expression in terms of trace can be found by substituting (\ref{free field start}-\ref{free field end}).

For $p=[3,2,1]$ and $[2,2,1,1]$ there are two BPS operators. For these, we have chosen the multiplicity space basis using the alternative orthogonalisation algorithm of section \ref{sec: alternative algorithm}, beginning with the choice of free field multiplicities in (\ref{free field [3,2,1] start}-\ref{free field [3,2,1] end}) and (\ref{free field [2,2,1,1] start}-\ref{free field [2,2,1,1] end}) respectively.

We present the operators starting from the longest partition $p=[2,1,1,1,1]$ and progressing to the shortest, $p = [4,2]$.
\fontsize{8pt}{0}
\begin{align}
	S^{BPS}_{
		[2, 1, 1, 1, 1]
	} & = \frac{1}{ 6 \sqrt{3 P_ 0 }} \left(
	{\left(N - 1\right)}  {\left(N - 3\right)} {\left(N - 4\right)} \left[ 3 \sqrt{3} {\left(N - 2\right)} \left\{  \, N 
	\mathcal{O}_{
		\ydiagram{5, 1}
	}
	- {\left(N + 4\right)} 
	\mathcal{O}_{
		\ydiagram{4, 2}
		, 2
	} \right\} \right. \right.
	\nonumber \\
	& \hspace{20pt} \left.
	- \sqrt{2} N {\left(N + 4\right)} \left\{ 2 \sqrt{5} 
	\mathcal{O}_{
		\ydiagram{4, 1, 1}
		, 1
	}
	- 11
	\mathcal{O}_{
		\ydiagram{4, 1, 1}
		, 2
	} \right\} 
	- 2 \sqrt{3} \, {\left(N + 4\right)} {\left(N + 3\right)} \left\{
	\mathcal{O}_{
		\ydiagram{3, 2, 1}
		, 1
	}
	+ 4 \, \sqrt{2}
	\mathcal{O}_{
		\ydiagram{3, 2, 1}
		, 3
	} \right\} \right]
	\nonumber \\
	& \hspace{20pt}
	+ \sqrt{2} N {\left(N + 4\right)} {\left(N + 3\right)} {\left(N - 1\right)} {\left(N - 4\right)} \left[  10 \, \sqrt{5}
	\mathcal{O}_{
		\ydiagram{3, 1, 1, 1}
		, 1
	}
	+ 29
	\mathcal{O}_{
		\ydiagram{3, 1, 1, 1}
		, 2
	} \right]
	\nonumber \\
	& \hspace{20pt}
	+ {\left(N + 4\right)} {\left(N + 3\right)} {\left(N + 2\right)} \left[ 10 \sqrt{3} {\left(N - 3\right)} {\left(N - 4\right)} 
	\mathcal{O}_{
		\ydiagram{2, 2, 2}
	}\right.
	\nonumber \\
	& \hspace{20pt}
	\left. \left. \begin{gathered}
	+ 3 \sqrt{3} \, {\left(N - 1\right)} {\left(N - 4\right)} \left\{ \begin{gathered} 2
	\mathcal{O}_{
		\ydiagram{2, 2, 1, 1}
		, 1
	}
	+ 13
	\mathcal{O}_{
		\ydiagram{2, 2, 1, 1}
		, 2
	} \end{gathered} \right\} 
	+ 65 \sqrt{3} N {\left(N - 1\right)} 
	\mathcal{O}_{
		\ydiagram{2, 1, 1, 1, 1}
	} \end{gathered} \right]
	\right)
\end{align}
\fontsize{12pt}{0}
where the normalisation polynomial is
\fontsize{8pt}{0}
\begin{align}
	P_{0} & = 195 \, N^{10} + 2298 \, N^{9} + 9767 \, N^{8} + 17008 \, N^{7} + 21041 \, N^{6} + 74974 \, N^{5} + 135005 \, N^{4} - 144704 \, N^{3} \nonumber \\ & \hspace{20pt} - 399936 \, N^{2} - 62976 \, N + 707328
\end{align}
\fontsize{12pt}{0}
For $p = [2,2,1,1]$ there is a two-dimensional multiplicity space. The first operator is
\fontsize{8pt}{0}
\begin{align}
	S^{BPS}_{
		[2, 2, 1, 1]
		, 1
	} & = \frac{1}{ 6 \sqrt{30 P_1 }} \left(
	- 20 \sqrt{3} N {\left(N + 1\right)} {\left(N - 2\right)} {\left(N - 3\right)} P_{1,1}
	\mathcal{O}_{
		\ydiagram{5, 1}
	}
	\right.
	\nonumber \\
	& \hspace{20pt}
	+ \sqrt{3} {\left(N - 3\right)} P_{1,2} \left[ 3 {\left(N + 1\right)} {\left(N - 2\right)} 
	\mathcal{O}_{
		\ydiagram{4, 2}
		, 1
	}
	+ 5 {\left(N + 3\right)} {\left(N - 2\right)}
	\mathcal{O}_{
		\ydiagram{3, 3}
	}
	- 12 \sqrt{2} {\left(N + 3\right)} {\left(N + 1\right)}
	\mathcal{O}_{
		\ydiagram{3, 2, 1}
		, 2
	} \right]
	\nonumber \\
	& \hspace{20pt}
	+ 2 \sqrt{3} {\left(N + 1\right)} {\left(N - 3\right)} P_{1,3} \left[ 3 {\left(N - 2\right)}
	\mathcal{O}_{
		\ydiagram{4, 2}
		, 2
	}
	+ 8 \, \sqrt{2} {\left(N + 3\right)}
	\mathcal{O}_{
		\ydiagram{3, 2, 1}
		, 3
	} \right]
	\nonumber \\
	& \hspace{20pt}
	+ \sqrt{10} N {\left(N + 1\right)} P_{1,4} \left[ {\left(N - 3\right)} 
	\mathcal{O}_{
		\ydiagram{4, 1, 1}
		, 1
	}
	- 5 {\left(N + 3\right)}
	\mathcal{O}_{
		\ydiagram{3, 1, 1, 1}
		, 1
	} \right]
	\nonumber \\
	& \hspace{20pt}
	- 2 \sqrt{2} {\left(N + 1\right)} \left[ 5 N {\left(N - 3\right)} P_{1,5}
	\mathcal{O}_{
		\ydiagram{4, 1, 1}
		, 2
	}
	+ \sqrt{6} {\left(N + 3\right)} {\left(N - 3\right)} P_{1,6}
	\mathcal{O}_{
		\ydiagram{3, 2, 1}
		, 1
	}
	+ 5 N {\left(N + 3\right)} P_{1,7}
	\mathcal{O}_{
		\ydiagram{3, 1, 1, 1}
		, 2
	} \right]
	\nonumber \\
	& \hspace{20pt}
	+ 5 \sqrt{3} {\left(N + 3\right)} {\left(N + 2\right)} {\left(N + 1\right)}  P_{1,8} \left[ 5 {\left(N - 3\right)}
	\mathcal{O}_{
		\ydiagram{2, 2, 2}
	}
	+ 3 {\left(N - 1\right)}
	\mathcal{O}_{
		\ydiagram{2, 2, 1, 1}
		, 1
	} \right]
	\nonumber \\
	& \hspace{20pt} \left.
	+ 10 \sqrt{3} {\left(N + 3\right)} {\left(N + 2\right)} {\left(N + 1\right)} \left[ 3 P_{1,9}
	\mathcal{O}_{
		\ydiagram{2, 2, 1, 1}
		, 2
	}
	- 26 N P_{1,10}
	\mathcal{O}_{
		\ydiagram{2, 1, 1, 1, 1}
	} \right]
	\right)
\label{[2,2,1,1] 1}
\end{align}
\fontsize{12pt}{0}
where the normalisation and coefficient polynomials are
\fontsize{8pt}{0}
	\begin{align}
	P_{1} & = 1254825 \, N^{16} + 25236900 \, N^{15} + 212913135 \, N^{14} + 949347864 \, N^{13} + 2265287922 \, N^{12} + 2296326096 \, N^{11} \nonumber \\ & \hspace{20pt} - 483268806 \, N^{10} - 64991400 \, N^{9} + 7717590681 \, N^{8} + 4250132076 \, N^{7} - 14563157385 \, N^{6} \nonumber \\ & \hspace{20pt} - 5596987632 \, N^{5} + 20300164460 \, N^{4} + 5660498272 \, N^{3} - 5514459136 \, N^{2} + 14594125824 \, N \nonumber \\ & \hspace{20pt} + 12396386304 \\
	P_{1,1} & = 78 \, N^{4} + 180 \, N^{3} - 411 \, N^{2} - 510 \, N + 788 \\
	P_{1,2} & = 195 \, N^{5} + 1149 \, N^{4} + 687 \, N^{3} - 3927 \, N^{2} - 1552 \, N + 4448 \\
	P_{1,3} & = 195 \, N^{5} + 1257 \, N^{4} + 801 \, N^{3} - 5871 \, N^{2} - 3656 \, N + 9024 \\
	P_{1,4} & = 975 \, N^{5} + 6177 \, N^{4} + 3891 \, N^{3} - 27411 \, N^{2} - 16176 \, N + 40544 \\
	P_{1,5} & = 507 \, N^{5} + 3225 \, N^{4} + 2037 \, N^{3} - 14487 \, N^{2} - 8664 \, N + 21632 \\
	P_{1,6} & = 195 \, N^{5} + 1041 \, N^{4} + 573 \, N^{3} - 1983 \, N^{2} + 552 \, N - 128 \\
	P_{1,7} & = 1443 \, N^{5} + 9129 \, N^{4} + 5745 \, N^{3} - 40335 \, N^{2} - 23688 \, N + 59456 \\
	P_{1,8} & = 117 \, N^{4} + 720 \, N^{3} + 1041 \, N^{2} + 240 \, N + 992  \\
	P_{1,9} & = 429 \, N^{5} + 2247 \, N^{4} + 1215 \, N^{3} - 3585 \, N^{2} + 2056 \, N - 2112 \\
	P_{1,10} & = 54 \, N^{3} + 273 \, N^{2} + 120 \, N - 572
\end{align}
\fontsize{12pt}{0}
The second operator is
\fontsize{8pt}{0}
\begin{align}
	S^{BPS}_{
		[2, 2, 1, 1]
		, 2
	} & = \frac{1}{ 3 \sqrt{6 P_ 2 }} \left(
	\sqrt{3} {\left(N - 2\right)} {\left(N - 3\right)} N P_{2,1}
	\mathcal{O}_{
		\ydiagram{5, 1}
	}
	\right.
	\nonumber \\
	& \hspace{20pt}
	- \sqrt{3} {\left(N - 3\right)} P_{2,2} \left[ 3 {\left(N + 1\right)} {\left(N - 2\right)}
	\mathcal{O}_{
		\ydiagram{4, 2}
		, 1
	}
	+ 5 {\left(N + 3\right)} {\left(N - 2\right)}
	\mathcal{O}_{
		\ydiagram{3, 3}
	}
	- 12 \sqrt{2} {\left(N + 3\right)} {\left(N + 1\right)}
	\mathcal{O}_{
		\ydiagram{3, 2, 1}
		, 2
	} \right]
	\nonumber \\
	& \hspace{20pt}
	+ \sqrt{3} {\left(N - 3\right)} P_{2,3} \left[ 3 {\left(N - 2\right)} 
	\mathcal{O}_{
		\ydiagram{4, 2}
		, 2
	}
	+ 8 \sqrt{2} {\left(N + 3\right)}
	\mathcal{O}_{
		\ydiagram{3, 2, 1}
		, 3
	} \right]
	\nonumber \\
	& \hspace{20pt}
	+  \sqrt{10} N P_{2,4} \left[ {\left(N - 3\right)} 
	\mathcal{O}_{
		\ydiagram{4, 1, 1}
		, 1
	}
	- 5 {\left(N + 3\right)}
	\mathcal{O}_{
		\ydiagram{3, 1, 1, 1}
		, 1
	} \right]
	- \sqrt{2} {\left(N - 3\right)} N P_{2,5}
	\mathcal{O}_{
		\ydiagram{4, 1, 1}
		, 2
	}
	\nonumber \\
	& \hspace{20pt}
	+ 2 \sqrt{3} {\left(N + 3\right)} {\left(N - 3\right)} P_{2,6}
	\mathcal{O}_{
		\ydiagram{3, 2, 1}
		, 1
	}
	+ \sqrt{2} N {\left(N + 3\right)} P_{2,7}
	\mathcal{O}_{
		\ydiagram{3, 1, 1, 1}
		, 2
	}
	\nonumber \\
	& \hspace{20pt} \left.
	- 5 \sqrt{3} {\left(N + 3\right)} {\left(N + 2\right)} {\left(N - 3\right)} P_{2,8}
	\mathcal{O}_{
		\ydiagram{2, 2, 2}
	}
	+ 3 \sqrt{3} P_{2,9}
	\mathcal{O}_{
		\ydiagram{2, 2, 1, 1}
		, 1
	}
	- 6 \sqrt{3} P_{2,10}
	\mathcal{O}_{
		\ydiagram{2, 2, 1, 1}
		, 2
	}
	- 2 \sqrt{3} N P_{2,11}
	\mathcal{O}_{
		\ydiagram{2, 1, 1, 1, 1}
	}
	\right)
\label{[2,2,1,1] 2}
\end{align}
\fontsize{12pt}{0}
where the normalisation and coefficient polynomials are
\fontsize{8pt}{0}
\begin{align}
	P_{2} & = 64575225 \, N^{16} + 1221543180 \, N^{15} + 9292923450 \, N^{14} + 34312809600 \, N^{13} + 49747071546 \, N^{12} \nonumber \\ & \hspace{20pt} - 49520811024 \, N^{11} - 212528733480 \, N^{10} + 81502221096 \, N^{9} + 872883407025 \, N^{8} + 609873915684 \, N^{7} \nonumber \\ & \hspace{20pt} - 949480261506 \, N^{6} - 778095650280 \, N^{5} + 986491220724 \, N^{4} + 591265527264 \, N^{3} - 532623199736 \, N^{2} \nonumber \\ & \hspace{20pt} - 150593123520 \, N + 181872634752 \\
	P_{2,1} & = 135 \, N^{5} + 423 \, N^{4} + 999 \, N^{3} + 1653 \, N^{2} + 1716 \, N + 74 \\
	P_{2,2} & = 351 \, N^{5} + 1485 \, N^{4} - 783 \, N^{3} - 3669 \, N^{2} + 5448 \, N - 1832 \\
	P_{2,3} & = 189 \, N^{6} + 903 \, N^{5} - 429 \, N^{4} - 4851 \, N^{3} - 1590 \, N^{2} + 98 \, N - 1320 \\
	P_{2,4} & = 27 \, N^{6} - 30 \, N^{5} - 1560 \, N^{4} - 5250 \, N^{3} - 4959 \, N^{2} - 3420 \, N - 808 \\
	P_{2,5} & = 675 \, N^{6} + 2589 \, N^{5} - 7527 \, N^{4} - 35553 \, N^{3} - 24606 \, N^{2} - 13386 \, N - 7192 \\
	P_{2,6} & = 1593 \, N^{6} + 8247 \, N^{5} + 2379 \, N^{4} - 22659 \, N^{3} + 5526 \, N^{2} + 14562 \, N - 8648 \\
	P_{2,7} & = 135 \, N^{6} + 3189 \, N^{5} + 23673 \, N^{4} + 69447 \, N^{3} + 74574 \, N^{2} + 55014 \, N + 8968 \\
	P_{2,8} & = 2835 \, N^{5} + 17493 \, N^{4} + 21549 \, N^{3} - 19317 \, N^{2} - 10044 \, N + 15464 \\
	P_{2,9} & = 10035 \, N^{8} + 100587 \, N^{7} + 320580 \, N^{6} + 201774 \, N^{5} - 613761 \, N^{4} - 529313 \, N^{3} + 665098 \, N^{2} + 243952 \, N \nonumber \\ & \hspace{20pt} - 359952 \\
	P_{2,10} & = 1131 \, N^{7} + 10440 \, N^{6} + 29667 \, N^{5} + 13182 \, N^{4} - 54074 \, N^{3} - 45886 \, N^{2} + 22026 \, N + 24264 \\
	P_{2,11} & = 2280 \, N^{6} + 24384 \, N^{5} + 95505 \, N^{4} + 166002 \, N^{3} + 120739 \, N^{2} + 22034 \, N - 11694
\end{align}
\fontsize{12pt}{0}
For $p = [3,1,1,1]$ the operator is
\fontsize{8pt}{0}
\begin{align}
	S^{BPS}_{
		[3, 1, 1, 1]
	} & = \frac{1}{ 18 \, \sqrt{2 P_ 3 }} \left(
	- 3 N {\left(N - 2\right)} {\left(N - 3\right)} P_{3,1}
	\mathcal{O}_{
		\ydiagram{5, 1}
	}
	\right.
	\nonumber \\
	& \hspace{20pt}
	+ 6 {\left(N - 3\right)} P_{3,2}  \left[ 3 {\left(N + 1\right)} {\left(N - 2\right)} 
	\mathcal{O}_{
		\ydiagram{4, 2}
		, 1
	}
	+ 5 \, {\left(N + 3\right)} {\left(N - 2\right)}
	\mathcal{O}_{
		\ydiagram{3, 3}
	}
	- 12 \, \sqrt{2} {\left(N + 3\right)} {\left(N + 1\right)} 
	\mathcal{O}_{
		\ydiagram{3, 2, 1}
		, 2
	} \right]
	\nonumber \\
	& \hspace{20pt}
	+ 3 {\left(N - 3\right)} P_{3,3}  \left[ 3 \, {\left(N - 2\right)} 
	\mathcal{O}_{
		\ydiagram{4, 2}
		, 2
	}
	+ 8 \, \sqrt{2} {\left(N + 3\right)}
	\mathcal{O}_{
		\ydiagram{3, 2, 1}
		, 3
	} \right]
	\nonumber \\
	& \hspace{20pt}
	+ {\left(N - 3\right)} \left[ 2 \, \sqrt{30}  N P_{3,4}
	\mathcal{O}_{
		\ydiagram{4, 1, 1}
		, 1
	}
	- \, \sqrt{6} N P_{3,5}
	\mathcal{O}_{
		\ydiagram{4, 1, 1}
		, 2
	}
	- 6 \, {\left(N + 3\right)} P_{3,6}
	\mathcal{O}_{
		\ydiagram{3, 2, 1}
		, 1
	} \right]
	\nonumber \\
	& \hspace{20pt}
	+ 2 \, \sqrt{30} P_{3,7}
	\mathcal{O}_{
		\ydiagram{3, 1, 1, 1}
		, 1
	}
	- \sqrt{6} P_{3,8}
	\mathcal{O}_{
		\ydiagram{3, 1, 1, 1}
		, 2
	}
	\nonumber \\
	& \hspace{20pt} \left.
	- 3 ( N + 2 ) \left[  10 \, {\left(N + 3\right)} {\left(N - 3\right)} P_{3,9}
	\mathcal{O}_{
		\ydiagram{2, 2, 2}
	}
	- 6 P_{3,10} 
	\mathcal{O}_{
		\ydiagram{2, 2, 1, 1}
		, 1
	}
	- 3 P_{3,11} 
	\mathcal{O}_{
		\ydiagram{2, 2, 1, 1}
		, 2
	}
	+ P_{3,12}
	\mathcal{O}_{
		\ydiagram{2, 1, 1, 1, 1}
	} \right]
	\right)
\end{align}
\fontsize{12pt}{0}
where the normalisation and coefficient polynomials are
\fontsize{8pt}{0}
\begin{align}
	P_{3} & = 93476025 \, N^{16} + 1612393695 \, N^{15} + 11013446394 \, N^{14} + 34526289987 \, N^{13} + 29660697936 \, N^{12} \nonumber \\ & \hspace{20pt} - 98498965581 \, N^{11} - 203072674968 \, N^{10} + 154945270125 \, N^{9} + 449766055695 \, N^{8} \nonumber \\ & \hspace{20pt} - 624364696710 \, N^{7} - 1246035300318 \, N^{6} + 1119952316004 \, N^{5} + 1953728842580 \, N^{4} \nonumber \\ & \hspace{20pt} - 1114329042600 \, N^{3} - 1086753482680 \, N^{2} + 1691309503680 \, N + 1297828640736 \\
	P_{3,1} & = 7155 \, N^{5} + 22752 \, N^{4} - 21231 \, N^{3} - 76512 \, N^{2} + 21066 \, N + 63020 \\
	P_{3,2} & = 270 \, N^{5} + 1728 \, N^{4} + 1287 \, N^{3} - 6762 \, N^{2} - 4278 \, N + 9380 \\
	P_{3,3} & = 2025 \, N^{6} + 14460 \, N^{5} + 19239 \, N^{4} - 46512 \, N^{3} - 80274 \, N^{2} + 42292 \, N + 71520 \\
	P_{3,4} & = 2295 \, N^{6} + 16458 \, N^{5} + 22254 \, N^{4} - 51987 \, N^{3} - 91314 \, N^{2} + 47394 \, N + 80900 \\
	P_{3,5} & = 24435 \, N^{6} + 175044 \, N^{5} + 235749 \, N^{4} - 555432 \, N^{3} - 971334 \, N^{2} + 506028 \, N + 861760 \\
	P_{3,6} & = 135 \, N^{6} + 1524 \, N^{5} + 4881 \, N^{4} + 2712 \, N^{3} - 8046 \, N^{2} - 1476 \, N + 3520 \\
	P_{3,7} & = 18630 \, N^{8} + 168027 \, N^{7} + 436488 \, N^{6} - 22071 \, N^{5} - 1221552 \, N^{4} - 330750 \, N^{3} + 1226756 \, N^{2} \nonumber \\ & \hspace{20pt} - 644796 \, N - 1298232 \\
	P_{3,8} & = 37260 \, N^{8} + 400629 \, N^{7} + 1309611 \, N^{6} + 387381 \, N^{5} - 4795443 \, N^{4} - 5190456 \, N^{3} + 4201270 \, N^{2} \nonumber \\ & \hspace{20pt} + 6554016 \, N + 1298232 \\
	P_{3,9} & = 135 \, N^{5} + 609 \, N^{4} - 1257 \, N^{3} - 9444 \, N^{2} - 12438 \, N - 5860 \\
	P_{3,10} & = 525 \, N^{6} + 3840 \, N^{5} + 6681 \, N^{4} - 8262 \, N^{3} - 49724 \, N^{2} - 91728 \, N - 41832 \\
	P_{3,11} & = 6825 \, N^{6} + 61005 \, N^{5} + 175743 \, N^{4} + 113439 \, N^{3} - 265222 \, N^{2} - 407694 \, N - 160596 \\
	P_{3,12} & = 22575 \, N^{6} + 198375 \, N^{5} + 553953 \, N^{4} + 307269 \, N^{3} - 994562 \, N^{2} - 1589994 \, N - 649116
\end{align}
\fontsize{12pt}{0}
For $p = [2,2,2]$ the operator is
\fontsize{8pt}{0}
\begin{align}
	S^{BPS}_{
		[2, 2, 2]
	} & = \frac{1}{ 36 \sqrt{P_ 4 }} \left(
	30 \, \sqrt{2} N {\left(N - 2\right)} P_{4,1}
	\mathcal{O}_{
		\ydiagram{5, 1}
	}
	\right.
	\nonumber \\
	& \hspace{20pt}
	- 15 \sqrt{2} P_{4,2} \left[ 3 {\left(N + 1\right)} {\left(N - 2\right)} 
	\mathcal{O}_{
		\ydiagram{4, 2}
		, 1
	}
	+ 5 {\left(N + 3\right)} {\left(N - 2\right)}
	\mathcal{O}_{
		\ydiagram{3, 3}
	}
	- 12 \sqrt{2} {\left(N + 3\right)} {\left(N + 1\right)} 
	\mathcal{O}_{
		\ydiagram{3, 2, 1}
		, 2
	} \right]
	\nonumber \\
	& \hspace{20pt}
	+ 30 \, \sqrt{2} P_{4,3} \left[ 3 {\left(N - 2\right)} 
	\mathcal{O}_{
		\ydiagram{4, 2}
		, 2
	}
	+ 8 \sqrt{2} (N + 3)
	\mathcal{O}_{
		\ydiagram{3, 2, 1}
		, 3
	} \right]
	+ 10 \, \sqrt{15} N P_{4,4}
	\mathcal{O}_{
		\ydiagram{4, 1, 1}
		, 1
	}
	- 20 \, \sqrt{3} N P_{4,5}
	\mathcal{O}_{
		\ydiagram{4, 1, 1}
		, 2
	}
	\nonumber \\
	& \hspace{20pt}
	+ 60 \, \sqrt{2} {\left(N + 3\right)} P_{4,6}
	\mathcal{O}_{
		\ydiagram{3, 2, 1}
		, 1
	}
	+ 10 \sqrt{3} N \left[ \sqrt{5} P_{4,7}
	\mathcal{O}_{
		\ydiagram{3, 1, 1, 1}
		, 1
	}
	+ 2 P_{4,8}
	\mathcal{O}_{
		\ydiagram{3, 1, 1, 1}
		, 2
	} \right]
	+ 3 \, \sqrt{2} P_{4,9}
	\mathcal{O}_{
		\ydiagram{2, 2, 2}
	}
	\nonumber \\
	& \hspace{20pt} \left.
	+ 45 \, \sqrt{2} P_{4,10}
	\mathcal{O}_{
		\ydiagram{2, 2, 1, 1}
		, 1
	}
	- 90 \, \sqrt{2} P_{4,11}
	\mathcal{O}_{
		\ydiagram{2, 2, 1, 1}
		, 2
	}
	- 30 \, \sqrt{2} N P_{4,12}
	\mathcal{O}_{
		\ydiagram{2, 1, 1, 1, 1}
	}
	\right)
\end{align}
\fontsize{12pt}{0}
where the normalisation and coefficient polynomials are
\fontsize{8pt}{0}
\begin{align}
	P_{4} & = 149226300 \, N^{14} + 2094533640 \, N^{13} + 10660893948 \, N^{12} + 20470965300 \, N^{11} - 2209082715 \, N^{10} \nonumber \\ & \hspace{20pt} - 23656646682 \, N^{9} + 108969897216 \, N^{8} + 185022077310 \, N^{7} - 186235972937 \, N^{6} \nonumber \\ & \hspace{20pt} - 216166001512 \, N^{5} + 413959581308 \, N^{4} + 246958572128 \, N^{3} - 287690109584 \, N^{2} \nonumber \\ & \hspace{20pt} - 143358681600 \, N + 276161485248 \\
	P_{4,1} & = 135 \, N^{5} - 18 \, N^{4} - 15 \, N^{3} - 1022 \, N^{2} - 76 \, N - 816 \\
	P_{4,2} & = 864 \, N^{5} + 2430 \, N^{4} - 5973 \, N^{3} - 6119 \, N^{2} + 17876 \, N - 12228 \\
	P_{4,3} & = 243 \, N^{6} + 924 \, N^{5} - 1152 \, N^{4} - 3670 \, N^{3} + 5307 \, N^{2} + 2256 \, N - 2988 \\
	P_{4,4} & = 108 \, N^{6} + 402 \, N^{5} - 1065 \, N^{4} - 2588 \, N^{3} + 9471 \, N^{2} + 3376 \, N + 276 \\
	P_{4,5} & = 945 \, N^{6} + 3576 \, N^{5} - 5586 \, N^{4} - 16186 \, N^{3} + 34863 \, N^{2} + 13520 \, N - 8412 \\
	P_{4,6} & = 1971 \, N^{6} + 7512 \, N^{5} - 8238 \, N^{4} - 27854 \, N^{3} + 28821 \, N^{2} + 13552 \, N - 27444 \\
	P_{4,7} & = 906 \, N^{5} + 7485 \, N^{4} + 18394 \, N^{3} + 9099 \, N^{2} - 14000 \, N - 13164 \\
	P_{4,8} & = 285 \, N^{5} + 5955 \, N^{4} + 34507 \, N^{3} + 63369 \, N^{2} + 19738 \, N + 7536 \\
	P_{4,9} & = 48060 \, N^{7} + 429834 \, N^{6} + 1227525 \, N^{5} + 919710 \, N^{4} - 762363 \, N^{3} - 208286 \, N^{2} + 1345500 \, N + 946584 \\
	P_{4,10} & = 3702 \, N^{6} + 29949 \, N^{5} + 68544 \, N^{4} + 7491 \, N^{3} - 79610 \, N^{2} + 45780 \, N + 78984 \\
	P_{4,11} & = 699 \, N^{6} + 5163 \, N^{5} + 8795 \, N^{4} - 9599 \, N^{3} - 20952 \, N^{2} + 21736 \, N + 26328 \\
	P_{4,12} & = 1605 \, N^{5} + 14460 \, N^{4} + 42159 \, N^{3} + 36288 \, N^{2} - 16754 \, N - 19428
\end{align}
\fontsize{12pt}{0}
For $p = [3,2,1]$ there is a two-dimensional multiplicity space. The first operator is
\fontsize{8pt}{0}
\begin{align}
	S^{BPS}_{
		[3, 2, 1]
		, 1
	} & = \frac{1}{ 45 \sqrt{P_ 5 }} \left(
	30 \, \sqrt{10} {\left(N - 2\right)} N P_{5,1}
	\mathcal{O}_{
		\ydiagram{5, 1}
	}
	\right.
	- 6 \sqrt{10} {\left(N - 2\right)} P_{5,2} \left[ 3 {\left(N + 1\right)} 
	\mathcal{O}_{
		\ydiagram{4, 2}
		, 1
	}
	+ 5 {\left(N + 3\right)}
	\mathcal{O}_{
		\ydiagram{3, 3}
	} \right]
	\nonumber \\
	& \hspace{20pt}
	- 18 \, \sqrt{10} {\left(N - 2\right)} P_{5,3}
	\mathcal{O}_{
		\ydiagram{4, 2}
		, 2
	}
	- 20 \, \sqrt{3} N \left[ P_{5,4}
	\mathcal{O}_{
		\ydiagram{4, 1, 1}
		, 1
	}
	- \sqrt{5} P_{5,5}
	\mathcal{O}_{
		\ydiagram{4, 1, 1}
		, 2
	} \right]
	- 12 \, \sqrt{10} P_{5,6}
	\mathcal{O}_{
		\ydiagram{3, 2, 1}
		, 1
	}
	\nonumber \\
	& \hspace{20pt}
	- 6 \, \sqrt{5} P_{5,7}
	\mathcal{O}_{
		\ydiagram{3, 2, 1}
		, 2
	}
	+ 3 \, \sqrt{5} P_{5,8}
	\mathcal{O}_{
		\ydiagram{3, 2, 1}
		, 3
	}
	- 20 \, \sqrt{3} N \left[ P_{5,9}
	\mathcal{O}_{
		\ydiagram{3, 1, 1, 1}
		, 1
	}
	- \sqrt{5} P_{5,10}
	\mathcal{O}_{
		\ydiagram{3, 1, 1, 1}
		, 2
	} \right]
	\nonumber \\
	& \hspace{20pt} \left.
	- 6  \, \sqrt{10} {\left(N + 2\right)} P_{5,11} \left[ 5
	\mathcal{O}_{
		\ydiagram{2, 2, 2}
	}
	+ 3
	\mathcal{O}_{
		\ydiagram{2, 2, 1, 1}
		, 1
	} \right]
	- 6 \, \sqrt{10} {\left(N + 2\right)} \left[ 3 P_{5,12}
	\mathcal{O}_{
		\ydiagram{2, 2, 1, 1}
		, 2
	}
	- 5 N P_{5,13}
	\mathcal{O}_{
		\ydiagram{2, 1, 1, 1, 1}
	} \right]
	\right)
\end{align}
\fontsize{12pt}{0}
where the normalisation and coefficient polynomials are
\fontsize{8pt}{0}
\begin{align}
	P_{5} & = 1329483780 \, N^{14} + 13761404280 \, N^{13} + 47552297508 \, N^{12} + 41944792356 \, N^{11} - 43156801080 \, N^{10} \nonumber \\ & \hspace{20pt} + 23239162764 \, N^{9} - 47497601127 \, N^{8} - 299164340106 \, N^{7} + 683116078397 \, N^{6} \nonumber \\ & \hspace{20pt} + 45647911732 \, N^{5} - 883683643044 \, N^{4} + 341394177280 \, N^{3} + 617090703216 \, N^{2} \nonumber \\ & \hspace{20pt} - 378227252672 \, N + 179121262144 \\
	P_{5,1} & = 3708 \, N^{5} + 2172 \, N^{4} - 19509 \, N^{3} + 17427 \, N^{2} + 9416 \, N - 13724 \\
	P_{5,2} & = 1656 \, N^{5} + 5619 \, N^{4} - 10194 \, N^{3} - 12393 \, N^{2} + 36194 \, N - 24704 \\
	P_{5,3} & = 5076 \, N^{6} + 23490 \, N^{5} - 14985 \, N^{4} - 85957 \, N^{3} + 116006 \, N^{2} + 32240 \, N - 75024 \\
	P_{5,4} & = 11808 \, N^{6} + 54255 \, N^{5} - 34545 \, N^{4} - 194501 \, N^{3} + 255813 \, N^{2} + 75970 \, N - 174752 \\
	P_{5,5} & = 12492 \, N^{6} + 57498 \, N^{5} - 36627 \, N^{4} - 207175 \, N^{3} + 274254 \, N^{2} + 80120 \, N - 184816 \\
	P_{5,6} & = 6462 \, N^{7} + 45960 \, N^{6} + 66285 \, N^{5} - 117979 \, N^{4} - 124298 \, N^{3} + 345680 \, N^{2} + 16132 \, N - 99960 \\
	P_{5,7} & = 16326 \, N^{7} + 99666 \, N^{6} + 144600 \, N^{5} - 54717 \, N^{4} - 52871 \, N^{3} + 89526 \, N^{2} + 116 \, N + 579336 \\
	P_{5,8} & = 165978 \, N^{7} + 1065798 \, N^{6} + 1496280 \, N^{5} - 1235731 \, N^{4} - 1949513 \, N^{3} + 8498 \, N^{2} + 679228 \, N + 177528 \\
	P_{5,9} & = 14061 \, N^{5} + 85773 \, N^{4} + 110389 \, N^{3} - 97565 \, N^{2} - 111014 \, N + 4008 \\
	P_{5,10} & = 15126 \, N^{5} + 66642 \, N^{4} - 35467 \, N^{3} - 293389 \, N^{2} + 81536 \, N + 220812 \\
	P_{5,11} & = 1137 \, N^{5} + 5721 \, N^{4} - 3097 \, N^{3} - 35915 \, N^{2} - 21618 \, N - 12024 \\
	P_{5,12} & = 6462 \, N^{5} + 40026 \, N^{4} + 56743 \, N^{3} - 30825 \, N^{2} - 44698 \, N + 8016 \\
	P_{5,13} & = 4332 \, N^{4} + 26304 \, N^{3} + 32807 \, N^{2} - 32861 \, N - 35466 
\end{align}
The second operator is	
\begin{align}
	S^{BPS}_{
		[3, 2, 1]
		, 2
	} & = \frac{1}{ 18 \sqrt{P_ 6 }} \left(
	- 3 \, {\left(N - 2\right)} N P_{6,1}
	\mathcal{O}_{
		\ydiagram{5, 1}
	}
	- 3 {\left(N - 2\right)} P_{6,2} \left[ 3 \, {\left(N + 1\right)} 
	\mathcal{O}_{
		\ydiagram{4, 2}
		, 1
	}
	+ 5 \, {\left(N + 3\right)}
	\mathcal{O}_{
		\ydiagram{3, 3}
	} \right] \right.
	\nonumber \\
	& \hspace{20pt}
	+ 9 ( N - 2) P_{6,3} 
	\mathcal{O}_{
		\ydiagram{4, 2}
		, 2
	}
	+ \sqrt{6} N \left[ \sqrt{5} P_{6,4}
	\mathcal{O}_{
		\ydiagram{4, 1, 1}
		, 1
	}
	- P_{6,5}
	\mathcal{O}_{
		\ydiagram{4, 1, 1}
		, 2
	} \right]
	- 6 P_{6,6}
	\mathcal{O}_{
		\ydiagram{3, 2, 1}
		, 1
	}
	\nonumber \\
	& \hspace{20pt}
	+ 12 \, \sqrt{2} P_{6,7}
	\mathcal{O}_{
		\ydiagram{3, 2, 1}
		, 2
	}
	- 3 P_{6,8} \left[ 8 \, \sqrt{2} {\left(N - 1\right)}
	\mathcal{O}_{
		\ydiagram{3, 2, 1}
		, 3
	}
	- 3 ( N + 2 ) 
	\mathcal{O}_{
		\ydiagram{2, 2, 1, 1}
		, 2
	} \right]
	+ \sqrt{30} N P_{6,9}
	\mathcal{O}_{
		\ydiagram{3, 1, 1, 1}
		, 1
	}
	\nonumber \\
	& \hspace{20pt} \left.
	- \sqrt{6} N P_{6,10}
	\mathcal{O}_{
		\ydiagram{3, 1, 1, 1}
		, 2
	}
	+ 3 ( N + 2 ) P_{6,11}  \left[ 5 
	\mathcal{O}_{
		\ydiagram{2, 2, 2}
	}
	+ 3
	\mathcal{O}_{
		\ydiagram{2, 2, 1, 1}
		, 1
	} \right]
	- 3 N {\left(N + 2\right)} P_{6,12}
	\mathcal{O}_{
		\ydiagram{2, 1, 1, 1, 1}
	}
	\right)
\end{align}
\fontsize{12pt}{0}
where the normalisation and coefficient polynomials are
\fontsize{8pt}{0}
\begin{align}
	P_{6} & = 433202580 \, N^{14} + 4164719976 \, N^{13} + 11536183026 \, N^{12} - 111051000 \, N^{11} - 29053464768 \, N^{10} \nonumber \\ & \hspace{20pt} + 10364014080 \, N^{9} + 31360792437 \, N^{8} - 51088773768 \, N^{7} + 37140544622 \, N^{6} - 29831349568 \, N^{5} \nonumber \\ & \hspace{20pt} + 55411748788 \, N^{4} - 79360524160 \, N^{3} + 66216685440 \, N^{2} - 31168716800 \, N + 6758052160 \\
	P_{6,1} & = 864 \, N^{5} + 7734 \, N^{4} - 29931 \, N^{3} + 34329 \, N^{2} - 2260 \, N - 8780 \\
	P_{6,2} & = 5724 \, N^{5} + 7509 \, N^{4} - 30912 \, N^{3} + 37572 \, N^{2} - 28648 \, N + 10144 \\
	P_{6,3} & = 4104 \, N^{6} + 12552 \, N^{5} - 15267 \, N^{4} - 24025 \, N^{3} + 50968 \, N^{2} - 18276 \, N - 4944 \\
	P_{6,4} & = 2484 \, N^{6} + 11871 \, N^{5} - 7131 \, N^{4} - 54710 \, N^{3} + 93012 \, N^{2} - 18048 \, N - 20032 \\
	P_{6,5} & = 22248 \, N^{6} + 85140 \, N^{5} - 74325 \, N^{4} - 290915 \, N^{3} + 524952 \, N^{2} - 127020 \, N - 94960 \\
	P_{6,6} & = 54810 \, N^{7} + 273690 \, N^{6} + 107331 \, N^{5} - 533633 \, N^{4} + 371678 \, N^{3} - 35050 \, N^{2} - 241112 \, N + 130440 \\
	P_{6,7} & = 7830 \, N^{7} + 45693 \, N^{6} + 36684 \, N^{5} - 104202 \, N^{4} - 1453 \, N^{3} + 78398 \, N^{2} - 91900 \, N + 46392 \\
	P_{6,8} & = 516 \, N^{5} + 5496 \, N^{4} + 12187 \, N^{3} - 4448 \, N^{2} - 12124 \, N + 4416 \\
	P_{6,9} & = 3717 \, N^{5} + 23415 \, N^{4} + 29524 \, N^{3} - 26804 \, N^{2} - 15272 \, N + 2208 \\
	P_{6,10} & = 16416 \, N^{5} + 68640 \, N^{4} - 93347 \, N^{3} - 401726 \, N^{2} + 197500 \, N + 132600 \\
	P_{6,11} & = 2685 \, N^{5} + 12423 \, N^{4} + 5150 \, N^{3} - 17908 \, N^{2} + 8976 \, N - 6624 \\
	P_{6,12} & = 6918 \, N^{4} + 41334 \, N^{3} + 46861 \, N^{2} - 49160 \, N - 18420 
\end{align}
For $p = [4,1,1]$ the operator is
\fontsize{8pt}{0}
\begin{align}
	S^{BPS}_{
		[4, 1, 1]
	} & = \frac{1}{ 6 \sqrt{P_ 7 }} \left(
	{\left(N - 2\right)} N P_{7,1}
	\mathcal{O}_{
		\ydiagram{5, 1}
	}
	- {\left(N - 2\right)} P_{7,2} \left[ 3 \, {\left(N + 1\right)} 
	\mathcal{O}_{
		\ydiagram{4, 2}
		, 1
	}
	+ 5 \, {\left(N + 3\right)}
	\mathcal{O}_{
		\ydiagram{3, 3}
	} \right] \right.
	\nonumber \\
	& \hspace{20pt}
	- 3 ( N - 2 ) P_{7,3}
	\mathcal{O}_{
		\ydiagram{4, 2}
		, 2
	}
	+ \sqrt{30} P_{7,4}
	\mathcal{O}_{
		\ydiagram{4, 1, 1}
		, 1
	}
	+ \sqrt{6} P_{7,5}
	\mathcal{O}_{
		\ydiagram{4, 1, 1}
		, 2
	}
	- 12 ( N + 3 ) \left[ P_{7,6}
	\mathcal{O}_{
		\ydiagram{3, 2, 1}
		, 1
	}
	+ \sqrt{2} P_{7,7}
	\mathcal{O}_{
		\ydiagram{3, 2, 1}
		, 2
	} \right]
	\nonumber \\
	& \hspace{20pt}
	+ {\left(N + 3\right)} \left[ 2 P_{7,8} \left\{ 8 \sqrt{2} {\left(N - 1\right)}
	\mathcal{O}_{
		\ydiagram{3, 2, 1}
		, 3
	}
	- 3 {\left(N + 2\right)}
	\mathcal{O}_{
		\ydiagram{2, 2, 1, 1}
		, 2
	} \right\}
	- \sqrt{30} N P_{7,9}
	\mathcal{O}_{
		\ydiagram{3, 1, 1, 1}
		, 1
	}
	+ 2 \sqrt{6} P_{7,10}
	\mathcal{O}_{
		\ydiagram{3, 1, 1, 1}
		, 2
	} \right]
	\nonumber \\
	& \hspace{20pt} \left.
	- {\left(N + 3\right)} {\left(N + 2\right)} P_{7,11} \left[ 5
	\mathcal{O}_{
		\ydiagram{2, 2, 2}
	}
	+ 3 
	\mathcal{O}_{
		\ydiagram{2, 2, 1, 1}
		, 1
	} \right]
	+ 2 \, {\left(N + 3\right)} {\left(N + 2\right)} P_{7,12} 
	\mathcal{O}_{
		\ydiagram{2, 1, 1, 1, 1}
	}
	\right)
\end{align}
\fontsize{12pt}{0}
where the normalisation and coefficient polynomials are
\fontsize{8pt}{0}
\begin{align}
	P_{7} & = 1691280 \, N^{14} + 14469840 \, N^{13} + 34933194 \, N^{12} - 15345720 \, N^{11} - 97734483 \, N^{10} + 108829584 \, N^{9} \nonumber \\ & \hspace{20pt} + 94236018 \, N^{8} - 365252412 \, N^{7} + 332214736 \, N^{6} + 23494544 \, N^{5} - 188670784 \, N^{4} + 59358800 \, N^{3} \nonumber \\ & \hspace{20pt} + 76067360 \, N^{2} - 55528000 \, N + 47136640 \\
	P_{7,1} & = 2052 \, N^{5} + 2592 \, N^{4} - 7293 \, N^{3} + 4232 \, N^{2} + 1320 \, N - 2240 \\
	P_{7,2} & = 108 \, N^{5} + 477 \, N^{4} - 348 \, N^{3} - 770 \, N^{2} + 1254 \, N - 772 \\
	P_{7,3} & = 612 \, N^{6} + 3210 \, N^{5} + 939 \, N^{4} - 7568 \, N^{3} + 5760 \, N^{2} + 692 \, N - 2472 \\
	P_{7,4} & = 1296 \, N^{7} + 6029 \, N^{6} + 1473 \, N^{5} - 11350 \, N^{4} + 9484 \, N^{3} - 4114 \, N^{2} - 3428 \, N + 3440 \\
	P_{7,5} & = 648 \, N^{7} + 4186 \, N^{6} + 1473 \, N^{5} - 12472 \, N^{4} + 8280 \, N^{3} + 6672 \, N^{2} - 4760 \, N - 3440 \\
	P_{7,6} & = 13 \, N^{5} - 7 \, N^{4} + 84 \, N^{3} - 129 \, N^{2} + 222 \, N - 40 \\
	P_{7,7} & = 15 \, N^{5} + 86 \, N^{4} - 194 \, N^{3} + 278 \, N^{2} - 178 \, N + 284 \\
	P_{7,8} & = 69 \, N^{4} + 222 \, N^{3} + 112 \, N^{2} - 193 \, N - 62 \\
	P_{7,9} & = 97 \, N^{4} + 316 \, N^{3} + 62 \, N^{2} - 306 \, N - 164 \\
	P_{7,10} & = 263 \, N^{5} + 410 \, N^{4} - 1216 \, N^{3} - 953 \, N^{2} + 1060 \, N + 860 \\
	P_{7,11} & = 15 \, N^{4} + 60 \, N^{3} - 262 \, N^{2} - 146 \, N - 244 \\
	P_{7,12} & = 222 \, N^{4} + 726 \, N^{3} + 74 \, N^{2} - 725 \, N - 430 
\end{align}
\fontsize{12pt}{0}
For $p = [3,3]$ the operator is
\fontsize{8pt}{0}
\begin{align}
	S^{BPS}_{
		[3, 3]
	} & = \frac{1}{ 18 \sqrt{2 P_ 8 }} \left(
	- 60 N P_{8,1}
	\mathcal{O}_{
		\ydiagram{5, 1}
	}
	+ 9 P_{8,2}
	\mathcal{O}_{
		\ydiagram{4, 2}
		, 1
	}
	- 18 P_{8,3}
	\mathcal{O}_{
		\ydiagram{4, 2}
		, 2
	} 
	- \sqrt{30} N P_{8,4}
	\mathcal{O}_{
		\ydiagram{4, 1, 1}
		, 1
	} \right.
	\nonumber \\
	& \hspace{20pt}
	- 10 \, \sqrt{6} N P_{8,5}
	\mathcal{O}_{
		\ydiagram{4, 1, 1}
		, 2
	}
	- 3 P_{8,6}
	\mathcal{O}_{
		\ydiagram{3, 3}
	}
	+ 12 P_{8,7}
	\mathcal{O}_{
		\ydiagram{3, 2, 1}
		, 1
	}
	- 12 \sqrt{2} P_{8,8}
	\mathcal{O}_{
		\ydiagram{3, 2, 1}
		, 2
	}
	\nonumber \\
	& \hspace{20pt}
	- 6 P_{8,9} \left[ 8 \sqrt{2} ( N - 1 )
	\mathcal{O}_{
		\ydiagram{3, 2, 1}
		, 3
	}
	- 3 {\left(N + 2\right)}
	\mathcal{O}_{
		\ydiagram{2, 2, 1, 1}
		, 2
	} \right]
	+ \sqrt{6} N \left[ \sqrt{5} P_{8,10}
	\mathcal{O}_{
		\ydiagram{3, 1, 1, 1}
		, 1
	}
	- 10  P_{8,11}
	\mathcal{O}_{
		\ydiagram{3, 1, 1, 1}
		, 2
	} \right]
	\nonumber \\
	& \hspace{20pt} \left.
	- 3 {\left(N + 2\right)} P_{8,12} \left[ 5
	\mathcal{O}_{
		\ydiagram{2, 2, 2}
	}
	+ 3
	\mathcal{O}_{
		\ydiagram{2, 2, 1, 1}
		, 1
	} \right]
	- 60 N {\left(N + 2\right)} P_{8,13}
	\mathcal{O}_{
		\ydiagram{2, 1, 1, 1, 1}
	}
	\right)
\end{align}
\fontsize{12pt}{0}
where the normalisation and coefficient polynomials are
\fontsize{8pt}{0}
\begin{align}
	P_{8} & = 64152 \, N^{12} + 209952 \, N^{11} - 137241 \, N^{10} - 640440 \, N^{9} + 908640 \, N^{8} - 322236 \, N^{7} - 116124 \, N^{6} \nonumber \\ & \hspace{20pt} - 675864 \, N^{5} + 2362028 \, N^{4} - 3013280 \, N^{3} + 2221520 \, N^{2} - 926400 \, N + 177280 \\
	P_{8,1} & = 27 \, N^{5} - 54 \, N^{4} + 39 \, N^{3} + 7 \, N^{2} - 36 \, N + 20 \\
	P_{8,2} & = 540 \, N^{6} + 765 \, N^{5} - 2364 \, N^{4} + 2098 \, N^{3} - 686 \, N^{2} - 592 \, N + 488 \\
	P_{8,3} & = 90 \, N^{6} + 75 \, N^{5} - 198 \, N^{4} + 156 \, N^{3} - 202 \, N^{2} + 216 \, N - 104 \\
	P_{8,4} & = 255 \, N^{4} - 78 \, N^{3} - 310 \, N^{2} + 218 \, N - 28 \\
	P_{8,5} & = 69 \, N^{4} - 156 \, N^{3} + 52 \, N^{2} + 178 \, N - 164 \\
	P_{8,6} & = 1188 \, N^{6} + 3519 \, N^{5} - 2868 \, N^{4} - 5066 \, N^{3} + 9654 \, N^{2} - 8560 \, N + 3000 \\
	P_{8,7} & = 87 \, N^{5} + 81 \, N^{4} - 302 \, N^{3} + 396 \, N^{2} - 214 \, N + 60 \\
	P_{8,8} & = 99 \, N^{5} - 12 \, N^{4} - 486 \, N^{3} + 706 \, N^{2} - 640 \, N + 168 \\
	P_{8,9} & = 21 \, N^{4} + 48 \, N^{3} - 36 \, N^{2} - 2 \, N - 12 \\
	P_{8,10} & = 51 \, N^{4} + 138 \, N^{3} - 86 \, N^{2} - 142 \, N - 12 \\
	P_{8,11} & = 33 \, N^{4} + 48 \, N^{3} - 164 \, N^{2} + 2 \, N + 156 \\
	P_{8,12} & = 33 \, N^{4} + 54 \, N^{3} - 58 \, N^{2} + 134 \, N - 36 \\
	P_{8,13} & = 3 \, N^{3} + 9 \, N^{2} - 5 \, N - 14 
\end{align}
\fontsize{12pt}{0}
For $p = [4,2]$ the operator is
\fontsize{8pt}{0}
\begin{align}
	S^{BPS}_{
		[4, 2]
	} & = \frac{1}{ 6 \sqrt{3 P_ 9 }} \left(
	- \sqrt{2} P_{9,1}
	\mathcal{O}_{
		\ydiagram{5, 1}
	}
	- 6 \, \sqrt{2} P_{9,2}
	\mathcal{O}_{
		\ydiagram{4, 2}
		, 1
	}
	- 3 \, \sqrt{2} P_{9,3}
	\mathcal{O}_{
		\ydiagram{4, 2}
		, 2
	}
	- 4 \, \sqrt{15} N P_{9,4}
	\mathcal{O}_{
		\ydiagram{4, 1, 1}
		, 1
	} \right.
	\nonumber \\
	& \hspace{20pt}
	- 2 \sqrt{3} P_{9,5}
	\mathcal{O}_{
		\ydiagram{4, 1, 1}
		, 2
	}
	+ {\left(N + 3\right)} \left[ - 10 \sqrt{2} {\left(N - 1\right)}  P_{9,6}
	\mathcal{O}_{
		\ydiagram{3, 3}
	}
	+ 6 \sqrt{2} P_{9,7} \left\{  N 
	\mathcal{O}_{
		\ydiagram{3, 2, 1}
		, 1
	} 
	+ 2 \sqrt{2} 
	\mathcal{O}_{
		\ydiagram{3, 2, 1}
		, 2
	} \right\} \right.
	\nonumber \\
	& \hspace{20pt}
	- \sqrt{2} P_{9,8} \left\{ 8 \sqrt{2} {\left(N - 1\right)} 
	\mathcal{O}_{
		\ydiagram{3, 2, 1}
		, 3
	}
	- 3 {\left(N + 2\right)}
	\mathcal{O}_{
		\ydiagram{2, 2, 1, 1}
		, 2
	} \right\}
	+ 4 \, \sqrt{15} {\left(N + 1\right)} {\left(N - 1\right)} N
	\mathcal{O}_{
		\ydiagram{3, 1, 1, 1}
		, 1
	}
	\nonumber \\
	& \hspace{20pt} \left. \left.
	- 2 \sqrt{3} P_{9,9}
	\mathcal{O}_{
		\ydiagram{3, 1, 1, 1} \, , \, 2
	}
	- \sqrt{2} {\left(N + 2\right)} \left\{ 10
	\mathcal{O}_{
		\ydiagram{2, 2, 2}
	}
	+ 6
	\mathcal{O}_{
		\ydiagram{2, 2, 1, 1} \, , \, 1
	}
	+ P_{9,10}
	\mathcal{O}_{
		\ydiagram{2, 1, 1, 1, 1}
	} \right\} \right]
	\right)
\end{align}
\fontsize{12pt}{0}
where the normalisation and coefficient polynomials are
\fontsize{8pt}{0}
\begin{align}
	P_{9} & = 297 \, N^{10} + 378 \, N^{8} - 1260 \, N^{7} + 390 \, N^{6} + 1080 \, N^{5} - 1256 \, N^{4} + 640 \, N^{3} + 760 \, N^{2} - 1920 \, N + 1440 \\
	P_{9,1} & = 81 \, N^{5} - 129 \, N^{4} + 51 \, N^{3} + 76 \, N^{2} - 130 \, N + 60 \\
	P_{9,2} & = 9 \, N^{5} + 3 \, N^{4} - 3 \, N^{3} - 13 \, N^{2} + 13 \, N - 6 \\
	P_{9,3} & = {\left(9 \, N^{3} + 13 \, N^{2} - 13 \, N + 6\right)} {\left(3 \, N^{2} - 2\right)} \\
	P_{9,4} & = 5 \, N^{3} - 2 \, N^{2} - 5 \, N + 3 \\
	P_{9,5} & = 29 \, N^{4} - 65 \, N^{3} + 24 \, N^{2} + 70 \, N - 60 \\
	P_{9,6} & = 3 \, N^{2} - 3 \, N + 2 \\
	P_{9,7} & = N^{2} - 2 \, N + 2 \\
	P_{9,8} & = 3 \, N^{2} - 2 \\
	P_{9,9} & = 11 \, N^{3} - 18 \, N^{2} - 10 \, N + 20 \\
	P_{9,10} & = 9 \, N^{2} - 10 
\end{align}
\fontsize{12pt}{0}

\subsection{Norms of operators with multiplicity}
\label{section: p=[2,2,1,1] and [3,2,1] norms}
	
As explained in section \ref{section: norms of BPS operators}, for $\Lambda, p$ with $\cM_{\Lambda, p} \geq 1$, the BPS norms of the operators are dependent on the choice of basis for the multiplicity space. In that section, we described a process to extract norm-like functions of $N$ that characterise the multiplicity space and are independent of the choice of basis.

In the $\Lambda = [4,2]$ sector, there are two partitions $p = [2,2,1,1]$ and $[3,2,1]$ with $\cM_{\Lambda, p} = 2$. For the first of these, we go through the process described in section \ref{section: norms of BPS operators} in some detail, while for the second we only give the results.

We begin by renormalising the BPS operators to have norm 1 in the physical inner product as given in \eqref{normalised by F inner product}. For $p = [2,2,1,1]$, this replaces $P_1$ and $P_2$ in the expansions \eqref{[2,2,1,1] 1} and \eqref{[2,2,1,1] 2} with
\fontsize{8pt}{0}
\begin{align}
\widehat{P}_1 & = 3 {\left(N + 3\right)} {\left(N + 2\right)} {\left(N + 1\right)}^{2} N^{2} {\left(N - 1\right)} {\left(N - 2\right)} {\left(N - 3\right)} Q_{mult} Q_{1} \\
\widehat{P}_2 & = 3 {\left(N + 1\right)} N^{2} {\left(N - 1\right)} {\left(N - 2\right)} {\left(N - 3\right)} Q_{mult} Q_{2}
\end{align}
\fontsize{12pt}{0}
where $Q_1$ and $Q_2$ are defined in \eqref{Q1} and \eqref{Q2} and
\fontsize{8pt}{0}
\begin{align}
Q_{mult} & = 2145 \, N^{8} + 21570 \, N^{7} + 69156 \, N^{6} + 44856 \, N^{5} - 130747 \, N^{4} - 117106 \, N^{3} + 138802 \, N^{2} \nonumber \\ & \hspace{20pt}  + 53280 \, N - 75456
\label{Qmult}
\end{align}
\fontsize{12pt}{0}
after normalising, the $S_n$ inner product matrix can be calculated, and is given by
\fontsize{8pt}{0}
\begin{equation}
A_{[2,2,1,1]} = 
\begin{pmatrix}
\left\langle \widehat{S}^{BPS}_{[2,2,1,1],1} | \widehat{S}^{BPS}_{[2,2,1,1],1} \right\rangle_{S_n} & \left\langle \widehat{S}^{BPS}_{[2,2,1,1],1} | \widehat{S}^{BPS}_{[2,2,1,1],2} \right\rangle_{S_n} \\
\left\langle \widehat{S}^{BPS}_{[2,2,1,1],2} | \widehat{S}^{BPS}_{[2,2,1,1],1} \right\rangle_{S_n} & \left\langle \widehat{S}^{BPS}_{[2,2,1,1],2} | \widehat{S}^{BPS}_{[2,2,1,1],2} \right\rangle_{S_n}
\end{pmatrix}
=
\begin{pmatrix}
\frac{A_{1,1}}{\widehat{P}_1} & \frac{A_{1,2}}{\sqrt{\widehat{P}_1 \widehat{P}_2}} \\
\frac{A_{1,2}}{\sqrt{\widehat{P}_1 \widehat{P}_2}} & \frac{A_{2,2}}{\widehat{P}_2}
\end{pmatrix}
\end{equation}
\fontsize{12pt}{0}
where
\fontsize{8pt}{0}
\begin{align}
A_{1,1} & = 1254825 \, N^{16} + 25236900 \, N^{15} + 212913135 \, N^{14} + 949347864 \, N^{13} + 2265287922 \, N^{12} + 2296326096 \, N^{11} \nonumber \\ & \hspace{20pt} - 483268806 \, N^{10} - 64991400 \, N^{9} + 7717590681 \, N^{8} + 4250132076 \, N^{7} - 14563157385 \, N^{6} \nonumber \\ & \hspace{20pt} - 5596987632 \, N^{5} + 20300164460 \, N^{4} + 5660498272 \, N^{3} - 5514459136 \, N^{2} + 14594125824 \, N \nonumber \\ & \hspace{20pt} + 12396386304 \\
A_{1,2} & = 2 \, \sqrt{5} \big( 394875 \, N^{15} + 7400484 \, N^{14} + 57527991 \, N^{13} + 231664914 \, N^{12} + 476892396 \, N^{11} + 249273666 \, N^{10} \nonumber \\ & \hspace{20pt} - 1301445666 \, N^{9} - 4474130634 \, N^{8} - 7919982621 \, N^{7} - 8401406142 \, N^{6} - 6257132757 \, N^{5} \nonumber \\ & \hspace{20pt} - 4801800696 \, N^{4} - 1575438250 \, N^{3} - 1395294808 \, N^{2} - 4205573568 \, N - 1295069184 \big) \\
A_{2,2} & = 64575225 \, N^{16} + 1221543180 \, N^{15} + 9292923450 \, N^{14} + 34312809600 \, N^{13} + 49747071546 \, N^{12} \nonumber \\ & \hspace{20pt} - 49520811024 \, N^{11} - 212528733480 \, N^{10} + 81502221096 \, N^{9} + 872883407025 \, N^{8} + 609873915684 \, N^{7} \nonumber \\ & \hspace{20pt} - 949480261506 \, N^{6} - 778095650280 \, N^{5} + 986491220724 \, N^{4} + 591265527264 \, N^{3} - 532623199736 \, N^{2} \nonumber \\ & \hspace{20pt} - 150593123520 \, N + 181872634752
\end{align}
\fontsize{12pt}{0}
\hspace{-3.5pt}We now take the trace of $A_{[2,2,1,1]}$, divide by $\cM_{\Lambda, p}$, and take the reciprocal. This gives the first $p = [2,2,1,1]$ invariant
\fontsize{8pt}{0}
\begin{align}
\frac{2}{ \tr A_{[2,2,1,1]}} = \frac{
	2 {\left(N + 3\right)} {\left(N + 2\right)} {\left(N + 1\right)}^{2} N^{2} {\left(N - 1\right)} {\left(N - 2\right)} {\left(N - 3\right)} Q_{1} Q_{2} }{D_1}
\end{align}
\fontsize{12pt}{0}
where the denominator is 
\fontsize{8pt}{0}
\begin{align}
D_1 & = 3913650 \, N^{16} + 78795855 \, N^{15} + 656781957 \, N^{14} + 2811679470 \, N^{13} + 5818416030 \, N^{12} + 1501757316 \, N^{11} \nonumber \\ & \hspace{20pt} - 15672370512 \, N^{10} - 14255947158 \, N^{9} + 42286367112 \, N^{8} + 71992040249 \, N^{7} - 32371301901 \, N^{6} \nonumber \\ & \hspace{20pt} - 121059621624 \, N^{5}  - 22843286488 \, N^{4} + 77152295508 \, N^{3} + 42542435352 \, N^{2} + 5036467584 \, N \nonumber \\ & \hspace{20pt} + 2255817600
\end{align}
\fontsize{12pt}{0}
We can also consider the trace of $A_{[2,2,1,1]}^2$. This leads to the second invariant
\fontsize{8pt}{0}
\begin{align}
\sqrt{\frac{2}{\tr A_{[2,2,1,1]}^2}} & = \frac{
	\sqrt{2} {\left(N + 3\right)} {\left(N + 2\right)} {\left(N + 1\right)}^2 N^{2} {\left(N - 1\right)} {\left(N - 2\right)} {\left(N - 3\right)} Q_{1} Q_{2} }{ \sqrt{ D_2 } }
\end{align}
\fontsize{12pt}{0}
where the denominator is
\fontsize{8pt}{0}
\begin{align}
D_2 & = 7658328161250 \, N^{32} + 308379397920750 \, N^{31}  + 5678040590961075 \, N^{30} + 62861883407800200 \, N^{29} 
\nonumber \\ & \hspace{20pt} 
+ 461553133569402069 \, N^{28} + 2323880655128992368 \, N^{27} + 7893896923770889320 \, N^{26} 
\nonumber \\ & \hspace{20pt} 
+ 16200841926037924512 \, N^{25} + 9738474984510581700 \, N^{24} - 43140893567922372492 \, N^{23} 
\nonumber \\ & \hspace{20pt}  
- 100830809456338189482 \, N^{22} + 66300678032545590264 \, N^{21} + 576422366985618028290 \, N^{20} 
\nonumber \\ & \hspace{20pt} 
+ 587496624365125252152 \, N^{19}  - 1266939757691694906384 \, N^{18} - 3370314414344723267400 \, N^{17} 
\nonumber \\ & \hspace{20pt} 
- 14422779155617085790 \, N^{16} + 8873284172309294711934 \, N^{15} + 9228283693975324117807 \, N^{14}
\nonumber \\ & \hspace{20pt} 
- 8309143471774592802944 \, N^{13} - 21871661389590847910159 \, N^{12} - 3725069874701998817592 \, N^{11}
\nonumber \\ & \hspace{20pt} 
+ 25451491117140266214976 \, N^{10} + 18757146605106723110568 \, N^{9} - 15395309506022451870416 \, N^{8}
\nonumber \\ & \hspace{20pt} 
- 22339442519546818907728 \, N^{7} + 3985689055612424950064 \, N^{6} + 16657691069689910952704 \, N^{5} 
\nonumber \\ & \hspace{20pt} 
+ 3604888800092578331072 \, N^{4} - 4775351642112978422784 \, N^{3} - 82696688563225374720 \, N^{2} 
\nonumber \\ & \hspace{20pt}
+ 2740871464097166655488 \, N + 1006239182315089379328
\end{align}
\fontsize{12pt}{0}	
For $p = [3,2,1]$, the same process produces
\fontsize{8pt}{0}
\begin{align}
\frac{2}{\tr A_{[3,2,1]}} &  =\frac{
	2 {\left(N + 2\right)} {\left(N + 1\right)} N^{2} {\left(N - 1\right)} {\left(N - 2\right)} Q_{4} Q_{5} }{ E_1 } \\
\sqrt{\frac{2}{\tr A_{[3,2,1]}^2}} & = \frac{
	\sqrt{2} {\left(N + 2\right)} {\left(N + 1\right)} N^{2} {\left(N - 1\right)} {\left(N - 2\right)}  Q_{4} Q_{5} }{ \sqrt{ E_2 } }
\end{align}
\fontsize{12pt}{0}
where the denominators are
\fontsize{8pt}{0}
\begin{align}
E_1 & = 41812200 \, N^{14} + 448198920 \, N^{13} + 1563219648 \, N^{12} + 1093147920 \, N^{11} - 3204936072 \, N^{10} \nonumber \\ & \hspace{20pt} - 1375066305 \, N^{9} + 4730520504 \, N^{8} - 3314823954 \, N^{7} + 4335640504 \, N^{6} - 6084970 \, N^{5} \nonumber \\ & \hspace{20pt} - 10209076192 \, N^{4} + 9690911824 \, N^{3} - 2443216896 \, N^{2} + 3777810528 \, N - 538272768 \\
E_2 & = 874130034420000 \, N^{28} + 18740182882824000 \, N^{27} + 166345754746996800 \, N^{26} \nonumber \\ & \hspace{20pt} + 753788224097235360 \, N^{25} + 1608498415010610504 \, N^{24} + 181361766700128024 \, N^{23} \nonumber \\ & \hspace{20pt} - 5390210561323512672 \, N^{22} - 4416942361725000252 \, N^{21} + 13381736971853528568 \, N^{20} \nonumber \\ & \hspace{20pt} + 14451638301852715944 \, N^{19} - 26479800963850159935 \, N^{18} - 31944440045187411534 \, N^{17} \nonumber \\ & \hspace{20pt} + 47031196210114852566 \, N^{16} + 73597499966176725312 \, N^{15} - 112834178522277863808 \, N^{14} \nonumber \\ & \hspace{20pt} - 122423268066628273308 \, N^{13} + 309535922049432602720 \, N^{12} + 10889533588200882344 \, N^{11} \nonumber \\ & \hspace{20pt} - 425645164341054775804 \, N^{10} + 196537079346722192144 \, N^{9} + 367648860348492413280 \, N^{8} \nonumber \\ & \hspace{20pt} - 423712842979380230656 \, N^{7} + 67969647225996116864 \, N^{6} + 101859033408413821440 \, N^{5} \nonumber \\ & \hspace{20pt} - 27177241919312392192 \, N^{4} - 38998356711672686592 \, N^{3} + 48696174595572179968 \, N^{2} \nonumber \\ & \hspace{20pt} - 18571148044644937728 \, N + 3609255644969508864
\end{align}
\fontsize{12pt}{0}

\subsection{Norms of BPS operators}
\label{section: [4,2] norms}
	
The physical norms of the BPS operators can be understood as characteristic functions of the pair $\Lambda, p$ and should be reproducible from stringy physics on the other side of the AdS/CFT duality. 

We give the norms for each of the BPS operators in the $\Lambda = [4,2]$ sector. For $p = [2,2,1,1]$ and $[3,2,1]$, we reproduce the invariants derived in the previous subsection in order to compare with other operators.
\fontsize{8pt}{0}
\begin{align}
	\left| S^{BPS}_{
		[2, 1, 1, 1, 1]
	} \right|^2 & = 
	\frac{
		\vspace{-5pt} {\left(N + 4\right)} {\left(N + 3\right)} {\left(N + 2\right)} {\left(N + 1\right)} N^{2} {\left(N - 1\right)}^{2} {\left(N - 2\right)} {\left(N - 3\right)} {\left(N - 4\right)} Q_{1}
	}{P_{0}} \\
\frac{2}{ \tr A_{[2,2,1,1]}} & = \frac{
	2 {\left(N + 3\right)} {\left(N + 2\right)} {\left(N + 1\right)}^{2} N^{2} {\left(N - 1\right)} {\left(N - 2\right)} {\left(N - 3\right)} Q_{1} Q_{2} }{D_1} \\
\sqrt{\frac{2}{\tr A_{[2,2,1,1]}^2}} & = \frac{
	\sqrt{2} {\left(N + 3\right)} {\left(N + 2\right)} {\left(N + 1\right)}^2 N^{2} {\left(N - 1\right)} {\left(N - 2\right)} {\left(N - 3\right)} Q_{1} Q_{2} }{ \sqrt{ D_2 } } \\
\left| S^{BPS}_{
	[3, 1, 1, 1]
} \right|^2 & = 
\frac{
	{\left(N + 2\right)} {\left(N + 1\right)} N {\left(N - 1\right)} {\left(N - 2\right)} {\left(N - 3\right)} Q_{2} Q_{3}
}{
	2 \, P_3 } \\
\left| S^{BPS}_{
	[2, 2, 2]
} \right|^2 & = 
\frac{{\left(N + 1\right)} N^{2} {\left(N - 1\right)} {\left(N - 2\right)} Q_{3} Q_{4} }{P_4}
\label{p=[2,2,2] norm} \\
\frac{2}{\tr A_{[3,2,1]}} &  =\frac{
	2 {\left(N + 2\right)} {\left(N + 1\right)} N^{2} {\left(N - 1\right)} {\left(N - 2\right)} Q_{4} Q_{5} }{ E_1 } \\
\sqrt{\frac{2}{\tr A_{[3,2,1]}^2}} & = \frac{
	\sqrt{2} {\left(N + 2\right)} {\left(N + 1\right)} N^{2} {\left(N - 1\right)} {\left(N - 2\right)}  Q_{4} Q_{5} }{ \sqrt{ E_2 } } \\
\left| S^{BPS}_{
	[4, 1, 1]
} \right|^2 & = 
\frac{ {\left(N + 3\right)} {\left(N + 2\right)} {\left(N + 1\right)} N {\left(N - 1\right)} {\left(N - 2\right)} Q_{5} Q_{6} }{P_{7}} \\
\left| S^{BPS}_{
	[3, 3]
} \right|^2 & = 
\frac{  {\left(N + 2\right)} {\left(N + 1\right)}  N^{2} {\left(N - 1\right)} Q_{6} Q_{7} }{P_8}
\label{p=[3,3] norm} \\
\left| S^{BPS}_{
	[4, 2]
} \right|^2 & = 
\frac{ {\left(N + 3\right)} {\left(N + 2\right)} {\left(N + 1\right)}  N^{4} {\left(N - 1\right)} \left( 3 \, N^{2} - 2 \right) Q_{7} }{P_9}
\end{align}
\fontsize{12pt}{0}
where the polynomials in the denominator have been defined in previous subsections and the polynomials in the numerator are
\fontsize{8pt}{0}
\begin{align}
	Q_{1} & = 195 \, N^{5} + 1149 \, N^{4} + 687 \, N^{3} - 3927 \, N^{2} - 1552 \, N + 4448 
	\label{Q1} \\
	Q_{2} & = 10035 \, N^{8} + 94914 \, N^{7} + 264876 \, N^{6} + 17268 \, N^{5} - 819309 \, N^{4} - 487830 \, N^{3} + 780722 \, N^{2} \nonumber \\ & \hspace{20pt} + 189568 \, N - 432744 
	\label{Q2} \\
	Q_{3} & = 18630 \, N^{8} + 160677 \, N^{7} + 371643 \, N^{6} - 204495 \, N^{5} - 1326729 \, N^{4} - 15804 \, N^{3} + 1726178 \, N^{2} \nonumber \\ & \hspace{20pt} - 442368 \, N - 1298232 \\
	Q_{4} & = 8010 \, N^{7} + 56214 \, N^{6} + 79800 \, N^{5} - 132315 \, N^{4} - 158273 \, N^{3} + 296994 \, N^{2} + 33500 \, N - 171336 
	\label{Q4} \\
	Q_{5} & = 2610 \, N^{7} + 12546 \, N^{6} + 3213 \, N^{5} - 25152 \, N^{4} + 20228 \, N^{3} - 5238 \, N^{2} - 8000 \, N + 5160 
	\label{Q5} \\
	Q_{6} & = 648 \, N^{7} + 2772 \, N^{6} + 51 \, N^{5} - 5484 \, N^{4} + 5438 \, N^{3} - 2026 \, N^{2} - 2000 \, N + 1720 \\
	Q_{7} & = 99 \, N^{6} + 162 \, N^{5} - 324 \, N^{4} + 102 \, N^{3} + 152 \, N^{2} - 260 \, N + 120
\end{align}
\fontsize{12pt}{0}
Comparing these norms with those in sections \ref{section: [3,2] norms} and \ref{section: norms of BPS operators}, we see a general pattern in the numerators. They typically contain a product of linear factors along with (in general) two complicated $Q$ polynomials. These $Q$ polynomials appear in two consecutive norms.

In \eqref{Qmult} we saw that the $Q$ polynomials appear in consecutive norms even in the non-physical multiplicity space. This suggests they are an artefact of the orthogonalisation process.

The linear factors are more interesting. Their presence is partially implied by SEP-compatibility, but there are generally more factors than would be sufficient for this purpose. The function $f_p$, defined in \eqref{omega in a representation}, that gives the free field norms, is a product of linear factors, and we can compare this with those found in the numerators of weak coupling BPS norms. In all but two ($p = [2,2,2]$ and $[3,3]$) of the examples we have calculated, the numerators contain $f_p$, while some partitions have considerably more factors. It would be interesting to enumerate the linear factors that appear in the numerator for general $\Lambda, p$.

\section{$\Lambda = [3,3]$ sector}
\label{appendix: lambda = [3,3]}
	
	\ytableausetup{boxsize=7pt}
	
	The final example we give here is the BPS basis for the $\Lambda = [3,3]$ sector at field content $(3,3)$.
	
	Throughout this section we will work with $\Lambda = [3,3]$ and $M_\Lambda = \begin{gathered}
	\fontsize{6pt}{0} \begin{ytableau}
	1 & 1 & 1 \\ 2 & 2 & 2
	\end{ytableau} \fontsize{12pt}{0} \end{gathered}$, so we will suppress this index in operator labels.
	
	\ytableausetup{boxsize=2pt}
	
	For each BPS operator, we will first present it as a sum over the free field basis \eqref{U(2) basis definition} and then as a sum over symmetrised traces and commutator traces, for which we use the covariant bases discussed in section \ref{section: covariant trace bases}. The covariant symmetrised trace basis is
	\begin{align}
	t_{[3,2,1]} & = \tr X^3 \tr Y \tr Y^2 
	- 2 \tr X^2 Y \tr X Y \tr Y 
	- \tr X \tr X^2 Y \tr Y^2 \nonumber \\ & \hspace{20pt}
	+ \tr X^2 \tr X Y^2 \tr Y 
	+ 2 \tr X \tr X Y \tr X Y^2 
	- \tr X \tr X^2 \tr Y^3 \\
	t_{[3,1,1,1]} & = \tr X^3 \left( \tr Y \right)^3 
	- 3 \tr X \tr X^2 Y \left( \tr Y \right)^2 
	+ 3 \left( \tr X \right)^2 \tr X Y^2 \tr Y \nonumber \\ & \hspace{20pt}
	+ \left( \tr X \right)^3 \tr Y^3
	\end{align}
	and the covariant commutator trace basis is 
	\begin{align}
	c_{[6]} & = \tr X^2 Y X Y^2 
	- \tr X^2 Y^2 X Y = \tr X^2 Y [X,Y] Y \\
	c_{[5,1]} & = \tr X^3 Y^2 \tr Y 
	- \tr X^2 Y X Y \tr Y 
	- \tr X \tr X^2 Y^3 
	+ \tr X \tr X Y X Y^2 \nonumber \\
	& = \tr X^2 [X,Y] Y \tr Y - \tr X \tr X [X,Y] Y^2
	\end{align}
	For these two bases, the partition label describes the cycle structure of the multi-traces.
	
	The free field operators can be written in terms of symmetrised and commutator traces
	\ytableausetup{boxsize=3pt}
	\begin{align}
	O_{
		\ydiagram{4,1,1}
	}
	& = \frac{\sqrt{3}}{36} \left(
	3t_{[3,2,1]} + t_{[3,1,1,1]} + 6 c_{[5,1]}
	\right) \\
	O_{
		\ydiagram{3,2,1} \, , \, \text{even}
	} 
	& = \frac{\sqrt{3}}{18} \left(
	t_{[3,1,1,1]} - 3 c_{[5,1]}
	\right) \\
	O_{
		\ydiagram{3,2,1} \, , \, \text{odd}
	}
	& = - \frac{1}{\sqrt{2}} c_{[6]} \\
	O_{
		\ydiagram{3,1,1,1}
	}
	& = \frac{\sqrt{3}}{36} \left(
	- 3 t_{[3,2,1]} + t_{[3,1,1,1]} + 6 c_{[5,1]}
	\right)
	\end{align}
	The odd/even labels for the $R=[3,2,1]$ multiplicity come from the odd/even permutations used to produce the respective traces. All other zero coupling operators are defined uniquely by $\Lambda$ and $R$.
	
	The BPS operators are	
	\begin{align}
	S^{BPS}_{[3,1,1,1]} & = \frac{1}{\sqrt{3 P_1}} \left( \begin{gathered}
	- N (N-3) O_{\ydiagram{4,1,1}}
	+ (N+3)(N-3) O_{\ydiagram{3,2,1} \, , \, \text{even}}
	+ 2 (N+3) N O_{\ydiagram{3,1,1,1}}
	\end{gathered}
	\right) \\
	& = \frac{1}{12\sqrt{P_1}} \left( - 3 N (N+1) t_{[3,2,1]} + (N^2 + 3N - 6) t_{[3,1,1,1]} + 18 (N+1) c_{[5,1]} \right) \\
	S^{BPS}_{[3,2,1]} & = \frac{1}{\sqrt{P_2}} \left( \begin{gathered}
	(N-1) O_{\ydiagram{4,1,1}}
	+ (N+1) O_{\ydiagram{3,2,1} \, , \, \text{even}}
	- O_{\ydiagram{3,1,1,1}}
	\end{gathered}
	\right) \\
	& = \frac{1}{4 \sqrt{3 P_2}} \left( N t_{[3,2,1]} + N t_{[3,1,1,1]} - 6 c_{[5,1]} \right) 
	\end{align}
	where the normalisation polynomials are
	\begin{align}
	P_1 & = 2 N^4 + 6 N^3 + 9 N^2 + 27 &
	P_2 & = 2N^2 + 3
	\end{align}
	The norms of the BPS operators are
	\begin{align}
	\left| S^{BPS}_{[3,1,1,1]} \right|^2 & = \frac{{\left(N + 3\right)} {\left(N + 2\right)} {\left(N + 1\right)} N^{2} {\left(N - 1\right)} {\left(N - 2\right)} {\left(N - 3\right)} Q }{P_1} \\
	\left| S^{BPS}_{[3,2,1]} \right|^2 & = \frac{ {\left(N + 2\right)} {\left(N + 1\right)} N^{2} {\left(N - 1\right)} {\left(N - 2\right)} Q }{P_2}
	\end{align}
	where
	\begin{equation}
	Q = 2 \, N^{2} + 3 \, N - 3
	\end{equation}
	
	\ytableausetup{boxsize=normal}

\end{document}